\documentclass[aps,prx,twocolumn,superscriptaddress,nofootinbib,longbibliography,notitlepage,floatfix,10pt]{revtex4-2}
\usepackage[dvipsnames]{xcolor}
\usepackage{times}
\usepackage{amsmath}   % provides gather, align, …
\usepackage{hyperref}  % hyperlinks (optional)
\usepackage{amssymb}
\usepackage{tikz}
\usepackage{xcolor}
\usetikzlibrary{3d}
\usetikzlibrary{decorations.pathmorphing}
\usepackage[english]{babel}
\usepackage{physics}
\usepackage{bm}%给希腊字母加粗
\usepackage{appendix}%加附录的包
\usepackage{setspace}
\usepackage{url}

\usepackage{pgfplots} % Required for 3D plots
\usepackage{tikz-3dplot} % Required for 3D plotting

\usepackage{graphicx}%插入图片的包
\graphicspath{{pics/}}%插入图片路径
\usepackage{wasysym}
\usepackage{color}
\usepackage{xcolor}
\usepackage{bm,wrapfig,cancel} 
\usepackage{float} 
\usepackage{subcaption}
\usepackage{amsthm,amssymb}
\usepackage{mathrsfs}
\usepackage[bottom, perpage]{footmisc}

\usepackage{scalerel}
\usetikzlibrary{svg.path}
\usepackage[justification=raggedright]{caption}

\usepackage{array}
\usepackage{multirow}
\definecolor{orcidlogocol}{HTML}{A6CE39}
\tikzset{
	orcidlogo/.pic={
		\fill[orcidlogocol] svg{M256,128c0,70.7-57.3,128-128,128C57.3,256,0,198.7,0,128C0,57.3,57.3,0,128,0C198.7,0,256,57.3,256,128z};
		\fill[white] svg{M86.3,186.2H70.9V79.1h15.4v48.4V186.2z}
		svg{M108.9,79.1h41.6c39.6,0,57,28.3,57,53.6c0,27.5-21.5,53.6-56.8,53.6h-41.8V79.1z M124.3,172.4h24.5c34.9,0,42.9-26.5,42.9-39.7c0-21.5-13.7-39.7-43.7-39.7h-23.7V172.4z}
		svg{M88.7,56.8c0,5.5-4.5,10.1-10.1,10.1c-5.6,0-10.1-4.6-10.1-10.1c0-5.6,4.5-10.1,10.1-10.1C84.2,46.7,88.7,51.3,88.7,56.8z};
	}
}

\newcommand\orcidicon[1]{\href{https://orcid.org/#1}{\mbox{\scalerel*{
				\begin{tikzpicture}[yscale=-1,transform shape]
					\pic{orcidlogo};
				\end{tikzpicture}
			}{|}}}}

\makeatletter
\protected\def\my@emoji@pic #1#2{\leavevmode@ifvmode
	\lower\dimexpr #1\p@*1/10\hbox{\includegraphics[height={#1\p@}]{#2}}}
\def\my@emoji@math #1{%
	\mathchoice
	{\my@emoji@pic\tf@size{#1}}{\my@emoji@pic\tf@size{#1}}
	{\my@emoji@pic\sf@size{#1}}{\my@emoji@pic\ssf@size{#1}}}
\protected\def\myemoji #1{{\ifmmode\my@emoji@math{#1}\else\my@emoji@pic\f@size{#1}\fi}}
\makeatother

\newcommand{\revsubset}{\mathchoice%
	{\rotatebox[origin=c]{180}{$\subset$}}% Display style
	{\rotatebox[origin=c]{180}{$\subset$}}% Text style
	{\rotatebox[origin=c]{180}{$\scriptstyle\subset$}}% Script style
	{\rotatebox[origin=c]{180}{$\scriptscriptstyle\subset$}}% Script script style
}

\usepackage{hyperref}
\hypersetup{
	colorlinks=true,
	linkcolor=blue,
	citecolor=blue,
	urlcolor=blue,
	pdfborder={0 0 0}
}

\begin{document}
	
	\title{Preparing    Code States via Seed-Entangler-Enriched Sequential Quantum Circuits: Application to Tetradigit Topological Error-Correcting Codes}
	\date{\today}

	\begin{abstract}
		Demonstrating how long-range entangled states are born from product states has gained much attention,  which is not only important for quantum technology but also provides an unconventional tool in characterizing and classifying exotic phases of matter.	In this paper, we introduce a unified and efficient framework of quantum circuits (i.e., a series of local unitary transformations), termed the \emph{Seed-Entangler-Enriched Sequential Quantum Circuit} (SEESQC) to construct long-range entangled states (i.e., code states) in code space of topological error-correcting codes. Specifically, we apply SEESQC to construct code states of tetradigit models—a broad class of long-range entangled stabilizer codes indexed by a four-digit parameter. These models are not rare but encompass Toric Codes across arbitrary dimensions and subsume the X-cube fracton code as special cases. Featuring a hierarchical structure of generalized entanglement renormalization group, many tetradigit models host spatially extended excitations (e.g., loops, membranes, and exotic non-manifold objects) with constrained mobility and deformability, and exhibit system-size-dependent ground state degeneracies that scale exponentially with a polynomial in linear sizes. In this work, we begin with graphical and algebraic demonstration of quantum circuits for computational basis states, before generalizing to broader cases. Central to this framework is a key ingredient termed the \emph{seed-entangler} acting on a small number of qubits termed \textit{seeds}, enabling a systematic scheme to achieve arbitrary code states. Remarkably, the number of available seeds equals the number of logical qubits for the constructed examples, which leaves plenty of room for future investigation in theoretical physics, mathematics and quantum information science. Beyond the critical limitation of prior state-engineering methodologies, which required entirely distinct, model-specific circuit designs for each class of topological order, this framework transcends spatial dimensions, bridges liquid and non-liquid states, and unifies gapped phases governed by distinct entanglement renormalization group schemes.  With experimental feasibility via synthetic dimensions in modern quantum simulators, the SEESQC framework offers a pathway toward engineering topological phases and manipulating logical qubits.
	\end{abstract}
	
	\author{Yu-Tao Hu}
	\affiliation{Guangdong Provincial Key Laboratory of Magnetoelectric Physics and Devices, State Key Laboratory of Optoelectronic Materials and Technologies, and School of Physics, Sun Yat-sen University, Guangzhou, 510275, China}
	\author{Meng-Yuan Li\orcidicon{0000-0001-8418-6372}}
	\affiliation{Institute for Advanced Study, Tsinghua University, Beijing, 100084, China}
	\author{Peng Ye\orcidicon{0000-0002-6251-677X}}
	\email{yepeng5@mail.sysu.edu.cn}
	\affiliation{Guangdong Provincial Key Laboratory of Magnetoelectric Physics and Devices, State Key Laboratory of Optoelectronic Materials and Technologies, and School of Physics, Sun Yat-sen University, Guangzhou, 510275, China}

	\maketitle

	%\tableofcontents
	
	%\hypertarget{catalogue}{}

	%	\newpage

	\section{Introduction}

	Searches for topological order~\cite{TO1,TO2} in   strongly correlated condensed matter systems remain vital for advancing our understanding of exotic quantum phases of matter beyond the Landau-Ginzburg-Wilson paradigm. Remarkably, driven by synergies between quantum many-body theory and quantum information science—as well as rapid progress in quantum technologies—researchers have simultaneously turned to   preparing these exotic states in qubit-based systems. Such prepared quantum states not only enable controlled manipulation and measurement but also bridge deep theoretical insights into topological order with practical advancements in quantum simulation technologies~\cite{quantum_simulation_overview_2021}.
	\begin{figure*}
		\centering
		\hspace*{-0.5cm}
		\begin{subfigure}[t]{0.3\textwidth}
			\centering
			\includegraphics[width=0.75\textwidth]{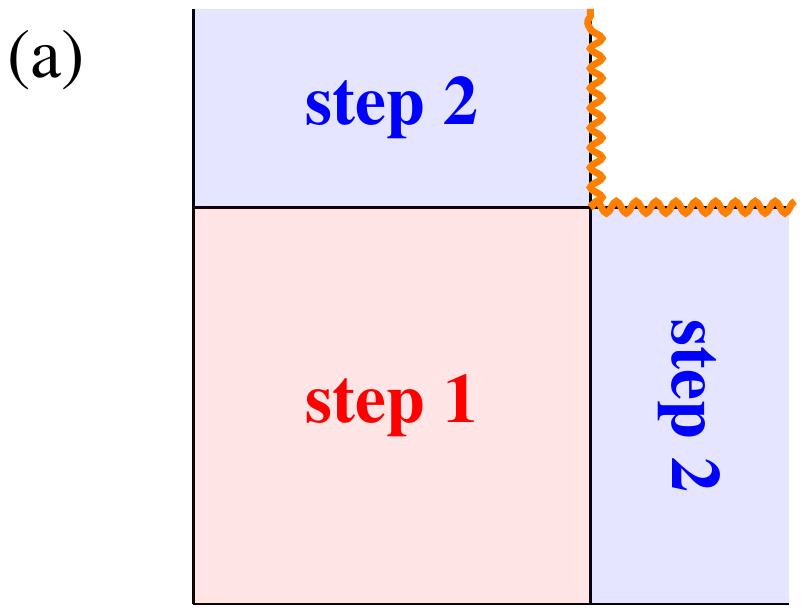}
		\end{subfigure}
		\begin{subfigure}{0.3\textwidth}
			\includegraphics[width=0.814\textwidth]{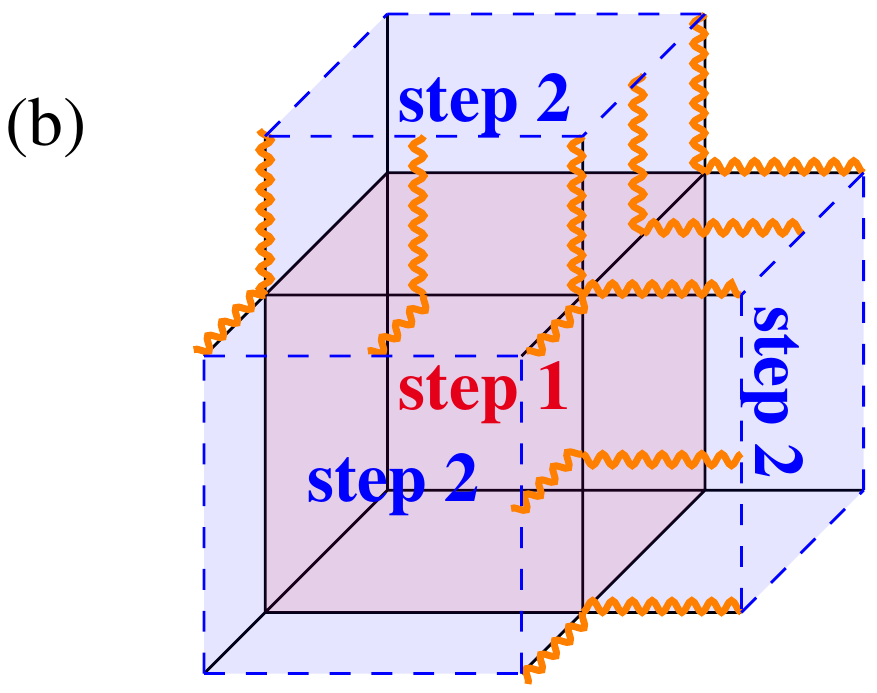}
		\end{subfigure}
		\begin{subfigure}{0.3\textwidth}
			\includegraphics[width=0.814\textwidth]{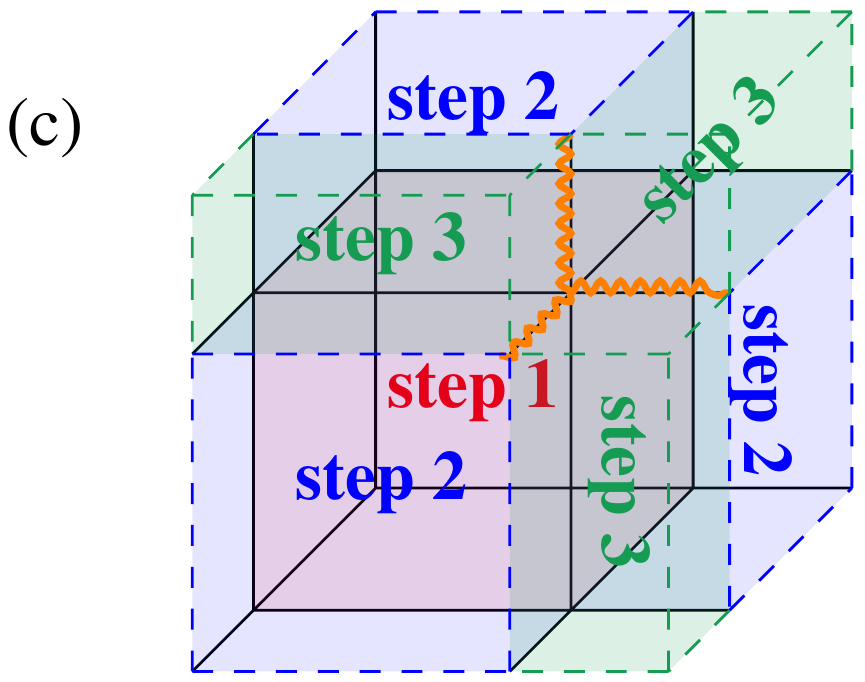}
		\end{subfigure}
		\vspace*{0.5cm}
		
		%\hspace*{-0.1cm}
		\begin{subfigure}{0.9\textwidth}
			\centering
			\begin{tikzpicture}
				\pgftext{\includegraphics[width=0.65\textwidth]{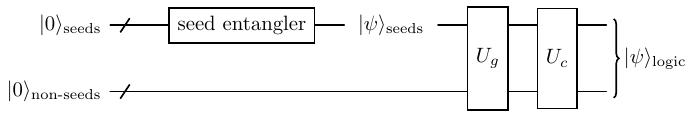}} at (0,0);
				\node[anchor=north west, xshift=-30,yshift=-4] at (current bounding box.north west) {(d)};
			\end{tikzpicture}
		\end{subfigure}
		\caption{\textbf{Schematic illustration of Seed-Entangler-Enriched Sequential Quantum Circuits (SEESQC).} All spins, including seeds and other spins, are initiated in product state $|00\cdots0\rangle$. (a)-(c) Unified linear-depth local unitary (LU) circuit $U_c$ for preparing computational basis states ($|0\rangle^{\otimes \log_2 \text{GSD}}_{\text{logic}}$) of logical qubits encoded by stabilizer code topological models: (a) Toric Code $[0,1,2,2]$, (b) X-cube $[0,1,2,3]$, (c) 3D Toric Code $[1,2,3,3]$. This circuit architecture can be extended to a large class of stabilizer code topological models called tetradigit (TD) models, which is reviewed in Sec.~\ref{section_TD_models}. (d) State preparation protocol: Initial product state undergoes seed entanglement, followed by $U_g$ (a series of CNOT gates) and $U_c$, mapping $|\psi\rangle_{\text{seeds}} \otimes |0\rangle_{\text{non-seeds}}$ to the code state $|\psi\rangle_{\text{logic}}$. Seeds (spins located at the center of each orange wavy line) are: edge-based in (a,b), plaquette-based in (c). When $|\psi\rangle_{\text{seeds}}$ is a Greenberger-Horne-Zeilinger (GHZ) state, $|\psi\rangle_{\text{logic}}$ is a logical GHZ state. For $[d-1,d,d+1,D]$ models (including all shown examples), the number of seeds equals the number of logical qubits.}
		\label{advertisement-1}
	\end{figure*}
	Along this line, local quantum circuits  have been applied to prepare interesting quantum states, which are also systematically categorized by their computational complexity---closely aligned with the classification of quantum phases in condensed matter physics. For gapped states (including topologically ordered phases and product states), two states connected by a finite-depth local unitary (LU) circuit belong to the same phase (when symmetry is not considered)~\cite{phase_transition_and_LU_circuit}, while connecting topologically distinct gapped phases requires at least a linear-depth LU circuit (depth proportional to system's linear size). Practically, quantum circuits for preparing quantum states initialize from product states of disentangled qubits on lattices. To prepare topologically ordered states that support at most area-law entanglement entropy, linear-depth quantum circuit must be properly designed to constrain entanglement growth---a constraint satisfied by sequential quantum circuits~\cite{2005_SQC_Cirac,2007_SQC_Cirac,2008_SQC_Cirac,Roushan_2021,Liu_Yu-Jie2021_SQC_string_net,Wei_Zhi-Yuan_2021_SQC_PEPS,SQC_time_evolution,chen2024sequential}.
	To realize quantum circuits between two topologically distinct gapped
	states, sequential quantum circuits as LU circuits are applied, where each qubit is acted by finite number of gates (independent of system size)~\cite{chen2024sequential}. Recent remarkable advances further leverage sequential quantum circuits to:  construct non-invertible symmetries~\cite{SQC_duality,2024arXiv240612962P,SQC_non_invertible_symmetry_2},   engineer Cheshire strings for charge condensation~\cite{SQC_Cheshire_string},  represent gauging procedures~\cite{SQC_triality},  and  implement non-Abelian anyon ribbon operators~\cite{SQC_ribbon_operator}.  
	While significant progress has been made in case-by-case basis, a unified and efficient framework for quantum circuit design remains elusive. A critical challenge lies in developing unified and efficient circuits capable of transcending spatial dimensions, bridging liquid and non-liquid states (e.g., fracton topological phases)~\cite{zeng2019quantum}, and unifying circuits for preparing gapped phases governed by distinct entanglement renormalization group (ERG) schemes~\cite{ERG_vidal,fracton16,dddd_ERG}. 
	
	Furthermore, in the presence of the robust ground state degeneracy (GSD), topological order discovered in condensed matter physics has also been widely recognized as a natural platform for topological quantum error-correcting codes. Topological error-correction reduces logical qubit error rates compared to physical qubits~\cite{toric_code,toric_code_error_correction_experiment}, and experiments have demonstrated the protection of logical qubit entanglement via topological error-correction protocols~\cite{logical_qubits_entanglement}. To implement quantum computation on error-correcting codes, universal logical gates are needed. For   prototypical topological error-correcting codes such as Toric Code, many protocols (e.g., lattice surgery~\cite{toric_code_lattice_surgery,fowlow_2018_lattice_surgery,game_lattice_surgery}) of implementing universal logical gate requires magic state injection, which in turn requires high-fidelity magic state preparation. For some other topological error-correcting codes, e.g., the X-cube model which is a typical non-liquid topological state,   it is still challenging to realize universal logical gate and arbitrary code state preparation protocol.  A systematic effort on quantum state engineering in code space remains a vibrant and critical research frontier at the intersection of quantum information, condensed matter physics, and mathematical physics.

	In this paper, we propose a unified circuit protocol, termed the \textbf{\emph{Seed-Entangler-Enriched Sequential Quantum Circuits}} (\textbf{SEESQC}) schematically illustrated in Fig.~\ref{advertisement-1},  which, beyond the case-by-case base, is able to systematically and efficiently realize arbitrary code states of a large class of topological error-correcting codes. Theoretically, the protocol unveils, in the context of quantum circuits, an elegant unification of distinct phases of matter across spatial dimensions, and dismantles the boundaries of liquid and non-liquid states that are governed by distinct ERG schemes~\cite{zeng2019quantum,ERG_vidal,fracton16,dddd_ERG}.  To demonstrate, we systematically apply the protocol to a broad class of long-range entangled states---tetradigit (TD) states---i.e., code states or ground states of TD models (to be introduced below shortly)~\cite{li2020fracton}. Notably, TD states are not rare; on the contrary,  TD states represent a wide spectrum of long-range entangled systems: Toric Codes in arbitrary spatial dimensions and X-cube stabilizer codes naturally emerge as special cases within this framework.  First, we show how to prepare the computational basis states of TD models' code space, using the circuit $U_c$ (conceptually illustrated in Fig.~\ref{advertisement-1} (a)-(c)). Then, to construct arbitrary code states of TD models, we employ the \emph{seed $\to$ logical qubit} scheme (conceptually illustrated in Fig.~\ref{advertisement-1} (d)). The \emph{seeds $\to$ logical qubit} scheme involves a ``seed-entangler'' that prepares ``seeds'' (a small fraction of all qubits marked by the orange wavy lines in Fig.~\ref{advertisement-1}) in state $|\psi\rangle_{\text{seeds}}$, and two linear-depth local unitary circuits called $U_g$ and $U_c$, sending the state of seeds $|\psi\rangle_{\text{seeds}}$ to the code state of logical qubits $|\psi\rangle_{\text{logic}}$. Intriguingly, for a large family of TD models and the unified circuit $U_c$, we find that the number of available seeds coincides with the number of encoded logical qubits.

	While a full technical analysis is reserved for the main text, it is beneficial to introduce TD models and their corresponding ground states---TD states, especially focusing on the main properties that are most relevant to the present paper:
	\begin{itemize}
		\item 	
		TD models were first introduced in Ref.~\cite{li2020fracton} and further explored in Refs.~\cite{li2021fracton,dddd_ERG}. These quantum spin-$1/2$ models are uniquely labeled by a four-digit index 
		$[d_n,d_s,d_l,D]$\footnote{Since this class of models was not assigned an official name in Refs.~\cite{li2020fracton,li2021fracton,dddd_ERG},  in this work we simply adopt the term ``tetradigit  models'' for the notational convenience. To ensure exact solvability and stabilizer-code ground states, the digits must satisfy specific algebraic constraints. For the sake of convenience, hereafter we use  the terms ``TD models'' and ``TD states''  for stabilizer code cases in which ground states serve as code states of error-correcting codes. These models are exactly solvable. All other cases will be specified by adding the prefix ``non-stabilizer code''. The latter cases are also of interest and will be  introduced in Sec.~\ref{section_non_stabilizer_code_TD_models}).}. The digit $D$ is just the spatial dimension of the hypercubic lattice hosting the model, while $d_s$ is the dimension of cell/cube that hosts a spin. The remaining digits—$d_n$ and $d_l$—characterize microscopic details of terms in Hamiltonian, which will be reviewed technically in Sec.~\ref{section_TD_models}. Notably, the $[0,1,2,2]$ model corresponds to the 2-dimensional Toric Code; the $[0,1,2,3]$ model describes the 3-dimensional X-cube fracton phase (see, e.g., Refs.~\cite{fractonorder1,fractonorder2,fractonorder3,fracton16,PhysRevLett.129.230502,fracton3,fracton2,fracton1,fracton19,fracton,fracton4,fracton17,li2020fracton,li2021fracton,dddd_ERG,fracton27,fracton53,fracton54,fracton15,PhysRevB.110.205108}); the $[1,2,3,3]$ model represents the 3-dimensional Toric Code. This broadness of long-range entangled states positions TD models as an ideal platform for systematic quantum circuit design, which is the main goal of this work.

		\item The primary motivation for constructing TD models in Ref.~\cite{li2020fracton} was  to extend fracton physics---traditionally focused on point-like topological excitations (e.g., fractons, lineons, planons)---to scenarios involving \textit{spatially extended excitations} such as loops and membranes. Actually, before the age of fracton physics,   excitations with nonlocal geometry have widely appeared in higher-dimensional topological phases of matter where   exotic braiding statistics~\cite{hansson_superconductors_2004,PRESKILL199050,PhysRevLett.62.1071,PhysRevLett.62.1221,ALFORD1992251,wang_levin1,PhysRevLett.114.031601,2016arXiv161209298P,string4,jian_qi_14,string5,PhysRevX.6.021015,Tiwari:2016aa,corbodism3,string6,3loop_ryu,PhysRevLett.121.061601,Kapustin:2014gua,bti2,PhysRevB.99.235137,cyaloop1,cyaloop2,cyaloop3,2024arXiv241201886K,zhang_compatible_2021,zhang_topological_2022,Zhang2023fusion,Zhang2023Continuum,Huang2023,huang2025diagrammatics}, topological response~\cite{lapa17,YW13a,ye16a,Ye:2017aa,bti6,RevModPhys.88.035001,PhysRevB.99.205120}, and symmetry fractionalization~\cite{Ning2018prb,ye16_set,2016arXiv161008645Y} go beyond two-dimensional   systems. 
		In conventional fracton systems like the X-cube model, fractons are point-like and immobile. Remarkably, introducing spatially extended objects necessitates considering not only \textit{mobility} but also \textit{deformability}, significantly broadening the scope of fracton topological order.  Furthermore, in the TD model series, excitations with non-manifold-like shapes---neither loops nor membranes---are discovered, exhibiting extreme diversity in both quantum many-body physics and quantum entanglement.  
		Notably, fracton physics has expanded to encompass many-body systems with Hamiltonians conserving higher moments, such as dipoles (i.e., center-of-mass)~\cite{fracton58,fracton59,Fractongauge,fracton55,fracton56,fracton57,fracton60,fracton20,fracton26,fracton5,diffusionofhigher-moment,Fractonicsuperfluids1,Fractonicsuperfluids2,wanghanxie,NSofFractonicsuperfluids,fracton23,Fractonicsuperfluidsdefect,reviewofFractonicsuperfluids,fracton28,fracton7,fracton21,fracton14,fracton22,fracton30,Lifshitzduality,fracton11,HMWT,DBHM,DBHM2,fracton31,fracton35,fracton44,fracton45,fracton6,fracton29,fracton24,fracton34,fracton36,fracton37,fracton33,fracton38,fracton39,fracton40,fracton9,fracton41,fracton32,fracton43,fracton48,fracton8,fracton42,fracton46,fracton25,fracton10,fracton12,fracton49,fracton50,fracton47,rank_two_TC,rank_two_TC_Z_N,effective_2d_fracton,2024PhRvB.109t5146D,2024arXiv240612962P}. These conserved quantities restrict the kinematics of constituent particles, leading to novel quantum phenomena in equilibrium and non-equilibrium settings. Consequently, the mobility and deformability constraints on spatially extended objects are anticipated to exhibit even more exotic behavior---a problem that remains unresolved.

		\item The series of TD models have been further explored. Ref.~\cite{li2021fracton} immediately revealed that their GSD formulas---expressed as $\text{log}_2\text{GSD}$---exhibit diverse polynomial dependencies on system size. This enriches the  linear dependence seen in the X-cube model and expands the counting of logical qubits. While a proof is demanded, the polynomial coefficients are conjectured to encode topological and geometric properties of both the quantum codes and the underlying lattice, as demonstrated for the X-cube model in Ref.~\cite{fracton16}.   In Ref.~\cite{dddd_ERG},   the origin of the polynomial dependence    is    connected to   a substantially generalized   ERG scheme, expanding the definitions of equivalence relations between gapped phases. While Toric Codes and X-cube states, as special examples of TD states, lack finite-temperature transitions, a subset of TD states ($[1,2,3,D]$ with $D \geq 4$) can host finite-temperature phase transitions~\cite{Shen_2022}. These exotic features and potential applications motivate methods to experimentally prepare TD states. In principle, any 4-dimensional TD model can be realized by synthetically engineering the extra dimension~\cite{synthetic_dimension_1,synthetic_dimension_2,synthetic_dimension_3}---a strategy extendable to higher-dimensional stabilizer code TD models.

	\end{itemize}

	Having briefly covered  TD models as well as TD states, we resume our exploration of the SEESQC  protocol that will be applied to TD models in this paper. In brief, we first explicitly propose a unified circuit, denoted by $U_c$, for preparing logical qubits' computational basis $|0\rangle_{\text{logic}}^{\otimes\text{log}_2\text{GSD}}$ of any stabilizer code TD model, and then add seed-entangler that enables the preparation of arbitrary code state of a large class of TD models. For computational basis of code states, first we prepare the \textbf{\emph{equal weight superposition}} of closed $D$-cube cage \textbf{\emph{configurations}} (EWSC), e.g., closed loop configurations when $D=2$, on a particular set of $D$-cubes. Then, we systematically study redundancy of stabilizers of TD models, by means of which, we show the prepared state is indeed the EWSC on all $D$-cubes, and thus a code state. Illustrative figures of the circuit $U_c$ preparing computational basis of code states are given in Fig.~\ref{advertisement-1} (a)-(c). 
	\begin{itemize}
		\item Fig.~\ref{advertisement-1}(a) illustrates the circuit $U_c$ for the $[0,1,2,2]$ model, where steps 1 and 2 are sequences of CNOT gates. Step 1 generates the EWSC in the ``bulk''; step 2 generates the EWSC on the ``boundary'' (note there is no physical boundary since the system is on a torus – here ``boundary'' refers to the perimeter of the planar graph); and the EWSC at ``corners'' automatically emerges when the circuit is completed.
		
		\item Fig.~\ref{advertisement-1}(b) illustrates the circuit $U_c$ for the $[0,1,2,3]$ model, where steps 1 and 2 are similarly sequences of CNOT gates. As in the $[0,1,2,2]$ case, step 1 generates the EWSC in the ``bulk'', step 2 generates the EWSC on the ``boundary'', and the EWSC at ``corners'' emerges automatically upon circuit completion. The key difference is that the $D$-cubes at ``corners'' scale with system size.
		
		\item Fig.~\ref{advertisement-1}(c) illustrates the circuit $U_c$ for the $[1,2,3,3]$ model, where steps 1–3 are sequences of CNOT gates. Step 1 generates the EWSC in the ``bulk'', step 2 generates the EWSC on the ``boundary'', and step 3 generates the EWSC at ``corners'', while the EWSC at ``corner squares'' automatically emerges when the circuit is completed.
	\end{itemize}
	This paradigm can be naturally generalized to arbitrary stabilizer code TD models. It is worth mentioning that among TD models, there exist different models with the same spatial dimension $D$ and distinct $d_s$ (i.e., the dimension of cubes that host spins). Consequently, the closed $D$-cube cage configurations and the EWSC depend not only on $D$, but also on $d_s$. For example, in the $[0,1,2,3]$ model, the closed cube cage configuration consists of edges on the boundaries of cubes, while in the $[1,2,3,3]$ model, it consists of plaquettes on the boundaries of cubes. Although our primary focus is on quantum circuits for periodic boundary conditions (PBC), we demonstrate that truncating these circuits (discussed in Sec.~\ref{section_unified_SQC}) immediately yields quantum circuits for open boundary conditions (OBC) and hybrid half-OBC-half-PBC (i.e., some directions under OBC while the others are under PBC). In the formula for the circuit's total depth (Eq.~(\ref{revise_19})), the so-called long-range entanglement (LRE) level~\cite{zeng2019quantum,ERG_vidal,fracton16,dddd_ERG} naturally emerges.

	On top of the unified circuit $U_c$ for preparing computational basis code states, we use the \emph{seed $\to$ logical qubit} scheme (see Fig.~\ref{advertisement-1} (d)), to prepare arbitrary code state of $[d-1,d,d+1,D]$ models, including Toric Code in all dimension and \textit{X-cube} model. Concretely, before the circuit $U_c$ for preparing computational basis of code states, we implement the \emph{seed-entangler} that maps seeds into any state $|\psi\rangle_{\text{seeds}}$, followed by a sequence of CNOT gates that make the seeds ``grow'', denoted by $U_g$. While the choice of seeds is model dependent and circuit dependent, it happens for $[0,1,2,2],[0,1,2,3],[1,2,3,3]$ models and circuit $U_c$, the choice of seeds can be written in a unified form, which gives (as a guess) a choice of seeds for any $[d-1,d,d+1,D]$ model. Believing in the equivalence of seeds number and logical qubits number, we obtain a formula of GSD for any $[d-1,d,d+1,D]$ model, which successfully returns back to the GSD of $[0,1,2,D]$ models and $[D-3,D-2,D-1,D]$ models~\cite{li2021fracton} when $d=1$ and $d=D-2$ are taken.

	This paper is organized as follows: In Sec.~\ref{section_TD_models}, we first introduce TD models, refining their definition beyond Ref.~\cite{li2020fracton}, and subsequently review their key properties. We then briefly discuss non-stabilizer code TD models, highlighting their distinctions from stabilizer-based cases. In Sec.~\ref{section_SQC}, we develop quantum circuit incrementally—starting with the 2-dimensional Toric Code, i.e., the code state of $[0,1,2,2]$ model, advancing to the code states of X-cube model (i.e., $[0,1,2,3]$ model), $[0,1,2,4]$ model and 3-dimensional Toric Code  model (i.e.,  $[1,2,3,3]$ model), and finally culminating in a unified quantum circuit framework applicable to generic TD models. In Sec.~\ref{section_any_GS}, we propose the so-called \emph{seed $\to$ logical qubit} scheme for preparing arbitrary code state of Toric Code, X-cube, 3-dimensional Toric Code, and discuss its generalization to any $[d-1,d,d+1,D]$ model. Finally, Sec.~\ref{section_summmary_outlook} concludes the paper and outlines future directions.

	\section{  TD models and TD states}\label{section_TD_models}
	\subsection{Notations}\label{section_TD_models_definition}
	
	The TD models constitute a family of spin lattice models defined on $D$-dimensional (hyper)cubic lattices~\cite{li2020fracton}. As their construction inherently involves higher-dimensional geometric constructs (e.g., membranes, volumes), we first establish dimension-agnostic notations. Building on the foundational framework of Refs.~\cite{li2020fracton,li2021fracton,dddd_ERG}, we refine the model definitions using terminology from algebraic topology and foliation theory. These additions not only streamline the characterization of spin placements and interactions but also generalize the lattice-dependent structure—enabling future extensions to non-cubic lattices. Crucially, this reformulation preserves the models’ definition while clarifying their geometric-topological underpinnings.
	
	We first review the geometric framework of TD models, following the notation and terminology established in Refs.~\cite{li2020fracton,li2021fracton,dddd_ERG}. Consider a $D$-dimensional (hyper)cubic lattice $\mathbb{Z}_{L_1} \times \mathbb{Z}_{L_2} \times \cdots \times \mathbb{Z}_{L_D}$ with lattice constant 1, subject to open (OBC), periodic (PBC), or half-OBC-half-PBC. The lattice spans $D$ orthogonal directions $\hat{x}_1, \hat{x}_2, \ldots, \hat{x}_D$.  
	An $n$-cube $\gamma_n$ denotes the $n$-dimensional analog of a cube:  
	\textit{(i)}    $0$-cube = vertex,  
	\textit{(ii)} $1$-cube = edge\footnote{Therefore, this ``edge''   does \textit{not} specifically denote the boundary of the bulk, but the link between two neighboring vertices, on the edge of some plaquette.},  
	\textit{(iii)}  $2$-cube = plaquette,  
	\textit{(iv)}   $3$-cube = 3-dimensional cube,  and 
	\textit{(v)}   $4$-cube = 4-dimensional hypercube,  
	and so forth. Each $n$-cube represents a unit  cell of the lattice. By fixing the origin at $(0,0,\ldots,0)$ in Cartesian coordinates, the geometric center of every $n$-cube $\gamma_n$ has coordinates where exactly $D-n$ components are integers and $n$ components are half-integers. This unique coordinate mapping provides a one-to-one identification of $n$-cubes.  
	A \textit{leaf} (or \textit{leaf space}) $l = \langle \hat{x}_{i_1}, \hat{x}_{i_2},   \cdots, \hat{x}_{i_{d_l}} \rangle$ is defined as a subspace extending along $d_l$ orthogonal directions among $\hat{x}_1,\hat{x}_2,\cdots,\hat{x}_D$~\cite{li2020fracton}. Leaves partition the lattice into lower-dimensional substructures critical for defining   TD models.

	\subsection{Notations explained via algebraic topology and foliation theory}
	The language introduced in the above subsection is natural for discussing TD models. However, some descriptions could be more precise. In the following several paragraphs, we aim to enhance precision by utilizing terminologies from algebraic topology and foliation theory of manifolds. Readers not interested in these details may skip them.
	
	We begin with a smooth $D$-dimensional manifold with corners $(X, \mathcal{A})$, where the topological manifold $X = T^m \times B^{D-m}$\footnote{Here, $T^m$ denotes an $m$-dimensional torus, and $B^{D-m}$ denotes a $(D-m)$-dimensional closed ball.} has a cubical cellular decomposition $\mathcal{E}$, with $n$-cells being $n$-cubes, and $\mathcal{A}$ is the atlas. As defined in \cite{manifolds_with_corners}, a manifold with boundary is a special case of a manifold with corners, while a manifold without boundary is a special case of a manifold with boundary. Unless specified otherwise, ``manifold'' hereafter refers to a manifold with corners. The cubical cellular decomposition $\mathcal{E}$ is the set of all (open) $n$-cubes for all $0 \leq n \leq D$. The pair $(X, \mathcal{E})$ is called a cell complex. Since all cells are $n$-cubes, $(X,\mathcal{E})$ is also referred to as a cubical complex. For simplicity, we sometimes say $X$ is a cell or cubical complex. Note that, in general, a cell complex $X$ need not be a manifold; it is sufficient for it to be a topological space. Physicists may intuitively consider the $D$-dimensional lattice as the set of all vertices $X^0$, or the union of all vertices and edges $X^1$. These two sets are known as the 0-skeleton $X^0$ and 1-skeleton $X^1$ of the cell or cubical complex $X$, respectively. For any $n$, the $n$-skeleton of $X$ is a subcomplex of $X$. A subcomplex of a cell complex $X$ is a closed subspace $A \subset X$ that is a union of cells of $X$.
	
	Now, we turn to leaves. As a smooth manifold (possibly with corners), we can define smooth foliations on $(X, \mathcal{A})$ \cite{foliation_1}. A $d_l$-dimensional smooth foliation on $(X, \mathcal{A})$ is an equivalence relation on $(X, \mathcal{A})$, with the equivalence classes being connected, disjoint $d_l$-dimensional smooth submanifolds called (smooth) leaves. Since we have defined manifolds to refer to manifolds with corners in general, it is worth emphasizing that leaves can only have boundary and corners at the boundary/corners of the original manifold $X$: cutting a complete leaf into two pieces with additional boundary/corners is forbidden. With this restriction, one can freely define different foliations to obtain different leaves. The leaves of interest in this paper are those $d_l$-dimensional (smooth) leaves that are simultaneously subcomplexes of $X$, which we call \emph{subcomplex leaves} hereafter. Subcomplex leaves are exactly those leaves defined in Ref.~\cite{li2020fracton}, here with a more rigorous and mathematical definition.

	\subsection{Hamiltonians in terms of mutually commuting stabilizers}\label{sec_notation_hamiltonian}

	Next, we review the definition of TD models. TD models (and their associated TD states) are labeled by four non-negative integers $[d_n, d_s, d_l, D]$, where $d_n \leq d_s \leq d_l \leq D$. The most distinguishing feature of this model family is that they host fracton orders in arbitrary spatial dimensions greater than two. Such high-dimensional fracton orders possess intriguing properties, such as spatially extended excitations with restricted mobility and deformability, and complex excitations with non-manifold-like shapes\footnote{To the best of our knowledge, there is no excitation with fractal support in the TD model series. What there are in the TD model series are branched strings/membranes-shaped excitation, which are not homeomorphic to any $\mathbb{R}^n$ locally at the branching points in the continuum limit, thus are called non-manifold-like excitation. For example, there are chairon (which is chair-like and has two 3-branch points), yuon in the $[1,2,3,4]$ model, see Fig. 8 in \cite{li2020fracton}.}~\cite{li2020fracton}.
	
\begin{figure}
		\centering
		\includegraphics[width=0.4\textwidth]{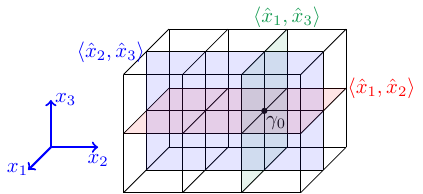}
		\caption{Leaves in the $[0,1,2,3]$ TD model (3-dimensional X-cube fracton phase) under open boundary conditions. The black dot denotes a $0$-dimensional node $\gamma_0$ at $(1,2,1)$. Three orthogonal leaves (2-dimensional planes) containing $\gamma_0$ are highlighted: $\langle \hat{x}_1, \hat{x}_2 \rangle$ (red, $x_3=1$ plane), $\langle \hat{x}_1, \hat{x}_3 \rangle$ (green, $x_2=2$ plane), and $\langle \hat{x}_2, \hat{x}_3 \rangle$ (blue, $x_1=1$ plane). Each leaf hosts $B$ terms formed by products of Pauli $Z$ operators on spins (qubits) within the respective plane. The axes $x_1$, $x_2$, $x_3$ are marked in blue. The blue dot stands for the original point $(0,0,0)$.}
		\label{leaf-illustrative-figure-Euclidean}
	\end{figure}

	The Hamiltonian of a TD model contains two kinds of stabilizer terms, called $A$ terms and $B$ terms. By our convention, an $A$ term is the product of Pauli $X$ operators, and a $B$ term is the product of Pauli $Z$ operators. In the label $[d_n, d_s, d_l, D]$, $d_n$ is the dimension of the cube/cell where we define a $B$ term (``n'' stands for ``node''), $d_s$ is the dimension of the cube/cell where we place a spin (``s'' stands for ``spin''), $d_l$ is the dimension of the leaf where a $B$ term is fully embedded (``l'' stands for ``leaf''), and $D$ simultaneously stands for the spatial dimension and the dimension of the cube/cell where we define an $A$ term.  The Hamiltonian is:
	\begin{equation}\label{paper_dddd_22}
		H_{[d_n, d_s, d_l, D]} = -\sum_{\gamma_D} A_{\gamma_D} - \sum_{\gamma_{d_n}} \sum_{l \revsubset \gamma_{d_n}} B_{\gamma_{d_n}, l},
	\end{equation}
	where
	\begin{equation}
		A_{\gamma_D} := \prod_{\gamma_{d_s} \subset \gamma_D} X_{\gamma_{d_s}}\label{revise_50}
	\end{equation}
	is the product of Pauli $X$ operators of all spins inside the (closed) $D$-cube $\gamma_D$, and
	\begin{equation}
		B_{\gamma_{d_n}, l} := \prod_{\substack{\gamma_{d_s} \\ \gamma_{d_n} \subset \gamma_{d_s} \subset l}} Z_{\gamma_{d_s}}
	\end{equation}
	is the product of Pauli $Z$ operators of all spins on the (closed) $d_s$-cube $\gamma_{d_s}$ that simultaneously contains the node $\gamma_{d_n}$ and is inside the leaf $l$. The leaves $l$ under summation in Eq.~(\ref{paper_dddd_22}) take all possible leaves of the form $\langle \hat{x}_{i_1}, \hat{x}_{i_2}, \dots, \hat{x}_{i_{d_l}} \rangle$ that simultaneously contain a straightforward $d_l$-dimensional sublattice and the $\gamma_{d_n}$, assuming the $D$-dimensional cubic lattice is a direct discretization of Euclidean space\footnote{This makes sense when OBC is taken, but lacks reasonable explanation when PBC is taken. Appendix~\ref{appendix_pbc} includes a discussion on leaf choice under PBC.}. For example, consider the $[0, 1, 2, 3]$ model under OBC, where the 3-dimensional cubic lattice is taken to be a regular lattice in Euclidean space. A leaf can be regarded as a truncated vector/Euclidean space, and in the Hamiltonian, $l$ takes through $\langle \hat{x}_1, \hat{x}_2 \rangle$, $\langle \hat{x}_1, \hat{x}_3 \rangle$, and $\langle \hat{x}_2, \hat{x}_3 \rangle$, each containing $\gamma_0$ in the summation $\sum_{l \revsubset \gamma_0}$. An illustration of this example is shown in Fig.~\ref{leaf-illustrative-figure-Euclidean}. In general, there are $C_{D-d_n}^{d_l-d_n}$ distinct $B_{\gamma_{d_n},l}$ terms attached to each node $\gamma_{d_n}$.
	
	In Ref.~\cite{li2020fracton}, it has been proven that when the labels satisfy
	\begin{equation}
		\label{dddd_3}
		C_{d_l - d_n}^{d_s - d_n} \mod 2 = 0,
	\end{equation}
	$[A_{\gamma_D}, B_{\gamma_{d_n}, l}] = 0$, the $[d_n, d_s, d_l, D]$ model is a stabilizer code model. In this paper, we primarily focus on the models where Eq.~(\ref{dddd_3}) is satisfied. In Sec.~\ref{section_non_stabilizer_code_TD_models}, we provide a brief discussion on TD models that are not stabilizer codes, which have been less studied previously.

	\subsection{Examples}
	
	\subsubsection{$[0,1,2,2]$ model (2-dimensional Toric Code )}

\begin{figure}[htbp]
		\centering
		\includegraphics[width=0.27\textwidth]{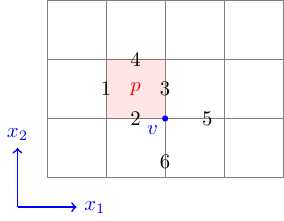}
		\caption{Illustration of $A$ terms and $B$ terms in the 2-dimensional Toric Code. The $A$ term associated with plaquette $p$ is $A_p = \prod_{e \subset p} X_e = X_1 X_2 X_3 X_4$. The $B$ term associated with vertex $v$ is $B_v = \prod_{e \revsubset v} Z_e = Z_2 Z_3 Z_5 Z_6$.}
		\label{2dTC-stabilizers}
	\end{figure}

	The 2-dimensional Toric Code is a prototypical model of pure topological order. The 2-dimensional Toric Code on a square lattice is the $[0,1,2,2]$ model in the TD model family; we show this equivalence now. Consider the 2-dimensional Toric Code on a square lattice, where each edge is assigned a qubit. The Hamiltonian is
	\begin{equation}
		H = -\sum_{p} A_p - \sum_{v} B_v = -\sum_{p} \prod_{e \subset p} X_e - \sum_{v} \prod_{e \revsubset v} Z_e,\label{revise_51}
	\end{equation}
	where $v$, $e$, and $p$ stand for vertex, edge, and plaquette, respectively. An illustration of the $A$ term and $B$ term is shown in Fig.~\ref{2dTC-stabilizers}. By definition, $v$, $e$, and $p$ correspond to $\gamma_0$, $\gamma_1$, and $\gamma_2$, respectively. Thus, $A_p$ can be written as $A_{\gamma_2}$, and $B_v$ can be written as $B_{\gamma_0,l}$, where $l$ has only one possible choice (since $d_l = 2$), i.e., $l$ spans the entire 2-dimensional cubical complex, so the index $l$ can be omitted.
	The dimension of $v$ (to which $B$ terms attach) is $0$, so $d_n = 0$. The dimension of $e$ (where qubits reside) is $1$, so $d_s = 1$. The leaf dimension $d_l = 2$. The dimension of $p$ (where $A$ terms reside) is 2, same to the lattice dimension, so $D=2$. Therefore, the 2-dimensional Toric Code on a square lattice is indeed the $[0,1,2,2]$ model.
	
	The 2-dimensional Toric Code model has two types of fundamental excitations:
	\begin{itemize}
		\item $e$ particles: Created by Pauli $Z$ operators on a string, at the endpoints of the string ($A_p=-1$).
		\item $m$ particles: Created by Pauli $X$ operators on a dual lattice string, at the endpoints of the string ($B_v=-1$).
	\end{itemize}
	$e,m$ particles have trivial (boson) self-statistics and non-trivial (-1 phase) relative statistics.

	\subsubsection{$[0,1,2,3]$ model (X-cube model)}

\begin{figure}[htbp]
		\centering
		\includegraphics[width=0.276\textwidth]{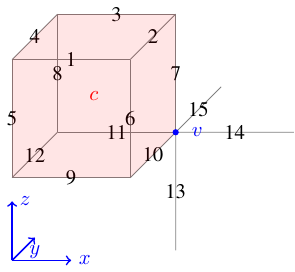}
		\caption{Illustration of $A$ terms and $B$ terms in the X-cube model. The $A$ term associated with cube $c$ is $A_c = X_1 X_2 \cdots X_{12}$. The three different $B$ terms attached to vertex $v$ are $B_{v,xy} = Z_{10}Z_{11}Z_{15}Z_{14}$, $B_{v,xz} = Z_7Z_{11}Z_{13}Z_{14}$, and $B_{v,yz} = Z_7Z_{10}Z_{13}Z_{15}$.}
		\label{X-cube-stabilizers}
	\end{figure}

	The X-cube model is a prototypical example of type-I fracton order~\cite{fuliang_xcube}. The X-cube model on a cubic lattice is the $[0,1,2,3]$ model in the TD model family; we demonstrate this equivalence now. Consider the X-cube model on a cubic lattice where each edge hosts a qubit. The Hamiltonian is:
	\begin{equation}\label{eq:x-cube-hamiltonian}
		H = -\sum_c A_c - \sum_v \left(B_{v,xy} + B_{v,xz} + B_{v,yz}\right),
	\end{equation}
	where $A_c = \prod_{e \subset c} X_e$ is the product of Pauli $X$ operators on all edges in cube $c$, and $B_{v,xy}$ denotes the product of four Pauli $Z$ operators forming an ``X'' in the $xy$-plane (similarly for $B_{v,xz}$ and $B_{v,yz}$). Fig.~\ref{X-cube-stabilizers} illustrates these operators.
	
	The vertex dimension $d_n = 0$ (to which $B$ terms attach), edge dimension $d_s = 1$ (where qubits reside), leaf dimension $d_l = 2$, and lattice dimension $D = 3$ (also the dimension of cubes where $A$ terms reside). Thus, the X-cube model corresponds to $[0,1,2,3]$, with Hamiltonian:
	\begin{align}
		H &= -\sum_{c} A_{c} - \sum_{v} \sum_{l \revsubset v} B_{v,l}.
	\end{align}
	
	The X-cube model exhibits a GSD of $2^{2L_x + 2L_y + 2L_z - 3}$ under PBC, encoding $2L_x + 2L_y + 2L_z - 3$ logical qubits. This extensive degeneracy significantly exceeds that of three-dimensional pure topological orders.
	
	The X-cube model has two fundamental excitations:
	\begin{itemize}
		\item  Fractons: Created by Pauli $Z$ operators on dual lattice rectangular planes, at the four corners of the rectangular plane ($A_c=-1$). Pairs of $A_c=-1$ fractons form mobile \textit{planons} confined within planes.
		\item  Lineons: Created by Pauli $X$ operators along straight lines, at the endpoints of the line, with three types:
		\begin{itemize}
			\item $l_x$ ($B_{v,xy}=B_{v,xz}=-1$): Moves along $x$-axis
			\item $l_y$ ($B_{v,xy}=B_{v,yz}=-1$): Moves along $y$-axis
			\item $l_z$ ($B_{v,xz}=B_{v,yz}=-1$): Moves along $z$-axis
		\end{itemize}
	\end{itemize}
	Lineons obey fusion rules $l_x \times l_y = l_z$~\cite{fuliang_xcube,li2020fracton,fracton_fusion}, demonstrating nontrivial interplay between fracton fusion and mobility.

	\subsubsection{$[0,1,2,4]$ model}

	The $[0,1,2,4]$ model is defined on a 4-dimensional hypercubic lattice where each edge hosts a qubit. The Hamiltonian is:
	\begin{equation}
		H = -\sum_{\gamma_4} A_{\gamma_4} - \sum_{\gamma_0} \sum_{l \revsubset \gamma_0} B_{\gamma_0,l},
	\end{equation}
	where $A_{\gamma_4} = \prod_{\gamma_1\subset \gamma_4} X_{\gamma_1}$ is the product of Pauli $X$ operators on all 1-cubes/edges in the 4-cube $\gamma_4$, and $B_{\gamma_0,l} = \prod_{\gamma_1 \revsubset \gamma_0, \gamma_1 \subset l} Z_{\gamma_1}$ acts on four edges containing vertex $\gamma_0$ within a 2-dimensional (smooth subcomplex) leaf $l$. Each vertex $\gamma_0$ has $C_4^2 = 6$ distinct leaves ($d_l=2$) in the $D=4$ lattice: $x_1x_2$, $x_1x_3$, $x_1x_4$, $x_2x_3$, $x_2x_4$, and $x_3x_4$ planes, each one hosts a $B$ term attached to $\gamma_0$. Fig.~\ref{0124-stabilizers} illustrates these operators.
	\begin{figure*}[htbp]
		\centering
		\begin{subfigure}{0.65\textwidth}
			\centering
			\begin{tikzpicture}[baseline=0ex]
				\draw[gray] (0,0,0)--(2,0,0)--(2,0,2)--(0,0,2)--cycle;
				\draw[gray] (0,2,0)--(2,2,0)--(2,2,2)--(0,2,2)--cycle;
				\draw[gray] 
				(0,0,0)--(0,2,0)
				(2,0,0)--(2,2,0)
				(2,0,2)--(2,2,2)
				(0,0,2)--(0,2,2);
				
				\draw[gray] (0+8,0+1.4,0)--(2+8,0+1.4,0)--(2+8,0+1.4,2)--(0+8,0+1.4,2)--cycle;
				\draw[gray] (0+8,2+1.4,0)--(2+8,2+1.4,0)--(2+8,2+1.4,2)--(0+8,2+1.4,2)--cycle;
				\draw[gray] 
				(0+8,0+1.4,0)--(0+8,2+1.4,0)
				(2+8,0+1.4,0)--(2+8,2+1.4,0)
				(2+8,0+1.4,2)--(2+8,2+1.4,2)
				(0+8,0+1.4,2)--(0+8,2+1.4,2);
				
				\draw[gray]
				(0,0,0)--(0+8,0+1.4,0)
				(0,0,2)--(0+8,0+1.4,2)
				(0,2,0)--(0+8,2+1.4,0)
				(0,2,2)--(0+8,2+1.4,2)
				(2,0,0)--(2+8,0+1.4,0)
				(2,0,2)--(2+8,0+1.4,2)
				(2,2,0)--(2+8,2+1.4,0)
				(2,2,2)--(2+8,2+1.4,2);
				
				\node at (1,0,0) {1};
				\node at (2,0,1) {2};
				\node at (1,0,2) {3};
				\node at (0,0,1) {4};
				\node at (0,1,0) {5};
				\node at (2,1,0) {6};
				\node at (2,1,2) {7};
				\node at (0,1,2) {8};
				\node at (1,2,0) {9};
				\node at (2,2,1) {10};
				\node at (1,2,2) {11};
				\node at (0,2,1) {12};
				\node at (0+4,0+0.7,0) {13};
				\node at (2+4,0+0.7,0) {14};
				\node at (0+4,0+0.7,2) {15};
				\node at (2+4,0+0.7,2) {16};
				\node at (0+4,2+0.7,0) {17};
				\node at (2+4,2+0.7,0) {18};
				\node at (0+4,2+0.7,2) {19};
				\node at (2+4,2+0.7,2) {20};
				\node at (1+8,0+1.4,0) {21};
				\node at (2+8,0+1.4,1) {22};
				\node at (1+8,0+1.4,2) {23};
				\node at (0+8,0+1.4,1) {24};
				\node at (0+8,1+1.4,0) {25};
				\node at (2+8,1+1.4,0) {26};
				\node at (2+8,1+1.4,2) {27};
				\node at (0+8,1+1.4,2) {28};
				\node at (1+8,2+1.4,0) {29};
				\node at (2+8,2+1.4,1) {30};
				\node at (1+8,2+1.4,2) {31};
				\node at (0+8,2+1.4,1) {32};
				
				\node[red] at (1+4,1+0.7,1) {$\gamma_4$};

				\draw[gray, dashed] (0+4,0+0.7,0)--(2+4,0+0.7,0)--(2+4,0+0.7,2)--(0+4,0+0.7,2)--cycle;
				\draw[gray, dashed] (0+4,2+0.7,0)--(2+4,2+0.7,0)--(2+4,2+0.7,2)--(0+4,2+0.7,2)--cycle;
				\draw[gray, dashed] 
				(0+4,0+0.7,0)--(0+4,2+0.7,0)
				(2+4,0+0.7,0)--(2+4,2+0.7,0)
				(2+4,0+0.7,2)--(2+4,2+0.7,2)
				(0+4,0+0.7,2)--(0+4,2+0.7,2);
			\end{tikzpicture}
			\caption{}
		\end{subfigure}
		\begin{subfigure}{0.34\textwidth}
			\centering
			\begin{tikzpicture}[baseline=5ex]
				\draw[gray] 
				(1.5,1.5,0)--(1.5,1.5,3)
				(0,1.5,1.5)--(3,1.5,1.5)
				(1.5,0,1.5)--(1.5,3,1.5)
				(1.5-2,1.5-0.35,1.5)--(1.5+2,1.5+0.35,1.5);
				
				\draw[->, blue, line width=1pt] (-0.5,3,0)--(-0.5,3,1);
				\draw[->, blue, line width=1pt] (-0.5,3,0)--(0.3,3,0);
				\draw[->, blue, line width=1pt] (-0.5,3,0)--(-0.5,3.8,0);
				\draw[->, blue, line width=1pt] (-0.5,3,0)--(2.7,3.56,0);
				\node[blue] at (-0.7,3,1) {$x_1$};
				\node[blue] at (0.3,2.75,0) {$x_2$};
				\node[blue] at (-0.25,3.8,0) {$x_3$};
				\node[blue] at (2.8,3.34,0) {$x_4$};

				\node at (1.5,2.5,1.5) {1};
				\node at (1.5,0.5,1.5) {2};
				\node at (0.5,1.5,1.5) {3};
				\node at (2.5,1.5,1.5) {4};
				\node at (1.5,1.5,2.5) {5};
				\node at (1.5,1.5,0.5) {6};
				\node at (1.5-1.5,1.5-0.2625,1.5) {7};
				\node at (1.5+1.5,1.5+0.2625,1.5) {8};
				
				\node[blue] at (1.3,1.65,1.5) {$\gamma_0$};
				\fill[blue] (1.5,1.5,1.5) circle (1.5pt);
			\end{tikzpicture}
			\caption{}
		\end{subfigure}
		\caption{Illustration of $A$ term and $B$ term in the $[0,1,2,4]$ model. The spins are on edges, and the numbers in this figure are placed at the center of each edge, representing the spin on it. (a) The $A$ term associated with 4-cube $\gamma_4$ is $A_{\gamma_4}=X_1X_2\cdots X_{32}$, supported by all the edges of $\gamma_4$. The dashed cube aids visualizing $x_4$-direction edges: the centers of edges spanning in $x_4$-direction coincide with the centers of vertices of the dashed cube. (b) Each $B$ term attached to vertex $v$ is supported by 4 edges on a cross containing $v$, e.g., $B_{\gamma_0,x_2x_3}=Z_1Z_2Z_3Z_4$, $B_{\gamma_0,x_3x_4}=Z_1Z_2Z_7Z_8$. There are 6 distinct $B$ terms attached to the vertex $v$ due to the combinatorial factor $C_{D-d_n}^{d_l-d_n}=C_{4-0}^{2-0}=6$, which are $B_{\gamma_0,x_1x_2},B_{\gamma_0,x_1x_3}, B_{\gamma_0,x_1x_4},B_{\gamma_0,x_2x_3},B_{\gamma_0,x_2x_4},B_{\gamma_0,x_3x_4}$.}
		\label{0124-stabilizers}
	\end{figure*}
	The GSD scales as:
	\begin{equation}
		\log_2 \text{GSD} = 2\sum_{i<j} L_iL_j - 3\sum_i L_i + 4
	\end{equation}
	for $L_1 \times L_2 \times L_3 \times L_4$ systems under PBC~\cite{li2021fracton}, enabling storage of $\mathcal{O}(L^2)$ logical qubits.
	
	The $[0,1,2,4]$ model has two fundamental excitations:
	\begin{itemize}
		\item  {Fractons}: Created by Pauli $Z$ operators on a 3-dimensional dual lattice cubical region, at the eight corners of the cubical region ($A_{\gamma_4}=-1$), immobile except as composite objects.
		\item  {Lineons} ($l_\mu$): Created by Pauli $X$ operators along straight lines, at the endpoints of the line, with four types ($\mu=1,2,3,4$):
		\begin{itemize}
			\item $l_\mu$ moves along $x_\mu$-axis only
			\item Fusion rules: $l_\mu \times l_\nu = f_{\mu\nu}$ (composite fracton) for $\mu \neq \nu$
		\end{itemize}
	\end{itemize}
	This mobility hierarchy demonstrates four-dimensional fracton phenomena~\cite{li2020fracton}.

	\subsubsection{$[1,2,3,3]$ model (3-dimensional Toric Code)}

\begin{figure*}[htbp]
		\centering
		\begin{subfigure}{0.4\textwidth}
			\centering
			\includegraphics[width=0.67\textwidth]{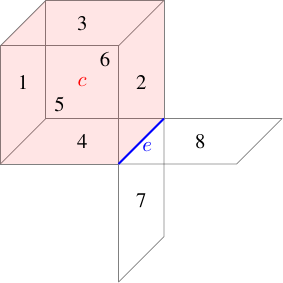}
			\caption{}
		\end{subfigure}
		\begin{subfigure}{0.4\textwidth}
			\centering
			\includegraphics[width=0.67\textwidth]{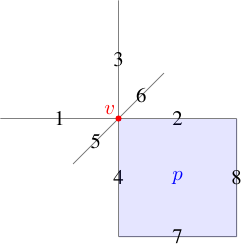}
			\caption{}
		\end{subfigure}
		\caption{Illustration of $A$ term and $B$ term of 3-dimensional toric code under two conventions. (a) The first convention, where spins are on plaquettes. The $A$ term supported on cube $c$ is $A_c=X_1X_2X_3X_4X_5X_6$. The $B$ term supported on edge $e$ is $B_e=Z_2Z_4Z_7Z_8$. (b) The second convention, where spins are on edges. The $A$ term supported on vertex $v$ is $A_v=X_1X_2X_3X_4X_5X_6$. The $B$ term supported on plaquette $p$ is $B_p=Z_2Z_4Z_7Z_8$. The first convention is unified in TD models as the $[1,2,3,3]$ model.}
		\label{3dtc-stabilizers}
	\end{figure*}

	The 3-dimensional Toric Code model represents a prototypical 3-dimensional topological order. Two equivalent conventions exist on cubic lattices, one corresponding to the $[1,2,3,3]$ model in the TD model family. They differ in qubit placement:
	\begin{align}
		H &= -\sum_c A_c - \sum_e B_e = -\sum_{c} \prod_{p \subset c} X_p - \sum_{e} \prod_{p \supset e} Z_p \label{eq:3dtc_hamiltonian} \\
		H &= -\sum_v A_v - \sum_p B_p = -\sum_{v} \prod_{e \supset v} X_e - \sum_{p} \prod_{e \subset p} Z_e \notag
	\end{align}
	where $v,e,p,c$ denote vertices, edges, plaquettes, and cubes respectively (dimensionally $\gamma_0$ through $\gamma_3$). Fig.~\ref{3dtc-stabilizers} illustrates both conventions.
	
	The two conventions are related by lattice duality, exchanging dimensions and codimensions. The plaquette-spin convention corresponds to $[1,2,3,3]$ through:
	\begin{itemize}
		\item $d_n=1$: $B$ terms attached to 1-dimensional edges
		\item $d_s=2$: Qubits on 2-dimensional plaquettes  
		\item $d_l=3$: $B$ terms span 3-dimensional leaves
		\item $D=3$: $A$ terms on 3-cubes $\&$ 3-dimensional lattice
	\end{itemize}
	
	The GSD under periodic boundary condition is $\text{GSD}=8$, encoding 3 logical qubits. 
	
	The $[1,2,3,3]$ model has two types of fundamental excitations:
	
	\begin{itemize}
		\item  $e$ particles: Created by Pauli $Z$ operators on dual lattice strings, at the endpoints of the string ($A_{c}=-1$). Two $e$ particles fuse to vacuum.
		\item $m$ loops: Created by Pauli $X$ operators on a membrane, at the edge of the membrane ($B_e=-1$). The energy of an $m$-loop is proportional to its length.
	\end{itemize}

	\subsubsection{$[1,2,3,4]$ model}
	
	The $[1,2,3,4]$ model is a 4-dimensional fracton ordered model~\cite{li2020fracton}, defined on a 4-dimensional hypercubic lattice where each plaquette hosts a $\frac{1}{2}$-spin.  The Hamiltonian is
	\begin{align}
		H&=-\sum_{\gamma_4}A_{\gamma_4} - \sum_{\gamma_1} \sum_{l\revsubset \gamma_1} B_{\gamma_1,l}\,,
	\end{align}
	where $A_{\gamma_4}=\prod_{\gamma_2\subset\gamma_4}X_{\gamma_2}$, and $B_{\gamma_1,l}$ is the product of Pauli $Z$ operators of four nearest spins to $\gamma_1$ in leaf $l$. The $A$ and $B$ terms are illustrated in Fig.~\ref{1234-stabilizers}.
	\begin{figure*}[htbp]
		\centering
		\begin{subfigure}{0.65\textwidth}
			\centering
			\begin{tikzpicture}[baseline=0ex]
				\draw[gray] (0,0,0)--(2,0,0)--(2,0,2)--(0,0,2)--cycle;
				\draw[gray] (0,2,0)--(2,2,0)--(2,2,2)--(0,2,2)--cycle;
				\draw[gray] 
				(0,0,0)--(0,2,0)
				(2,0,0)--(2,2,0)
				(2,0,2)--(2,2,2)
				(0,0,2)--(0,2,2);
				
				\draw[gray] (0+8,0+1.4,0)--(2+8,0+1.4,0)--(2+8,0+1.4,2)--(0+8,0+1.4,2)--cycle;
				\draw[gray] (0+8,2+1.4,0)--(2+8,2+1.4,0)--(2+8,2+1.4,2)--(0+8,2+1.4,2)--cycle;
				\draw[gray] 
				(0+8,0+1.4,0)--(0+8,2+1.4,0)
				(2+8,0+1.4,0)--(2+8,2+1.4,0)
				(2+8,0+1.4,2)--(2+8,2+1.4,2)
				(0+8,0+1.4,2)--(0+8,2+1.4,2);
				
				\draw[gray]
				(0,0,0)--(0+8,0+1.4,0)
				(0,0,2)--(0+8,0+1.4,2)
				(0,2,0)--(0+8,2+1.4,0)
				(0,2,2)--(0+8,2+1.4,2)
				(2,0,0)--(2+8,0+1.4,0)
				(2,0,2)--(2+8,0+1.4,2)
				(2,2,0)--(2+8,2+1.4,0)
				(2,2,2)--(2+8,2+1.4,2);
				
				\node at (1,0,1) {1};
				\node at (1,1,0) {2};
				\node at (2,1,1) {3};
				\node at (1,1,2) {4};
				\node at (0,1,1) {5};
				\node at (1,2,1) {6};
				\node at (0+4,1+0.7,0) {11};
				\node at (2+4,1+0.7,0) {12};
				\node at (2+4,1+0.7,2) {13};
				\node at (0+4,1+0.7,2) {14};
				\node at (1+4,2+0.7,0) {15};
				\node at (2+4,2+0.7,1) {16};
				\node at (1+4,2+0.7,2) {17};
				\node at (0+4,2+0.7,1) {18};
				\node at (1+4,0+0.7,0) {7};
				\node at (2+4,0+0.7,1) {8};
				\node at (1+4,0+0.7,2) {9};
				\node at (0+4,0+0.7,1) {10};
				\node at (1+8,0+1.4,1) {19};
				\node at (1+8,1+1.4,0) {20};
				\node at (2+8,1+1.4,1) {21};
				\node at (1+8,1+1.4,2) {22};
				\node at (0+8,1+1.4,1) {23};
				\node at (1+8,2+1.4,1) {24};
				
				\node[red] at (1+4,1+0.7,1) {$\gamma_4$};

				\draw[gray, dashed] (0+4,0+0.7,0)--(2+4,0+0.7,0)--(2+4,0+0.7,2)--(0+4,0+0.7,2)--cycle;
				\draw[gray, dashed] (0+4,2+0.7,0)--(2+4,2+0.7,0)--(2+4,2+0.7,2)--(0+4,2+0.7,2)--cycle;
				\draw[gray, dashed] 
				(0+4,0+0.7,0)--(0+4,2+0.7,0)
				(2+4,0+0.7,0)--(2+4,2+0.7,0)
				(2+4,0+0.7,2)--(2+4,2+0.7,2)
				(0+4,0+0.7,2)--(0+4,2+0.7,2);
			\end{tikzpicture}
			\caption{$A_{\gamma_4}=X_1X_2\cdots X_{24}$.}
		\end{subfigure}
		\begin{subfigure}{0.34\textwidth}
			\centering
			\begin{tikzpicture}[baseline=5ex]
				\draw[gray] 
				(0,-1.5,0)--(0,1.5,0)--(0,1.5,1.5)--(0,-1.5,1.5)--cycle;
				\draw[gray] 
				(-1.5,0,0)--(1.5,0,0)--(1.5,0,1.5)--(-1.5,0,1.5)--cycle;
				\draw[gray] (0,0,0)--(0,0,1.5);
				\draw[gray] 
				(0-2.5,0-0.4375,0)--(0+2.5,0+0.4375,0)--(0+2.5,0+0.4375,1.5)--(0-2.5,0-0.4375,1.5)--cycle;
				
				\draw[->, blue, line width=1pt] (-1.5,2,0)--(-1.5,2,1);
				\draw[->, blue, line width=1pt] (-1.5,2,0)--(-0.7,2,0);
				\draw[->, blue, line width=1pt] (-1.5,2,0)--(-1.5,2.8,0);
				\draw[->, blue, line width=1pt] (-1.5,2,0)--(1.7,2.56,0);
				\node[blue] at (-1.7,2,1) {$x_1$};
				\node[blue] at (-0.7,1.75,0) {$x_2$};
				\node[blue] at (-1.25,2.8,0) {$x_3$};
				\node[blue] at (1.8,2.34,0) {$x_4$};

				\draw[blue] (0,0,0)--(0,0,1.5);
				\node[blue] at (-0.13,0.05,0.75) {$\gamma_1$};
				
				\node at (0,0.75,0.75) {1};
				\node at (0,-0.75,0.75) {2};
				\node at (-0.75,0,0.75) {3};
				\node at (0.75,0,0.75) {4};
				\node at (0-1.25,0-0.25,0.75) {5};
				\node at (0+1.25,0+0.25,0.75) {6};
			\end{tikzpicture}
			\caption{}
		\end{subfigure}
		\caption{Illustration of $A$ term and $B$ term in the $[1,2,3,4]$ model. The spins are on plaquettes, and the numbers in this figure are placed at the center of each plaquette, representing the spin on it. (a) The $A$ term associated with 4-cube $\gamma_4$ is $A_{\gamma_4}=X_1X_2\cdots X_{24}$, supported by all the plaquettes in $\gamma_4$. The dashed cube aids visualizing plaquettes spanning in $x_4$-direction: the centers of plaquettes spanning in $x_4$-direction coincide with the centers of edges of the dashed cube. (b) Each $B$ term attached to edge $e$ is supported by 4 plaquettes containing $e$, e.g., $B_{\gamma_1,x_1x_2x_3}=Z_1Z_2Z_3Z_4$, $B_{\gamma_1,x_1x_3x_4}=Z_1Z_2Z_5Z_6$ There are 3 $B$ terms attached to $\gamma_1$ in due to the combinatorial factor $C_{D-d_n}^{d_l-d_n}=C_{4-1}^{3-1}=3$, which are $B_{\gamma_1,x_1x_2x_3},B_{\gamma_1,x_1x_2x_4}, B_{\gamma_1,x_1x_3x_4}$.}
		\label{1234-stabilizers}
	\end{figure*} 
	For each $\gamma_1$, there are $C_{D-d_n}^{d_l-d_n}=C_{4-1}^{3-1}=3$ different $B_{\gamma_1,l}$ attached to it. 
	The GSD of $[1,2,3,4]$ model satisfies $\log_2 GSD = 3(L_1 + L_2 +L_3 +L_4) -6$ on a lattice of size $L_1\times L_2\times L_3\times L_4$ under PBC~\cite{li2021fracton}, encoding $3(L_1 + L_2 + L_3 + L_4) - 6$ logical qubits.

	The $[1,2,3,4]$ model has two fundamental types of excitations:
	\begin{itemize}
		\item Fractons: Created by Pauli $Z$ operators on a dual lattice rectangular plane, at the four corners of the plane ($A_{\gamma_4}=-1$). Two nearby fractons combine to a ``volumeon'' that is mobile within a 3-dimensional leaf.
		\item $m$-loops constraint on a plane: Created by Pauli $X$ operators in an open flat membrane, at the edge of the flat membrane ($B_{\gamma_2}=-1$). The energy of an $m$-loop is proportional to its length. Intriguingly, a curved membrane creates additional string-like excitations along its crease lines, which prevent the $m$-loop from moving or deforming freely. This phenomenon indicates that curved membranes generate non-manifold-like excitations, collectively referred to as ``complex excitations'' in Ref.~\cite{li2020fracton}.
	\end{itemize}

	\subsubsection{$[2,3,4,4]$ model (4-dimensional Toric Code)}

	The $[2,3,4,4]$ model represents a 4-dimensional pure topological order, serving as a $(1,3)$-type 4-dimensional Toric Code with 1-dimensional $Z$-type and 3-dimensional $X$-type logical operators. While not unique among 4-dimensional Toric Codes --- other variants include $(2,2)$-type~\cite{Kitaev_topo_quantum_memory} and octaplex-tessellated models~\cite{octaplex_4d_toric_code} --- its Hamiltonian on a 4-dimensional hypercubic lattice (qubits on 3-cubes) is:
	\begin{align}
		H&=-\sum_{\gamma_4}A_{\gamma_4}-\sum_{\gamma_2}B_{\gamma_2}\notag\\
		&=-\sum_{\gamma_4}\prod_{\gamma_3\subset\gamma_4}X_{\gamma_3} - \sum_{\gamma_2}\prod_{\gamma_3\supset \gamma_2}Z_{\gamma_3}\,.
	\end{align}
	Like 2-dimensional Toric Code and 3-dimensional Toric Code, the only 4-dimensional leaf $l$ contains the whole lattice, so the index $l$ is neglected. As an $(1,3)$-type 4-dimensional Toric Code model, in $[2,3,4,4]$ model we have two kinds of logical operators which are respectively 1-dimensional and 3-dimensional. An 1-dimensional logical operator here is the product of Pauli $Z$ on cubes along a non-contractible dual string $S^1$, and a 3-dimensional logical operator here is the product of Pauli $X$ on cubes along a 3-dimensional torus $T^3$. By counting the number of independent logical operators, we can obtain that under PBC the GSD of $[2,3,4,4]$ model is 16, encoding $4$ logical qubits. 
	
	The two types of fundamental excitations are:
	\begin{itemize}
		\item $e$ particles: Created by Pauli $Z$ operators on dual lattice strings, at the endpoints of the strings ($A_{\gamma_4}=-1$). Two $e$ particles fuse to vacuum.
		\item $m$-membranes: Created by Pauli $X$ operators in an open 3-dimensional region, at the boundary of the region. The energy of an $m$-membrane is proportional to its area, in which sense it is called a spatially extended topological excitation.
	\end{itemize}

	\subsection{Hierarchy of long-range entanglement, nested leaf Structure, and ground state degeneracy}\label{subsec:ERG_hier}

	It is generally established that topological orders are distinguished by their nontrivial long-range entanglement (LRE) patterns, whereas fracton orders exhibit intricate LRE structures revealed by entanglement renormalization group (ERG) transformations~\cite{fracton16,chen_2018_foliated_fracton_phase_classification,dddd_ERG}. The X-cube model ($[0,1,2,3]$) serves as a fixed point under ERG transformations, which involve two key steps: first, inserting layers of 2-dimensional Toric Code ($[0,1,2,2]$) ground states into the cubic lattice, followed by application of finite-depth  LU  circuits.
	A hierarchical classification framework of long range entanglement (LRE) for TD models has been established~\cite{dddd_ERG}, organizing these models into distinct levels where lower-level TD states function as building blocks in ERG transformations of higher-level counterparts. Specifically:
	\begin{itemize}
		\item Level-0: Short-range entangled states or disentangled product states
		\item Level-1: $[0,1,2,2]$ ground states requiring spin-line insertion/removal in product states
		\item Level-2: $[0,1,2,3]$ ground states involving membrane insertion/removal of level-1 states
	\end{itemize}
	This hierarchy generalizes through the relation $[d-1,d,d+1,D] \rightarrow \text{level-}(D-d)$ LRE states~\cite{dddd_ERG}, extending foliated fracton theory. For instance,  3-dimensional leaves added/removed in the ERG of $[0,1,2,4]$ model are in $[0,1,2,3]$ ground states, which also has a leaf structure under ERG with LRE states on the leaves.    Such nested leaf architectures distinguish general TD models from conventional 3-dimensional foliated fracton systems. The hierarchical structure of ERG among TD states with different LRE-levels suggests LRE-level quantifies LRE complexity.
	
	Analogous to the Toric Code, any TD model exhibits unique ground state under OBC, while demonstrating nontrivial GSD under PBC. Reference~\cite{li2021fracton} quantifies the GSD for two classes of these models: level-1 TD models display constant GSD under PBC, whereas higher-level counterparts exhibit $\log_2\mathrm{GSD}$ scaling polynomially with system size under equivalent boundary conditions.

	\subsection{Non-stabilizer-code TD models: A pathway to transverse field Ising models and beyond}\label{section_non_stabilizer_code_TD_models}
	
	We briefly discuss TD models that are not stabilizer codes, where $A$ and $B$ terms are non-commuting. These   models exhibit interesting properties in the context of (quantum) statistical physics. The 1-dimensional transverse field Ising model serves as prototype $[0,1,1,1]$ TD model with edge spins:
	\begin{align}
		H = -\sum_{e}A_e - \sum_{v}B_v= -\sum_{e}X_e - \sum_{v}\prod_{e \supset v}Z_e.
	\end{align}
	For general cases with $d_n=d_s=d_l$, the Hamiltonian reduces to transverse field Ising-type:
	\[
	B_{\gamma_{d_n}} = Z_{\gamma_{d_n}} = Z_{\gamma_{d_s}},
	\]
	yielding non-stabilizer codes. Notable examples include:
	\begin{itemize}
		\item $[0,0,0,2]$ and $[1,1,1,2]$: Transverse field plaquette Ising models
		\item $[0,0,0,3]$: Transverse field cubic Ising model
	\end{itemize}
	Non-stabilizer TD models exhibit rich critical behavior. The paradigmatic 1-dimension transverse-field Ising model with tunable parameters:
	\begin{align}
		H = -h\sum_{e}A_e - J\sum_{v}B_v= -h\sum_{e}X_e - J\sum_{v}\prod_{e \supset v}Z_e
	\end{align}
	shows a critical line at $h=J$ with gapless CFT description~\cite{cardy1984conformal,oshikawa2019universal},  $\mathbb{Z}_2$-symmetric phase ($h>J$)
	and symmetry-broken phase ($h<J$). 
	Key open questions include:  
	existence of critical lines/intermediate phases in general non-stabilizer TD models, ERG connections between different critical systems, and universality class classification for higher-dimensional analogs. 
	These directions represent promising avenues for exploring critical phenomenon across dimension.

	The unified form and broadness (across dimension, liquid and non-liquid states) of TD models make it an ideal platform for searching quantum circuits mapping among high-dimensional topological order, fracton, and trivial phases. In Sec.~\ref{section_SQC}, we construct a unified local unitary circuit $U_c$, for preparing computational basis code states of any stabilizer code TD model from trivial product state.

	\section{Preparation of computational basis of code states}\label{section_SQC}
	
	\subsection{Overall strategy}\label{section_strate}
	
	For simplicity, we denote $d_s$ by $d$ throughout this section. Given any stabilizer code TD model $[d_n,d,d_l,D]$ under OBC, PBC, or half-OBC-half-PBC, a specific code state (the computational basis of code states $|0\rangle_{\text{logic}}^{\otimes\text{log}_2\text{GSD}}$) can be expressed as:
	\begin{align}
		\label{paper_dddd_27}
		c\prod_{\text{all }\gamma_D}\left(1 + A_{\gamma_D}\right)|00\cdots0\rangle\,,
	\end{align}
	where $|0\rangle:=|Z=1\rangle$, $c$ is the normalization factor, which is explicitly calculated in Sec.~\ref{section_unified_SQC}, and $A_{\gamma_D}$ is defined in Eq.~(\ref{revise_50}). In this section, we provide a linear-depth local unitary quantum circuit $U_c$ for preparing such a computational basis of code states (i.e., computational basis of logical qubits) of any stabilizer code TD model, i.e., a circuit mapping the initial state $|00\cdots0\rangle$, to Eq.~(\ref{paper_dddd_27}). Our circuit is applicable to OBC, PBC, and half-OBC-half-PBC: while the circuit for PBC consists of $d+2$ mutually non-commutable steps, the first two steps (namely, step 0 and step 1) prepare the TD state under OBC, and the circuit for half-OBC-half-PBC can be obtained by truncating the circuit for PBC. Therefore, we focus on circuit for PBC first, and after introducing the unified circuit for any stabilizer code TD model, we show how to truncate the circuit for PBC to obtain the circuit for half-OBC-half-PBC. From now on, if not specified, we discuss the circuit for PBC by default.

	The overall strategy of our circuit is:\\
	\begin{enumerate}
		\item Prepare the system in the initial state $|00\cdots0\rangle$.\\
		\item  Choose a set of $D$-cubes, denoted by $\Gamma_D\backslash R_D$\footnote{$\backslash$ stands for set difference.}, which supports a minimal complete set of $A$ terms. $\Gamma_D$ is the set of all $D$-cubes in lattice, and $R_D\subset\Gamma_D$. The choice of $R_D$ is model dependent, and will be discussed in detail later.\\
		\item  Choose a \textbf{\emph{representative spin}} (or $\gamma_d$) in/for each $\gamma_D\in\Gamma_D\backslash R_D$, such that there is a one-to-one correspondence between each representative spin and the represented $\gamma_D\in\Gamma_D\backslash R_D$. Apply Hadamard gates on all representative spins. Hereafter, we use the symbol   $\mathcal{R}(\gamma_D)$ to denote the representative spin of $\gamma_D$.\\
		\item  Apply CNOT gates, with the representative spins being control qubits, and all other spins in the represented $D$-cube being target qubits, in an order that no representative spin plays the role of target qubit before it has finished playing the role of control qubit.
	\end{enumerate}	
	
	The readers may wonder why not choosing a representative spin in every $D$-cube and apply CNOT gates, which seems much more straightforward. The answer is under PBC, this is impossible: whether the constraint of order stated in the overall strategy would be broken, or if one does it anyway, the result state will not be a code state. We will see the reason soon, in the first example.
	
	How arbitrarily can $\Gamma_D\backslash R_D$ and representative spins be chosen is not studied in this paper; instead, we provide a concrete scheme. In our scheme, the first layer of circuit is the Hadamard gates that are applied to all the representative spins. We call the first layer \textbf{step}  $\mathbf{0}$, for convenience. Apart from the Hadamard gates, the circuit $U_c$ for preparing $[d_n,d,d_l,D]$ model code state consists of $d+1$ steps, which are mutually non-commutable, namely, \textbf{step} $\mathbf{1, 2, \cdots, d+1}$. \textbf{Step} $\bf{k+1}$ has $C_D^k$ mutually commutable parts, which can be applied simultaneously. The concrete form of any part of any step of $U_c$ will be presented in the rest of this section, starting from concrete examples. While the examples show useful step-by-step details, all these examples can be obtained from the unified circuit introduced in Sec.~\ref{section_unified_SQC}.
	
	Before we start, as the concept ``layer'' appears frequently in this section, we clarify here that a \textit{\textbf{layer} (of circuit presented in this paper) consists of mutually commutable (rather than disjoint) local unitary gates by default in this paper}. As long as $D\not\to\infty$, each degree of freedom has finite connectivity. Thus, the depth of a LU circuit in the sense of disjoint gates differs from the depth of a LU circuit in the sense of commuting gates up to multiplying a constant.
	
	Another massively used terminology deserving a precise definition is the \textbf{\emph{equal weight superposition}} of closed $D$-cube cage \textbf{\emph{configurations}} (\textbf{EWSC}). For \textit{any union of $D$-cubes $\Omega$ in lattice}, we define the EWSC on $\Omega$ as the normalized state
	\begin{equation}\label{revise_53}
		\mathcal{F}(\Omega)=c\prod_{\gamma_D\subset\Omega}\Big(1+A_{\gamma_D}\Big)\bigotimes_{\gamma_d\subset\Omega}|0\rangle_{\gamma_d}\otimes|\zeta\rangle_{\{\gamma_d\nsubseteq\Omega\}}\,,
	\end{equation}
	with	
	\begin{equation}
		|\zeta\rangle_{\{\gamma_d\nsubseteq\Omega\}} = \prod_{\gamma_D\nsubseteq\Omega}\text{Had}_{\mathcal{R}(\gamma_D)}\bigotimes_{\gamma_d\nsubseteq\Omega}|0\rangle_{\gamma_d}\,,\label{revise_39}
	\end{equation} 
	where  $\mathcal{R}(\gamma_D)$ denotes the representative spin of $\gamma_D$.  $\text{Had}_a$ refers to the Hadamard gate on spin $a$.
	In addition, since we will meet $\mathcal{F}(\Omega)$ with plenty of different $\Omega$, we stick to the following convention in Sec.~\ref{section_SQC} for simplicity:
	\textit{The symbol $\Omega$ is treated as a variable, such that $\mathcal{F}(\Omega)$ always refers to the current quantum state (the state after some step or some part of a step is finished)}.

	%	revise_47

	\subsection{$[0,1,2,2]$ model}\label{section_toric_code_sqc}
	As illustrated in Sec.~\ref{section_TD_models}, the $[0,1,2,2]$ model is Toric Code (on square lattice). In this subsection, we construct a circuit $U_c$ for preparing computational basis of the $[0,1,2,2]$ model code states, where $d=1,D=2$. We start from graphic illustration, then algebraically write the circuit $U_c$, find a stabilizer redundancy, by using which we show the prepared state is the EWSC on all plaquettes, and thus a code state.
	
	\subsubsection{Graphical approach on a $4\times 4$ square lattice}

	Let us start with the graphic illustration. Consider a $4\times4$ square lattice under PBC, where each edge has a $\frac{1}{2}$-spin on it. Prepare the system in product state $|00\cdots0\rangle$, where $|0\rangle=|Z=1\rangle$ is a Pauli $Z$ eigenstate. Then, apply Hadamard gates to each spin on solid orange edges shown below:
	\begin{equation}\label{revise_28}
		\begin{tikzpicture}[baseline=-3ex]
			\foreach \j in {0,1,2,3}{
				\draw (0,\j)--(4,\j);}
			\foreach \i in {0,1,2,3}{
				\draw (\i,4)--(\i,0);}

			\draw[->, blue, line width=1pt] (-0.5,-0.5)--(-0.5,1);
			\draw[->, blue, line width=1pt] (-0.5,-0.5)--(1,-0.5);

			\node[blue] at (-0.75,1) {$x_1$};
			\node[blue] at (1,-0.75) {$x_2$};

			\draw[orange, line width=1pt]
			(0,0)--(4,0)
			(0,1)--(4,1)
			(0,2)--(4,2)
			(0,3)--(0,4)
			(1,3)--(1,4)
			(2,3)--(2,4);
			
			\fill[blue] (0,0) circle (2pt);
			
			\foreach \i in {0.5,1.5,2.5,3.5}{
				\draw[orange, dashed, line width=1pt]
				(\i,0)--(\i,0.5)
				(\i,1)--(\i,1.5)
				(\i,2)--(\i,2.5);}
			\draw[orange, dashed, line width=1pt]
			(0,3.5)--(0.5,3.5)
			(1,3.5)--(1.5,3.5)
			(2,3.5)--(2.5,3.5);
		\end{tikzpicture}\ ,
	\end{equation}
	after which the spins on solid orange lines are in state $|+\rangle=|X=1\rangle$. Those solid orange edges are the representative spins (i.e., $\mathcal{R}(\gamma_D)$ with $D=2$ defined in Sec.~\ref{section_strate}) in the overall strategy, with the dashed orange lines connecting the representative spins to the centers of their represented plaquettes. The blue dot stands for the original point $(0,0)$. 
	
	From now on, we use black edges to represent spins in state $|0\rangle$, and orange edges to represent spins in state $|+\rangle$ in this subsection, so that Eq.~(\ref{revise_28}) represents the state after applying step 0 (i.e., Hadamard gates on representative spins). After step 0, apply $d+1=2$ non-commutable steps of CNOT gates, namely, step 1 and step 2, which shall be applied with the order 1,2. Denote 
	\begin{equation}\label{revise_12}
		\begin{tikzpicture}[baseline=2.8ex]
			\draw (0,0)--(0,1)--(1,1)--(1,0)--cycle;
			\draw[orange, line width=1pt] (0,0)--(1,0);
			\draw[->, blue, line width=1pt] (0.5,0)--(0,0.5);
			\draw[->, blue, line width=1pt] (0.5,0)--(1,0.5);
			\draw[->, blue, line width=1pt] (0.5,0)--(0.5,1);
		\end{tikzpicture}
		\ =\ 
		\begin{tikzpicture}[baseline=2.8ex]
			\draw (0,0)--(0,1)--(1,1)--(1,0)--cycle;
			\draw[orange, line width=1pt] (0,0)--(1,0);
			\draw[blue, line width=1pt] (0.5,0)--(0.5,0.2);
			\draw[blue, line width=1pt] (0.5,0.5) circle (0.3);
		\end{tikzpicture}\ .
	\end{equation} 
	On the left hand side of Eq.~(\ref{revise_12}), each arrow represents a CNOT gate, with the tail being the control qubit, and the head being the target qubit. The right hand side of Eq.~(\ref{revise_12}) is just a symbolic simplification. Step 1 and step 2 are:\\
	\begin{itemize}
		\item \textbf{Step 1} (consisting of $C_2^0=1$ part):
		\begin{equation}\label{paper_2dTC_1}
			\begin{tikzpicture}[baseline=-3ex]
				\foreach \j in {0,1,2,3}{
					\draw (0,\j)--(4,\j);}
				\foreach \i in {0,1,2,3}{
					\draw (\i,4)--(\i,0);}

				\draw[->, blue, line width=1pt] (-0.5,-0.5)--(-0.5,1);
				\draw[->, blue, line width=1pt] (-0.5,-0.5)--(1,-0.5);

				\node[blue] at (-0.75,1) {$x_1$};
				\node[blue] at (1,-0.75) {$x_2$};

				\node[blue] at (0.5,0.5) {3};
				\node[blue] at (1.5,0.5) {3};
				\node[blue] at (2.5,0.5) {3};
				\node[blue] at (0.5,1.5) {2};
				\node[blue] at (1.5,1.5) {2};
				\node[blue] at (2.5,1.5) {2};
				\node[blue] at (0.5,2.5) {1};
				\node[blue] at (1.5,2.5) {1};
				\node[blue] at (2.5,2.5) {1};

				\foreach \i in {0,1,2}{
					\foreach \j in {0,1,2}{
						\draw[blue, line width=1pt] (\i+0.5,\j)--(\i+0.5,\j+0.2);
						\draw[blue, line width=1pt] (\i+0.5,\j+0.5) circle (0.3);
				}}

				\draw[orange, line width=1pt]
				(0,0)--(4,0)
				(0,1)--(4,1)
				(0,2)--(4,2)
				(0,3)--(0,4)
				(1,3)--(1,4)
				(2,3)--(2,4);
				
				\fill[blue] (0,0) circle (2pt);
			\end{tikzpicture}\ ,
		\end{equation}
		where the numbers label the layers defined in Sec.~\ref{section_strate}. That is to say, gates labeled by the same number are mutually commutable, and are applied simultaneously; gates labeled by smaller number are applied first, then follow the gates with larger numbers.
		\item \textbf{Step 2} consists of $C_2^1=2$ mutually commutable parts, denoted by part $\{1\}$ and part $\{2\}$. \emph{The reason for using a set to denote a part is that a set of $k$ distinct numbers in $\{1:D\}:=\{1,2,\cdots,D\}$ has exactly $C_D^k$ possibilities, which perfectly labels the $C_D^k$ mutually commutable parts of step $k+1$.} Later, when the circuit is written algebraically, it will be easily seen the set $S_k$ labeling a part has a clear meaning.
		\begin{itemize}
			\item \textbf{Step 2 part $\{1\}$}:
			\begin{equation}
				\label{paper_dddd_28}\begin{tikzpicture}[baseline=-3ex]
					\foreach \j in {0,1,2,3}{
						\draw (0,\j)--(4,\j);}
					\foreach \i in {0,1,2,3}{
						\draw (\i,4)--(\i,0);}

					\draw[->, blue, line width=1pt] (-0.5,-0.5)--(-0.5,1);
					\draw[->, blue, line width=1pt] (-0.5,-0.5)--(1,-0.5);

					\node[blue] at (-0.75,1) {$x_1$};
					\node[blue] at (1,-0.75) {$x_2$};

					\node[blue] at (2.5,3.5) {4};
					\node[blue] at (1.5,3.5) {5};
					\node[blue] at (0.5,3.5) {6};

					\foreach \i in {0,1,2}{
						\foreach \j in {3}{
							\draw[blue, line width=1pt] (\i,\j+0.5)--(\i+0.2,\j+0.5);
							\draw[blue, line width=1pt] (\i+0.5,\j+0.5) circle (0.3);
					}}

					\draw[orange, line width=1pt]
					(0,4)--(0,3)
					(1,4)--(1,3)
					(2,4)--(2,3)
					(3,0)--(4,0)
					(3,1)--(4,1)
					(3,2)--(4,2);
					
					\fill[blue] (0,0) circle (2pt);
				\end{tikzpicture}\ .
			\end{equation}
			\item \textbf{Step 2 part $\{2\}$}:
			\begin{equation}
				\begin{tikzpicture}[baseline=-3ex]
					\foreach \j in {0,1,2,3}{
						\draw (0,\j)--(4,\j);}
					\foreach \i in {0,1,2,3}{
						\draw (\i,4)--(\i,0);}

					\draw[->, blue, line width=1pt] (-0.5,-0.5)--(-0.5,1);
					\draw[->, blue, line width=1pt] (-0.5,-0.5)--(1,-0.5);

					\node[blue] at (-0.75,1) {$x_1$};
					\node[blue] at (1,-0.75) {$x_2$};

					\node[blue] at (3.5,2.5) {4};
					\node[blue] at (3.5,1.5) {5};
					\node[blue] at (3.5,0.5) {6};

					\foreach \i in {3}{
						\foreach \j in {0,1,2}{
							\draw[blue, line width=1pt] (\i+0.5,\j)--(\i+0.5,\j+0.2);
							\draw[blue, line width=1pt] (\i+0.5,\j+0.5) circle (0.3);
					}}

					\draw[orange, line width=1pt]
					(0,4)--(0,3)
					(1,4)--(1,3)
					(2,4)--(2,3)	
					(3,0)--(4,0)
					(3,1)--(4,1)
					(3,2)--(4,2);
					
					\fill[blue] (0,0) circle (2pt);
				\end{tikzpicture}\label{paper_dddd_29}\ .
			\end{equation}
		\end{itemize}
		Step 2 part $\{1\}$ and step 2 part $\{2\}$ are mutually commutable, thus can be applied simultaneously. Step 2 must be applied after step 1.
	\end{itemize}
	Steps 1,2 prepare the EWSC on all plaquettes, except the one on the right top corner\footnote{In this example, the plaquette in the right top corner itself forms the plaquette set $R_D$ defined in Sec.~\ref{section_strate}.}. However, the prepared state equals to the EWSC on all plaquettes, out of the $A$ term redundancy, i.e., the $A$ term supported on the plaquette in the top right corner equals to the product of all other $A$ terms. Semantically, the term ``redundancy'' suggests removing the redundant stabilizer from Hamiltonian does not change the ground space. The formal definition of stabilizer redundancy is given in Sec.~\ref{section_X-cube_SQC_algebraic}, and will be frequently used throughout this section.
	
	\subsubsection{Algebraic approach on a general square lattice}
	
	To give a precise description of why the missing configurations exist in the prepared state, writing the circuit algebraically is necessary. Let us first represent specific $n$-cubes algebraically, and then write the circuit. For any dimensional cubical lattice, the geometric center of $\gamma_n$ ($0\leq n\leq D$) has a one-to-one correspondence with $\gamma_n$, no matter what $n$, thus we can use the coordinates of geometric center to represent $\gamma_n$. Take the lattice constant to be 1, and some vertex to be at $(0,0,\cdots,0)$, the coordinates of $\gamma_n$'s geometric center $(x_1,x_2,\cdots,x_D)$ will contain $n$ half-integers and $(D-n)$ integers. Use the coordinates of geometric center to represent any $n$-cube $\gamma_n$,\footnote{Not to be confused with the TD model index $[d_n,d,d_l,D]$ when $D=4$. The two different meanings of the symbol are distinguishable from context.}
	\begin{equation}\label{geometric_notation_TD_model}
		\gamma_n\Big(\text{center at }(x_1,x_2,\cdots,x_D)\Big):=[x_1,x_2,\cdots,x_D]\,.
	\end{equation}
	According to Eq.~(\ref{geometric_notation_TD_model}), we can use for example $[2,3]$ to represent the vertex at $(2,3)$, $[5,3+\frac{1}{2}]$ to represent the edge whose center is at $(5,3+\frac{1}{2})$, $[\frac{1}{2},L_2-\frac{3}{2}]$ to represent the plaquette whose center is at $(\frac{1}{2},L_2-\frac{3}{2})$. Using this notation, one can not only represent $n$-cubes, but also represent stabilizer redundancies with complex structure conveniently (see Eqs.~(\ref{paper_redundancy_0})-(\ref{revise_37}) and Fig.~\ref{*-intuition} for definitions and the example for X-cube). Using this notation, we can write the circuit $U_c$ for $[0,1,2,2]$ model as following:
	
	The initial state is $|00\cdots0\rangle$. The state after step 0 is:
	\begin{equation}\label{revise_52}
		\begin{split}
			&\Bigg(\prod_{\substack{x_1\in\mathbb{Z}_{L_1-1}\\x_2\in\mathbb{Z}_{L_2-1}}}\text{Had}_{[x_1,x_2+\frac{1}{2}]}\Bigg)\Bigg(\prod_{x_2\in\mathbb{Z}_{L_2-1}}\text{Had}_{[-\frac{1}{2},x_2]}\Bigg)\\
			&\quad\quad\quad\quad\quad\quad\quad\Bigg(\prod_{x_1\in\mathbb{Z}_{L_1-1}}\text{Had}_{[x_1,-\frac{1}{2}]}\Bigg)|00\cdots0\rangle\,,
		\end{split}
	\end{equation}
	where $L_1$ and $L_2$ are the length of the lattice in $x_1$- and $x_2$-directions, respectively. The huge parentheses are added here to restrict the effective range of indices under $\sum$, i.e., ($x_1,x_2$). $\mathbb{Z}_{N}:=\{0,1,\cdots,N-1\}$ is the $N$-th order cyclic group.
	$		x_i=\big(x_i\ \text{mod}\ L_i\big)
	$	is applied because of PBC, so that $\left[-\frac12,x_2\right]=\left[L_1-\frac12,x_2\right]$, $\left[x_1,-\frac{1}{2}\right]=\left[x_1,L_2-\frac12\right]$. The state in Eq.~(\ref{revise_28}) is the state in Eq.~(\ref{revise_52}) with $L_1=L_2=4$. 
	\begin{itemize}
		\item \textbf{Step 1}:
		\begin{equation}\label{paper_revise_1}
			\prod_{n=0}^{L_1-2}\prod_{\substack{x_i\in\mathbb{Z}_{L_i-1},\,i=1,2\\x_1=n}}\prod_{\substack{\gamma_1\subset[x_1+\frac{1}{2},x_2+\frac{1}{2}]\\\gamma_1\neq[x_1,x_2+\frac{1}{2}]}}\text{CNOT}_{[x_1,x_2+\frac{1}{2}],\gamma_1}\,,
		\end{equation}
		where $\text{CNOT}_{a,b}$ is the Controlled-NOT gate with $a$ being the control qubit and $b$ being the target qubit. Recall that for the $[0,1,2,2]$ model, spins are placed on edges (or 1-cubes) $\gamma_1$, and $\left[x_1+\frac12,x_2+\frac12\right]$ are closed plaquettes (or 2-cubes), so $\gamma_1\subset\left[x_1+\frac12,x_2+\frac12\right]$ refers to the edges on the boundary of $\left[x_1+\frac12,x_2+\frac12\right]$. $\prod_{n=0}^{L_1-2}$ is the product in order, where the operator with $n=0$ is placed on left most, and the operator with $n=L_1-2$ is placed on the right most. Formally, for any set of operators $\{O_i\}$ and $m,n\in\mathbb{R}$, where $m-n\in\mathbb{Z}$,
		\begin{equation}
			\prod_{i=m}^n O_i=\left\{\begin{array}{l}
				O_m O_{m+1}\cdots O_n,\quad m\leq n\\
				O_m O_{m-1}\cdots O_n,\quad m\geq n
			\end{array}\right.\,.\label{revise_35}
		\end{equation}
		The readers may feel confused at   first glance to the conditions under the second $\prod$ in Eq.~(\ref{paper_revise_1}): $x_1\in\mathbb{Z}_{L_1-1}$ can be derived from $x_1=n$, why writing the redundant condition $x_1\in\mathbb{Z}_{L_1-1}$? The answer is to be consistent with the unified form of circuit. Just like all other $\prod$ with multi-conditions, only the indices satisfying all conditions under $\prod$ is taken to support the product. Step 1 adds all the plaquettes with no coordinate being $-\frac12$ into $\Omega$ (see the convention below Eq.~(\ref{revise_39})). The state before step 1 is $\mathcal{F}(\emptyset)$, while the state after step 1 is $\mathcal{F}(\Omega_1)$, where $\Omega_1$ is the union of all the plaquettes with no coordinate being $-\frac{1}{2}$. The circuit in Eq.~(\ref{paper_2dTC_1}) is the circuit in Eq.~(\ref{paper_revise_1}) with $L_1=L_2=4$.
		\item \textbf{Step 2}:
		\begin{itemize}
			\item \textbf{Step 2 part $\{1\}$}:
			\begin{equation}
				\prod_{n=0}^{L_2-2}\prod_{\substack{x_2=n}}\,\prod_{\substack{\gamma_1\subset[-\frac{1}{2},x_2+\frac{1}{2}]\\\gamma_1\neq[-\frac{1}{2},x_2]}}\text{CNOT}_{[-\frac{1}{2},x_2],\gamma_1}\,.
			\end{equation}
			Step 2 part $\{1\}$ adds the plaquettes with only $x_1$ but no other coordinate being $-\frac12$ into $\Omega$, i.e., the state after step 1, step 2 part $\{1\}$ is $\mathcal{F}(\Omega_1\cup\Omega_{2,\{1\}})$, where $\Omega_{2,\{1\}}$ is the union of all plaquettes with only $x_1$ but no other coordinate being $-\frac12$.
			\item \textbf{Step 2 part $\{2\}$}:
			\begin{equation}
				\prod_{n=0}^{L_1-2}\prod_{\substack{x_1=n}}\,\prod_{\substack{\gamma_1\subset[x_1+\frac{1}{2},-\frac{1}{2}]\\\gamma_1\neq[x_1,-\frac{1}{2}]}}\text{CNOT}_{[x_1,-\frac{1}{2}],\gamma_1}
				\,.
			\end{equation}
			Step 2 part $\{2\}$ adds the plaquettes with only $x_2$ but no other coordinate being $-\frac12$ into $\Omega$, i.e., the state after step 1, step 2 part $\{1\}$, step 2 part $\{2\}$ is $\mathcal{F}(\Omega_1\cup\Omega_{2,\{1\}}\cup\Omega_{2,\{2\}})$, where $\Omega_{2,\{2\}}$ is the union of all plaquettes with only $x_2$ but no other coordinate being $-\frac12$.
		\end{itemize}
		In total, step 2 adds the plaquettes with exactly 1 coordinate being $-\frac{1}{2}$ into $\Omega$, i.e., the state after steps 1,2 is $\mathcal{F}(\Omega_1\cup\Omega_2)$, where $\Omega_2$ is the union of all plaquettes with exactly 1 coordinate being $-\frac12$.
	\end{itemize}
	As illustrated above, the state after circuit $U_c$ is $\mathcal{F}(\Omega_1\cup\Omega_2)$. Note that there is no edge not in $\Omega_1\cup\Omega_2$, so by definition (i.e., Eq.~(\ref{revise_53})), $\mathcal{F}(\Omega_1\cup\Omega_2)$ can be written as 
	\begin{equation}\label{paper_2dTC_2}
		\begin{split}
			&\Bigg(\prod_{x_i\in\mathbb{Z}_{L_i-1},\,i=1,2}\frac{1}{\sqrt{2}}\Big(1+A_{[x_1+\frac{1}{2},x_2+\frac12]}\Big)\Bigg)\\
			&\Bigg(\prod_{x_2\in\mathbb{Z}_{L_2-1}}\frac{1}{\sqrt{2}}\Big(1+A_{[-\frac{1}{2},x_2+\frac{1}{2}]}\Big)\Bigg)\\
			&\Bigg(\prod_{x_1\in\mathbb{Z}_{L_1-1}}\frac{1}{\sqrt{2}}\Big(1+A_{[x_1+\frac{1}{2},-\frac{1}{2}]}\Big)\Bigg)|00\cdots0\rangle\,,
		\end{split}
	\end{equation}
	where $A_p$ are defined in Eq.~(\ref{revise_51}). Under PBC, there is a redundant $A$ term in Hamiltonian, i.e., there is a $A$ term, which equals to the multiplication of other $A$ terms, so that the ground space is invariant under deleting or adding that $A$ term from or into Hamiltonian. We define such an equation
	\begin{equation}\label{paper_2dTC_3}
		A_{p'} = \prod_{p\in \Gamma_2\backslash\{p'\}}A_p\,.
	\end{equation}
	as an $A$ term redundancy, as a type of stabilizer redundancy, where $\Gamma_2$ is the set of all 2-cubes or plaquettes. Specifically, we can choose $p'$ to be $\left[-\frac{1}{2},-\frac{1}{2}\right]$ in Eq.~(\ref{paper_2dTC_3}). Note that $\frac{1}{\sqrt{2}}\big(1+A_p\big)$ of all plaquettes, except $p'=\left[-\frac{1}{2},-\frac{1}{2}\right]$, explicitly appear in Eq.~(\ref{paper_2dTC_2}), so Eq.~(\ref{paper_2dTC_2}) can be written as
	\begin{equation}\label{paper_2dTC_2_1}
		\prod_{p\in\Gamma_2\backslash\{p'\}}\frac{1}{\sqrt{2}}\Big(1+A_{p}\Big)|00\cdots0\rangle\,.
	\end{equation}
	Using Eq.~(\ref{paper_2dTC_3}) with $p'=\left[-\frac12,-\frac12\right]$, we can write
	\begin{align}
		A_{p'}\prod_{p\in\Gamma_2\backslash\{p'\}}\frac{1}{\sqrt{2}}\Big(1+A_p\Big)
		=&\prod_{p\in\Gamma_2\backslash\{p'\}}A_p\frac{1}{\sqrt{2}}\Big(1+A_p\Big)\notag\\
		=&\prod_{p\in\Gamma_2\backslash\{p'\}}\frac{1}{\sqrt{2}}\Big(1+A_p\Big)\label{paper_2dTC_4}\,,
	\end{align}
	thus Eq.~(\ref{paper_2dTC_2_1}) equals to
	\begin{equation}
		\begin{split}
			&\frac{1}{2}\Big(1+A_{p'}\Big)\prod_{p\in\Gamma_2\backslash\{p'\}}\frac{1}{\sqrt{2}}\Big(1+A_p\Big)|00\cdots0\rangle\\
			&=\frac{1}{\sqrt{2}}\prod_{p\in\Gamma_2}\frac{1}{\sqrt{2}}\Big(1+A_p\Big)|00\cdots0\rangle\,.
		\end{split}
	\end{equation}
	This is a ground state (or code state) of 2-dimensional Toric Code, since for any $A_{p'},B_{v'}$,
	\begin{gather}
		A_{p'}\prod_{p\in\Gamma_2}\Big(1+A_p\Big)|00\cdots0\rangle = \prod_{p\in\Gamma_2}\Big(1+A_p\Big)|00\cdots0\rangle\,,\\
		\begin{split}
			B_{v'}\prod_{p\in\Gamma_2}\Big(1+A_p\Big)|00\cdots0\rangle&=\prod_{p\in\Gamma_2}\Big(1+A_p\Big)B_{v'}|00\cdots0\rangle\\
			&=\prod_{p\in\Gamma_2}\Big(1+A_p\Big)|00\cdots0\rangle\,,
		\end{split}
	\end{gather}
	and the eigenvalues of $A_{p'},B_{v'}$ can only be $\pm1$. Note that the convention is $B_v=\prod_{e\revsubset v}Z_e$.
	
	Finally, it is worthy noting that after the circuit, further applying Hadamard and CNOT gates inside the right top plaquette or the plaquette $\left[-\frac{1}{2},-\frac{1}{2}\right]$ would make the state no longer a code state of Toric Code. The existence of such a plaquette is topological, since it is directly related to the $A$ term redundancy, which exists on any 2-manifold without boundary (e.g. two-torus $T^2$, i.e., under PBC), and does not exist on any 2-manifold with boundary (e.g., 2-ball $B^2$, i.e., under OBC; or cylindrical surface $S^1\times B^1$, i.e., under hybrid half-OBC-half-PBC).

	\subsection{$[0,1,2,3]$ model}\label{section_x-cube_sqc}

	As illustrated in Sec.~\ref{section_TD_models}, the $[0,1,2,3]$ model is X-cube (on cubic lattice). In this subsection, we construct a circuit $U_c$ for preparing computational basis of the $[0,1,2,3]$ model code states, where $d=1,D=3$. Again, we start from graphic illustration, then algebraically write the circuit, find a set of stabilizer redundancies, by using which we show the prepared state is the EWSC on all cubes, and thus a code state. Unlike the $[0,1,2,2]$ model, which has only one $A$ term redundancy equivalence class\footnote{For a formal definition, see Eq.~(\ref{paper_redundancy_1})}, the $[0,1,2,3]$ model has multiple $A$ term redundancy classes with intersecting structure, which we will discuss soon.

	\subsubsection{Graphical approach on a $2\times 3\times 3$ cubic lattice}

	Let us start with the graphic illustration again. Consider a $2\times3\times3$ cubic lattice under PBC, where each edge has a $\frac{1}{2}$-spin on it. Prepare the system in product state $|00\cdots0\rangle$, where $|0\rangle=|Z=1\rangle$ is a Pauli $Z$ eigenstate. Then, apply Hadamard gates to each spin on orange edges shown below:
	\begin{equation}\label{revise_29}
		\begin{tikzpicture}[baseline=-3ex]
			\foreach \j in {0,1,2}{\foreach \k in {0,1}{\draw(0,\j,\k)--(3,\j,\k);}}
			\foreach \i in {0,1,2}{\foreach \j in {0,1,2}{\draw(\i,\j,0)--(\i,\j,2);}}
			\foreach \i in {0,1,2}{\foreach \k in {0,1}{\draw(\i,0,\k)--(\i,3,\k);}}
			
			\draw[->, blue, line width=1pt] (-1.5,0,0)--(-1.5,0,1);
			\draw[->, blue, line width=1pt] (-1.5,0,0)--(-0.7,0,0);
			\draw[->, blue, line width=1pt] (-1.5,0,0)--(-1.5,0.8,0);
			\node[blue] at (-1.68,0,1) {$x_1$};
			\node[blue] at (-0.7,-0.22,0) {$x_2$};
			\node[blue] at (-1.25,0.8,0) {$x_3$};
			
			\draw[orange, line width=1pt] 
			(0,0,0)--(0,3,0)
			(1,0,0)--(1,3,0)
			(0,1,1)--(0,1,2)
			(1,1,1)--(1,1,2)
			(0,0,1)--(0,0,2)
			(1,0,1)--(1,0,2)
			(2,0,0)--(3,0,0)
			(2,1,0)--(3,1,0);
			
			\fill[blue] (0,0,0) circle (2pt);
		\end{tikzpicture}\ ,
	\end{equation}
	after which the spins on orange lines are in state $|+\rangle=|X=1\rangle$. The spins on those orange lines are the representative spins in the overall strategy introduced in Sec.~\ref{section_strate}; their represented cubes can be read from Eqs.~(\ref{paper_X-cube_1}, \ref{revise_41}, \ref{revise_eqn_step22},  \ref{revise_42}), or Eq.~(\ref{revise_45}). The blue dot stands for the original point $(0,0,0)$. From now on, we use black edges to represent spins in state $|0\rangle$, and orange edges to represent spins in state $|+\rangle$ in this subsection, so that Eq.~(\ref{revise_29}) represents the state after step 0 (i.e., Hadamard gates on representative spins). After step 0, apply $d+1=2$ non-commutable steps of CNOT gates, namely, step 1 and step 2, which shall be applied with the order 1,2. Denote
	\begin{equation}
		\begin{tikzpicture}[baseline=2ex]
			\draw (0,0,0)--(1,0,0)--(1,0,1)--(0,0,1)--cycle;
			\draw (0,1,0)--(1,1,0)--(1,1,1)--(0,1,1)--cycle;
			\draw[orange, line width=1pt]
			(0,0,1)--(0,1,1);
			\draw (1,0,0)--(1,1,0)
			(1,0,1)--(1,1,1)
			(0,0,0)--(0,1,0);
			
			\draw[->, blue, line width=1pt] (0,0.5,1)--(0,0.4,0);
			\draw[->, blue, line width=1pt] (0,0.5,1)--(1,0.5,0);
			\draw[->, blue, line width=1pt] (0,0.5,1)--(1,0.5,1);
			
			\draw[->, blue, line width=1pt] (0,0.5,1)--(0.5,1,0);
			\draw[->, blue, line width=1pt] (0,0.5,1)--(0.6,1,1);
			\draw[->, blue, line width=1pt] (0,0.5,1)--(0,1,0.5);
			\draw[->, blue, line width=1pt] (0,0.5,1)--(1,1,0.5);
			
			\draw[->, blue, line width=1pt] (0,0.5,1)--(0.5,0,0);
			\draw[->, blue, line width=1pt] (0,0.5,1)--(0.5,0,1);
			\draw[->, blue, line width=1pt] (0,0.5,1)--(0,0,0.5);
			\draw[->, blue, line width=1pt] (0,0.5,1)--(1,0,0.5);
		\end{tikzpicture}
		\ =\ 
		\begin{tikzpicture}[baseline=2ex]
			\draw (0,0,0)--(1,0,0)--(1,0,1)--(0,0,1)--cycle;
			\draw (0,1,0)--(1,1,0)--(1,1,1)--(0,1,1)--cycle;
			\draw
			(0,0,0)--(0,1,0)
			(1,0,0)--(1,1,0)
			(1,0,1)--(1,1,1)
			(0,0,1)--(0,1,1);
			\draw[orange, line width=1pt] (0,0,1)--(0,1,1);

			\draw[blue, line width=1pt] 
			(0.25,0.25,0.25)--(0.75,0.25,0.25)--(0.75,0.75,0.25)--(0.25,0.75,0.25)--cycle;
			\draw[blue, line width=1pt] 
			(0.25,0.25,0.75)--(0.75,0.25,0.75)--(0.75,0.75,0.75)--(0.25,0.75,0.75)--cycle;
			\draw[blue, line width=1pt]
			(0.25,0.25,0.25)--(0.25,0.25,0.75)
			(0.75,0.25,0.25)--(0.75,0.25,0.75)
			(0.25,0.75,0.25)--(0.25,0.75,0.75)
			(0.75,0.75,0.25)--(0.75,0.75,0.75);
			
			\draw[blue, line width=1pt] (0,0.5,1)--(0.25,0.5,0.75);
		\end{tikzpicture}\ .
		\label{revise_11}
	\end{equation}
	On the left hand side of Eq.~(\ref{revise_11}), each arrow represents a CNOT gate, as before, and the right hand side is just a symbolic simplification. Step 1 is:
	\begin{itemize}
		\item 
		\textbf{Step 1} (consisting of $C_3^0=1$ part):
		\begin{equation}\label{paper_X-cube_1}
			\begin{tikzpicture}[baseline=-3ex]
				\foreach \j in {0,1,2}{\foreach \k in {0,1}{\draw(0,\j,\k)--(3,\j,\k);}}
				\foreach \i in {0,1,2}{\foreach \j in {0,1,2}{\draw(\i,\j,0)--(\i,\j,2);}}
				\foreach \i in {0,1,2}{\foreach \k in {0,1}{\draw(\i,0,\k)--(\i,3,\k);}}
				
				\draw[->, blue, line width=1pt] (-1.5,0,0)--(-1.5,0,1);
				\draw[->, blue, line width=1pt] (-1.5,0,0)--(-0.7,0,0);
				\draw[->, blue, line width=1pt] (-1.5,0,0)--(-1.5,0.8,0);
				\node[blue] at (-1.68,0,1) {$x_1$};
				\node[blue] at (-0.7,-0.22,0) {$x_2$};
				\node[blue] at (-1.25,0.8,0) {$x_3$};
				
				\draw[orange, line width=1pt] 
				(0,0,0)--(0,2,0)
				(1,0,0)--(1,2,0);
				
				\foreach \i in {0,1}{\foreach \j in {0,1}{\foreach \k in {0}{
							\draw[blue, line width=1pt] 
							(\i+0.25,\j+0.25,\k+0.25)--(\i+0.75,\j+0.25,\k+0.25)--(\i+0.75,\j+0.75,\k+0.25)--(\i+0.25,\j+0.75,\k+0.25)--cycle;
							\draw[blue, line width=1pt] 
							(\i+0.25,\j+0.25,\k+0.75)--(\i+0.75,\j+0.25,\k+0.75)--(\i+0.75,\j+0.75,\k+0.75)--(\i+0.25,\j+0.75,\k+0.75)--cycle;
							\draw[blue, line width=1pt]
							(\i+0.25,\j+0.25,\k+0.25)--(\i+0.25,\j+0.25,\k+0.75)
							(\i+0.75,\j+0.25,\k+0.25)--(\i+0.75,\j+0.25,\k+0.75)
							(\i+0.25,\j+0.75,\k+0.25)--(\i+0.25,\j+0.75,\k+0.75)
							(\i+0.75,\j+0.75,\k+0.25)--(\i+0.75,\j+0.75,\k+0.75);
							
							\draw[blue, line width=1pt] (\i,\j+0.35,\k)--(\i+0.25,\j+0.45,\k+0.25);
				}}}

				\node[blue] at (1.5,0.5,0.5) {1};
				\node[blue] at (1.5,1.5,0.5) {1};
				\node[blue] at (0.5,0.5,0.5) {2};
				\node[blue] at (0.5,1.5,0.5) {2};
			\end{tikzpicture}\ .
		\end{equation}
	\end{itemize}
	Note that in Eq.~(\ref{paper_X-cube_1}), we do not draw the orange lines that are not involved in step 1 and the original point, for visual clearness. The number labels different layers, as before. Unfortunately, a $2\times3\times3$ lattice is not large enough to display the full pattern of step 1, but a larger 3-dimensional figure is not friendly to read, so we draw a projective 2-dimensional figure below:
	\begin{equation}\label{paper_X-cube_2}
		\begin{tikzpicture}
			\foreach \i in {0,1,2,3,4}{
				\draw (\i,0)--(\i,4);}
			\foreach \j in {1,2,3,4}{
				\draw (0,\j)--(5,\j);}
			
			\foreach \i in {0,1,2,3}{\foreach \j in {1,2,3}{
					\fill[orange] (\i,\j+1) circle (2pt);
					\draw[blue, line width=1pt] (\i,\j+1)--(\i+0.25,\j+0.75);
					\draw[blue, line width=1pt] (\i+0.25,\j+0.25)--(\i+0.75,\j+0.25)--(\i+0.75,\j+0.75)--(\i+0.25,\j+0.75)--cycle;}}
			
			\node[blue] at (3.5,3.5) {3};
			\node[blue] at (2.5,3.5) {4};
			\node[blue] at (3.5,2.5) {2};
			\node[blue] at (1.5,3.5) {5};
			\node[blue] at (2.5,2.5) {3};
			\node[blue] at (3.5,1.5) {1};
			\node[blue] at (1.5,2.5) {4};
			\node[blue] at (2.5,1.5) {2};
			\node[blue] at (0.5,3.5) {6};
			\node[blue] at (1.5,1.5) {3};
			\node[blue] at (0.5,2.5) {5};
			\node[blue] at (0.5,1.5) {4};
			
			\draw[->, blue, line width=1pt] (-0.5,4.5)--(1,4.5);
			\draw[->, blue, line width=1pt] (-0.5,4.5)--(-0.5,3);
			\draw[blue, line width=1pt] (-0.5,4.5) circle (0.1);
			\fill[blue] (-0.5,4.5) circle (1.5pt);
			
			\node[blue] at (-0.7,3) {$x_1$};
			\node[blue] at (1,4.7) {$x_2$};
			\node[blue] at (-0.7,4.7) {$x_3$};
		\end{tikzpicture}\ .
	\end{equation}
	The figure in Eq.~(\ref{paper_X-cube_2}) should be recognized as the top view of Eq.~(\ref{paper_X-cube_1}) from $x_3$-direction, with larger $L_1,L_2$.
	
	If OBC is applied, step 1 is enough to create a code state of X-cube. Under PBC, where ground state is not unique, additional circuit, i.e., step 2, needs to be applied. Step 2 is:
	\begin{itemize}
		\item \textbf{Step 2} consists of $C_3^1=3$ mutually commutable parts, denoted by parts $\{1\},\{2\},\{3\}$.
		\begin{itemize}
			\item \textbf{Step 2 part $\{1\}$} (following Eq.~(\ref{paper_X-cube_1})):
			\begin{equation}\label{revise_41}
				\begin{tikzpicture}[baseline=-3ex]
					\foreach \j in {0,1,2}{\foreach \k in {0,1}{\draw(0,\j,\k)--(3,\j,\k);}}
					\foreach \i in {0,1,2}{\foreach \j in {0,1,2}{\draw(\i,\j,0)--(\i,\j,2);}}
					\foreach \i in {0,1,2}{\foreach \k in {0,1}{\draw(\i,0,\k)--(\i,3,\k);}}
					
					\draw[->, blue, line width=1pt] (-1.5,0,0)--(-1.5,0,1);
					\draw[->, blue, line width=1pt] (-1.5,0,0)--(-0.7,0,0);
					\draw[->, blue, line width=1pt] (-1.5,0,0)--(-1.5,0.8,0);
					\node[blue] at (-1.68,0,1) {$x_1$};
					\node[blue] at (-0.7,-0.22,0) {$x_2$};
					\node[blue] at (-1.25,0.8,0) {$x_3$};
					
					\draw[orange, line width=1pt] 
					(0,0,2)--(0,0,1)
					(0,1,2)--(0,1,1)
					(1,0,2)--(1,0,1)
					(1,1,2)--(1,1,1);
					
					\foreach \i in {0,1}{\foreach \j in {0,1}{\foreach \k in {1}{
								\draw[blue, line width=1pt] 
								(\i+0.25,\j+0.25,\k+0.25)--(\i+0.75,\j+0.25,\k+0.25)--(\i+0.75,\j+0.75,\k+0.25)--(\i+0.25,\j+0.75,\k+0.25)--cycle;
								\draw[blue, line width=1pt] 
								(\i+0.25,\j+0.25,\k+0.75)--(\i+0.75,\j+0.25,\k+0.75)--(\i+0.75,\j+0.75,\k+0.75)--(\i+0.25,\j+0.75,\k+0.75)--cycle;
								\draw[blue, line width=1pt]
								(\i+0.25,\j+0.25,\k+0.25)--(\i+0.25,\j+0.25,\k+0.75)
								(\i+0.75,\j+0.25,\k+0.25)--(\i+0.75,\j+0.25,\k+0.75)
								(\i+0.25,\j+0.75,\k+0.25)--(\i+0.25,\j+0.75,\k+0.75)
								(\i+0.75,\j+0.75,\k+0.25)--(\i+0.75,\j+0.75,\k+0.75);
								
								\draw[blue, line width=1pt] (\i,\j,\k+0.5)--(\i+0.25,\j+0.5,\k+0.75);
					}}}

					\node[blue] at (1.5,0.5,1.5) {4};
					\node[blue] at (1.5,1.5,1.5) {3};
					\node[blue] at (0.5,0.5,1.5) {5};
					\node[blue] at (0.5,1.5,1.5) {4};
				\end{tikzpicture}\ .
			\end{equation}
			\item  \textbf{Step 2 part $\{2\}$}:
			\begin{equation}\label{revise_eqn_step22}
				\begin{tikzpicture}[baseline=-3ex]
					\foreach \j in {0,1,2}{\foreach \k in {0,1}{\draw(0,\j,\k)--(3,\j,\k);}}
					\foreach \i in {0,1,2}{\foreach \j in {0,1,2}{\draw(\i,\j,0)--(\i,\j,2);}}
					\foreach \i in {0,1,2}{\foreach \k in {0,1}{\draw(\i,0,\k)--(\i,3,\k);}}
					
					\draw[->, blue, line width=1pt] (-1.5,0,0)--(-1.5,0,1);
					\draw[->, blue, line width=1pt] (-1.5,0,0)--(-0.7,0,0);
					\draw[->, blue, line width=1pt] (-1.5,0,0)--(-1.5,0.8,0);
					\node[blue] at (-1.68,0,1) {$x_1$};
					\node[blue] at (-0.7,-0.22,0) {$x_2$};
					\node[blue] at (-1.25,0.8,0) {$x_3$};
					
					\draw[orange, line width=1pt] 
					(2,0,0)--(3,0,0)
					(2,1,0)--(3,1,0);
					
					\foreach \i in {2}{\foreach \j in {0,1}{\foreach \k in {0}{
								\draw[blue, line width=1pt] 
								(\i+0.25,\j+0.25,\k+0.25)--(\i+0.75,\j+0.25,\k+0.25)--(\i+0.75,\j+0.75,\k+0.25)--(\i+0.25,\j+0.75,\k+0.25)--cycle;
								\draw[blue, line width=1pt] 
								(\i+0.25,\j+0.25,\k+0.75)--(\i+0.75,\j+0.25,\k+0.75)--(\i+0.75,\j+0.75,\k+0.75)--(\i+0.25,\j+0.75,\k+0.75)--cycle;
								\draw[blue, line width=1pt]
								(\i+0.25,\j+0.25,\k+0.25)--(\i+0.25,\j+0.25,\k+0.75)
								(\i+0.75,\j+0.25,\k+0.25)--(\i+0.75,\j+0.25,\k+0.75)
								(\i+0.25,\j+0.75,\k+0.25)--(\i+0.25,\j+0.75,\k+0.75)
								(\i+0.75,\j+0.75,\k+0.25)--(\i+0.75,\j+0.75,\k+0.75);
								
								\draw[blue, line width=1pt]
								(\i+0.36,\j,\k)--(\i+0.45,\j+0.25,\k+0.25);
					}}}

					\node[blue] at (2.5,1.5,0.5) {3};
					\node[blue] at (2.5,0.5,0.5) {4};
				\end{tikzpicture}\ .
			\end{equation}
			\item \textbf{Step 2 part $\{3\}$}:
			\begin{equation}\label{revise_42}
				\begin{tikzpicture}[baseline=-3ex]
					\foreach \j in {0,1,2}{\foreach \k in {0,1}{\draw(0,\j,\k)--(3,\j,\k);}}
					\foreach \i in {0,1,2}{\foreach \j in {0,1,2}{\draw(\i,\j,0)--(\i,\j,2);}}
					\foreach \i in {0,1,2}{\foreach \k in {0,1}{\draw(\i,0,\k)--(\i,3,\k);}}
					
					\draw[->, blue, line width=1pt] (-1.5,0,0)--(-1.5,0,1);
					\draw[->, blue, line width=1pt] (-1.5,0,0)--(-0.7,0,0);
					\draw[->, blue, line width=1pt] (-1.5,0,0)--(-1.5,0.8,0);
					\node[blue] at (-1.68,0,1) {$x_1$};
					\node[blue] at (-0.7,-0.22,0) {$x_2$};
					\node[blue] at (-1.25,0.8,0) {$x_3$};
					
					\draw[orange, line width=1pt] 
					(0,2,0)--(0,3,0)
					(1,2,0)--(1,3,0);
					
					\foreach \i in {0,1}{\foreach \j in {2}{\foreach \k in {0}{
								\draw[blue, line width=1pt] 
								(\i+0.25,\j+0.25,\k+0.25)--(\i+0.75,\j+0.25,\k+0.25)--(\i+0.75,\j+0.75,\k+0.25)--(\i+0.25,\j+0.75,\k+0.25)--cycle;
								\draw[blue, line width=1pt] 
								(\i+0.25,\j+0.25,\k+0.75)--(\i+0.75,\j+0.25,\k+0.75)--(\i+0.75,\j+0.75,\k+0.75)--(\i+0.25,\j+0.75,\k+0.75)--cycle;
								\draw[blue, line width=1pt]
								(\i+0.25,\j+0.25,\k+0.25)--(\i+0.25,\j+0.25,\k+0.75)
								(\i+0.75,\j+0.25,\k+0.25)--(\i+0.75,\j+0.25,\k+0.75)
								(\i+0.25,\j+0.75,\k+0.25)--(\i+0.25,\j+0.75,\k+0.75)
								(\i+0.75,\j+0.75,\k+0.25)--(\i+0.75,\j+0.75,\k+0.75);
								
								\draw[blue, line width=1pt] (\i,\j+0.35,\k)--(\i+0.25,\j+0.45,\k+0.25);
					}}}

					\node[blue] at (0.5,2.5,0.5) {3};
					\node[blue] at (1.5,2.5,0.5) {4};
				\end{tikzpicture}\ .
			\end{equation}
			Like for step 1, we do not draw the orange lines that are not involved in the present part of step 2, for visual clearness. The number labels different layers, as before. Again, a $2\times3\times3$ lattice is not large enough to display the full pattern of parts $\{2\},\{3\}$ of step 2, but parts $\{1\},\{2\},\{3\}$ have the same pattern, so the full pattern of parts $\{2\},\{3\}$ can be inferred. Different parts of step 2 are mutually commutable, thus can be applied simultaneously.
		\end{itemize}
	\end{itemize}
	The two steps prepare the EWSC on all cubes, except the red cubes $c_1,c_2,c_3,c_4,c_5,c_6$ shown below:
	\begin{equation}\label{paper_X-cube_3}
		\includegraphics[width=0.3\textwidth]{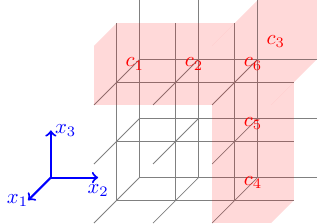}\ .
	\end{equation}
	Denote the set of all $\gamma_3$(cubes) by $\Gamma_3$, and the set of all $\gamma_3$ without a representative spin as $R_3$\footnote{This $R_3$ is the set $R_D$ in the overall strategy in Sec.~\ref{section_strate}.}. Under the setting of Eq.~(\ref{paper_X-cube_3}), $R_3=\{c_1,c_2,c_3,c_4,c_5,c_6\}$. The state after applying steps 1,2 is
	\begin{equation}\label{paper_X-cube_6}
		\prod_{c\in\Gamma_3\backslash R_3}\frac{1}{\sqrt{2}}\Big(1+A_c\Big)|00\cdots0\rangle\,.
	\end{equation}
	This resulting state is an EWSC on all cubes and is therefore a code state, which is explained in detail in Sec.~\ref{section_X-cube_SQC_algebraic}.
	
	\subsubsection{Algebraic approach on a  general cubic lattice and the notion of stabilizer redundancy equivalence class (RE-class)}\label{section_X-cube_SQC_algebraic}
	
	Next, we describe the above unitary circuit algebraically before explaining why the state in Eq.~(\ref{paper_X-cube_6}) is the EWSC on all cubes and is therefore a code state. For this purpose, we switch to a  general cubic lattice. The initial state is $|00\cdots0\rangle$. Then we apply Hadamard gates on representative spins (step 0), after which the state becomes
	\begin{equation}
		\begin{split}
			&\Bigg(\prod_{x_i\in\mathbb{Z}_{L_i-1},\,i=1,2,3}\text{Had}_{[x_1,x_2,x_3+\frac{1}{2}]}\Bigg)\\
			&\Bigg(\prod_{j=1}^3\  \prod_{x_i\in\mathbb{Z}_{L_i-1},\,i\neq j}\text{Had}_{[-\frac{1}{2},x_i|i\neq j]}\Bigg)|00\cdots0\rangle\,,
		\end{split}
	\end{equation}
	where the subscript $[-\frac{1}{2},x_i|i\neq j]$ stands for $[-\frac{1}{2},x_2,x_3]$ when $j=1$, $[x_1,-\frac{1}{2},x_3]$ when $j=2$, and $[x_1,x_2,-\frac{1}{2}]$ when $j=3$.

\begin{figure}[t]
		\centering
		\hspace*{-1cm}
		\begin{subfigure}{0.28\textwidth}
			\includegraphics[width=1\textwidth]{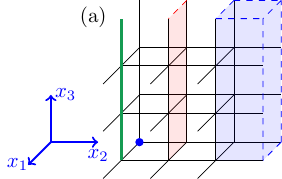}
		\end{subfigure}
		\begin{subfigure}{0.18\textwidth}
			\includegraphics[width=1.15\textwidth]{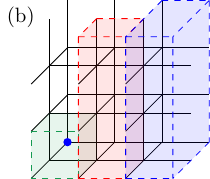}
		\end{subfigure}
		\caption{Illustration of $*$ notation. The original point is dotted blue in the two figures, and the lattice constant is taken to be 1. (a) The green line is a circle $S^1$, which can be denoted by $\left[1,0,*\right]$, and viewed as the trajectory (uncontractible loop) of dragging the vertex $[1,0,0]$ along $x_3$-direction. The red surface is a spanned circle $S^1\times B^1$, which can be denoted by $\left[\frac12,1,*\right]$, and viewed as the trajectory (uncontractible loop) of dragging the edge $\left[\frac12,1,0\right]$ along $x_3$-direction. Similarly, the blue volume is a spanned circle $S^1\times B^2$, which can be denoted by $\left[\frac12,\frac52,*\right]$, and viewed as the trajectory (uncontractible loop) of dragging the plaquette $\left[\frac12,\frac52,0\right]$ along $x_3$-direction. (b) The green cube, red volume and blue volume can be denoted by $\left[\frac32,\frac12,\frac12\right],\ \left[\frac32,\frac32,*\right]$ and $\left[*,\frac52,*\right]$, respectively. Topologically, they are $B^3$, $T^1\times B^2$ and $T^2\times B^1$, respectively. $T^1=S^1$. For a $D$-dimensional cubic lattice, when $m$ inputs are half-integers and $n$ inputs are taken to be $*$, the output is topologically $T^n\times B^m$.}
		\label{*-intuition}
	\end{figure}

	Define a symbol called \textbf{\textit{group-CNOT}} to simplify the formula:
	\begin{equation}\label{paper_dddd_18}
		\text{GCNOT}_{s}^{\xi}:=\prod_{s'\in(\xi\backslash \{s\})}\text{CNOT}_{s,s'}\,,
	\end{equation}
	i.e., the product of CNOT gates with the same control qubit $s$, and a set of target qubits $\xi\backslash\{s\}$. The control qubit is automatically excluded from the target qubits set, to meet with the definition of CNOT gate as a two qubit operation. Note that the CNOT gates with the same control qubit are commutable, group-CNOT just collects some of those commutable CNOT gates by multiplying them together. The idea is to represent the CNOT gates inside the same cube as a whole.
	\begin{itemize}
		\item \textbf{Step 1}:
		\begin{equation}\label{paper_X-cube_5}
			\prod_{n=0}^{\sum_{i=1,2}(L_i-2)}\prod_{\substack{x_i\in\mathbb{Z}_{L_i-1},\,i=1:3\\\sum_{i=1,2}x_i=n}}\text{GCNOT}_{[x_1,x_2,x_3+\frac{1}{2}]}^{\delta_d[x_1+\frac{1}{2},x_2+\frac12,x_3+\frac12]}\,,
		\end{equation}
		where the symbol $\delta_d\gamma_D$ is defined as:
		\begin{equation}\label{revise_17}
			\delta_d\gamma_D:=\big\{\gamma_d\big|\gamma_d\subset \gamma_D\big\}\,.
		\end{equation}
		It is the set of all $d$-cubes inside the closed $D$-cube $\gamma_D$. When $d<D$, $\delta_d\gamma_D$ is the set of all $\gamma_d$ on the boundary or corners of $\gamma_D$. When $d=D$, $\delta_d\gamma_D=\{\gamma_D\}$. When $d>D$, $\delta_d\gamma_D=\emptyset$. For example, when $d=1$ and $D=3$, $\delta_d\gamma_D$ is the set of all 1-cubes or edges on the boundary of the cube $\gamma_3$. Step 1 adds cubes with no coordinate being $-\frac12$ into $\Omega$ (see the convention below Eq.~(\ref{revise_39})). The state before step 1 is $\mathcal{F}(\emptyset)$, while the state after step 1 is $\mathcal{F}(\Omega_1)$, where $\Omega_1$ is the union of all the cubes with no coordinate being $-\frac12$.
		\item \textbf{Step 2}:
		\begin{itemize}
			\item \textbf{Step 2 part $\{1\}$}:
			\begin{equation}\label{revise_43}
				\prod_{n=0}^{\sum_{i=2,3}(L_i-2)}\prod_{\substack{x_i\in\mathbb{Z}_{L_i-1},\,i=2,3\\\sum_{i=2,3}x_i=n}}\text{GCNOT}_{[-\frac{1}{2},x_2,x_3]}^{\delta_d[-\frac{1}{2},x_2+\frac{1}{2},x_3+\frac12]}\,.
			\end{equation} 
			Step 2 part $\{1\}$ adds all the cubes with $x_1$ but no other coordinates being $-\frac12$ into $\Omega$, i.e., the state after step 1, step 2 part $\{1\}$ is $\mathcal{F}(\Omega_1\cup\Omega_{2,\{1\}})$, where $\Omega_{2,\{1\}}$ is the union of all cubes with $x_1$ but no other coordinates being $-\frac12$.
			\item \textbf{Step 2 part $\{2\}$}:
			\begin{equation}
				\prod_{n=0}^{\sum_{i=1,3}(L_i-2)}\prod_{\substack{x_i\in\mathbb{Z}_{L_i-1},\,i=1,3\\\sum_{i=1,3}x_i=n}}\text{GCNOT}_{[x_1,-\frac{1}{2},x_3]}^{\delta_d[x_1+\frac12,-\frac{1}{2},x_3+\frac{1}{2}]}\,.
			\end{equation}
			Step 2 part $\{2\}$ adds all the cubes with $x_2$ but no other coordinates being $-\frac12$ into $\Omega$, i.e., the state after step 1, step 2 part $\{1\}$, step 2 part $\{2\}$ is $\mathcal{F}(\Omega_1\cup\Omega_{2,\{1\}}\cup\Omega_{2,\{2\}})$, where $\Omega_{2,\{2\}}$ is the union of all cubes with $x_2$ but no other coordinates being $-\frac12$.
			\item \textbf{Step 2 part $\{3\}$}:
			\begin{equation}\label{revise_44}
				\prod_{n=0}^{\sum_{i=1,2}(L_i-2)}\prod_{\substack{x_i\in\mathbb{Z}_{L_i-1},\,i=1,2\\\sum_{i=1,2}x_i=n}}\text{GCNOT}_{[x_1,x_2,-\frac{1}{2}]}^{\delta_d[x_1+\frac12,x_2+\frac12,-\frac{1}{2}]}\,.
			\end{equation}
			Step 2 part $\{3\}$ adds all the cubes with $x_3$ but no other coordinates being $-\frac12$ into $\Omega$, i.e., the state after step 1, step 2 part $\{1\}$, step 2 part $\{2\}$, step 2 part $\{3\}$ is $\mathcal{F}(\Omega_1\cup\Omega_{2,\{1\}}\cup\Omega_{2,\{2\}}\cup\{2,\{3\}\})$, where $\Omega_{2,\{3\}}$ is the union of all cubes with $x_3$ but no other coordinates being $-\frac12$.
		\end{itemize} 
		In total, step 2 adds all the cubes with exactly 1 coordinate being $-\frac12$ into $\Omega$, i.e., the state after steps 1,2 is $\mathcal{F}(\Omega_1\cup\Omega_2)$, where $\Omega_2$ is the union of all cubes with exactly 1 coordinate being $-\frac12$.
	\end{itemize}
	The representative spin and its represented cube pairs can be read from GCNOT gates in Eqs.~(\ref{paper_X-cube_5}),(\ref{revise_43})-(\ref{revise_44}) as
	\begin{equation}\label{revise_45}
		\text{GCNOT}^{\delta_d\gamma_D}_{\mathcal{R}(\gamma_D)}\ ,\quad \gamma_D\subset\Omega_1\cup\Omega_2\,,
	\end{equation}
	where $\mathcal{R}(\gamma_D)$ refers to the representative spin of $\gamma_D$. As illustrated above, the state after circuit $U_c$ is $\mathcal{F}(\Omega_1\cup\Omega_2)$. Note that there is no edge not in $\Omega_1\cup\Omega_2$, so $\mathcal{F}(\Omega_1\cup\Omega_2)$ can be written as
	\begin{equation}
		\begin{split}
			&\Bigg(\prod_{x_i\in\mathbb{Z}_{L_i-1},\,i=1,2,3}\frac{1}{\sqrt{2}}\Big(1+A_{[x_1+\frac{1}{2},x_2+\frac12,x_3+\frac12]}\Big)\Bigg)\\
			&\Bigg(\prod_{j=1}^3\ \prod_{x_i\in\mathbb{Z}_{L_i-1},\,i\neq j}\frac{1}{\sqrt{2}}\Big(1+A_{[-\frac{1}{2},x_i+\frac{1}{2}|i\neq j]}\Big)\Bigg)|00\cdots0\rangle\,.\label{revise_40}
		\end{split}
	\end{equation}
	Here, $\big[-\frac12,x_i+\frac12\big|i\neq  j\big]$ stands for $\big[-\frac12,x_2+\frac12,x_3+\frac12\big]$ when $j=1$, $\big[x_1+\frac12,-\frac12,x_3+\frac12\big]$ when $j=2$ and  $\big[x_1+\frac12,x_2+\frac12,-\frac12\big]$ when $j=3$. Now we discuss why the state in Eq.~(\ref{revise_40}) is the EWSC of all cubes, because of the existence of stabilizer redundancies. At the end of Sec.~\ref{section_toric_code_sqc}, we wrote a stabilizer redundancy, here we define this concept formally. Use the symbol $s$ or $s'$ to represent an arbitrary stabilizer, define an equation of the form
	\begin{equation}\label{paper_redundancy_0}
		s'=\prod_{s\in\xi\backslash\{s'\}}s
	\end{equation}
	as a stabilizer redundancy, where $\xi$ is a stabilizer set containing $s'$, and define an equation of the form
	\begin{equation}\label{paper_redundancy_1}
		r^1:\ \prod_{s\in\xi}s=1
	\end{equation}
	as a \textbf{{\textit{stabilizer redundancy equivalence class}}}, abbreviated as a \textbf{\textit{RE-class}} (symbolically represented by $r^1$) for the notational convenience. For simplicity, we restrict the discussion with stabilizers satisfying
	\begin{equation}
		s^2=1\quad\&\quad [s,s']=0\,.
	\end{equation} 
	The RE-class in Eq.~(\ref{paper_redundancy_1}) is equivalent to $|\xi|$ mutually equivalent stabilizer redundancies. We can write the following $A$ term RE-classes of the X-cube:
	\begin{gather}
		r^1_{[x_1+\frac{1}{2},*,*]}:\ \prod_{x_i\in\mathbb{Z}_{L_i},\,i=2,3}A_{[x_i+\frac{1}{2}|i=1:3]}=1,\ x_1\in\mathbb{Z}_{L_1}\,,\notag\\
		r^1_{[*,x_2+\frac{1}{2},*]}:\ \prod_{x_i\in\mathbb{Z}_{L_i},\,i=1,3}A_{[x_i+\frac{1}{2}|i=1:3]}=1,\ x_2\in\mathbb{Z}_{L_2}\,,\notag\\
		r^1_{[*,*,x_3+\frac{1}{2}]}:\ \prod_{x_i\in\mathbb{Z}_{L_i},\,i=1,2}A_{[x_i+\frac{1}{2}|i=1:3]}=1,\ x_3\in\mathbb{Z}_{L_3}\,,\label{paper_X-cube_4}
	\end{gather}
	where 
	\begin{align}
		\left[x_1+\frac{1}{2},*,*\right]&\equiv\bigcup_{x_i\in\frac12\mathbb{Z},\,i=2,3}\left[x_1+\frac12,x_2,x_3\right]\,,\label{revise_31}
	\end{align}
	i.e., $*$ takes through all possible legal inputs, with the output being the union of all possible output set. Formally, for any set-valued function 
	\begin{equation}
		g:X\to Y,\ x\mapsto y=g(x)\,,
	\end{equation}
	define 
	\begin{align}
		g(*)\equiv\bigcup_{x\in X}g(x)\,.\label{revise_36}
	\end{align}
	$\left[x_1+\frac{1}{2},*,*\right],\left[*,x_2+\frac12,*\right],\left[*,*,x_3+\frac12\right]$ all follow the definition in Eq.~(\ref{revise_36}). The notation $*$ will be frequently used in this paper, and it has a straightforward intuition in our context, as illustrated in Fig.~\ref{*-intuition}.
	Notice that since $\left[x_1+\frac12,x_2+\frac12,x_3+\frac12\right]$ are closed 3-cubes, 
	\begin{equation}\label{revise_37}
		\left[x_1+\frac{1}{2},*,*\right]=\bigcup_{x_i\in\mathbb{Z}_{L_i},\,i=2,3}\left[x_1+\frac{1}{2},x_2+\frac12,x_3+\frac12\right]\,.
	\end{equation}
	Therefore, all the $A$ terms associated with the 3-cubes in $\left[x_1+\frac12,*,*\right]$ multiply to 1, or $\left[x_1+\frac12,*,*\right]$ support an order-1 RE class, denoted by $r^1_{\left[x_1+\frac12,*,*\right]}$. Similarly, each $[*,x_2+\frac12,*]$ or $[*,*,x_3+\frac12]$ supports an order-1 RE-class, as written in Eq.~(\ref{paper_X-cube_4}). There are in total $L_1+L_2+L_3$ RE-classes in Eq.~(\ref{paper_X-cube_4}). All other $A$ term RE-classes can be generated by multiplying RE-classes in Eq.~(\ref{paper_X-cube_4}). 
	
	For concreteness, we keep the setting of Eq.~(\ref{paper_X-cube_3}), i.e., $L_1=2,L_2=3,L_3=3$. The only $A$ term appearing in $r^1_{[*,\frac{1}{2},*]}$ but not appearing in Eq.~(\ref{paper_X-cube_6}) is the $A$ term associated with $c_1$ drawn in Eq.~(\ref{paper_X-cube_3}), so based on the same reason as Eqs.~(\ref{paper_2dTC_3}, \ref{paper_2dTC_4}),
	\begin{equation}
		\begin{split}
			&\frac{1}{2}\Big(1+A_{c_1}\Big)\prod_{c\in\Gamma_3\backslash R_3}\frac{1}{\sqrt{2}}\Big(1+A_c\Big)|00\cdots0\rangle\\
			=&\prod_{c\in\Gamma_3\backslash R_3}\frac{1}{\sqrt{2}}\Big(1+A_c\Big)|00\cdots0\rangle\,,
		\end{split}
	\end{equation}
	i.e., the EWSC on $\Omega_1\cup\Omega_2$\footnote{Note that $\Omega_1\cup\Omega_2=\bigcup_{c\in\Gamma_3\backslash R_3}c$.} equals to the EWSC on $\Omega_1\cup\Omega_2\cup c_1$. $c_1$ can be directly added into $\Omega$, out of the $A$ term redundancy $r^1_{[*,\frac{1}{2},*]}$. Similarly, $c_2$ can be directly added into $\Omega$ out of the $A$ term redundancy $r^1_{[*,\frac{3}{2},*]}$, while $c_3$ out of $r^1_{[\frac{1}{2},*,*]}$, $c_4$ out of $r^1_{[*,*,\frac{1}{2}]}$, $c_5$ out of $r^1_{[*,*,\frac{3}{2}]}$. Therefore,
	\begin{equation}\label{paper_X-cube_6_2}
		\begin{split}
			&\prod_{c\in\Gamma_3\backslash R_3}\frac{1}{\sqrt{2}}\Big(1+A_c\Big)|00\cdots0\rangle\\
			=&\prod_{i=1,2,\cdots,5}\frac{1}{2}\Big(1+A_{c_i}\Big)\prod_{c\in\Gamma_3\backslash R_3}\frac{1}{\sqrt{2}}\Big(1+A_c\Big)|00\cdots0\rangle\,.
		\end{split}
	\end{equation}
	Then, the only $A$ term appearing in $r^1_{[*,\frac{5}{2},*]}$ but not appearing on the right hand side of Eq.~(\ref{paper_X-cube_6_2}) is $A_{c_6}$, so based on the same reason as Eqs.~(\ref{paper_2dTC_3}, \ref{paper_2dTC_4}),
	\begin{equation}\label{paper_X-cube_8}
		\begin{split}
			&\prod_{i=1,2,\cdots,5}\frac{1}{2}\Big(1+A_{c_i}\Big)\prod_{c\in\Gamma_3\backslash R_3}\frac{1}{\sqrt{2}}\Big(1+A_c\Big)|00\cdots0\rangle\\
			=&\prod_{i=1,2,\cdots,6}\frac{1}{2}\Big(1+A_{c_i}\Big)\prod_{c\in\Gamma_3\backslash R_3}\frac{1}{\sqrt{2}}\Big(1+A_c\Big)|00\cdots0\rangle\,.
		\end{split}
	\end{equation}
	There are three different order 1 RE-classes in Eq.~(\ref{paper_X-cube_4}) that allows the direct addition of $c_6$ into $\Omega$, which are $r^1_{[*,\frac{5}{2},*]},r^1_{[\frac{3}{2},*,*]},r^1_{[*,*,\frac{5}{2}]}$. This is because the order 1 RE-classes listed in Eq.~(\ref{paper_X-cube_4}) are not independent, the dependency is described by higher order redundancy. We will talk about it in another paper in the future.
	
	The right hand side of Eq.~(\ref{paper_X-cube_8}) is just
	\begin{equation}
		\frac{1}{\sqrt{2}^{|R_3|}}\prod_{c\in\Gamma_3}\frac{1}{\sqrt{2}}\Big(1+A_c\Big)|00\cdots0\rangle\,,\label{any_GS_19}
	\end{equation}
	which is a computational basis of the X-cube model code states. The set order $|R_3|=L_1+L_2+L_3-2$, which is a special case of Eq.~(\ref{general_R_D_order}).
	
	The strategy of showing the result state after circuit $U_c$ is actually the EWSC on all cubes is: (1) showing the state after circuit $U_c$ is the EWSC of all cubes with at most $d=1$ coordinate being $-\frac12$; (2) showing all the cubes with exactly 2 coordinates being $-\frac12$ can be directly added to $\Omega$, out of stabilizer redundancies; (3) showing all the cubes with exactly 3 coordinates being $-\frac12$ can be directly added to $\Omega$, out of stabilizer redundancies. This strategy can be straightforwardly generalized to the case of general stabilizer code TD model.

	\subsection{$[0,1,2,4]$ model}\label{section_0124_sqc}
	
	In this subsection, we construct a circuit $U_c$ for preparing computational basis of the $[0,1,2,4]$ model code states, where $d=1,D=4$. Since graphic illustration for 4-dimensional models is no longer straightforward, we illustrate the circuit alternatively with figures and formulas. The readers are expected to have become familiar with the algebraic description of circuit for X-cube, in order to read this subsection.

	We start with the algebraic description here. Consider a 4-dimensional hypercubic lattice under PBC, with each edge having a $\frac{1}{2}$-spin on it. Prepare the initial product state $|00\cdots0\rangle$. Then apply Hadamard gates on all the representative spins (step 0), after which the state becomes
	\begin{equation}
		\begin{split}
			&\Bigg(\prod_{x_i\in\mathbb{Z}_{L_i-1},\,i=1,2,3,4}\text{Had}_{[x_1,x_2,x_3,x_4+\frac{1}{2}]}\Bigg)\\
			&\Bigg(\prod_{j=1}^4\  \prod_{x_i\in\mathbb{Z}_{L_i-1},\,i\neq j}\text{Had}_{[-\frac{1}{2},x_i|i\neq j]}\Bigg)|00\cdots0\rangle\,,
		\end{split}
	\end{equation}
	where $\left[\left.-\frac    {1}{2},x_i\right|i\neq 1\right]=\left[-\frac{1}{2},x_2,x_3,x_4\right]$, $\left[\left.-\frac    {1}{2},x_i\right|i\neq 2\right]=\left[x_1,-\frac{1}{2},x_3,x_4\right]$ and similarly when $j=3,4$. After step 0, apply $d+1=2$ non-commutable steps of CNOT gates, namely, step 1 and step 2, which shall be applied in the order 1,2. \\\textbf{Step 1} is
	\begin{equation}\label{revise_14}
		\prod_{n=0}^{\sum_{i\neq4}(L_i-2)}\prod_{\substack{x_i\in\mathbb{Z}_{L_i-1},\,i=1:4\\\sum_{i\neq4}x_i=n}}\text{GCNOT}_{[x_1,x_2,x_3,x_4+\frac{1}{2}]}^{\delta_d[x_i+\frac{1}{2}|i=1:4]}\,,
	\end{equation}
	where $\left[\left.x_i+\frac12\right|i=1,2,3,4\right]=\left[x_1+\frac12,\cdots,x_4+\frac12\right]$, and $\delta_d\big[x_i+\frac{1}{2}\big|i=1,2,3,4\big]$ is defined in Eq.~(\ref{revise_17}). Step 1 adds all the 4-cubes with no coordinate being $-\frac12$ into $\Omega$ (see the convention below Eq.~(\ref{revise_39})). The state before step 1 is $\mathcal{F}(\emptyset)$ while the state after step 1 is $\mathcal{F}(\Omega_1)$, where $\Omega_1$ is the union of all 4-cubes with no coordinate being $-\frac12$.
	
	Now we illustrate step 1 graphically by considering a $3\times3\times3\times3$ hypercubic lattice under PBC. Denote
	\begin{equation}\label{paper_dddd_2}
		\begin{tikzpicture}
			\draw (0,0,0)--(1,0,0)--(1,0,1)--(0,0,1)--cycle;
			\draw (0,1,0)--(1,1,0)--(1,1,1)--(0,1,1)--cycle;
			\draw
			(0,0,0)--(0,1,0)
			(1,0,0)--(1,1,0)
			(0,0,1)--(0,1,1)
			(1,0,1)--(1,1,1);
			
			\draw (0+4,0+0.7,0)--(1+4,0+0.7,0)--(1+4,0+0.7,1)--(0+4,0+0.7,1)--cycle;
			\draw (0+4,1+0.7,0)--(1+4,1+0.7,0)--(1+4,1+0.7,1)--(0+4,1+0.7,1)--cycle;
			\draw
			(0+4,0+0.7,0)--(0+4,1+0.7,0)
			(1+4,0+0.7,0)--(1+4,1+0.7,0)
			(0+4,0+0.7,1)--(0+4,1+0.7,1)
			(1+4,0+0.7,1)--(1+4,1+0.7,1);
			
			\draw
			(0,0,0)--(0+4,0+0.7,0)
			(1,0,0)--(1+4,0+0.7,0)
			(0,0,1)--(0+4,0+0.7,1)
			(1,0,1)--(1+4,0+0.7,1)
			(0,1,0)--(0+4,1+0.7,0)
			(1,1,0)--(1+4,1+0.7,0)
			(0,1,1)--(0+4,1+0.7,1)
			(1,1,1)--(1+4,1+0.7,1);

			\draw[->, blue, line width=1pt] 
			(0,0.5,0)--(0,0,0.5);
			\draw[->, blue, line width=1pt] 
			(0,0.5,0)--(1,0,0.5);
			\draw[->, blue, line width=1pt] 
			(0,0.5,0)--(0.5,0,0);
			\draw[->, blue, line width=1pt] 
			(0,0.5,0)--(0.5,0,1);
			
			\draw[->, blue, line width=1pt] 
			(0,0.5,0)--(0,1,0.5);
			\draw[->, blue, line width=1pt] 
			(0,0.5,0)--(1,1,0.5);
			\draw[->, blue, line width=1pt] 
			(0,0.5,0)--(0.5,1,0);
			\draw[->, blue, line width=1pt] 
			(0,0.5,0)--(0.5,1,1);
			
			\draw[->, blue, line width=1pt] 
			(0,0.5,0)--(0+2,0+0.35,0);
			\draw[->, blue, line width=1pt] 
			(0,0.5,0)--(1,0.5,0);
			\draw[->, blue, line width=1pt] 
			(0,0.5,0)--(0,0.5,1);
			\draw[->, blue, line width=1pt] 
			(0,0.5,0)--(1,0.5,1);

			\draw[->, blue, line width=1pt] 
			(0,0.5,0)--(0+4,0+0.7,0.5);
			\draw[->, blue, line width=1pt] 
			(0,0.5,0)--(1+4,0+0.7,0.5);
			\draw[->, blue, line width=1pt] 
			(0,0.5,0)--(0.5+4,0+0.7,0);
			\draw[->, blue, line width=1pt] 
			(0,0.5,0)--(0.5+4,0+0.7,1);
			
			\draw[->, blue, line width=1pt] 
			(0,0.5,0)--(0+4,1+0.7,0.5);
			\draw[->, blue, line width=1pt] 
			(0,0.5,0)--(1+4,1+0.7,0.5);
			\draw[->, blue, line width=1pt] 
			(0,0.5,0)--(0.5+4,1+0.7,0);
			\draw[->, blue, line width=1pt] 
			(0,0.5,0)--(0.5+4,1+0.7,1);
			
			\draw[->, blue, line width=1pt] 
			(0,0.5,0)--(0+4,0.5+0.7,0);
			\draw[->, blue, line width=1pt] 
			(0,0.5,0)--(1+4,0.5+0.7,0);
			\draw[->, blue, line width=1pt] 
			(0,0.5,0)--(0+4,0.5+0.7,1);
			\draw[->, blue, line width=1pt] 
			(0,0.5,0)--(1+4,0.5+0.7,1);

			\draw[->, blue, line width=1pt] 
			(0,0.5,0)--(1+2,0+0.35,0);
			\draw[->, blue, line width=1pt] 
			(0,0.5,0)--(0+2,1+0.35,0);
			\draw[->, blue, line width=1pt] 
			(0,0.5,0)--(1+2,1+0.35,0);
			\draw[->, blue, line width=1pt] 
			(0,0.5,0)--(0+2,0+0.35,1);
			\draw[->, blue, line width=1pt] 
			(0,0.5,0)--(1+2,0+0.35,1);
			\draw[->, blue, line width=1pt] 
			(0,0.5,0)--(0+2,1+0.35,1);
			\draw[->, blue, line width=1pt] 
			(0,0.5,0)--(1+2,1+0.35,1);

			\draw[orange, line width=1pt] (0,0,0)--(0,1,0);
		\end{tikzpicture}
	\end{equation}
	by
	\begin{equation}\label{paper_dddd_1}
		\includegraphics[width=0.30186\textwidth]{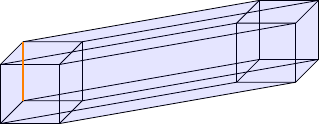}\ ,
	\end{equation}
	where the black, orange edges represent spins in state $|0\rangle,|+\rangle$, respectively, and the blue arrows represent CNOT gates, as before. Eq.~(\ref{paper_dddd_2}) or Eq.~(\ref{paper_dddd_1}) represents the group-CNOT gate with one spin (the orange one) in a closed 4-cube $\gamma_4$ being control qubit, and all other spins inside $\gamma_4$ being target qubits. Based on Eq.~(\ref{paper_dddd_1}), denote
	\begin{equation}\label{paper_dddd_3}
		\includegraphics[width=0.3813\textwidth]{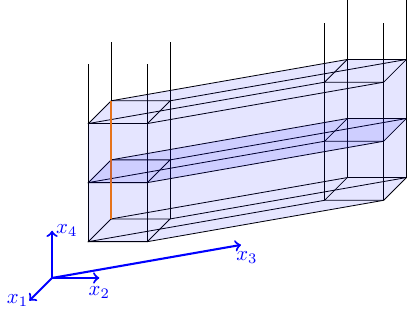}
	\end{equation}
	by
	\begin{equation}\label{paper_dddd_4}
		\includegraphics[width=0.248\textwidth]{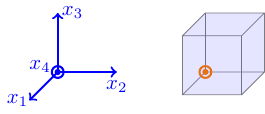}\ ,
	\end{equation}
	where Eq.~(\ref{paper_dddd_4}) is the planform of Eq.~(\ref{paper_dddd_3}), viewed from $x_4$-direction. The lines along $x_4$-direction degenerate to a dot (i.e., the circled dot in Eq.~(\ref{paper_dddd_4})). The orange line, together with the blue dye, extends only between $[0,L_4-1]$ in $x_4$-direction, as shown in Eq.~(\ref{paper_dddd_3}), where $L_4=3$. With the notation in Eq.~(\ref{paper_dddd_4}), we can graphically illustrate step 1 as following: \\
	\textbf{Step 1 layer 1}:
	\begin{equation}\label{paper_dddd_5}
		\includegraphics[width=0.36\textwidth]{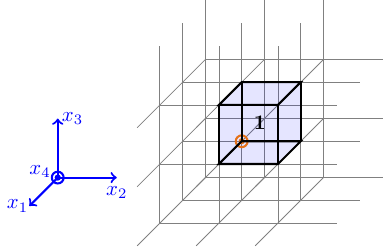}\ ,
	\end{equation}
	where the newly added $\bf{1}$ stands for layer 1, and the black bold lines are added to stress the edge of dyed blue cube (just for visual clearness). 
	\\\textbf{Step 1 layer 2}:
	\begin{equation}\label{paper_dddd_6}
		\includegraphics[width=0.36\textwidth]{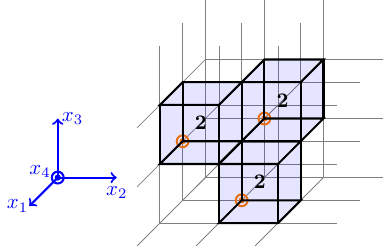}\ .
	\end{equation}
	\textbf{Step 1 layer 3}:
	\begin{equation}\label{paper_dddd_7}
		\includegraphics[width=0.36\textwidth]{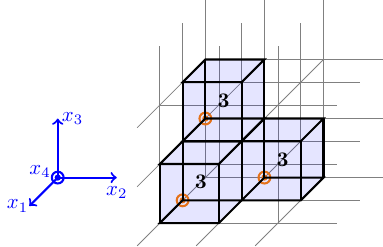}\ .
	\end{equation}
	\textbf{Step 1 layer 4}:
	\begin{equation}\label{paper_dddd_8}
		\includegraphics[width=0.36\textwidth]{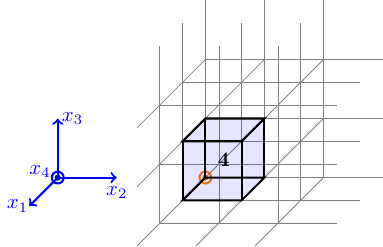}\ .
	\end{equation}
	Now we move on to step 2.\\ \textbf{Step 2} consists of $C_4^1$ mutually commutable parts, denoted by parts $\{j\} (j=1,2,3,4)$. Part $\{j\}$ is
	\begin{equation}\label{revise_13}
		\prod_{n=0}^{\sum_{i\neq j}(L_i-2)}\prod_{\substack{x_i\in\mathbb{Z}_{L_i-1},\,i\neq j\\\sum_{i\neq j}x_i=n}}\text{GCNOT}_{[-\frac{1}{2},x_i|i\neq j]}^{\delta_d[-\frac{1}{2},x_i+\frac{1}{2}|i\neq j]}\,,
	\end{equation}
	where $\left[-\frac12,x_i|i\neq j\right]$ stands for $\left[-\frac12,x_2,x_3,x_4\right]$ when $j=1$, $\left[x_1,-\frac12,x_3,x_4\right]$ when $j=2$, and similarly when $j=3,4$. $\delta_d\gamma_D$ is defined in Eq.~(\ref{revise_17}). Step 2 part $\{j\}$ adds all 4-cubes with $x_j$ but no other coordinate being $-\frac12$ into $\Omega$. In total, step 2 adds all the 4-cubes with exactly 1 coordinate being $-\frac12$ into $\Omega$. The state after step 2 is $\mathcal{F}(\Omega_1\cup\Omega_2)$, where $\Omega_2$ is the union of all 4-cubes with exactly 1 coordinate being $-\frac12$.
	
	Eq.~(\ref{revise_14}) (i.e., step 1) and Eq.~(\ref{revise_13}) (i.e., any part of step 2) have much similarity. In fact, if we visualize part $\{j\}$ of step 2, the same figures as Eqs.~(\ref{paper_dddd_5}, \ref{paper_dddd_6}, \ref{paper_dddd_7}, \ref{paper_dddd_8}) will be obtained, only with two changes:
	\begin{enumerate}
		\item Exchange the identity of $x_j,x_4$ (and $L_j,L_4$ simultaneously). Of course, if $j=4$, this exchange does nothing.
		\item The orange lines (i.e., the orange dots in planforms), together with the blue dye, extend between $[L_j-1,L_j]$, rather than $[0,L_j-1]$, i.e., step 2 part $\{j\}$ adds 4-cubes with $x_j=L_j-\frac{1}{2}$, rather than $x_j=\frac{1}{2},\frac{3}{2},\cdots,L_j-\frac{3}{2}$, into $\Omega$.
	\end{enumerate}
	The representative spin and its represented 4-cube pairs can be read from Eqs.~(\ref{revise_14})(\ref{revise_14}) as
	\begin{equation}
		\text{GCNOT}^{\delta_d\gamma_D}_{\mathcal{R}(\gamma_D)}\ ,\quad \gamma_D\subset\Omega_1\cup\Omega_2\,,
	\end{equation}
	just as in the X-cube case, where $\mathcal{R}(\gamma_D)$ refers to the representative spin of $\gamma_D$. Like the former circuits introduced for preparing Toric Code and X-cube model, the state after circuit $U_c$ is $\mathcal{F}(\Omega_1\cup\Omega_2)$. During the application of circuit $U_c$, all the representative spins have played the role of control qubits, 4-cubes not in $\Omega_1\cup\Omega_2$ do not have representative spins. So, for $\mathcal{F}(\Omega_1\cup\Omega_2)$, $|\zeta\rangle_{\gamma_1\nsubseteq\Omega_1\cup\Omega_2}=\bigotimes_{\gamma_1\nsubseteq\Omega_1\cup\Omega_2}|0\rangle_{\gamma_1}$. Therefore, $\mathcal{F}(\Omega_1\cup\Omega_2)$ can be written as
	\begin{equation}\label{paper_dddd_10}
		\begin{split}
			&\Bigg(\prod_{x_i\in\mathbb{Z}_{L_i-1},\,i=1,2,3,4}\frac{1}{\sqrt{2}}\Big(1+A_{[x_i+\frac{1}{2}|i=1:4]}\Big)\Bigg)\\
			&\Bigg(\prod_{j=1}^4\ \prod_{x_i\in\mathbb{Z}_{L_i-1},\,i\neq j}\frac{1}{\sqrt{2}}\Big(1+A_{[-\frac{1}{2},x_i+\frac{1}{2}|i\neq j]}\Big)\Bigg)|00\cdots0\rangle\,,
		\end{split}
	\end{equation}
	where $\left.\left[x_i+\frac12\right|i=1:4\right]$ stands for $\left[x_1+\frac12,\cdots,x_4+\frac12\right]$, $\left.\left[-\frac12,x_i\right|i\neq j\right]$ stands for $\left[-\frac12,x_2,x_3,x_4\right]$ when $j=1$, $\left[x_1,-\frac12,x_3,x_4\right]$ when $j=2$, and similarly when $j=3,4$. Next, we find a set of $A$ term redundancies, by using which we show the state $\mathcal{F}(\Omega_1\cup\Omega_2)$ is the EWSC on all 4-cubes in lattice. We can write the following $A$ term redundancy equivalence classes (RE-classes, in short) of $[0,1,2,4]$ model:
	\begin{align}
		&r^1_{[x_1+\frac{1}{2},x_2+\frac{1}{2},*,*]}:\ \prod_{x_i\in\mathbb{Z}_{L_i},\,i=3,4}A_{[x_i+\frac{1}{2}|i=1:4]}=1\,,\notag\\
		&r^1_{[x_1+\frac{1}{2},*,x_3+\frac{1}{2},*]}:\ \prod_{x_i\in\mathbb{Z}_{L_i},\,i=2,4}A_{[x_i+\frac{1}{2}|i=1:4]}=1\,,\notag\\
		&r^1_{[x_1+\frac{1}{2},*,*,x_4+\frac{1}{2}]}:\ \prod_{x_i\in\mathbb{Z}_{L_i},\,i=2,3}A_{[x_i+\frac{1}{2}|i=1:4]}=1\,,\notag\\
		&r^1_{[*,x_2+\frac{1}{2},x_3+\frac{1}{2},*]}:\ \prod_{x_i\in\mathbb{Z}_{L_i},\,i=1,4}A_{[x_i+\frac{1}{2}|i=1:4]}=1\,,\notag\\
		&r^1_{[*,x_2+\frac{1}{2},*,x_4+\frac{1}{2}]}:\ \prod_{x_i\in\mathbb{Z}_{L_i},\,i=1,3}A_{[x_i+\frac{1}{2}|i=1:4]}=1\,,\notag\\
		&r^1_{[*,*,x_3+\frac{1}{2},x_4+\frac{1}{2}]}:\ \prod_{x_i\in\mathbb{Z}_{L_i},\,i=1,2}A_{[x_i+\frac{1}{2}|i=1:4]}=1\,,\label{paper_dddd_9}
	\end{align}
	where $x_i\in\mathbb{Z}_{L_i},\,i=1,2,3,4$ when they are free indices, and the subscripts of $r^1$ are defined similarly as Eq.~(\ref{revise_31}), e.g.,
	\begin{align}
		&\left[*,x_2+\frac12,x_3+\frac12,*\right]\notag\\
		\equiv&\bigcup_{x_i\in\frac12\mathbb{Z},\,i=1,4}\left[x_1,x_2+\frac12,x_3+\frac12,x_4\right]\notag\\
		=&\bigcup_{x_i\in\mathbb{Z}_{L_i},\,i=1,4}\left[\left.x_i+\frac12\right|i=1:4\right]\,.
	\end{align} 
	The set-valued function with $*$ input is defined in Eq.~(\ref{revise_36}), with an intuitive explanation given in Fig.~\ref{*-intuition}. Here, the set $[*,x_2+\frac{1}{2},x_3+\frac12,*]$ is a manifold $T^2\times B^2$, which is extensive in $x_1,x_4$-directions, and spans between $[x_2,x_2+1],[x_3,x_3+1]$ in $x_2,x_3$-directions, respectively. There are in total $L_1L_2+L_1L_3+L_1L_4+L_2L_3+L_2L_4+L_3L_4$ RE-classes in Eq.~(\ref{paper_dddd_9}). All other $A$ term RE-classes of the $[0,1,2,4]$ model can be generated by multiplying RE-classes in Eq.~(\ref{paper_dddd_9}).
	
	We follow the strategy used in the X-cube case to show the state $\mathcal{F}(\Omega_1\cup\Omega_2)$ is the EWSC on all 4-cubes. First, we show all the 4-cubes with exactly 2 coordinates being $-\frac12$ can be added to $\Omega$ and then those with three coordinates being $-\frac{1}{2}$, and finally the one with four coordinates being $-\frac{1}{2}$.
	
	Similar to the X-cube case, denote the set of all 4-cubes as $\Gamma_4$ and denote the set of all 4-cubes without a representative spin by $R_4$. Note that all the 4-cubes appearing in Eq.~(\ref{paper_dddd_10}) form $\Gamma_4\backslash R_4$, so the state in Eq.~(\ref{paper_dddd_10}) can be written as
	\begin{equation}\label{paper_dddd_10_1}
		\mathcal{F}(\Omega_1\cup\Omega_2)=\prod_{\gamma_4\in\Gamma_4\backslash R_4}\frac{1}{\sqrt{2}}\Big(1+A_{\gamma_4}\Big)|00\cdots0\rangle\,.
	\end{equation}
	Without losing generality, consider a 4-cube $\gamma'_4$, with $x_3(\gamma'_4),x_4(\gamma'_4)=-\frac{1}{2}$, and $x_1(\gamma'_4)=x'_1+\frac{1}{2}\neq-\frac{1}{2},\ x_2(\gamma'_4)=x'_2+\frac{1}{2}\neq-\frac{1}{2}$, where $x_i(\gamma'_4)$ is the $x_i$-coordinate of $\gamma'_4$. Since $\gamma'_4$ is the only 4-cube, which is in $\left[x'_1+\frac{1}{2},x'_2+\frac{1}{2},*,*\right]$ and has 2 coordinates being $-\frac{1}{2}$, the $A$ term on $\gamma'_4$ is the only $A$ term that is involved in $r^1_{[x'_1+\frac{1}{2},x'_2+\frac{1}{2},*,*]}$, but not appearing in Eq.~(\ref{paper_dddd_10_1}), so based on the same reason as Eqs.~(\ref{paper_2dTC_3}, \ref{paper_2dTC_4}),
	\begin{equation}
		\begin{split}
			&\prod_{\gamma_4\in\Gamma_4\backslash R_4}\frac{1}{\sqrt{2}}\Big(1+A_{\gamma_4}\Big)|00\cdots0\rangle\\
			=&\frac{1}{2}\Big(1+A_{\gamma'_4}\Big)\prod_{\gamma_4\in\Gamma_4\backslash R_4}\frac{1}{\sqrt{2}}\Big(1+A_{\gamma_4}\Big)|00\cdots0\rangle\,,
		\end{split}
	\end{equation}
	or equivalently,
	\begin{equation}
		\mathcal{F}(\Omega_1\cup\Omega_2)=\mathcal{F}(\Omega_1\cup\Omega_2\cup\gamma'_4)\,.
	\end{equation}
	Similarly, other 4-cubes with exactly two coordinates being $-\frac{1}{2}$ can be directly added to $\Omega$, out of RE-classes in Eq.~(\ref{paper_dddd_9}). For convenience, denote the set of all $\gamma_4$ with exactly $d+i$ coordinates being $-\frac{1}{2}$ as $R_{4,i}$, $R_4=\bigcup_{i=1}^3 R_{4,i}$. With this notation, and the analysis above, Eq.~(\ref{paper_dddd_10_1}) equals to
	\begin{equation}\label{paper_dddd_10_2}
		\frac{1}{\sqrt{2}^{|R_{4,1}|}}\prod_{\gamma_4\in\Gamma_4\backslash \bigcup_{i=2}^3 R_{4,i}}\frac{1}{\sqrt{2}}\Big(1+A_{\gamma_4}\Big)|00\cdots0\rangle\,.
	\end{equation}
	Now we further show the 4-cubes with exactly 3 coordinates being $-\frac{1}{2}$, i.e., 4-cubes in $R_{4,2}$, can be directly added to $\Omega$, based on Eq.~(\ref{paper_dddd_10_2}). Without losing generality, consider a 4-cube $\gamma''_4$, with $x_2(\gamma''_4),x_3(\gamma''_4),x_4(\gamma''_4)=-\frac{1}{2}$, and $x_1(\gamma''_4)=x''_1+\frac{1}{2}\neq-\frac{1}{2}$. Since $\gamma''_4$ is the only 4-cube, which is in $[x''_1+\frac{1}{2},-\frac{1}{2},*,*]$ and has 3 coordinates being $-\frac{1}{2}$, the $A$ term on $\gamma''_4$ is the only $A$ term that is involved in $r^1_{[x''_1+\frac{1}{2},-\frac{1}{2},*,*]}$, but not appearing in Eq.~(\ref{paper_dddd_10_2}), so based on the same reason as Eqs.~(\ref{paper_2dTC_3}, \ref{paper_2dTC_4}),
	\begin{equation}\label{paper_dddd_11}
		\begin{split}
			&\prod_{\gamma_4\in\Gamma_4\backslash R_{4,1}}\frac{1}{\sqrt{2}}\Big(1+A_{\gamma_4}\Big)|00\cdots0\rangle\\
			=&\frac{1}{2}\Big(1+A_{\gamma''_4}\Big)\prod_{\gamma_4\in\Gamma_4\backslash R_{4,1}}\frac{1}{\sqrt{2}}\Big(1+A_{\gamma_4}\Big)|00\cdots0\rangle\,.
		\end{split}
	\end{equation}
	Similarly, other 4-cubes with exactly 3 coordinates being $-\frac{1}{2}$ can be directly added to $\Omega$, out of RE-classes in Eq.~(\ref{paper_dddd_9}). Therefore, Eq.~(\ref{paper_dddd_10_2}) equals to
	\begin{align}\label{revise_38}
		\frac{1}{\sqrt{2}^{|\bigcup_{i=1}^2 R_{4,i}|}}\prod_{\gamma_4\in\Gamma_4\backslash R_{4,3}}\frac{1}{\sqrt{2}}\Big(1+A_{\gamma_4}\Big)|00\cdots0\rangle\,.
	\end{align}
	The difference here from ``generating'' configurations on 4-cubes with 2 coordinates being $-\frac{1}{2}$ is: here 3 different RE-classes in Eq.~(\ref{paper_dddd_9}) can be used for each 4-cube with 3 coordinates being $-\frac{1}{2}$. For example, $r^1_{[x''_1+\frac{1}{2},-\frac{1}{2},*,*]},\ r^1_{[x''_1+\frac{1}{2},*,-\frac{1}{2},*]},\ r^1_{[x''_1+\frac{1}{2},*,*,-\frac{1}{2}]}$ can be used for the $\gamma''_4$ in Eq.~(\ref{paper_dddd_11}). 
	
	Now we further show the 4-cubes with exactly 4 coordinates being $-\frac{1}{2}$, i.e., the 4-cube in $R_{4,3}$, can be directly added to $\Omega$, based on Eq.~(\ref{paper_dddd_10_2}). The 4-cube with 4 coordinates being $-\frac{1}{2}$ is unique, denote it by $\gamma'''_4$ now. Since $\gamma'''_4$ is the only 4-cube, which is in $[-\frac{1}{2},-\frac{1}{2},*,*]$ and has 4 coordinates being $-\frac{1}{2}$, the $A$ term on $\gamma'''_4$ is the only $A$ term that is involved in $r^1_{[-\frac{1}{2},-\frac{1}{2},*,*]}$, but not appearing in Eq.~(\ref{revise_38}), so based on the same reason as Eqs.~(\ref{paper_2dTC_3}, \ref{paper_2dTC_4}),
	\begin{equation}\label{revise_15}
		\begin{split}
			&\prod_{\gamma_4\in\Gamma_4\backslash R_{4,2}}\frac{1}{\sqrt{2}}\Big(1+A_{\gamma_4}\Big)|00\cdots0\rangle\\
			=&\frac{1}{2}\Big(1+A_{\gamma'''_4}\Big)\prod_{\gamma_4\in\Gamma_4\backslash R_{4,2}}\frac{1}{\sqrt{2}}\Big(1+A_{\gamma_4}\Big)|00\cdots0\rangle\,.
		\end{split}
	\end{equation}
	Here $C_4^2=6$ different RE-classes in Eq.~(\ref{paper_dddd_9}) can be used for showing $\gamma'''_4$ can be directly added to $\Omega$, which are $r^1_{[-\frac{1}{2},-\frac{1}{2},*,*]},\ r^1_{[-\frac{1}{2},*,-\frac{1}{2},*]},\ r^1_{[-\frac{1}{2},*,*,-\frac{1}{2}]},\\ r^1_{[*,-\frac{1}{2},-\frac{1}{2},*]},\ r^1_{[*,-\frac{1}{2},*,-\frac{1}{2}]},\ r^1_{[*,*,-\frac{1}{2},-\frac{1}{2}]}$. On the right hand side of Eq.~(\ref{revise_15}), all the 4-cubes with at most 4 coordinates being $-\frac{1}{2}$, i.e., all 4-cubes, appear in the product, therefore, Eq.~(\ref{revise_15}) equals to
	\begin{equation}
		\frac{1}{\sqrt{2}^{|R_4|}}\prod_{\gamma_4\in\Gamma_4}\frac{1}{\sqrt{2}}\Big(1+A_{\gamma_4}\Big)|00\cdots0\rangle\,,
	\end{equation}
	which is a computational basis of the $[0,1,2,4]$ model code states. The factor $1/\sqrt{2}^{|R_4|}$ is obtained from Eqs.~(\ref{revise_38}, \ref{revise_15}), and $|R_4|=\left|\bigcup_{i=1}^3 R_{4,i}\right|=\left|\bigcup_{i=1}^2R_{4,i}\right|+1$. $|R_4|$ is a special case of Eq.~(\ref{general_R_D_order}), and can be represented in $L_i$ by using Eq.~(\ref{general_R_D_order}).

	\subsection{$[1,2,3,3]$ model}\label{section_3dTC_sqc}
	
	As illustrated in Sec.~\ref{section_TD_models}, the $[1,2,3,3]$ model is 3-dimensional Toric Code (on cubic lattice). In this subsection, we construct a circuit $U_c$ for preparing computational basis of the $[1,2,3,3]$ model code states, where $d=2,D=3$. We start from graphic illustration, then algebraically write the circuit $U_c$, find a stabilizer redundancy, by using which we show the prepared state is the EWSC on all cubes, and thus a code state. Though the graphic illustration of circuit for $[1,2,3,3]$ model is much easier than that of $[0,1,2,4]$ model, the algebraic description for $[1,2,3,3]$ model requires more sophisticated skills, which is why we put $[1,2,3,3]$ model after $[0,1,2,4]$ model. 
	
	Let us start with the graphic illustration. Consider a $3\times3\times2$ cubic lattice under PBC, where each plaquette has a $\frac{1}{2}$-spin on it. Prepare the system in product state $|00\cdots0\rangle$. Then, apply Hadamard gates to each spin on orange plaquettes shown below:
	\begin{equation}\label{revise_20}
		\includegraphics[width=0.3276\textwidth]{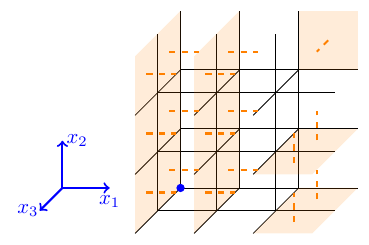}\ ,
	\end{equation}
	after which the spins on orange plaquettes are in state $|+\rangle=|X=1\rangle$. The spins on those orange plaquettes are the representative spins in the overall strategy, and the dashed orange lines link the representative spins with their represented cubes. The blue dot stands for the original poing $(0,0,0)$. From now on, we use blank plaquettes to represent spins in state $|0\rangle$, and orange plaquettes to represent spins in state $|+\rangle$ in this subsection, so that Eq.~(\ref{revise_20}) represents the state after step 0 (i.e., Hadamard gates on all representative spins). After step 0, apply $d+1=3$ non-commutable steps of CNOT gates, namely, step 1, step 2 and step 3, which shall be applied with the order 1,2,3. Denote
	\begin{equation}\label{revise_16_a}
		\includegraphics[width=0.07742\textwidth]{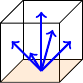}
	\end{equation}
	by
	\begin{equation}\label{revise_16_b}
		\includegraphics[width=0.07742\textwidth]{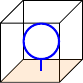}\ .
	\end{equation}
	In Eq.~(\ref{revise_16_a}), each arrow represents a CNOT gate, with the tail being the control qubit, and the head being the target qubit. Eq.~(\ref{revise_16_b}) is just a symbolic simplification of the 5 CNOT gates. Steps 1,2,3 are:
	\begin{itemize}
		\item \textbf{Step 1}:
		\begin{equation}\label{revise_21}
			\includegraphics[width=0.3276\textwidth]{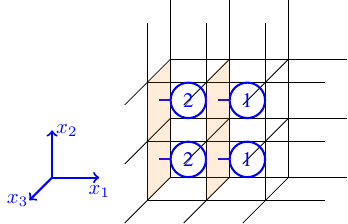}\ .
		\end{equation}
		Step 1 adds all cubes with no coordinate being $-\frac12$ into $\Omega$ (see the convention below Eq.~(\ref{revise_39})). The state after step 1 is $\mathcal{F}(\Omega_1)$, where $\Omega_1$ is the union of all cubes with no coordinate being $-\frac12$. As before, the numbers in the figure refer to different layers. We do not draw the original point here for visual clearness.
		\item \textbf{Step 2} consists of $C_3^1=3$ mutually commutable parts, denoted by parts $\{1\},\{2\},\{3\}$. 
		\begin{itemize}
			\item \textbf{Step 2 part \{1\}}:
			\begin{equation}\label{revise_22}
				\includegraphics[width=0.3276\textwidth]{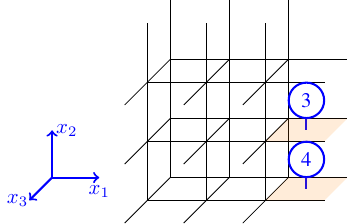}\ .
			\end{equation}
			Step 2 part $\{1\}$ adds all cubes with $x_1$ but no other coordinate being $-\frac12$ into $\Omega$.
			\item \textbf{Step 2 part $\{2\}$}:
			\begin{equation}\label{revise_23}
				\includegraphics[width=0.3276\textwidth]{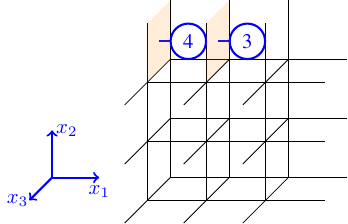}\ .
			\end{equation}
			Step 2 part $\{1\}$ adds all cubes with $x_1$ but no other coordinate being $-\frac12$ into $\Omega$.
			\item \textbf{Step 2 part $\{3\}$}:
			\begin{equation}\label{revise_24}
				\includegraphics[width=0.3276\textwidth]{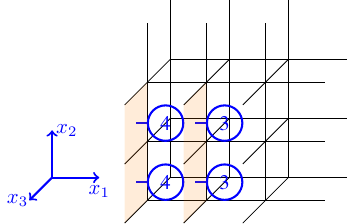}\ .
			\end{equation}
			Step 2 part $\{1\}$ adds all cubes with $x_1$ but no other coordinate being $-\frac12$ into $\Omega$.
		\end{itemize}
		In total, Step 2 adds all cubes with exactly 1 coordinate being $-\frac12$ into $\Omega$. The state after step 2 is $\mathcal{F}(\Omega_1\cup\Omega_2)$, where $\Omega_2$ is the union of all cubes with exactly 1 coordinate being $-\frac12$.
		\item \textbf{Step 3} consists of $C_3^2=3$ mutually commutable parts, denoted by part $\{1,2\}$, part $\{1,3\}$ and part $\{2,3\}$.
		\begin{itemize}
			\item \textbf{Step 3 part $\{1,2\}$}:
			\begin{equation}\label{revise_25}
				\includegraphics[width=0.3276\textwidth]{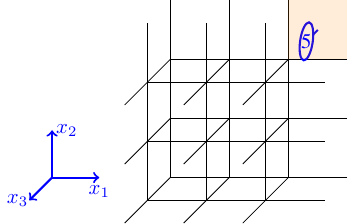}\ .
			\end{equation}
			Step 3 part $\{1,2\}$ adds all cubes with $x_1,x_2$ but no other coordinate being $-\frac12$ into $\Omega$.
			\item \textbf{Step 3 part $\{1,3\}$}:
			\begin{equation}\label{revise_26}
				\includegraphics[width=0.3276\textwidth]{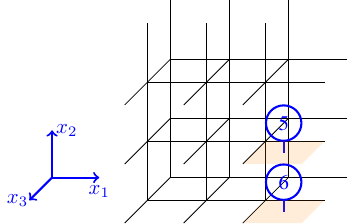}\ .
			\end{equation}
			Step 3 part $\{1,3\}$ adds all cubes with $x_1,x_3$ but no other coordinate being $-\frac12$ into $\Omega$.
			\item \textbf{Step 3 part $\{2,3\}$}:
			\begin{equation}\label{revise_27}
				\includegraphics[width=0.3276\textwidth]{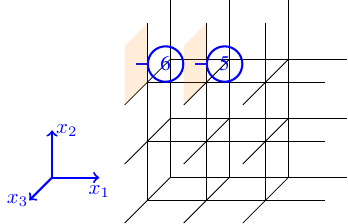}\ .
			\end{equation}
			Step 3 part $\{2,3\}$ adds all cubes with $x_2,x_3$ but no other coordinate being $-\frac12$ into $\Omega$.
		\end{itemize}
		In total, step 3 adds all cubes with exactly 2 coordinates being $-\frac12$ into $\Omega$. The state after step 3 is $\mathcal{F}(\Omega_1\cup\Omega_2\cup\Omega_3)$.
	\end{itemize}
	The three steps prepare the EWSC on all cubes, except the red cube $c_1$ shown below:
	\begin{equation}\label{paper_dddd_19}
		\includegraphics[width=0.3276\textwidth]{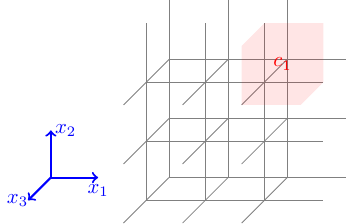}\ .
	\end{equation}
	Denote the set of all cubes by $\Gamma_3$, and the set of the cube without a representative spin as $R_3$. $R_3=\{c_1\}$ under the setting of Eq.~(\ref{paper_dddd_19}). The state after applying steps 1,2,3 is
	\begin{equation}\label{revise_46}
		\prod_{c\in\Gamma_3\backslash R_3}\frac{1}{\sqrt{2}}\Big(1+A_{c}\Big)|00\cdots0\rangle\,.
	\end{equation}
	Now we describe the circuit $U_c$ algebraically, and discuss why the state in Eq.~(\ref{revise_46}) is the EWSC on all cubes, and thus a code state.
	
	To write a form convenient for generalizing to any stabilizer code TD models, it is worthy to introduce some notations here. Denote
	\begin{equation}\label{revise_34}
		S_k\subset\{1,2,3\}
	\end{equation}
	as a $k$-element subset of $\{1,2,3\}$. The only $S_0$ is $\emptyset$, and there are $C_3^1=3$ possible values of $S_1$, i.e., $\{1\},\{2\},\{3\}$, and there are $C_3^2$ possible values of $S_2$, i.e., $\{1,2\},\{1,3\},\{2,3\}$. Denote the complement set ($\{1,2,3\}$ being the total set) of $S_k$ as $S_k^c:=\{1,2,3\}\backslash S_k$. For example, when $S_1=\{2\}$, $S_1^c=\{1,3\}$. With these notations, the state after step 0 can be written as
	\begin{align}
		&\Bigg(\prod_{x_i\in\mathbb{Z}_{L_i-1},\,i\in S_0^c}\text{Had}_{f\left([x_i+\frac{1}{2}|i\in S_0^c],S_0\right)}\Bigg)\notag\\
		&\Bigg(\prod_{S_1}\ \prod_{x_i\in\mathbb{Z}_{L_i-1},\,i\in S_1^c}\text{Had}_{f\left([-\frac{1}{2},x_i+\frac{1}{2}|i\in S_1^c],S_1\right)}\Bigg)\notag\\
		&\Bigg(\prod_{S_2}\ \prod_{x_i\in\mathbb{Z}_{L_i-1},\,i\in S_2^c}\text{Had}_{f\left([-\frac{1}{2},-\frac{1}{2},x_i+\frac{1}{2}|i\in S_2^c],S_2\right)}\Bigg)|00\cdots0\rangle\,,
	\end{align}
	where $f\big(\gamma_3,S_k\big)$ is the following map
	\begin{align}\label{equation_map_f}
		f\big(\gamma_3,S_k\big)=\big[x_i(\gamma_3)-x'_i\big|i=1:3\big]\,,\  x'_i=\left\{\begin{array}{l}
			\frac{1}{2},\ i=\text{min}(S_k^c)  \\
			0,\ i\neq\text{min}(S_k^c) 
		\end{array}\right.
	\end{align}
	$\text{min}(S_k^c)$ is the smallest element in $S_k^c$ in common sense.
	\begin{itemize}
		\item \textbf{Step 1}:
		\begin{align}
			\prod_{n=0}^{L_1-2}\,\prod_{\substack{x_i\in\mathbb{Z}_{L_i-1},\,i\in S_0^c\\x_1=n}}\text{GCNOT}_{f\left([x_i+\frac{1}{2}|i\in S_0^c],S_0\right)}^{\delta_d[x_i+\frac{1}{2}|i\in S_0^c]}\,,
		\end{align}
		where $\delta_d[x_i+\frac{1}{2}|i\in S_0^c]$ is defined in Eq.~(\ref{revise_17}). \\Step 1 adds all cubes with no coordinate being $-\frac12$ into $\Omega$.
		\item \textbf{Step 2}:
		\begin{itemize}
			\item \textbf{Step 2 part $\{1\}$}:
			\begin{equation}
				\prod_{n=0}^{L_2-2}\ \prod_{\substack{x_i\in\mathbb{Z}_{L_i-1},\,i\in \{1\}^c\\x_2=n}}\text{GCNOT}_{f\left([-\frac{1}{2},x_i+\frac{1}{2}|i\in \{1\}^c],\{1\}\right)}^{\delta_d[-\frac{1}{2},x_i+\frac{1}{2}|i\in \{1\}^c]}\,.
			\end{equation}
			Step 2 part $\{1\}$ adds all cubes with $x_1$ but no other coordinate being $-\frac12$ into $\Omega$.
			\item \textbf{Step 2 part $\{2\}$}:
			\begin{equation}
				\prod_{n=0}^{L_1-2}\ \prod_{\substack{x_i\in\mathbb{Z}_{L_i-1},\,i\in \{2\}^c\\x_1=n}}\text{GCNOT}_{f\left([-\frac{1}{2},x_i+\frac{1}{2}|i\in \{2\}^c],\{2\}\right)}^{\delta_d[-\frac{1}{2},x_i+\frac{1}{2}|i\in \{2\}^c]}\,.
			\end{equation}
			Step 2 part $\{2\}$ adds all cubes with $x_2$ but no other coordinate being $-\frac12$ into $\Omega$.
			\item \textbf{Step 2 part $\{3\}$}:
			\begin{equation}
				\prod_{n=0}^{L_1-2}\ \prod_{\substack{x_i\in\mathbb{Z}_{L_i-1},\,i\in \{3\}^c\\x_1=n}}\text{GCNOT}_{f\left([-\frac{1}{2},x_i+\frac{1}{2}|i\in \{3\}^c],\{3\}\right)}^{\delta_d[-\frac{1}{2},x_i+\frac{1}{2}|i\in \{3\}^c]}\,.
			\end{equation}
			Step 2 part $\{3\}$ adds all cubes with $x_3$ but no other coordinate being $-\frac12$ into $\Omega$.
		\end{itemize} 
		In total, step 2 adds all cubes with exactly 1 coordinate being $-\frac12$ into $\Omega$.
		\item \textbf{Step 3}:
		\begin{itemize}
			\item \textbf{Step 3 part $\{1,2\}$}:
			\begin{equation}
				\prod_{n=0}^{L_3-2}\ \prod_{\substack{x_i\in\mathbb{Z}_{L_i-1},\,i\in\{1,2\}^c\\x_3=n}}\text{GCNOT}_{f\left([-\frac{1}{2},-\frac{1}{2},x_i+\frac{1}{2}|i\in \{1,2\}^c],\{1,2\}\right)}^{\delta_d[-\frac{1}{2},-\frac{1}{2},x_i+\frac{1}{2}|i\in \{1,2\}^c]}\,.
			\end{equation}
			Step 3 part $\{1,2\}$ adds all cubes with $x_1,x_2$ but no other coordinate being $-\frac12$ into $\Omega$.
			\item \textbf{Step 3 part $\{1,3\}$}:
			\begin{equation}
				\prod_{n=0}^{L_2-2}\ \prod_{\substack{x_i\in\mathbb{Z}_{L_i-1},\,i\in\{1,3\}^c\\x_2=n}}\text{GCNOT}_{f\left([-\frac{1}{2},-\frac{1}{2},x_i+\frac{1}{2}|i\in\{1,3\}^c],\{1,3\}\right)}^{\delta_d[-\frac{1}{2},-\frac{1}{2},x_i+\frac{1}{2}|i\in\{1,3\}^c]}\,.
			\end{equation}
			Step 3 part $\{1,3\}$ adds all cubes with $x_1,x_3$ but no other coordinate being $-\frac12$ into $\Omega$.
			\item \textbf{Step 3 part $\{2,3\}$}:
			\begin{equation}
				\prod_{n=0}^{L_1-2}\ \prod_{\substack{x_i\in\mathbb{Z}_{L_i-1},\,i\in\{2,3\}^c\\x_1=n}}\text{GCNOT}_{f\left([-\frac{1}{2},-\frac{1}{2},x_i+\frac{1}{2}|i\in\{2,3\}^c],\{2,3\}\right)}^{\delta_d[-\frac{1}{2},-\frac{1}{2},x_i+\frac{1}{2}|i\in\{2,3\}^c]}\,.
			\end{equation}
			Step 3 part $\{2,3\}$ adds all cubes with $x_2,x_3$ but no other coordinate being $-\frac12$ into $\Omega$.
		\end{itemize}
		In total, step 3 adds all cubes with exactly 2 coordinates being $-\frac12$ into $\Omega$.
	\end{itemize}
	As illustrated in Eq.~(\ref{paper_dddd_19}), the output state $\mathcal{F}(\Omega_1\cup\Omega_2\cup\Omega_3)$ after circuit $U_c$ is the EWSC on all cubes except $c_1$. There is no spin in $c_1$ that is still in product state, no representative spin can be chosen for $c_1$. There is a unique $A$ term RE-class in three-dimensional Toric Code model, i.e., the product of all $A$ terms equal to 1. Apparently, based on the same reason as Eqs.~(\ref{paper_2dTC_3},\ref{paper_2dTC_4}), the output state $\mathcal{F}(\Omega_1\cup\Omega_2\cup\Omega_3)$ is the EWSC on all cubes in lattice. Till here we shall have accumulated enough intuition, conventions and notations to go for a circuit for all stabilizer code TD models, unifying the examples introduced in this section before.

	\subsection{General quantum circuits for computational basis of the code statess of all TD models}\label{section_unified_SQC}
	
	In this subsection, we provide a LU circuit $U_c$ for preparing computational basis of the code states of general stabilizer code TD model, which is a natural generalization of the circuit introduced in Secs.~\ref{section_toric_code_sqc},\ref{section_x-cube_sqc},\ref{section_0124_sqc},\ref{section_3dTC_sqc}. After writing down the circuit, we discuss the truncation of the circuit $U_c$: while this circuit is constructed for preparing TD states under PBC, by truncating (i.e., not applying specific parts/steps) $U_c$, circuits for preparing TD states under OBC or hybrid half-OBC-half-PBC can be obtained. Then we calculate the number of layers for this circuit, and compare it with the so called long range entanglement (LRE) level introduced in Ref.~\cite{dddd_ERG}. It turns out the LRE level appears as a factor in the number of layers, no matter under PBC or OBC.
	
	Consider an $L_1\times L_2\times \cdots\times L_D$ $D$-dimensional cubic lattice under PBC, with each $d$-cube having a $\frac{1}{2}$-spin on it. The whole preparing process is: First preparing the initial state $|00\cdots0\rangle$, then apply Hadamard gates (namely, step 0) to all the representative spins, then apply $d+1$ non-commutable steps (namely, steps $1,2,\cdots,d+1$) of CNOT gates. For any $k=0,1,\cdots,d$, step $k+1$ consists of $C_D^k$ mutually commutable parts, with each part labeled by an order-$k$ subset 
	\begin{equation}
		S_k\subset\{1:D\}
	\end{equation}
	of the total set $\{1:D\}=\{1,2,\cdots,D\}$. As the circuit introduced in Secs.~\ref{section_toric_code_sqc},\ref{section_x-cube_sqc},\ref{section_0124_sqc},\ref{section_3dTC_sqc}, step $k+1$ part $S_k$ adds $D$-cubes with all and only indices within $S_k$ being $-\frac{1}{2}$ into $\Omega$. In total, step $k+1$ adds all $D$-cubes with exactly $k$ coordinates being $-\frac{1}{2}$ into $\Omega$. For any $k=0:d$, the state after step $k+1$ is $\mathcal{E}\left(\bigcup_{i=1}^{k+1}\Omega_i\right)$, where $\Omega_i$ is the union of all $D$-cubes with exactly $i-1$ coordinates being $-\frac12$. To express the set of all representative spins, the CNOT gates in steps $1,2,\cdots,d+1$, as well as the result state, we define the following notations:
	\begin{itemize} 
		\item Denote the complement set of $S_k$ by $S_k^c:=\{1:D\}\backslash S_k$.
		\item Denote $D^*:=D-d$.
		\item Denote the depth of step $k+1$ part $S_k$ as
		\begin{equation}
			\text{depth}(S_k)=\sum_{i\in\text{min}_{D^*}(S_k^c)}(L_i-2)+1\,,
		\end{equation}
		where $\text{min}_{D^*}(S_k^c)$ is the set of $D^*$ minimal elements of $S_k^c$.
		\item Define
		\begin{widetext}
			\begin{gather}
				\label{revise_18}\Gamma_D(k+1,S_k,\alpha):=\Bigg\{\Big[-\frac{1}{2},\cdots,-\frac{1}{2},x_i+\frac{1}{2}\Big|i\in S_k^c\Big]\Bigg|\Big(x_i\in\mathbb{Z}_{L_i-1},\,i\in S_k^c\Big)\ \ \&\sum_{i\in\text{min}_{D^*}(S_k^c)}x_i=\text{depth}(S_k)-\alpha\Bigg\}\,,\\
				f(\gamma_D,S_k)=\left[x_1(\gamma_D)-x'_1,\cdots,x_D(\gamma_D)-x'_D\right]\ ,\quad x'_i=\left\{\begin{array}{ll}
					\frac{1}{2}\ ,&i\in\text{min}_{D^*}(S_k^c)\\
					0\ ,&i\notin\text{min}_{D^*}(S_k^c)
				\end{array}\right.\,,\label{any_GS_22}
			\end{gather}
		\end{widetext}
		where $\Gamma_D(k+1,S_k,\alpha)$ is the set of $D$-cubes, which are added to $\Omega$ (see the convention below Eq.~(\ref{revise_39})) by the layer $\alpha$ of part $S_k$ of step $k+1$. On the other hand, $f(\gamma_D,S_k)$ is the representative spin of the $D$-cube $\gamma_D$, given
		\begin{equation}
			\forall i\in S_k,\ x_i(\gamma_D)=-\frac12\quad\&\quad\forall j\notin S_k,\ x_j(\gamma_D)\neq-\frac12\,.
		\end{equation}
		Note that $f$ is dependent on $d,D$.
	\end{itemize}
	With these notations, we can write the set of all representative spins as
	\begin{equation}
		\bigcup_{k=0}^d\bigcup_{S_k}\bigcup_{\alpha=1}^{\text{depth}(S_k)}\bigcup_{\gamma_D\in\Gamma_D(k+1,S_k,\alpha)}\Big\{f(\gamma_D,S_k)\Big\}\,,
	\end{equation}
	so after step 0, i.e., Hadamard gates on all representative spins, the state becomes
	\begin{equation}
		\prod_{k=0}^d \prod_{S_k} \prod_{\alpha=1}^{\text{depth}(S_k)}\prod_{\gamma_D\in\Gamma_D(k+1,S_k,\alpha)}\text{Had}_{f(\gamma_D,S_k)}|00\cdots0\rangle\,,
	\end{equation}
	where $S_k$ is summed over all possible order-$k$ subset of $\{1:D\}$. After step 0, apply steps $1,2,\cdots,d+1$ in order, where the layer $\alpha$ of part $S_k$ of step $k+1$ is
	\begin{gather}
		\prod_{\gamma_D\in\Gamma_D(k+1,S_k,\alpha)}\text{GCNOT}_{f(\gamma_D,S_k)}^{\delta_d\gamma_D}\label{paper_dddd_14}\,.
	\end{gather}
	Within each part, apply layers with the order $1,2,\cdots,\text{depth}(S_k)$, and different parts within the same step can be implemented simultaneously. After all parts of a step are finished, apply the next step. For any $k=0,1,\cdots,d$, step $k+1$ adds all $D$-cubes with exactly $k$ coordinates being $-\frac12$ into $\Omega$, so after step $d+1$ is finished, the result state is
	\begin{widetext}
		\begin{equation}\label{revise_32}
			\prod_{k=0}^d \prod_{S_k} \prod_{x_i\in\mathbb{Z}_{L_i-1},\,i\in S_k^c}\frac{1}{\sqrt{2}}\Big(1+A_{[-\frac{1}{2},\cdots,-\frac{1}{2},x_i+\frac{1}{2}|i\in S_k^c]}\Big)\Big|00\cdots0\Big\rangle\,,
		\end{equation}
	\end{widetext}
	where $\left[\left.-\frac12,\cdots,-\frac12,x_i+\frac12\right|i\in S_k^c\right]$ is the $D$-cube with the $i$-th coordinate being $-\frac12$ if $i\notin S_k^c$ and $x_i+\frac12$ if $i\in S_k^c$. Because of the existence of $A$ term redundancies, the state in Eq.~(\ref{revise_32}) equals to
	\begin{equation}\label{revise_33}
		\frac{1}{\sqrt{2}^{|R_D|}}\prod_{\gamma_D}\big(1+A_{\gamma_D}\big)|00\cdots0\rangle\,,
	\end{equation}
	where $R_D$ is the set of $D$-cubes without a representative spin.
	
	$|R_D|$ can be explicitly calculated in terms of $L_i$, as follows. The number of $D$-cubes with all and only $x_{j_1},\cdots,x_{j_p}$-coordinates being $-\frac{1}{2}$ is 
	\begin{equation}
		\prod_{i\in\{j_1,\cdots,j_p\}^c}(L_i-1)\,,\label{any_GS_28}
	\end{equation}
	so, denoting an order-$p$ subset of $\{1:D\}$ as $S_p$ (like $S_k$ defined earlier),
	\begin{equation}\label{general_R_D_order}
		|R_D| = \sum_{p=d+1}^D \sum_{S_p}\prod_{i\in S_p^c}(L_i-1) = \sum_{p=0}^{D^*-1}\sum_{S_p}\prod_{i\in S_p}(L_i-1)\,,
	\end{equation}
	where $\sum_{S_p}$ sums over all possible order-$p$ subsets of $\{1:D\}$, $S_0=\emptyset$, $\prod_{i\in S_0}(L_i-1)=1$. $|R_D|$ also equals to the number of independent $A$ term redundancies.
	
	Now we discuss how to truncate the circuit to obtain the circuit for OBC or half-OBC-half-PBC. To obtain a circuit for OBC, the truncation is straightforward: step 0 and step 1 create a TD state under OBC. For half-OBC-half-PBC, suppose among $D$ directions, $x_{j_1},\cdots,x_{j_n}$-directions are under OBC, and the rest directions are under PBC, then truncating all parts $S_k$ of the circuit satisfying 
	\begin{align}
		S_k\cap\{j_1,\cdots,j_n\}\neq\emptyset\label{paper_dddd_29_two}
	\end{align}
	yields a circuit for this half-OBC-half-PBC. Here truncating a part simply means not applying the part. Note that $S_k$ is enough to specify a part \textit{and} the step it belongs to: one can count the order of $S_k$ to obtain the step to which the part belongs. As an example of truncated circuit, consider 2-dimensional Toric Code, with $x_1$ under OBC, $x_2$ under PBC, then according to Eq.~(\ref{paper_dddd_29_two}), only part $\{1\}$ is truncated, the truncated circuit creates a 2-dimensional Toric Code with $x_1$ under OBC, $x_2$ under PBC. However, it is worth noting that the truncated circuit for half-OBC-half-PBC may have easy method to reduce depth, for instance, in the 2-dimensional Toric Code truncated circuit just mentioned, step 1 and the untruncated step 2 part $\{2\}$ are commutable, thus can be applied simultaneously.
	
	If we set $L_1=L_2=\cdots=L_D=L$, then for any part $S_k$,
	\begin{equation}
		\text{depth}(S_k)=D^*(L-2)+1=(D-d)(L-2)+1\,,
	\end{equation}
	so the total depth (excluding step 0) of this circuit for PBC is
	\begin{equation}\label{revise_19}
		[(D-d)(L-2)+1](d+1)\,,
	\end{equation}
	while for OBC, the total depth (excluding step 0) is
	\begin{equation}
		(D-d)(L-2)+1\,.
	\end{equation} 
	Whether PBC or OBC is applied, the total depth of circuit is proportional to $D-d$, which suggests the long range entanglement (LRE) complexity grows with $D-d$. The factor $D-d$ coincides with the so called LRE level introduced in Ref.~\cite{dddd_ERG}, which is also a quantity that was argued to quantify LRE complexity. The factor $d+1$, which is the number of extensive directions of manifolds that support a simple $A$ term redundancy class under PBC, implies topological order under PBC, where non-trivial GSD exist, is intrinsically harder to prepare than under OBC. For the PBC case, a natural question after constructing the circuit $U_c$ for preparing computational basis code states is: can we intentionally prepare arbitrary code state? In Sec.~\ref{section_any_GS}, we try to answer this question.

	\section{Preparation of arbitrary code state}\label{section_any_GS}
	
	In Sec.~\ref{section_SQC}, we proposed a circuit $U_c$ to prepare computational basis of the code states of general stabilizer code TD model, which, although being a long-range entangled state, is a specific product state of logical qubits $|00\cdots0\rangle_{\text{logic}}$. In this section, we propose a scheme, namely, \emph{seed $\to$ logical qubit} scheme, to prepare arbitrary code state, including magic states and states with entangled logical qubits, of $[0,1,2,2],[0,1,2,3],[1,2,3,3]$ models under PBC, basing on the circuit introduced in Sec.~\ref{section_SQC}. As any $[d-1,d,d+1,D]$ model share a common property --- logical Pauli $X$ operators are supported on $d$-dimensional torus under PBC, the method should be generalizable to any $[d-1,d,d+1,D]$ model. Furthermore, inspired by the circuit in this section, we make a guess for the GSD of $[d-1,d,d+1,D]$ models, which reduces to the GSD of $[0,1,2,D]$ and $[D-3,D-2,D-1,D]$~\cite{li2021fracton} when $d=1$ and $d=D-2$ are taken, respectively. The circuit presented in this section is largely based on $U_c$ introduced in Sec.~\ref{section_SQC}, only with a \emph{pre-step} added before $U_c$.
	
	Roughly speaking, the pre-step consists of two procedures: (1) choosing a \textit{seed}, which is a spin/physical qubit, for each logical qubit, and use the seed-entangler to prepare seeds into any desired state $|\psi\rangle_{\text{seeds}}$; (2) applying a set of CNOT gates to make each seed grow to the corresponding logical qubit, but without topological protection since the topological order is not yet prepared. After the pre-step, apply the circuit $U_c$ introduced in Sec.~\ref{section_SQC}, then the TD state $|\psi\rangle_{\text{logic}}$ is prepared. It will be shown that the seed qubits and logical qubits are in the same state, e.g., if two seed qubits are in a Bell state, after finishing the circuit, the two corresponding logical qubits will also be in the same Bell state.
	As in Sec.~\ref{section_SQC}, we use $d$ to represent the dimension of cubic cell where we put a spin, and $D$ to represent the dimension of lattice in this section.

	\subsection{$[0,1,2,2]$ model}\label{section_any_GS_TC}
	
	As stated above, first we choose a seed for each logical qubit. Toric Code has two logical qubits under PBC, and we choose the two seeds as the two spins on the wavy orange edges in the following figure:
	\begin{equation}
		\begin{tikzpicture}
			\foreach \i in {0,1,2}{\draw (\i,0)--(\i,3);}
			\foreach \j in {0,1,2}{\draw (0,\j)--(3,\j);}
			\draw[decorate, decoration={snake, amplitude=0.3mm, segment length=1mm}, orange, line width=1pt] 
			(2,2) -- (3,2)
			(2,2) -- (2,3);
			\node[orange] at (1.85,2.5) {1};
			\node[orange] at (2.5,1.8) {2};
			\draw[->, blue, line width=1pt] (-1.3,0)--(-0.5,0);
			\draw[->, blue, line width=1pt] (-1.3,0)--(-1.3,0.8);
			\node[blue] at (-0.3,0) {$x_2$};
			\node[blue] at (-1.3,0.94) {$x_1$};
			
			\fill[blue] (0,0) circle (2pt);
		\end{tikzpicture}\ .
	\end{equation}
	We denote the two seeds as seed 1 and seed 2, as shown above. The blue dot stands for the original point $(0,0)$. Next, use the seed-entangler\footnote{For the $[0,1,2,2]$ model, a local unitary acting on seeds 1,2 is enough to be the general seed-entangler.} to prepare seeds 1,2 in state
	\begin{align}
		&a_{00}|00\rangle_{12}+a_{01}|01\rangle_{12}+a_{10}|10\rangle_{12}+a_{11}|11\rangle_{12}\notag\\
		=&\Big(a_{00}+a_{01}X_2+a_{10}X_1+a_{11}X_1X_2\Big)|00\rangle_{12}\,,\label{any_GS_1}
	\end{align}
	while the other spins still being in state $|0\rangle$, disentangled with seeds 1,2. $|ij\rangle_{12}$ is the abbreviation of $|i\rangle_1\otimes|j\rangle_2$. Using black and red edges to represent spins in states $|0\rangle$ and $|1\rangle$ respectively, denote the state in Eq.~(\ref{any_GS_1}) as
	\begin{equation}
		a_{00}\begin{tikzpicture}[baseline=4ex]
			\foreach \i in {0,0.5,1}{\draw (\i,0)--(\i,1.5);}
			\foreach \j in {0,0.5,1}{\draw (0,\j)--(1.5,\j);}
		\end{tikzpicture} + 
		a_{01}\begin{tikzpicture}[baseline=4ex]
			\foreach \i in {0,0.5,1}{\draw (\i,0)--(\i,1.5);}
			\foreach \j in {0,0.5,1}{\draw (0,\j)--(1.5,\j);}
			\draw[red, line width=1pt] (1,1)--(1.5,1);
		\end{tikzpicture} + 
		a_{10}\begin{tikzpicture}[baseline=4ex]
			\foreach \i in {0,0.5,1}{\draw (\i,0)--(\i,1.5);}
			\foreach \j in {0,0.5,1}{\draw (0,\j)--(1.5,\j);}
			\draw[red, line width=1pt] (1,1)--(1,1.5);
		\end{tikzpicture} + 
		a_{11}\begin{tikzpicture}[baseline=4ex]
			\foreach \i in {0,0.5,1}{\draw (\i,0)--(\i,1.5);}
			\foreach \j in {0,0.5,1}{\draw (0,\j)--(1.5,\j);}
			\draw[red, line width=1pt] (1,1)--(1.5,1);
			\draw[red, line width=1pt] (1,1)--(1,1.5);
		\end{tikzpicture}\ .\label{any_GS_2}
	\end{equation}
	Next, apply the following CNOT sequence (namely, $U_g$, making the seeds grow to their corresponding logical qubits) to the state in Eq.~(\ref{any_GS_2}):
	\begin{equation}
		U_g=\begin{tikzpicture}[baseline=8ex]
			\foreach \i in {0,1,2}{\draw (\i,0)--(\i,3);}
			\foreach \j in {0,1,2}{\draw (0,\j)--(3,\j);}
			\foreach \j in {2.45,1.45}{
				\draw[blue, line width=1pt, ->] (2.05,\j)--(2.05,\j-0.9);}
			\foreach \i in {2.45,1.45}{
				\draw[blue, line width=1pt, ->]
				(\i,2.05)--(\i-0.9,2.05);}
			\node[blue] at (2.15,2.2) {1};
			\node[blue] at (1.15,2.2) {2};
			\node[blue] at (2.17,1.15) {2};
			
			\draw[blue, ->, line width=1pt] (-1.3,0)--(-0.5,0);
			\draw[blue, ->, line width=1pt] (-1.3,0)--(-1.3,0.8);
			\node[blue] at (-0.3,0) {$x_2$};
			\node[blue] at (-1.3,0.95) {$x_1$};
			
			\fill[blue] (0,0) circle (2pt);
		\end{tikzpicture}\ ,\label{any_GS_3}
	\end{equation}
	where each blue arrow stands for a CNOT gate with the tail being the control qubit, and the head being the target qubit, as before. The numbers in the figure are the label of layers. Layer with smaller number should be applied first. After the CNOT sequence $U_g$ applied, the state becomes
	\begin{equation}
		a_{00}\begin{tikzpicture}[baseline=4ex]
			\foreach \i in {0,0.5,1}{\draw (\i,0)--(\i,1.5);}
			\foreach \j in {0,0.5,1}{\draw (0,\j)--(1.5,\j);}
		\end{tikzpicture} + 
		a_{01}\begin{tikzpicture}[baseline=4ex]
			\foreach \i in {0,0.5,1}{\draw (\i,0)--(\i,1.5);}
			\foreach \j in {0,0.5,1}{\draw (0,\j)--(1.5,\j);}
			\draw[red, line width=1pt] (0,1)--(1.5,1);
		\end{tikzpicture} + 
		a_{10}\begin{tikzpicture}[baseline=4ex]
			\foreach \i in {0,0.5,1}{\draw (\i,0)--(\i,1.5);}
			\foreach \j in {0,0.5,1}{\draw (0,\j)--(1.5,\j);}
			\draw[red, line width=1pt] (1,0)--(1,1.5);
		\end{tikzpicture} + 
		a_{11}\begin{tikzpicture}[baseline=4ex]
			\foreach \i in {0,0.5,1}{\draw (\i,0)--(\i,1.5);}
			\foreach \j in {0,0.5,1}{\draw (0,\j)--(1.5,\j);}
			\draw[red, line width=1pt] (0,1)--(1.5,1);
			\draw[red, line width=1pt] (1,0)--(1,1.5);
		\end{tikzpicture}\ .\label{any_GS_4}
	\end{equation}
	Using the coordinate system introduced in Sec.~\ref{section_SQC}, denote
	\begin{align}
		\tilde{X}_{1}^{\text{logic}}:=\prod_{x_2\in\mathbb{Z}_{L_2}}X_{[-1,x_2+\frac{1}{2}]}\,,\ 
		\tilde{X}_{2}^{\text{logic}}:=\prod_{x_1\in\mathbb{Z}_{L_1}}X_{[x_1+\frac{1}{2},-1]}\,,\label{any_GS_12}
	\end{align}
	by using which we can write the state in Eq.~(\ref{any_GS_4}) as
	\begin{equation}
		\Big(a_{00}+a_{01}\tilde{X}_{2}^{\text{logic}}+a_{10}\tilde{X}_{1}^{\text{logic}}+a_{11}\tilde{X}_{1}^{\text{logic}}\tilde{X}_{2}^{\text{logic}}\Big)|00\cdots0\rangle\,.\label{any_GS_5}
	\end{equation}
	Note that $\tilde{X}_i^{\text{logic}}$ is not an arbitrary uncontractible Pauli $X$ Wilson loop along $x_i$-direction, but the specific Wilson loop defined in Eq.~(\ref{any_GS_12}).
	
	Recall the circuit $U_c$ for preparing computational basis of the $[0,1,2,2]$ model code states. As illustrated in Fig.~\ref{TC-rep-qubits-and-logical-X}, 
	\begin{figure}[htbp]
		\centering
		\begin{tikzpicture}
			\foreach \i in {0,1,2}{\draw (\i,0)--(\i,3);}
			\foreach \j in {0,1,2}{\draw (0,\j)--(3,\j);}
			
			\draw[red, line width=1pt] (0,2)--(3,2) (2,0)--(2,3);
			\draw[black, line width=1.5pt] (0,0)--(3,0) (0,1)--(3,1) (0,2)--(0,3) (1,2)--(1,3);
			
			\draw[blue, ->, line width=1pt] (-1.3,0)--(-0.5,0);
			\draw[blue, ->, line width=1pt] (-1.3,0)--(-1.3,0.8);
			\node[blue] at (-0.3,0) {$x_1$};
			\node[blue] at (-1.3,0.95) {$x_2$};
		\end{tikzpicture}
		\caption{Illustration of representative spins of $U_c$ (on bold black edges) and support spins of $\tilde{X}^{\text{logic}}_{i}$ (on red edges) for the $[0,1,2,2]$ model. There is no overlap between representative spins and $\tilde{X}^{\text{logic}}_{i}$.}
		\label{TC-rep-qubits-and-logical-X}
	\end{figure}
	no representative spins of $U_c$ are involved in $\tilde{X}_{i}^{\text{logic}}$, so\footnote{Because $[\text{CNOT}_{12},X_2]=[\text{CNOT}_{12},X_3]=0$, and Hadamard gates in $U_c$ do not overlap with $\tilde{X}_{i}^{\text{logic}}$.}
	\begin{equation}
		\left[U_c,\tilde{X}_{i}^{\text{logic}}\right]=0,\quad i=1,2\,.
	\end{equation}
	Therefore, applying $U_c$ on the state in Eq.~(\ref{any_GS_5}), one obtains
	\begin{align}
		&a_{00}|00\rangle_{\text{logic}}+a_{01}|01\rangle_{\text{logic}}+a_{10}|10\rangle_{\text{logic}}+a_{11}|11\rangle_{\text{logic}}\notag\\
		=&\Big(a_{00}+a_{01}\tilde{X}_2^{\text{logic}}+a_{10}\tilde{X}_1^{\text{logic}}+a_{11}\tilde{X}_1^{\text{logic}}\tilde{X}_2^{\text{logic}}\Big)|00\rangle_{\text{logic}}\,,\label{any_GS_6}
	\end{align}
	where
	\begin{align}
		|00\rangle_{\text{logic}}:=U_c|00\cdots0\rangle=\frac{1}{\sqrt{2}}\prod_p\frac{1}{\sqrt{2}}\Big(1+A_p\Big)|00\cdots0\rangle\,.\label{any_GS_9}
	\end{align}
	Apparently, the state in Eq.~(\ref{any_GS_6}) is an arbitrary code state of $[0,1,2,2]$ model. 
	The CNOT sequence $U_g$ and circuit $U_c$ map the state in Eq.~(\ref{any_GS_1}) to the state in Eq.~(\ref{any_GS_9}), i.e.,
	\begin{align}
		&\Big(a_{00}+a_{01}X_2+a_{10}X_1+a_{11}X_1X_2\Big)|00\rangle_{12}\otimes|00\cdots0\rangle\notag\\
		\mapsto&\Big(a_{00}+a_{01}\tilde{X}_{2}^{\text{logic}}+a_{10}\tilde{X}_{1}^{\text{logic}}+a_{11}\tilde{X}_{1}^{\text{logic}}\tilde{X}_{2}^{\text{logic}}\Big)|00\rangle_{\text{logic}}\,.
	\end{align}
	Pictorially, the CNOT sequence $U_g$ and circuit $U_c$ map $X_i$ to $\tilde{X}_i^{\text{logic}}$, and each seed to its corresponding logical qubit.

	\subsection{$[0,1,2,3]$ model}\label{section_any_GS_X_cube}
	
	It has been a challenge to prepare arbitrary code state of $[0,1,2,3]$ model, i.e., X-cube model. With the circuit introduced in this section, as long as arbitrary state of seeds can be prepared, arbitrary code state of X-cube can be prepared. The physical picture is simple and clear: the circuit in this section is a generalization of the circuit in Sec.~\ref{section_any_GS_TC}, with more seeds and logical qubits. X-cube model has $2(L_1+L_2+L_3)-3$ logical qubits on cubic lattice under PBC, we choose $2(L_1+L_2+L_3)-3$ seeds as the qubits on orange wavy edges in the following figure:
	\begin{equation}
		\begin{tikzpicture}
			\foreach \i in {0,1,2}{\foreach \j in {0,1,2}{\draw (\i,\j,0)--(\i,\j,2);}}
			\foreach \i in {0,1,2}{\foreach \k in {0,1}{\draw (\i,0,\k)--(\i,3,\k);}}
			\foreach \j in {0,1,2}{\foreach \k in {0,1}{\draw (0,\j,\k)--(3,\j,\k);}}
			
			\draw[->, blue, line width=1pt] (-2,0,0)--(-2,0,1);
			\draw[->, blue, line width=1pt] (-2,0,0)--(-1.2,0,0);
			\draw[->, blue, line width=1pt] (-2,0,0)--(-2,0.8,0);
			\node[blue] at (-2.18,0,1) {$x_1$};
			\node[blue] at (-1.2,-0.22,0) {$x_2$};
			\node[blue] at (-1.75,0.8,0) {$x_3$};
			
			\draw[decorate, decoration={snake, amplitude=0.3mm, segment length=1mm}, orange, line width=1pt] 
			(0,2,1) -- (0,3,1)
			(0,2,1) -- (0,2,2)
			(1,2,1) -- (1,3,1)
			(1,2,1) -- (1,2,2)
			(2,2,1) -- (2,3,1)
			(2,2,1) -- (2,2,2)
			(2,2,0) -- (3,2,0)
			(2,2,0) -- (2,3,0)
			(2,2,1) -- (3,2,1)
			(2,1,1) -- (3,1,1)
			(2,1,1) -- (2,1,2)
			(2,0,1) -- (3,0,1)
			(2,0,1) -- (2,0,2);
			
			\fill[blue] (0,0,0) circle (2pt);
		\end{tikzpicture}\,.\label{any_GS_7}
	\end{equation}
	There are in total
	\begin{equation}
		2(L_1-1)+2(L_2-1)+2(L_3-1)+3
	\end{equation}
	seeds in Eq.~(\ref{any_GS_7}), equaling to the number of logical qubits of X-cube. Denote the set of seeds as $\Gamma_d^{\text{seed}}$, use the seed-entangler\footnote{For the $[0,1,2,3]$ model, since seeds are not in a local region, the most general seed-entangler is a non-local unitary acting on all seeds. However, for some cases, linear-depth SQC is enough (e.g., the SQC illustrated in Fig.~\ref{seeds-GHZ}). For some cases, constant-depth LU circuit together with on-site measurement is enough\cite{any_single_qubit_logical_state,long_range_entanglement_from_measuring_SPT}.} to prepare an arbitrary state of seeds, i.e.,
	\begin{align}
		&\sum_{i_{\gamma_d}=0,1|\gamma_d\in\Gamma_d^{\text{seed}}}a_{\{i_{\gamma_d}\}}\bigotimes_{\gamma_d\in\Gamma_d^{\text{seed}}}|i_{\gamma_d}\rangle\notag\\
		=&\sum_{i_{\gamma_d}=0,1|\gamma_d\in\Gamma_d^{\text{seed}}}a_{\{i_{\gamma_d}\}}\bigotimes_{\gamma_d\in\Gamma_d^{\text{seed}}}X_{\gamma_d}^{i_{\gamma_d}}|0\rangle\,,\label{any_GS_8}
	\end{align}
	while other spins stay in $|0\rangle$, disentangled with seeds. Here $|i_{\gamma_d}\rangle$ is the computational basis of the spin on $\gamma_d$, $\bigotimes_{\gamma_d\in\Gamma_d^{\text{seed}}}|i_{\gamma_d}\rangle$ is the computational basis of all seeds, $X_{\gamma_d}$ is the Pauli $X$ operator of seed $\gamma_d$, and $a_{\{i_{\gamma_d}\}}$ is the amplitude of configuration $\{i_{\gamma_d}\}$.

	The entanglement among seeds will finally be transformed to the entanglement among logical qubits by a CNOT sequence (introduced soon) and circuit $U_c$, akin to Eq.~(\ref{any_GS_6}). For example, if one intends to prepare a GHZ state of logical qubits, the state in Eq.~(\ref{any_GS_8}) should be a GHZ state of seeds, which can be straightforwardly prepared by a Hadamard gate followed by a sequence of CNOT among seeds, as shown in Fig.~\ref{seeds-GHZ}. 
	\begin{figure}
		\begin{tikzpicture}
			\foreach \i in {0,1,2}{\foreach \j in {0,1,2}{\draw (\i,\j,0)--(\i,\j,2);}}
			\foreach \i in {0,1,2}{\foreach \k in {0,1}{\draw (\i,0,\k)--(\i,3,\k);}}
			\foreach \j in {0,1,2}{\foreach \k in {0,1}{\draw (0,\j,\k)--(3,\j,\k);}}
			
			\draw[->, blue, line width=1pt] (-2,0,0)--(-2,0,1);
			\draw[->, blue, line width=1pt] (-2,0,0)--(-1.2,0,0);
			\draw[->, blue, line width=1pt] (-2,0,0)--(-2,0.8,0);
			\node[blue] at (-2.18,0,1) {$x_1$};
			\node[blue] at (-1.2,-0.22,0) {$x_2$};
			\node[blue] at (-1.75,0.8,0) {$x_3$};
			
			\draw[red, line width=1pt] (0,2,1)--(0,2,2);
			\draw[blue, ->, line width=1pt] (0,2,1.5)--(0,2.5,1);
			\draw[blue, ->, line width=1pt] (0,2,1.5)--(1,2,1.5);
			\draw[blue, ->, line width=1pt] (0,2.5,1)--(1,2.5,1);
			\draw[blue, ->, line width=1pt] (1,2,1.5)--(2,2,1.5);
			\draw[blue, ->, line width=1pt] (1,2.5,1)--(2,2.5,1);
			\draw[blue, ->, line width=1pt] (2,2.5,1)--(2,2.5,0);
			\draw[blue, ->, line width=1pt] (2,2,1.5)--(2.5,2,1);
			\draw[blue, ->, line width=1pt] (2.5,2,1)--(2.5,2,0);
			\draw[blue, ->, line width=1pt] (2,2,1.5)--(2,1,1.5);
			\draw[blue, ->, line width=1pt] (2,1,1.5)--(2,0,1.5);
			\draw[blue, ->, line width=1pt] (2.5,2,1)--(2.5,1,1);
			\draw[blue, ->, line width=1pt] (2.5,1,1)--(2.5,0,1);
			
			\fill[blue] (0,0,0) circle (2pt);
		\end{tikzpicture}\,.
		\caption{A straightforward preparation of GHZ state of seeds via a circuit. This circuit is what we call the seed-entangler previously. Red edge stands for spin in state $|+\rangle$, while black edge stands for $|0\rangle$. The blue arrows stands for CNOT gates. The CNOT gates should be applied in the order that spins play the role of control qubit only after it has been entangled with the spin on red edge.}
		\label{seeds-GHZ}
	\end{figure}
	If one intends to prepare a Toric Code ground state of logical qubits with some synthetically defined connectivity, the state in Eq.~(\ref{any_GS_8}) should be a Toric Code ground state, which can be prepared by the circuit introduced in Sec.~\ref{section_toric_code_sqc}. To prepare a generic state of seeds, a general unitary on seeds is needed, which, in principle, can be approximated by a finite sequence of universal quantum gate (e.g., a set of universal quantum gate is $\{\text{CNOT},\text{Had},T\}$).

	After the state in Eq.~(\ref{any_GS_8}) is prepared, apply the following CNOT sequence, namely, $U_g$, making the seeds grow to the corresponding logical qubits (for visual clearness, we depart $U_g$ into 3 mutually commutable parts):
	\begin{itemize}
		\item Part 1:
		\begin{equation}
			\begin{tikzpicture}
				\foreach \i in {0,1,2}{\foreach \j in {0,1,2}{\draw (\i,\j,0)--(\i,\j,2);}}
				\foreach \i in {0,1,2}{\foreach \k in {0,1}{\draw (\i,0,\k)--(\i,3,\k);}}
				\foreach \j in {0,1,2}{\foreach \k in {0,1}{\draw (0,\j,\k)--(3,\j,\k);}}
				
				\draw[->, blue, line width=1pt] (-2,0,0)--(-2,0,1);
				\draw[->, blue, line width=1pt] (-2,0,0)--(-1.2,0,0);
				\draw[->, blue, line width=1pt] (-2,0,0)--(-2,0.8,0);
				\node[blue] at (-2.18,0,1) {$x_1$};
				\node[blue] at (-1.2,-0.22,0) {$x_2$};
				\node[blue] at (-1.75,0.8,0) {$x_3$};
				
				\draw[decorate, decoration={snake, amplitude=0.3mm, segment length=1mm}, orange, line width=1pt] 
				(0,2,1) -- (0,3,1)
				(0,2,1) -- (0,2,2)
				(1,2,1) -- (1,3,1)
				(1,2,1) -- (1,2,2)
				(2,2,1) -- (2,3,1)
				(2,2,1) -- (2,2,2)
				(2,2,0) -- (3,2,0)
				(2,2,0) -- (2,3,0)
				(2,2,1) -- (3,2,1)
				(2,1,1) -- (3,1,1)
				(2,1,1) -- (2,1,2)
				(2,0,1) -- (3,0,1)
				(2,0,1) -- (2,0,2);
				
				\draw[blue, line width=1pt, ->] (0,2,1.45)--(0,2,0.55);
				\draw[blue, line width=1pt, ->] (1,2,1.45)--(1,2,0.55);
				\draw[blue, line width=1pt, ->] (2,2,1.45)--(2,2,0.55);
				\draw[blue, line width=1pt, ->] (2,1,1.45)--(2,1,0.55);
				\draw[blue, line width=1pt, ->] (2,0,1.45)--(2,0,0.55);
			\end{tikzpicture}\ .
		\end{equation}
		Part 1 has $L_1-1$ layer, which is just 1 under the setting in figure. Part 1 maps each Pauli $X$ of a seed spanning along $x_1$-direction to the corresponding logical Pauli $X$ supported on the line\footnote{Strictly speaking, since $T^3$ is curved, it is not a straight line but a 1-dimensional smooth subcomplex leaf.} containing the seed. 
		\item Part 2:
		\begin{equation}
			\begin{tikzpicture}
				\foreach \i in {0,1,2}{\foreach \j in {0,1,2}{\draw (\i,\j,0)--(\i,\j,2);}}
				\foreach \i in {0,1,2}{\foreach \k in {0,1}{\draw (\i,0,\k)--(\i,3,\k);}}
				\foreach \j in {0,1,2}{\foreach \k in {0,1}{\draw (0,\j,\k)--(3,\j,\k);}}
				
				\draw[->, blue, line width=1pt] (-2,0,0)--(-2,0,1);
				\draw[->, blue, line width=1pt] (-2,0,0)--(-1.2,0,0);
				\draw[->, blue, line width=1pt] (-2,0,0)--(-2,0.8,0);
				\node[blue] at (-2.18,0,1) {$x_1$};
				\node[blue] at (-1.2,-0.22,0) {$x_2$};
				\node[blue] at (-1.75,0.8,0) {$x_3$};
				
				\draw[decorate, decoration={snake, amplitude=0.3mm, segment length=1mm}, orange, line width=1pt] 
				(0,2,1) -- (0,3,1)
				(0,2,1) -- (0,2,2)
				(1,2,1) -- (1,3,1)
				(1,2,1) -- (1,2,2)
				(2,2,1) -- (2,3,1)
				(2,2,1) -- (2,2,2)
				(2,2,0) -- (3,2,0)
				(2,2,0) -- (2,3,0)
				(2,2,1) -- (3,2,1)
				(2,1,1) -- (3,1,1)
				(2,1,1) -- (2,1,2)
				(2,0,1) -- (3,0,1)
				(2,0,1) -- (2,0,2);
				
				\draw[blue, line width=1pt, ->] 
				(2.45,2.05,0)--(1.55,2.05,0);
				\draw[blue, line width=1pt, ->] 
				(1.45,2.05,0)--(0.55,2.05,0);
				\draw[blue, line width=1pt, ->] 
				(2.45,2.05,1)--(1.55,2.05,1);
				\draw[blue, line width=1pt, ->] 
				(1.45,2.05,1)--(0.55,2.05,1);
				\draw[blue, line width=1pt, ->] 
				(2.45,1.05,1)--(1.55,1.05,1);
				\draw[blue, line width=1pt, ->] 
				(1.45,1.05,1)--(0.55,1.05,1);
				\draw[blue, line width=1pt, ->] 
				(2.45,0.05,1)--(1.55,0.05,1);
				\draw[blue, line width=1pt, ->] 
				(1.45,0.05,1)--(0.55,0.05,1);
				
				\node[blue] at (2.15,2.2,0) {1};
				\node[blue] at (2.15,2.2,1) {1};
				\node[blue] at (2.15,1.2,1) {1};
				\node[blue] at (2.15,0.2,1) {1};
				\node[blue] at (1.15,2.2,0) {2};
				\node[blue] at (1.15,2.2,1) {2};
				\node[blue] at (1.15,1.2,1) {2};
				\node[blue] at (1.15,0.2,1) {2};
			\end{tikzpicture}\ .
		\end{equation}
		Part 2 has $L_2-1$ layers, which is 2 under the setting in figure. The numbers 1,2 are labels of layers. Part 2 maps each Pauli $X$ of a seed spanning along $x_2$-direction to the corresponding logical Pauli $X$ supported on the line containing the seed.
		\item Part 3:
		\begin{equation}
			\begin{tikzpicture}
				\foreach \i in {0,1,2}{\foreach \j in {0,1,2}{\draw (\i,\j,0)--(\i,\j,2);}}
				\foreach \i in {0,1,2}{\foreach \k in {0,1}{\draw (\i,0,\k)--(\i,3,\k);}}
				\foreach \j in {0,1,2}{\foreach \k in {0,1}{\draw (0,\j,\k)--(3,\j,\k);}}
				
				\draw[->, blue, line width=1pt] (-2,0,0)--(-2,0,1);
				\draw[->, blue, line width=1pt] (-2,0,0)--(-1.2,0,0);
				\draw[->, blue, line width=1pt] (-2,0,0)--(-2,0.8,0);
				\node[blue] at (-2.18,0,1) {$x_1$};
				\node[blue] at (-1.2,-0.22,0) {$x_2$};
				\node[blue] at (-1.75,0.8,0) {$x_3$};
				
				\draw[decorate, decoration={snake, amplitude=0.3mm, segment length=1mm}, orange, line width=1pt] 
				(0,2,1) -- (0,3,1)
				(0,2,1) -- (0,2,2)
				(1,2,1) -- (1,3,1)
				(1,2,1) -- (1,2,2)
				(2,2,1) -- (2,3,1)
				(2,2,1) -- (2,2,2)
				(2,2,0) -- (3,2,0)
				(2,2,0) -- (2,3,0)
				(2,2,1) -- (3,2,1)
				(2,1,1) -- (3,1,1)
				(2,1,1) -- (2,1,2)
				(2,0,1) -- (3,0,1)
				(2,0,1) -- (2,0,2);
				
				\draw[blue, line width=1pt, ->] 
				(0.05,2.45,1) -- (0.05,1.55,1);
				\draw[blue, line width=1pt, ->] 
				(0.05,1.45,1) -- (0.05,0.55,1);
				\draw[blue, line width=1pt, ->] 
				(1.05,2.45,1) -- (1.05,1.55,1);
				\draw[blue, line width=1pt, ->] 
				(1.05,1.45,1) -- (1.05,0.55,1);
				\draw[blue, line width=1pt, ->] 
				(2.05,2.45,1) -- (2.05,1.55,1);
				\draw[blue, line width=1pt, ->] 
				(2.05,1.45,1) -- (2.05,0.55,1);
				\draw[blue, line width=1pt, ->] 
				(2.05,2.45,0) -- (2.05,1.55,0);
				\draw[blue, line width=1pt, ->] 
				(2.05,1.45,0) -- (2.05,0.55,0);
				
				\node[blue] at (0.15,2,1) {1};
				\node[blue] at (1.15,2,1) {1};
				\node[blue] at (2.15,2,1) {1};
				\node[blue] at (2.15,2,0) {1};
				\node[blue] at (0.15,1,1) {2};
				\node[blue] at (1.15,1,1) {2};
				\node[blue] at (2.15,1,1) {2};
				\node[blue] at (2.15,1,0) {2};
			\end{tikzpicture}\ .
		\end{equation}
		Part 3 has $L_3-1$ layers, which is 2 under the setting in figure. The numbers 1,2 are labels of layers. Part 3 maps each Pauli $X$ of a seed spanning along $x_3$-direction to the corresponding logical Pauli $X$ supported on the line containing the seed.
	\end{itemize}
	Formally, denote the corresponding logical Pauli $X$ of seed $\gamma_d$ as $\tilde{X}^{\text{logic}}_{\gamma_d}$, we can write
	\begin{equation}
		\tilde{X}^{\text{logic}}_{\gamma_d}=\prod_{\gamma'_d\subset\text{ext}(\gamma_d)}X_{\gamma'_d}\,,
	\end{equation}
	where $\text{ext}(\gamma_d)$\footnote{Semantically, the symbol ext stands for \textit{extension}. Each $\gamma_d$ is mapped to a unique leaf by ext.} is the $d$-dimensional smooth subcomplex leaf containing $\gamma_d$. For the $[0,1,2,3]$ model, $d=1$. Pictorially, $\text{ext}(\gamma_1)$ is the line extended from $\gamma_1$. Analogous to Eqs.~(\ref{any_GS_1})-(\ref{any_GS_5}), $U_g$ maps each $X_{\gamma_d}$ to its corresponding $\tilde{X}_{\gamma_d}^{\text{logic}}$, so the result state of applying $U_g$ to the state in Eq.~(\ref{any_GS_8}) is
	\begin{align}\label{revise_48}
		\sum_{i_{\gamma_d}=0,1|\gamma_d\in\Gamma_d^{\text{seed}}}a_{\{i_{\gamma_d}\}}\bigotimes_{\gamma_d\in\Gamma_d^{\text{seeds}}}\left(\tilde{X}^{\text{logic}}_{\gamma_d}\right)^{i_{\gamma_d}}|00\cdots0\rangle\,,
	\end{align}
	where $|00\cdots0\rangle$ is the product state of all spins. As illustrated in Fig.~\ref{any-GS-10}, for any seed $\gamma_d$, $\tilde{X}^{\text{logic}}_{\gamma_d}$ does not overlap with any representative spin of $U_c$, so $[U_c,\tilde{X}^{\text{logic}}_{\gamma_d}]=0$\footnote{Because $[\text{CNOT}_{12},X_2]=[\text{CNOT}_{12},X_3]=0$, and Hadamard gates in $U_c$ do not overlap with $\tilde{X}_{i}^{\text{logic}}$.}.
	\begin{figure}[htbp]
		\centering
		\begin{tikzpicture}
			\foreach \i in {0,1,2}{\foreach \j in {0,1,2}{\draw (\i,\j,0)--(\i,\j,2);}}
			\foreach \i in {0,1,2}{\foreach \k in {0,1}{\draw (\i,0,\k)--(\i,3,\k);}}
			\foreach \j in {0,1,2}{\foreach \k in {0,1}{\draw (0,\j,\k)--(3,\j,\k);}}
			
			\draw[->, blue, line width=1pt] (-2,0,0)--(-2,0,1);
			\draw[->, blue, line width=1pt] (-2,0,0)--(-1.2,0,0);
			\draw[->, blue, line width=1pt] (-2,0,0)--(-2,0.8,0);
			\node[blue] at (-2.18,0,1) {$x_1$};
			\node[blue] at (-1.2,-0.22,0) {$x_2$};
			\node[blue] at (-1.75,0.8,0) {$x_3$};
			
			\draw[red, line width=1pt]
			(0,0,1) -- (0,3,1)
			(0,2,0) -- (0,2,2)
			(1,0,1) -- (1,3,1)
			(1,2,0) -- (1,2,2)
			(2,0,1) -- (2,3,1)
			(2,2,0) -- (2,2,2)
			(0,2,0) -- (3,2,0)
			(2,0,0) -- (2,3,0)
			(0,2,1) -- (3,2,1)
			(0,1,1) -- (3,1,1)
			(2,1,0) -- (2,1,2)
			(0,0,1) -- (3,0,1)
			(2,0,0) -- (2,0,2);
			
			\draw[black, line width=1.5pt]
			(0,0,0)--(0,3,0)
			(1,0,0)--(1,3,0)
			(0,0,1)--(0,0,2)
			(1,0,1)--(1,0,2)
			(0,1,1)--(0,1,2)
			(1,1,1)--(1,1,2)
			(2,0,0)--(3,0,0)
			(2,1,0)--(3,1,0); 
		\end{tikzpicture}\ .
		\caption{The logical Pauli $X$ operators created by the pre-step are supported on the red lines, and the representative spins in $U_c$ are supported on the bold black lines. There is no overlap between red edges and bold black edges, so the logical Pauli $X$ operators created by the pre-step are commutable with $U_c$.}
		\label{any-GS-10}
	\end{figure}
	Therefore, applying $U_c$ on the state in Eq.~(\ref{revise_48}), one obtains
	\begin{align}
		&\sum_{i_{\gamma_d}=0,1|\gamma_d\in\Gamma_d^{\text{seed}}}a_{\{i_{\gamma_d}\}}\bigotimes_{\gamma_d\in\Gamma_d^{\text{seed}}}\left(\tilde{X}^{\text{logic}}_{\gamma_d}\right)^{i_{\gamma_d}}U_c|00\cdots0\rangle\notag\\
		=&\sum_{i_{\gamma_d}=0,1|\gamma_d\in\Gamma_d^{\text{seed}}}a_{\{i_{\gamma_d}\}}\bigotimes_{\gamma_d\in\Gamma_d^{\text{seed}}}\left(\tilde{X}^{\text{logic}}_{\gamma_d}\right)^{i_{\gamma_d}}|00\cdots0\rangle_{\text{logic}}\label{any_GS_11}\,,
	\end{align}
	where\footnote{See Eq.~(\ref{any_GS_19}).}
	\begin{align}
		&|00\cdots0\rangle_{\text{logic}}:=U_c|00\cdots0\rangle\notag\\&=\frac{1}{\sqrt{2}^{|R_3|}}\prod_{c\in\Gamma_3}\frac{1}{\sqrt{2}}\Big(1+A_c\Big)|00\cdots0\rangle\,.
	\end{align}
	Again, pictorially, the CNOT sequence $U_g$ and circuit $U_c$ map $X_{\gamma_d}$ to $\tilde{X}_{\gamma_d}^{\text{logic}}$ ($\gamma_d\in\Gamma_d^{\text{seed}}$), and each seed to its corresponding logical qubit.
	
	It is not hard to see that all the uncontractible Pauli $X$ Wilson loops along $x_1$-direction can be obtained by multiplying logical $X$ operators created by part 1 of the pre-step and $A_c$ terms. Similarly, all the uncontractible Pauli $X$ Wilson loops along $x_2,x_3$-directions can be obtained by multiplying logical $X$ created by parts 2,3 of $U_g$ and $A_c$ terms, respectively. Therefore, all the logical Pauli $X$ are obtainable by applying the pre-step and $U_c$, the states in Eq.~(\ref{any_GS_11}) form the whole ground space. The number of seeds or $\tilde{X}_{\gamma_d}^{\text{logic}}$ in Eq.~(\ref{any_GS_11}) equaling to $\text{log}_2\text{GSD}$ further shows that $\{\tilde{X}_{\gamma_d}^{\text{logic}}\}$ are mutually independent, i.e., no $\tilde{X}_{\gamma_d}^{\text{logic}}$ equals to the multiplication of other $\tilde{X}_{\gamma_d}^{\text{logic}}$ and $A_c$ terms.

	\subsection{$[1,2,3,3]$ model}\label{section_any_GS_3dTC}
	
	$[1,2,3,3]$ model has three logical qubits under PBC, and we choose the three seeds as the three qubits on the plaquettes with orange wavy edges in the following figure:
	\begin{equation}
		\begin{tikzpicture}
			\foreach \i in {0,1,2}{\foreach \j in {0,1,2}{\draw (\i,\j,0)--(\i,\j,2);}}
			\foreach \i in {0,1,2}{\foreach \k in {0,1}{\draw (\i,0,\k)--(\i,3,\k);}}
			\foreach \j in {0,1,2}{\foreach \k in {0,1}{\draw (0,\j,\k)--(3,\j,\k);}}
			
			\draw[->, blue, line width=1pt] (-2,0,0)--(-2,0,1);
			\draw[->, blue, line width=1pt] (-2,0,0)--(-1.2,0,0);
			\draw[->, blue, line width=1pt] (-2,0,0)--(-2,0.8,0);
			\node[blue] at (-2.18,0,1) {$x_1$};
			\node[blue] at (-1.2,-0.22,0) {$x_2$};
			\node[blue] at (-1.75,0.8,0) {$x_3$};
			
			\draw[decorate, decoration={snake, amplitude=0.3mm, segment length=1mm}, orange, line width=1pt] 
			(2,2,1)--(3,2,1)
			(3,2,1)--(3,3,1)
			(3,3,1)--(2,3,1)
			(2,3,1)--(2,2,1)
			(2,2,1)--(2,2,2)
			(2,2,2)--(2,3,2)
			(2,3,2)--(2,3,1)
			(2,2,1)--(3,2,1)
			(3,2,1)--(3,2,2)
			(3,2,2)--(2,2,2);
			
			\node[orange] at (2.5,2.5,1) {$\bm{1}$};
			\node[orange] at (1.97,2.5,1.5) {$\bm{2}$};
			\node[orange] at (2.5,1.97,1.5) {$\bm{3}$};
			
			\fill[blue] (0,0,0) circle (2pt);
		\end{tikzpicture}\ .
	\end{equation}
	The blue dot stands for the original point $(0,0,0)$. Using the coordinate notation introduced in Sec.~\ref{section_SQC}, seeds 1,2,3 can be written as $\left[-1,-\frac{1}{2},-\frac{1}{2}\right],\left[-\frac{1}{2},-1,-\frac{1}{2}\right],\left[-\frac{1}{2},-\frac{1}{2},-1\right]$ respectively. Denote the set of three seeds as $\Gamma_d^{\text{seed}}$. Use the seed-entangler to prepare an arbitrary state of the three seeds, which can always be written as
	\begin{align}
		&\sum_{i_{\gamma_d}=0,1|\gamma_d\in\Gamma_d^{\text{seed}}}a_{\{i_{\gamma_d}\}}\otimes_{\gamma_d\in\Gamma_d^{\text{seed}}}|i_{\gamma_d}\rangle\notag\\
		=&\sum_{i_{\gamma_d}=0,1|\gamma_d\in\Gamma_d^{\text{seed}}}a_{\{i_{\gamma_d}\}}\otimes_{\gamma_d\in\Gamma_d^{\text{seed}}}X^{i_{\gamma_d}}_{\gamma_d}|0\rangle\label{any_GS_13}\,,
	\end{align}
	while other spins stay in $|0\rangle$, disentangled with seeds. Just like in the X-cube case, here $\bigotimes_{\gamma_d\in\Gamma_d^{\text{seed}}}$ is the computational basis of all seeds, $X_{\gamma_d}$ is the Pauli $X$ of seed $\gamma_d$, and $a_{\{i_{\gamma_d}\}}$ is the amplitude of configuration $\{i_{\gamma_d}\}$. Eq.~(\ref{any_GS_13}) has exactly the same form as Eq.~(\ref{any_GS_8}), which opens the way to general $[d-1,d,d+1,D]$ model.
	
	Next, starting from seeds 1,2,3, one applies the CNOT sequence $U_g$ on the state in Eq.~(\ref{any_GS_13}), making the seeds growing to their corresponding logical qubits. $U_g$ consists of three parts, shown as follows:
	\begin{itemize}
		\item Starting from seed 1, or $[-1,-\frac{1}{2},-\frac{1}{2}]$, part 1:
		\begin{equation}
			\begin{tikzpicture}
				\foreach \i in {0,1,2}{\foreach \j in {0,1,2}{\draw (\i,\j,0)--(\i,\j,2);}}
				\foreach \i in {0,1,2}{\foreach \k in {0,1}{\draw (\i,0,\k)--(\i,3,\k);}}
				\foreach \j in {0,1,2}{\foreach \k in {0,1}{\draw (0,\j,\k)--(3,\j,\k);}}
				
				\draw[->, blue, line width=1pt] (-2,0,0)--(-2,0,1);
				\draw[->, blue, line width=1pt] (-2,0,0)--(-1.2,0,0);
				\draw[->, blue, line width=1pt] (-2,0,0)--(-2,0.8,0);
				\node[blue] at (-2.18,0,1) {$x_1$};
				\node[blue] at (-1.2,-0.22,0) {$x_2$};
				\node[blue] at (-1.75,0.8,0) {$x_3$};
				
				\draw[decorate, decoration={snake, amplitude=0.3mm, segment length=1mm}, orange, line width=1pt] 
				(2,2,1)--(3,2,1)
				(3,2,1)--(3,3,1)
				(3,3,1)--(2,3,1)
				(2,3,1)--(2,2,1);
				
				\foreach \i in {0.5,1.5,2.5}{\foreach \j in {0.5,1.5,2.5}{
						\draw[blue, dashed, opacity=0.5] (\i+0.5,\j+0.5,1)--(\i-0.5,\j-0.5,1)
						(\i+0.5,\j-0.5,1)--(\i-0.5,\j+0.5,1);}}
				
				\draw[blue, line width=1pt, ->] (2.45,2.5,1)--(1.55,2.5,1);
				\draw[blue, line width=1pt, ->] (1.45,2.5,1)--(0.55,2.5,1);
				\node[blue] at (2.15,2.65,1) {1};
				\node[blue] at (1.15,2.65,1) {2};
				\draw[blue, line width=1pt, ->] (2.5,2.45,1)--(2.5,1.55,1);
				\draw[blue, line width=1pt, ->] (1.5,2.45,1)--(1.5,1.55,1);
				\draw[blue, line width=1pt, ->] (0.5,2.45,1)--(0.5,1.55,1);
				\draw[blue, line width=1pt, ->] (2.5,1.45,1)--(2.5,0.55,1);
				\draw[blue, line width=1pt, ->] (1.5,1.45,1)--(1.5,0.55,1);
				\draw[blue, line width=1pt, ->] (0.5,1.45,1)--(0.5,0.55,1);
				\node[blue] at (2.65,2,1) {3};
				\node[blue] at (1.65,2,1) {3};
				\node[blue] at (0.65,2,1) {3};
				\node[blue] at (2.65,1,1) {4};
				\node[blue] at (1.65,1,1) {4};
				\node[blue] at (0.65,1,1) {4};
			\end{tikzpicture}\ .\label{any_GS_16}
		\end{equation}
		The blue dashed lines are added to help visualize the plaquettes supporting control and target qubits of CNOT gates. The blue numbers are labels of layers.
		\item Starting from seed 2, or $[-\frac{1}{2},-1,-\frac{1}{2}]$, part 2:
		\begin{equation}
			\begin{tikzpicture}
				\foreach \i in {0,1,2}{\foreach \j in {0,1,2}{\draw (\i,\j,0)--(\i,\j,2);}}
				\foreach \i in {0,1,2}{\foreach \k in {0,1}{\draw (\i,0,\k)--(\i,3,\k);}}
				\foreach \j in {0,1,2}{\foreach \k in {0,1}{\draw (0,\j,\k)--(3,\j,\k);}}
				
				\draw[->, blue, line width=1pt] (-2,0,0)--(-2,0,1);
				\draw[->, blue, line width=1pt] (-2,0,0)--(-1.2,0,0);
				\draw[->, blue, line width=1pt] (-2,0,0)--(-2,0.8,0);
				\node[blue] at (-2.18,0,1) {$x_1$};
				\node[blue] at (-1.2,-0.22,0) {$x_2$};
				\node[blue] at (-1.75,0.8,0) {$x_3$};
				
				\draw[decorate, decoration={snake, amplitude=0.3mm, segment length=1mm}, orange, line width=1pt] 
				(2,2,1)--(2,3,1)
				(2,2,1)--(2,2,2)
				(2,2,2)--(2,3,2)
				(2,3,2)--(2,3,1);
				
				\foreach \k in {0.5,1.5}{\foreach \j in {0.5,1.5,2.5}{
						\draw[blue, dashed, opacity=0.5] (2,\j+0.5,\k+0.5)--(2,\j-0.5,\k-0.5)
						(2,\j+0.5,\k-0.5)--(2,\j-0.5,\k+0.5);}}
				
				\draw[blue, line width=1pt, ->] (2,2.5,1.45)--(2,2.5,0.6);
				\node[blue] at (1.87,2.65,1) {1};
				\draw[blue, line width=1pt, ->] (2,2.45,1.5)--(2,1.55,1.5);
				\draw[blue, line width=1pt, ->] (2,2.45,0.5)--(2,1.55,0.5);
				\draw[blue, line width=1pt, ->] (2,1.45,1.5)--(2,0.55,1.5);
				\draw[blue, line width=1pt, ->] (2,1.45,0.5)--(2,0.55,0.5);
				\node[blue] at (2.12,2,1.5) {2};
				\node[blue] at (2.12,2,0.5) {2};
				\node[blue] at (2.12,1,1.5) {3};
				\node[blue] at (2.12,1,0.5) {3};
			\end{tikzpicture}\ .\label{any_GS_17}
		\end{equation}
		\item Starting from seed 3, or $[-\frac{1}{2},-\frac{1}{2},-1]$, part 3:
		\begin{equation}
			\begin{tikzpicture}
				\foreach \i in {0,1,2}{\foreach \j in {0,1,2}{\draw (\i,\j,0)--(\i,\j,2);}}
				\foreach \i in {0,1,2}{\foreach \k in {0,1}{\draw (\i,0,\k)--(\i,3,\k);}}
				\foreach \j in {0,1,2}{\foreach \k in {0,1}{\draw (0,\j,\k)--(3,\j,\k);}}
				
				\draw[->, blue, line width=1pt] (-2,0,0)--(-2,0,1);
				\draw[->, blue, line width=1pt] (-2,0,0)--(-1.2,0,0);
				\draw[->, blue, line width=1pt] (-2,0,0)--(-2,0.8,0);
				\node[blue] at (-2.18,0,1) {$x_1$};
				\node[blue] at (-1.2,-0.22,0) {$x_2$};
				\node[blue] at (-1.75,0.8,0) {$x_3$};
				
				\draw[decorate, decoration={snake, amplitude=0.3mm, segment length=1mm}, orange, line width=1pt] 
				(2,2,1)--(2,2,2)
				(2,2,1)--(3,2,1)
				(3,2,1)--(3,2,2)
				(3,2,2)--(2,2,2);
				
				\foreach \i in {0.5,1.5,2.5}{\foreach \k in {0.5,1.5}{\draw[blue, dashed, opacity=0.5] (\i+0.5,2,\k+0.5)--(\i-0.5,2,\k-0.5)
						(\i+0.5,2,\k-0.5)--(\i-0.5,2,\k+0.5);}}
				
				\draw[blue, line width=1pt, ->] (2.5,2,1.45)--(2.5,2,0.6);
				\node[blue] at (2.75,2,1.5) {1};
				\draw[blue, line width=1pt, ->] (2.45,2,1.5)--(1.55,2,1.5);
				\draw[blue, line width=1pt, ->] (2.45,2,0.5)--(1.55,2,0.5);
				\draw[blue, line width=1pt, ->] (1.45,2,1.5)--(0.55,2,1.5);
				\draw[blue, line width=1pt, ->] (1.45,2,0.5)--(0.55,2,0.5);
				\node[blue] at (2,2.15,0.5) {2};
				\node[blue] at (2,2.15,1.5) {2};
				\node[blue] at (1,2.15,0.5) {3};
				\node[blue] at (1,2.15,1.5) {3};
			\end{tikzpicture}\ .\label{any_GS_18}
		\end{equation}
	\end{itemize}
	For the $[1,2,3,3]$ model, $d=2$. Notice that for each seed $\gamma_d\in\Gamma_d^{\text{seed}}$, $U_g$ maps $X_{\gamma_d}$ to the logical Pauli $X$ supported on the $d$-dimensional smooth subcomplex leaf containing $\gamma_d$. Therefore, we can use the same notation as in Sec.~\ref{section_any_GS_X_cube}, i.e.,
	\begin{equation}
		\tilde{X}^{\text{logic}}_{\gamma_d}:=\prod_{\gamma'_d\subset\text{ext}(\gamma_d)}X_{\gamma'_d}\,.
	\end{equation}
	Intuitively, here ext lengthens each seed on plaquette $\gamma_2$ in the $2$ directions where $\gamma_2$ spans, resulting in a $2$-dimensional torus $T^2$.
	\begin{figure}
		\centering
		\includegraphics[width=0.3288\textwidth]{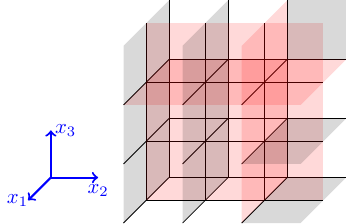}
		\caption{The logical Pauli $X$ operators created by the pre-step are supported on the red plaquettes, and the representative spins in $U_c$ are supported on the black plaquettes. There is no overlap between red plaquettes and black plaquettes, so the logical Pauli $X$ operators created by the pre-step are commutable with $U_c$.}
		\label{any-GS-20}
	\end{figure}
	Analogous to Eqs.~(\ref{any_GS_1})-(\ref{any_GS_5}), $U_g$ maps each $X^{i_{\gamma_d}}_{\gamma_d}$ to its corresponding $\tilde{X}_{\gamma_d}^{i_{\gamma_d}}$, the result state of applying $U_g$ to the state in Eq.~(\ref{any_GS_13}) is
	\begin{align}
		\sum_{i_{\gamma_d}=0,1|\gamma_d\in\Gamma_d^{\text{seed}}}a_{\{i_{\gamma_d}\}}\otimes_{\gamma_d\in\Gamma_d^{\text{seed}}}\left(\tilde{X}^{\text{logic}}_{\gamma_d}\right)^{i_{\gamma_d}}|00\cdots0\rangle\,,\label{revise_49}
	\end{align}
	where $|00\cdots0\rangle$ is the product state of all spins. Again, as illustrated in Fig.~\ref{any-GS-20}, the three $\tilde{X}^{\text{logic}}_{\gamma_d}$ do not overlap with the representative spins of $U_c$, so $\left[U_c,\tilde{X}^{\text{logic}}_{\gamma_d}\right]=0$. Therefore, $U_c$ maps the state in Eq.~(\ref{revise_49}) to
	\begin{align}
		&\sum_{i_{\gamma_d}=0,1|\gamma_d\in\Gamma_d^{\text{seed}}}a_{\{i_{\gamma_d}\}}\otimes_{\gamma_d\in\Gamma_d^{\text{seed}}}\left(\tilde{X}^{\text{logic}}_{\gamma_d}\right)^{i_{\gamma_d}}U_c|00\cdots0\rangle\notag\\
		=&\sum_{i_{\gamma_d}=0,1|\gamma_d\in\Gamma_d^{\text{seed}}}a_{\{i_{\gamma_d}\}}\otimes_{\gamma_d\in\Gamma_d^{\text{seed}}}\left(\tilde{X}^{\text{logic}}_{\gamma_d}\right)^{i_{\gamma_d}}|000\rangle_{\text{logic}}\,,\label{any_GS_21}
	\end{align}
	where
	\begin{align}
		&|000\rangle_{\text{logic}}:=U_c|00\cdots0\rangle\notag\\
		=&\frac{1}{\sqrt{2}^3}\prod_{c\in\Gamma_3}\Big(1+A_c\Big)|00\cdots0\rangle\,.
	\end{align}
	Apparently, the state in Eq.~(\ref{any_GS_21}) is an arbitrary code state of 3-dimensional Toric Code.

	\subsection{$[d-1,d,d+1,D]$ model}\label{section_general_SESQC}

	One might wonder how general is the \emph{seed $\to$ logical qubit} scheme for preparing arbitrary code state? While a detailed step-by-step illustration lacks, we believe this scheme can be used for any $[d-1,d,d+1,D]$ model, even with no change of formal notations, because all $[d-1,d,d+1,D]$ models share a common property: for any $\gamma_d$, $\text{ext}(\gamma_d)$ supports a logical Pauli $X$. Whether this scheme can be generalized to prepare arbitrary code state of other stabilizer code models remains to be studied. For $[d-1,d,d+1,D]$ model, we choose the seeds to be
	\begin{widetext}
		\begin{align}
			\Gamma_d^{\text{seed}}:=\Big\{f(\gamma_D,S_d)\Big|(\gamma_D,S_d):\big|ES(\gamma_D)\big|\geq d+1,\,S_d\subset ES(\gamma_D)\Big\}\,,\label{any_GS_23}
		\end{align}
	\end{widetext}
	where $ES(\gamma_D)$ is the complete set of coordinate indices satisfying
	\begin{align}
		\forall i\in ES(\gamma_D),\ x_i(\gamma_D)=-\frac{1}{2}\,,
	\end{align}
	and $f$ is defined in Eq.~(\ref{any_GS_22}). Let us give Eq.~(\ref{any_GS_23}) some explanation: 
	\begin{itemize}
		\item The $\gamma_D$ satisfying $\big|ES(\gamma_D)\big|\geq d+1$ are those $D$-cubes with at least $d+1$ coordinates being $-\frac{1}{2}$, i.e., $D$-cubes without a representative spin (see Sec.~\ref{section_SQC} for details).
		\item Define $+$ and $-$ between $n$-cubes as
		\begin{align}
			\gamma_n\pm\gamma'_{n'}&=[x_1,\cdots,x_D]\pm[x'_1,\cdots,x'_D]\notag\\&=[x_1\pm x'_1,\cdots,x_D\pm x'_D]\,.\label{any_GS_14}
		\end{align}
		Two cubes with different dimension can be added. Using Eq.~(\ref{any_GS_14}), we can write $f(\gamma_D,S_d)=\gamma_D-\gamma'_{D^*}$, where
		\begin{align}
			x_i(\gamma'_{D^*})=\left\{\begin{array}{ll}
				\frac{1}{2},\ i\in\text{min}_{D^*}(S_d^c)=S_d^c\\
				0,\ i\notin\text{min}_{D^*}(S_d^c)=S_d^c
			\end{array}\right.\,.
		\end{align}
		As the pair $(\gamma_D,S_d)$ takes through all possibilities, $f(\gamma_D,S_d)$ takes through all the $d$-cubes in $\gamma_D$ satisfying $\big|ES(\gamma_D)\big|\geq d+1$ that do not support a representative spin.
	\end{itemize}	
	As for former examples, to prepare an arbitrary code state $|\psi\rangle_{\text{logic}}$ of general $[d-1,d,d+1,D]$ model, the first step is applying the seed-entangler to $|00\cdots0\rangle$, resulting in $|\psi\rangle_{\text{seeds}}\otimes|00\cdots0\rangle_{\text{non-seeds}}$. Then, apply the CNOT sequence $U_g$, making the seeds grow to their corresponding logical qubits. To express $U_g$ for general $[d-1,d,d+1,D]$ model, define $hS(\gamma_n)$ as the complete set of coordinate indices satisfying
	\begin{align}
		\forall i\in hS(\gamma_n),\ x_i(\gamma_n)\text{ is a half-integer}.
	\end{align}
	$\forall\alpha\in\mathbb{Z}$, define
	\begin{align}
		\alpha\gamma_n=\alpha[x_1,\cdots,x_D]=[\alpha x_1,\cdots,\alpha x_D]\,.
	\end{align}
	For any $i\in\{1:D\}$, denote $\hat{v}_i$ as the vertex where the $x_j$-coordinate of $\hat{v}_i$ is
	\begin{align}
		x_j(\hat{v}_i)=\delta_{ij}\,.
	\end{align}
	Note that we need a CNOT sequence mapping $X_{\gamma_d}$ ($\gamma_d\in\Gamma_d^{\text{seed}}$) to $\tilde{X}_{\gamma_d}^{\text{logic}}$ supported on $\text{ext}(\gamma_d)$, i.e.,
	\begin{align}
		\tilde{X}^{\text{logic}}_{\gamma_d}=\prod_{\gamma'_d\subset\text{ext}(\gamma_d)} X_{\gamma'_d} \,.
	\end{align} 
	With the notations defined above, we can write a choice of the required CNOT sequence $U_g$ for $[d-1,d,d+1,D]$ models as follows: 
	
	Supposing for $\gamma_d\in\Gamma_d^{\text{seed}}$, $hS(\gamma_d)=\{i_1,i_2,\cdots,i_d\}$\footnote{The order of $i_1,i_2,\cdots,i_d$ can be arbitrarily chosen.}, a CNOT sequence mapping $X_{\gamma_d}$ to $\tilde{X}_{\gamma_d}^{\text{logic}}$ is
	\begin{widetext}
		\begin{gather}
			\iota(\gamma_d)=\prod_{t=d}^1\Bigg(\prod_{x_{i_t}=L_{i_t}-2}^0\Bigg(\prod_{x_{i_{t'}}\in\mathbb{Z}_{L_{i_{t'}}},\,t'=1:t-1}\text{CNOT}_{\text{control},\text{target}}\Bigg)\Bigg)\,,\notag\\
			\text{control}=\gamma_d-\sum_{t'=1}^{t-1}x_{i_{t'}}\hat{v}_{i_{t'}}-x_{i_t}\hat{v}_{i_t}\quad,\quad\text{target}=\gamma_d-\sum_{t'=1}^{t-1}x_{i_{t'}}\hat{v}_{i_{t'}}-(x_{i_t}+1)\hat{v}_{i_t}\,,
		\end{gather}
	\end{widetext}
	where $\prod_{t=d}^1$ and $\prod_{x_{i_t}=L_{i_t}-2}^0$ are products in order, defined in Eq.~(\ref{revise_35}), and the rest two products are products without order (the order does not matter since the multiplied operators are commutable). For two different seeds $\gamma_d,\gamma'_d\in\Gamma_d^{\text{seed}}$, $\iota(\gamma_d)$ and $\iota(\gamma'_d)$ are commutable, thus can be applied in parallel. In all,
	\begin{align}
		U_g=\prod_{\gamma_d\in\Gamma_d^{\text{seed}}}\iota(\gamma_d)\label{any_GS_30}
	\end{align}
	is a choice of the required CNOT sequence mapping all $X_{\gamma_d}$ ($\gamma_d\in\Gamma_d^{\text{seed}}$) to $\tilde{X}_{\gamma_d}^{\text{logic}}$. The seed set and $U_g$ for $[0,1,2,2],[0,1,2,3],[1,2,3,3]$ models introduced earlier in this section are special cases of Eq.~(\ref{any_GS_23}) and Eq.~(\ref{any_GS_30}), respectively.

	Finally, we would like to discuss the fact that for all the three models $[0,1,2,2],[0,1,2,3],[1,2,3,3]$, the number of available $\left.\left(\gamma_d,\tilde{X}^{\text{logic}}_{\gamma_d}\right)\right|_{\gamma_d\in\Gamma_d^{\text{seed}}}$ pairs exactly equals to the number of logical qubits, which can be seen from Figs.~\ref{TC-rep-qubits-and-logical-X},\ref{any-GS-10},\ref{any-GS-20}, respectively. More precisely, in each logical $X$ equivalence class $\left\{\left.X_i^{\text{logic}}s_X\right|s_X\text{ is }X\text{-type stabilizer}\right\}$\footnote{For simplicity, we restrict the equivalence class of logical $X$ operator within multiplying $X$-type stabilizer (i.e., stabilizers that are pure Pauli $X$ product) here. Under this setting, logical $X$ operators are always pure Pauli $X$ products.}, there is a unique element $\tilde{X}_i^{\text{logic}}$, such that $\text{supp}(\tilde{X}_i^{\text{logic}})$ has no intersection with any representative spin in the circuit $U_c$, and we can find a seed for $\tilde{X}_i^{\text{logic}}$. This statement actually works for any Calderbank-Shor-Steane (CSS) code\footnote{For general CSS code, we need to allow logical qubit reidentification to ensure the existence of seeds, see details in Appendix~\ref{proof_for_CSS_code_and_Had_CNOT_type_circuit}.} and a Hadamard + CNOT type preparation circuit (a direct generalization of $U_c$), we illustrate it in Appendix~\ref{proof_for_CSS_code_and_Had_CNOT_type_circuit}.
	
	Back to TD models and the circuit $U_c$, using the result in Appendix~\ref{proof_for_CSS_code_and_Had_CNOT_type_circuit}, when the structure of representative spins and $\tilde{X}_i^{\text{logic}}$ is known, the number of logical qubits or GSD can be counted by counting seeds, since
	\begin{equation}
		\#\ \text{logical qubits}\ =\ \#\ \tilde{X}_i^{\text{logic}}\ =\ \#\ \text{seeds}\,.
	\end{equation}
	The number of seeds for $U_c$ and $[d-1,d,d+1,D]$ model, i.e., the order of $\Gamma_d^{\text{seed}}$ given in Eq.~(\ref{any_GS_23}) is
	\begin{equation}
		\sum_{p=d+1}^{D}\sum_{S_p}C_p^d\sum_{i\in S_p^c}(L_i-1)=\sum_{p=0}^{D^*-1}C_{D-p}^d\sum_{S_p}\sum_{i\in S_p}(L_i-1)\,,\label{any_GS_29}
	\end{equation}
	which reduces to $\#\,\text{logical qubit}$ of $[0,1,2,D]$ and $[D-3,D-2,D-1,D]$ models (calculated in Ref.~\cite{li2021fracton}) when $d=1$ and $d=D-2$ are taken. This suggests the $\#\,\text{logical qubit}$ or $\text{log}_2\text{GSD}$ of any $[d-1,d,d+1,D]$ models might be given by Eq.~(\ref{any_GS_29}).

	\section{Summary and outlook}\label{section_summmary_outlook}
	
	In this paper, we have established the SEESQC protocol as a unified and efficient framework for preparing the  code states of topological error-correcting codes, including the computational basis states and their superpositions. By successfully constructing quantum circuits for the diverse tetradigit models—encompassing arbitrary-dimensional Toric codes and fracton phases like the X-cube model—we demonstrate SEESQC's capability to generate a broad class of stabilizer code states, including those states supporting spatially extended excitations and system-size-dependent degeneracies. 
	To reach arbitrary states in code space, a key step lies in finding seeds and their growth, which is both model dependent and circuit dependent. Significantly, our analysis reveals that for a large class of topological stabilize code models, the number of available seeds exactly equals to the number of logical qubits, a fundamental observation warranting deeper exploration in the future.

	Several interesting questions and promising directions  emerge from this work:	\begin{enumerate}
		\item As raised in Sec.~\ref{section_general_SESQC} and Appendix~\ref{proof_for_CSS_code_and_Had_CNOT_type_circuit}, for any CSS code and Hadamard + CNOT type preparation circuit (including the SQC $U_c$ constructed in this paper), enough number of seeds can always be found to prepare the whole code space. The SQC constructed in Ref.~\cite{chen2024sequential} for preparing string-net models is similar to the Hadamard + CNOT type preparation circuit, can it be managed to be a SEESQC? What about other preparation circuits for other models? Can they be seed-entangler-enriched?
		
		\item From Figs.~\ref{TC-rep-qubits-and-logical-X},\ref{any-GS-10},\ref{any-GS-20}, we see that for a topological ordered state (Toric Code), seeds can be chosen in a way that all seeds belong to a local area (when the system size becomes infinite). However, for a fracton state (X-cube), there is no available choice of seeds that all seeds belong to a local area. Is this a universal property that could distinguish topological order and fracton states?
		
		\item Since a unitary circuit is an automorphism of the total Hilbert space, so for any unitary circuit, if we admit seeds in a more general sense as a subspace with the same dimension as the code space, there must be equal number of seeds as logical qubits. In this context, two interesting questions are: (1) When are seeds local? More precisely, when is each seed a subspace of some tensor product Hilbert space of degrees of freedom (DOF) within a local area? (2) When are seeds just physical DOF? 
		
		\item During the process of circuits, it is interesting to implement subsystem symmetry via, e.g., higher-order cellular automata, which may lead to exotic long-range entangled states enriched by various types of subsystem symmetries~\cite{PRXquantum.5.030342,SSPT1,zhang2025programmable}. In such context, designing strange order correlators to probe symmetry fractionalization / enrichment may be of great interest~\cite{PRXquantum.5.030342,SSPT1}.  Further generalizing the present TD models by extending local Hilbert space (e.g., qudits in Ref.~\cite{PRXquantum.5.030328}) could lead to a unified circuit with more emerging structures.
		
		\item Experimental realization of higher-dimensional TD models presents intriguing possibilities. The $[0,1,2,2]$ model/Toric Code, which requires no synthetic dimension in 2-dimensional quantum simulator, has already been experimentally prepared and examined on programmable superconducting qubits~\cite{2019Nature_programmable_superconducting_qubits,Roushan_2021} and Rydberg atom arrays~\cite{Rydberg_atom_toric_code_1,Rydberg_atom_toric_code_2,Rydberg_atom_toric_code_3}. As dimension is defined by connectivity, synthetic dimensions can be realized by engineering the connectivity of degrees of freedom~\cite{synthetic_dimension_1,synthetic_dimension_2,synthetic_dimension_3}. Recent quantum simulators on the platforms of neutral atoms~\cite{quantum_simulation_with_movable_qubits_2,quantum_simulation_with_movable_qubits_3,quantum_simulation_with_movable_qubits_4} or ion traps~\cite{quantum_simulation_with_movable_qubits_1} with non-local interaction based on movable qubits provides the possibility of simulating high-dimensional TD models. Photonic systems present another promising avenue for synthetic dimension engineering~\cite{photon_synthetic_dimension_1,photon_synthetic_dimension_2,photon_synthetic_dimension_3,photon_synthetic_dimension_4}, which might be used to simulate high-dimensional TD models.

	\end{enumerate}

	\acknowledgements  We thank Xie Chen and Jing-Yu Zhao for the helpful explanation of quantum circuits during the initial stage of this work and thank Zhi-Yuan Wei for the helpful discussion on quantum simulation. This work was in part supported by National Natural Science Foundation of China (NSFC)  under Grants No. 12474149, Research Center for Magnetoelectric Physics of Guangdong Province under Grant No. 2024B0303390001, and Guangdong Provincial Key Laboratory of Magnetoelectric Physics and Devices under Grant No. 2022B1212010008.

	\appendix

	\bigskip
	
	\section{Technical details about lattice under PBC}\label{appendix_pbc}
	
	When PBC is applied, the lattice and leaves are no longer Euclidean. Therefore, an alternative description is required to specify leaves containing a straightforward $d_l$-dimensional sublattice, analogous to the OBC case.
	
	A suitable description that accommodates both OBC and PBC for the desired leaves is smooth subcomplex leaves, as introduced in Sec.~\ref{section_TD_models_definition}. Specifically, the leaves $l$ in Eq.~(\ref{paper_dddd_22}) under summation run through all possible $d_l$-dimensional subcomplex leaves containing the node $\gamma_{d_n}$, obtainable from smooth foliations of $X$. This approach avoids the need to identify tangent vectors at different points, which is only reasonable for Euclidean spaces.
	
	The requirement that a leaf $l$ is simultaneously a subcomplex excludes the following undesirable situations:
	\begin{itemize}
		\item $l$ does not contain a $d_l$-dimensional sublattice (a subset of the $0$-skeleton $X^0$),
		\item $l$ contains a slashed $d_l$-dimensional sublattice,
		\item $l$ exhibits undulations and contains a $d_l$-dimensional sublattice in a checkerboard or other pattern with missing vertices,
		\item $l$ contains only vertices of a $d_l$-dimensional sublattice but not all the $n$-cubes ($n \leq d_l$) ``in'' the sublattice.
	\end{itemize}
	
\begin{figure}
		\centering
		\includegraphics[width=0.4\textwidth]{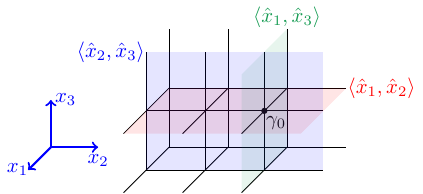}
		\caption{All the three smooth 2-dim subcomplex leaves containing $\gamma_0$ in 3-dim cubical lattice/complex under PBC, where each leaf $l$ is assigned with a $B$ term of the $[0,1,2,3]$ model, denoted by $B_{\gamma_0,l}$. The blue dot stands for the original point $(0,0,0)$.}
		\label{leaf-illustrative-figure-PBC}
	\end{figure}

	As an example, consider again the $[0,1,2,3]$ model under PBC. For each node $\gamma_0$, there are still three distinct $B_{\gamma_0,l}$ terms associated with it, each embedded in a 2-dimensional smooth subcomplex leaf, as shown in Fig.~\ref{leaf-illustrative-figure-PBC}.
	Although Fig.~\ref{leaf-illustrative-figure-PBC} appears similar to Fig.~\ref{leaf-illustrative-figure-Euclidean}, the underlying manifold in Fig.~\ref{leaf-illustrative-figure-PBC} is now a curved space ($T^3$). Therefore, interpreting leaves as truncated vector/Euclidean spaces is no longer applicable. A proper description for the desired leaves is smooth subcomplex leaves, which do not exhibit any angular discontinuities and contain full closed $n$-cubes ($n\leq d_l$) within them. We continue to use the symbol $\langle \hat{x}_{i_1}, \hat{x}_{i_2}, \dots, \hat{x}_{i_{d_l}} \rangle$ to represent subcomplex leaves, but they are no longer considered truncated vector/Euclidean spaces.
	
	The foliation and cellulation structure introduced in this subsection were utilized to construct the X-cube model on general 3-dimensional manifolds \cite{fracton16} and the foliated fracton phase classification theory \cite{chen_2018_foliated_fracton_phase_classification}. Later, it was observed that the long-range entanglement pattern of TD models extends beyond the description of foliated fracton orders, due to the nested foliation structure in TD models \cite{dddd_ERG}. For a more detailed discussion, see Sec.~\ref{subsec:ERG_hier}.

	\bigskip
	
	\bigskip
	
	\section{Proof of $\#$ seeds = $\#$ logical qubits for CSS code and Hadamard + CNOT type circuit}\label{proof_for_CSS_code_and_Had_CNOT_type_circuit}

	In this appendix, we prove that for any $[[n,k]]$ CSS code and any Hadamard + CNOT type preparation circuit $U'_c$, there exist $k$ seeds (physical qubits) and a growing circuit $U'_g$ (a product of mutually commuting CNOT gates) that together prepare the entire code space. Furthermore, for a fixed identification of the logical qubits, each logical $X$ operator that does not intersect the representative spins is unique. We begin by recalling the background and terminology used in the proof.

	The TD model series are special cases of CSS codes. An $[[n,k]]$ CSS code is a $2^k$-dimensional subspace (the code space) $\mathcal{H}_{\text{code}}$ of the $2^n$-dimensional Hilbert space of $n$ physical qubits. Such a code can be specified by two sets of independent stabilizer generators $S_X,S_Z$, where $s_X\in S_X$ are pure Pauli-$X$ products and $s_Z\in S_Z$ are pure Pauli-$Z$ products; the code space is the common $+1$ eigenspace of $S_X$ and $S_Z$. The union $S_X\cup S_Z$ generates the stabilizer group $\mathcal{S}$; any state $|\psi\rangle\in\mathcal{H}_{\text{code}}$ is stabilized by all elements of $\mathcal{S}$, meaning that $\forall s\in\mathcal{S}$, $s|\psi\rangle=|\psi\rangle$. Let $r=|S_X|$; then $|S_Z|=n-k-r$. Encode $S_X$ and $S_Z$ as an $r\times n$ matrix $G_X$ and an $(n-k-r)\times n$ matrix $G_Z$, respectively, where a row $v=(v_1, \dots, v_n)$ in $G_X$ corresponds to an operator $\prod_{i=1}^n X_i^{v_i} \in S_X$, and similarly for $G_Z$ and $S_Z$. For simplicity, we henceforth do not distinguish Pauli operator products from their corresponding row vectors. Logical operators are those that commute with all stabilizers. Since stabilizers act trivially on code states, any stabilizer is a logical identity operator, and logical operators that differ by multiplication by stabilizers belong to the same equivalence class (i.e., they have the same action on states in $\mathcal{H}_{\text{code}}$). We adopt the usual convention that the logical basis state
	\begin{equation}\label{invariant_base}
		|00\cdots0\rangle_{\text{logic}}=\prod_{s_X\in S_X}\frac{1}{\sqrt{2}}\Big(1+s_X\Big)|00\cdots0\rangle\,,
	\end{equation}
	and we take the logical operators $X^{\text{logic}}_{\alpha}$ ($Z^{\text{logic}}_{\alpha}$) to be pure $X$ ($Z$) products, $\alpha=1,\cdots,k$.

	A Hadamard + CNOT type preparation circuit for a CSS code with independent $X$ stabilizer generators $s_{X,1},\cdots,s_{X,r}$ is defined as follows: (1) choose a representative spin $\mathcal{R}(s_{X,i})$ for each $s_{X,i}$, such that for $i\neq j$, $\mathcal{R}(s_{X,i})\neq\mathcal{R}(s_{X,j})$; (2) apply $\prod_{i=1}^r \text{Had}_{\mathcal{R}(s_{X,i})}$ to the initial state $|00\cdots0\rangle$; (3) apply $\{\text{GCNOT}_{\mathcal{R}(s_{X,i})}^{\text{supp}(s_{X,i})}\mid i=1,\cdots,r\}$ in an order such that no representative spin acts as a target qubit before it has finished acting as a control qubit. The circuits $U_c$ constructed in Sec.~\ref{section_SQC} are Hadamard + CNOT type preparation circuits for TD models. In this appendix we fix the growing circuit $U'_g$ as a product of mutually commuting CNOT gates: each seed of a logical $X$ controls the remaining qubits in its support. $U'_g$ is not guaranteed to be sequential and local as $U_g$ in  Sec.~\ref{section_any_GS} (generic CSS codes may not admit locality). For TD models and the preparation circuit $U_c$, one can obtain $U'_g$ from $U_g$ by replacing, within each $\text{supp}(\tilde{X}_\alpha^{\text{logic}})$, the control qubits of the CNOT gates by the seed of $\tilde{X}_\alpha^{\text{logic}}$.

	Now fix an $[[n,k]]$ CSS code with $X$ stabilizer generators $s_{X,1},\cdots,s_{X,r}$, and a Hadamard + CNOT type preparation circuit $U'_c$ with representative spins $\mathcal{R}(s_{X,i})=c(i)\in\{1,\cdots,n\}$. By definition, $(s_{X,i})_{c(i)}=1$ and, for $i\neq j$, $c(i)\neq c(j)$. The circuit $U'_c$ applies a layer of Hadamard gates on $\{c(i)\mid i=1,\cdots,r\}$, followed by an ordered sequence of CNOT gates $\left(\text{GCNOT}_{c(i_1)}^{\text{supp}(s_{X,i_1})},\cdots,\text{GCNOT}_{c(i_r)}^{\text{supp}(s_{X,i_r})}\right)$, where $\{i_1,\cdots,i_r\}=\{1,\cdots,r\}$. In the Hadamard + CNOT type preparation circuit, no representative spin acts as a target qubit before it finishes acting as a control, which implies $(s_{X,i_j})_{c(i_{j'})}=0$ for all $j'>j$. Equivalently, the matrix $K$ with entries $K_{j,j'}=(s_{X,i_j})_{c(i_{j'})}$ is a unit (i.e., it has $1$ on the main diagonal)\footnote{The unit condition follows from $c(i)\in\text{supp}(s_{X,i})$, in other words, $(s_{X,i})_{c(i)}=1$.} lower triangular matrix. Particularly, the $r\times r$ submatrix $K^{(0)}$ of $G_X$, formed by columns $c(1),\cdots,c(r)$ of $G_X$, can be transformed into the unit lower triangular matrix $K$ by joint row and column swaps,\footnote{Suppose the matrix $K^{(0)}$ has entries
		$$\left(K^{(0)}\right)_{j,j'}=(s_{X,i'_j})_{c(i'_{j'})}\,,$$
		where $\{i'_1,\cdots,i'_r\}=\{1,\cdots,r\}$ specifies an ordered sequence $\left(\text{GCNOT}_{c(i'_1)}^{\text{supp}(s_{X,i'_1})},\cdots,\text{GCNOT}_{c(i'_r)}^{\text{supp}(s_{X,i'_r})}\right)$ that may not satisfy $(s_{X,i'_j})_{c(i_{j'})}=0$ for all $j'>j$. Any permutation of the ordered sequence decomposes into transpositions of $i'_j$ and $i'_{j'}$; exchanging $i'_j$ and $i'_{j'}$ implements a $j,j'$ row swap followed by a $j,j'$ column swap on $K^{(0)}$. This precisely describes the joint row–column swap.}
	which preserves rank; hence $K^{(0)}$ has the same rank as $K$ and is therefore full rank.

	Since $K^{(0)}$ is full rank, by performing row reductions on $G_X$ we obtain another generating matrix $G'_X$ in which the $c(i)$-th column has a single $1$ at row $i$. Denote the rows of $G'_X$ by $s'_{X,1},\cdots,s'_{X,r}$. Clearly, $\{s'_{X,i}\}$ are stabilizers. Since the $c(i)$-th column of $G'_X$ has a single $1$ at row $i$, the $c(i)$-th entry of the row vector representing $X^{\text{logic}}_{\alpha}$ can be zeroed by adding $s'_{X,i}$ (i.e., add it iff $(X^{\text{logic}}_{\alpha})_{c(i)}=1$). That is, for every $\alpha\in\{1,\cdots,k\}$ there exists
	$
	\tilde{X}^{\text{logic}}_{\alpha}=X^{\text{logic}}_{\alpha}+\sum_{i=1}^r b_{\alpha i}s'_{X,i}
	$
	such that
	\begin{equation}\label{no_rep_condition}
		(\tilde{X}^{\text{logic}}_{\alpha})_{c(i)}=0,\ \text{for all } i=1,\cdots,r\,.
	\end{equation}
	Note that the identification of logical qubits is not unique; for example, one may perform the following reidentification:
%	\begin{widetext}
  %	\end{widetext}
	\begin{align}
		Z^{\,\text{logic}}_{1}&\mapsto Z'^{\,\text{logic}}_{1}=Z^{\,\text{logic}}_{1}\notag\\
		X^{\text{logic}}_{1}&\mapsto X'^{\,\text{logic}}_{1}=X^{\text{logic}}_{1}X^{\text{logic}}_{2}\notag\\
		Z^{\text{logic}}_{2}&\mapsto Z'^{\,\text{logic}}_{2}=Z^{\text{logic}}_{1}Z^{\text{logic}}_{2}\notag\\
		X^{\text{logic}}_{2}&\mapsto X'^{\,\text{logic}}_{2}=X^{\text{logic}}_{2}\label{logical_qubit_reidentification_1}\,,
	\end{align}
	while at the same time
	\begin{align}
		|00\rangle_{\text{logic}}&\mapsto|00\rangle'_{\text{logic}}=|00\rangle_{\text{logic}}\notag\\
		|01\rangle_{\text{logic}}&\mapsto|01\rangle'_{\text{logic}}=|01\rangle_{\text{logic}}\notag\\
		|10\rangle_{\text{logic}}&\mapsto|10\rangle'_{\text{logic}}=|11\rangle_{\text{logic}}\notag\\
		|11\rangle_{\text{logic}}&\mapsto|11\rangle'_{\text{logic}}=|10\rangle_{\text{logic}}\,.
	\end{align}
	Let $\tilde{J}_X$ be the $k\times n$ matrix whose $\alpha$-th row corresponds to $\tilde{X}_\alpha^{\text{logic}}$ (each satisfying (\ref{no_rep_condition})). The transformation in Eq.~(\ref{logical_qubit_reidentification_1}) is precisely a row addition on $\tilde{J}_X$ (left-multiplication by an elementary matrix). Over $\mathbb{F}_2$, row additions alone generate all elementary row operations—row scaling is trivial and a row swap can be implemented by three row additions—so iterating such reidentifications implements an arbitrary row reduction while preserving (\ref{no_rep_condition}). Since $\tilde{J}_X$ has rank $k$, there exists a row-reduced form $\tilde{J}'_X$ that contains a $k\times k$ identity submatrix $\mathbb{I}_k$. Moreover, the rows of $\tilde{J}'_X$ (denoted by $\tilde{X}'^{\,\text{logic}}_\alpha$) also satisfy (\ref{no_rep_condition}). If the $q(1),\cdots,q(k)$-th columns of $\tilde{J}'_X$ form $\mathbb{I}_k$, then $\{q(\alpha)\mid \alpha=1,\cdots,k\}$ is a set of seeds,
	\begin{align}
		U'_cU'_g\bigotimes_{\alpha=1}^k|\sigma_{q(\alpha)}\rangle\bigotimes_{\text{non-seeds}}|0\rangle=\bigotimes_{\alpha=1}^k|\sigma_{q(\alpha)}\rangle_{\text{logic}}\,,
	\end{align}
	where the growing circuit is
	\begin{align}
		U'_g=\prod_{\alpha=1}^k\text{GCNOT}_{q(\alpha)}^{\text{supp}(\tilde{X}'^{\text{logic}}_\alpha)}\,.
	\end{align}
	Recall that $s_{X,i}$ are row vectors in our convention. Let $s^c_{X,i}$ denote the $1\times r$ row vector obtained by restricting $s_{X,i}$ to the columns $c(1),\cdots,c(r)$; equivalently, the $s^c_{X,i}$ are the rows of $K^{(0)}$. As shown above, $K^{(0)}$ is full rank, so there is no nonzero $v\in\mathbb{F}_2^r$ such that $\sum_{i=1}^r v_i s_{X,i}^c=0$. Consequently, since $\tilde{X}'^{\,\text{logic}}_\alpha$ satisfies (\ref{no_rep_condition}), there is no nonzero $v\in\mathbb{F}_2^r$ such that $\tilde{X}''^{\,\text{logic}}_\alpha=\tilde{X}'^{\,\text{logic}}_\alpha+\sum_{i=1}^r v_i s_{X,i}$ also satisfies (\ref{no_rep_condition}). Equivalently (since $\{s_{X,i}\}$ generate all $X$ stabilizers), there is no $X$ stabilizer $s_X$ with $\tilde{X}''^{\,\text{logic}}_\alpha=\tilde{X}'^{\,\text{logic}}_\alpha+s_X$ satisfying (\ref{no_rep_condition}). In other words, $\tilde{X}'^{\,\text{logic}}_\alpha$ is the unique logical $X$ in its equivalence class that satisfies (\ref{no_rep_condition}). Finally, we would like to point out that logical qubit reidentification of the form in Eq.~(\ref{logical_qubit_reidentification_1}) never changes $|00\cdots0\rangle_{\text{logic}}$, so Eq.~(\ref{invariant_base}) always holds in our construction.


\begin{thebibliography}{178}%
\makeatletter
\providecommand \@ifxundefined [1]{%
 \@ifx{#1\undefined}
}%
\providecommand \@ifnum [1]{%
 \ifnum #1\expandafter \@firstoftwo
 \else \expandafter \@secondoftwo
 \fi
}%
\providecommand \@ifx [1]{%
 \ifx #1\expandafter \@firstoftwo
 \else \expandafter \@secondoftwo
 \fi
}%
\providecommand \natexlab [1]{#1}%
\providecommand \enquote  [1]{``#1''}%
\providecommand \bibnamefont  [1]{#1}%
\providecommand \bibfnamefont [1]{#1}%
\providecommand \citenamefont [1]{#1}%
\providecommand \href@noop [0]{\@secondoftwo}%
\providecommand \href [0]{\begingroup \@sanitize@url \@href}%
\providecommand \@href[1]{\@@startlink{#1}\@@href}%
\providecommand \@@href[1]{\endgroup#1\@@endlink}%
\providecommand \@sanitize@url [0]{\catcode `\\12\catcode `\$12\catcode
  `\&12\catcode `\#12\catcode `\^12\catcode `\_12\catcode `\%12\relax}%
\providecommand \@@startlink[1]{}%
\providecommand \@@endlink[0]{}%
\providecommand \url  [0]{\begingroup\@sanitize@url \@url }%
\providecommand \@url [1]{\endgroup\@href {#1}{\urlprefix }}%
\providecommand \urlprefix  [0]{URL }%
\providecommand \Eprint [0]{\href }%
\providecommand \doibase [0]{https://doi.org/}%
\providecommand \selectlanguage [0]{\@gobble}%
\providecommand \bibinfo  [0]{\@secondoftwo}%
\providecommand \bibfield  [0]{\@secondoftwo}%
\providecommand \translation [1]{[#1]}%
\providecommand \BibitemOpen [0]{}%
\providecommand \bibitemStop [0]{}%
\providecommand \bibitemNoStop [0]{.\EOS\space}%
\providecommand \EOS [0]{\spacefactor3000\relax}%
\providecommand \BibitemShut  [1]{\csname bibitem#1\endcsname}%
\let\auto@bib@innerbib\@empty
%</preamble>
\bibitem [{\citenamefont {Wen}(1990)}]{TO1}%
  \BibitemOpen
  \bibfield  {author} {\bibinfo {author} {\bibfnamefont {X.~G.}\ \bibnamefont
  {Wen}},\ }\bibfield  {title} {\bibinfo {title} {{Topological Order in Rigid
  States}},\ }\href {https://doi.org/10.1142/S0217979290000139} {\bibfield
  {journal} {\bibinfo  {journal} {Int. J. Mod. Phys. B}\ }\textbf {\bibinfo
  {volume} {4}},\ \bibinfo {pages} {239} (\bibinfo {year} {1990})}\BibitemShut
  {NoStop}%
\bibitem [{\citenamefont {{Wen}}(2017)}]{TO2}%
  \BibitemOpen
  \bibfield  {author} {\bibinfo {author} {\bibfnamefont {X.-G.}\ \bibnamefont
  {{Wen}}},\ }\bibfield  {title} {\bibinfo {title} {{Colloquium: Zoo of
  quantum-topological phases of matter}},\ }\href
  {https://doi.org/10.1103/RevModPhys.89.041004} {\bibfield  {journal}
  {\bibinfo  {journal} {Reviews of Modern Physics}\ }\textbf {\bibinfo {volume}
  {89}},\ \bibinfo {eid} {041004} (\bibinfo {year} {2017})},\ \Eprint
  {https://arxiv.org/abs/1610.03911} {arXiv:1610.03911 [cond-mat.str-el]}
  \BibitemShut {NoStop}%
\bibitem [{\citenamefont {Altman}\ \emph {et~al.}(2021)\citenamefont {Altman},
  \citenamefont {Brown}, \citenamefont {Carleo}, \citenamefont {Carr},
  \citenamefont {Demler}, \citenamefont {Chin}, \citenamefont {DeMarco},
  \citenamefont {Economou}, \citenamefont {Eriksson}, \citenamefont {Fu},
  \citenamefont {Greiner}, \citenamefont {Hazzard}, \citenamefont {Hulet},
  \citenamefont {Koll\'ar}, \citenamefont {Lev}, \citenamefont {Lukin},
  \citenamefont {Ma}, \citenamefont {Mi}, \citenamefont {Misra}, \citenamefont
  {Monroe}, \citenamefont {Murch}, \citenamefont {Nazario}, \citenamefont {Ni},
  \citenamefont {Potter}, \citenamefont {Roushan}, \citenamefont {Saffman},
  \citenamefont {Schleier-Smith}, \citenamefont {Siddiqi}, \citenamefont
  {Simmonds}, \citenamefont {Singh}, \citenamefont {Spielman}, \citenamefont
  {Temme}, \citenamefont {Weiss}, \citenamefont {Vu\ifmmode \check{c}\else
  \v{c}\fi{}kovi\ifmmode~\acute{c}\else \'{c}\fi{}}, \citenamefont
  {Vuleti\ifmmode~\acute{c}\else \'{c}\fi{}}, \citenamefont {Ye},\ and\
  \citenamefont {Zwierlein}}]{quantum_simulation_overview_2021}%
  \BibitemOpen
  \bibfield  {author} {\bibinfo {author} {\bibfnamefont {E.}~\bibnamefont
  {Altman}}, \bibinfo {author} {\bibfnamefont {K.~R.}\ \bibnamefont {Brown}},
  \bibinfo {author} {\bibfnamefont {G.}~\bibnamefont {Carleo}}, \bibinfo
  {author} {\bibfnamefont {L.~D.}\ \bibnamefont {Carr}}, \bibinfo {author}
  {\bibfnamefont {E.}~\bibnamefont {Demler}}, \bibinfo {author} {\bibfnamefont
  {C.}~\bibnamefont {Chin}}, \bibinfo {author} {\bibfnamefont {B.}~\bibnamefont
  {DeMarco}}, \bibinfo {author} {\bibfnamefont {S.~E.}\ \bibnamefont
  {Economou}}, \bibinfo {author} {\bibfnamefont {M.~A.}\ \bibnamefont
  {Eriksson}}, \bibinfo {author} {\bibfnamefont {K.-M.~C.}\ \bibnamefont {Fu}},
  \bibinfo {author} {\bibfnamefont {M.}~\bibnamefont {Greiner}}, \bibinfo
  {author} {\bibfnamefont {K.~R.}\ \bibnamefont {Hazzard}}, \bibinfo {author}
  {\bibfnamefont {R.~G.}\ \bibnamefont {Hulet}}, \bibinfo {author}
  {\bibfnamefont {A.~J.}\ \bibnamefont {Koll\'ar}}, \bibinfo {author}
  {\bibfnamefont {B.~L.}\ \bibnamefont {Lev}}, \bibinfo {author} {\bibfnamefont
  {M.~D.}\ \bibnamefont {Lukin}}, \bibinfo {author} {\bibfnamefont
  {R.}~\bibnamefont {Ma}}, \bibinfo {author} {\bibfnamefont {X.}~\bibnamefont
  {Mi}}, \bibinfo {author} {\bibfnamefont {S.}~\bibnamefont {Misra}}, \bibinfo
  {author} {\bibfnamefont {C.}~\bibnamefont {Monroe}}, \bibinfo {author}
  {\bibfnamefont {K.}~\bibnamefont {Murch}}, \bibinfo {author} {\bibfnamefont
  {Z.}~\bibnamefont {Nazario}}, \bibinfo {author} {\bibfnamefont {K.-K.}\
  \bibnamefont {Ni}}, \bibinfo {author} {\bibfnamefont {A.~C.}\ \bibnamefont
  {Potter}}, \bibinfo {author} {\bibfnamefont {P.}~\bibnamefont {Roushan}},
  \bibinfo {author} {\bibfnamefont {M.}~\bibnamefont {Saffman}}, \bibinfo
  {author} {\bibfnamefont {M.}~\bibnamefont {Schleier-Smith}}, \bibinfo
  {author} {\bibfnamefont {I.}~\bibnamefont {Siddiqi}}, \bibinfo {author}
  {\bibfnamefont {R.}~\bibnamefont {Simmonds}}, \bibinfo {author}
  {\bibfnamefont {M.}~\bibnamefont {Singh}}, \bibinfo {author} {\bibfnamefont
  {I.}~\bibnamefont {Spielman}}, \bibinfo {author} {\bibfnamefont
  {K.}~\bibnamefont {Temme}}, \bibinfo {author} {\bibfnamefont {D.~S.}\
  \bibnamefont {Weiss}}, \bibinfo {author} {\bibfnamefont {J.}~\bibnamefont
  {Vu\ifmmode \check{c}\else \v{c}\fi{}kovi\ifmmode~\acute{c}\else
  \'{c}\fi{}}}, \bibinfo {author} {\bibfnamefont {V.}~\bibnamefont
  {Vuleti\ifmmode~\acute{c}\else \'{c}\fi{}}}, \bibinfo {author} {\bibfnamefont
  {J.}~\bibnamefont {Ye}},\ and\ \bibinfo {author} {\bibfnamefont
  {M.}~\bibnamefont {Zwierlein}},\ }\bibfield  {title} {\bibinfo {title}
  {Quantum simulators: Architectures and opportunities},\ }\href
  {https://doi.org/10.1103/PRXQuantum.2.017003} {\bibfield  {journal} {\bibinfo
   {journal} {PRX Quantum}\ }\textbf {\bibinfo {volume} {2}},\ \bibinfo {pages}
  {017003} (\bibinfo {year} {2021})}\BibitemShut {NoStop}%
\bibitem [{\citenamefont {Chen}\ \emph {et~al.}(2010)\citenamefont {Chen},
  \citenamefont {Gu},\ and\ \citenamefont
  {Wen}}]{phase_transition_and_LU_circuit}%
  \BibitemOpen
  \bibfield  {author} {\bibinfo {author} {\bibfnamefont {X.}~\bibnamefont
  {Chen}}, \bibinfo {author} {\bibfnamefont {Z.-C.}\ \bibnamefont {Gu}},\ and\
  \bibinfo {author} {\bibfnamefont {X.-G.}\ \bibnamefont {Wen}},\ }\bibfield
  {title} {\bibinfo {title} {Local unitary transformation, long-range quantum
  entanglement, wave function renormalization, and topological order},\ }\href
  {https://doi.org/10.1103/PhysRevB.82.155138} {\bibfield  {journal} {\bibinfo
  {journal} {Phys. Rev. B}\ }\textbf {\bibinfo {volume} {82}},\ \bibinfo
  {pages} {155138} (\bibinfo {year} {2010})}\BibitemShut {NoStop}%
\bibitem [{\citenamefont {Sch\"on}\ \emph {et~al.}(2005)\citenamefont
  {Sch\"on}, \citenamefont {Solano}, \citenamefont {Verstraete}, \citenamefont
  {Cirac},\ and\ \citenamefont {Wolf}}]{2005_SQC_Cirac}%
  \BibitemOpen
  \bibfield  {author} {\bibinfo {author} {\bibfnamefont {C.}~\bibnamefont
  {Sch\"on}}, \bibinfo {author} {\bibfnamefont {E.}~\bibnamefont {Solano}},
  \bibinfo {author} {\bibfnamefont {F.}~\bibnamefont {Verstraete}}, \bibinfo
  {author} {\bibfnamefont {J.~I.}\ \bibnamefont {Cirac}},\ and\ \bibinfo
  {author} {\bibfnamefont {M.~M.}\ \bibnamefont {Wolf}},\ }\bibfield  {title}
  {\bibinfo {title} {Sequential generation of entangled multiqubit states},\
  }\href {https://doi.org/10.1103/PhysRevLett.95.110503} {\bibfield  {journal}
  {\bibinfo  {journal} {Phys. Rev. Lett.}\ }\textbf {\bibinfo {volume} {95}},\
  \bibinfo {pages} {110503} (\bibinfo {year} {2005})}\BibitemShut {NoStop}%
\bibitem [{\citenamefont {Sch\"on}\ \emph {et~al.}(2007)\citenamefont
  {Sch\"on}, \citenamefont {Hammerer}, \citenamefont {Wolf}, \citenamefont
  {Cirac},\ and\ \citenamefont {Solano}}]{2007_SQC_Cirac}%
  \BibitemOpen
  \bibfield  {author} {\bibinfo {author} {\bibfnamefont {C.}~\bibnamefont
  {Sch\"on}}, \bibinfo {author} {\bibfnamefont {K.}~\bibnamefont {Hammerer}},
  \bibinfo {author} {\bibfnamefont {M.~M.}\ \bibnamefont {Wolf}}, \bibinfo
  {author} {\bibfnamefont {J.~I.}\ \bibnamefont {Cirac}},\ and\ \bibinfo
  {author} {\bibfnamefont {E.}~\bibnamefont {Solano}},\ }\bibfield  {title}
  {\bibinfo {title} {Sequential generation of matrix-product states in cavity
  qed},\ }\href {https://doi.org/10.1103/PhysRevA.75.032311} {\bibfield
  {journal} {\bibinfo  {journal} {Phys. Rev. A}\ }\textbf {\bibinfo {volume}
  {75}},\ \bibinfo {pages} {032311} (\bibinfo {year} {2007})}\BibitemShut
  {NoStop}%
\bibitem [{\citenamefont {Ba\~nuls}\ \emph {et~al.}(2008)\citenamefont
  {Ba\~nuls}, \citenamefont {P\'erez-Garc\'{\i}a}, \citenamefont {Wolf},
  \citenamefont {Verstraete},\ and\ \citenamefont {Cirac}}]{2008_SQC_Cirac}%
  \BibitemOpen
  \bibfield  {author} {\bibinfo {author} {\bibfnamefont {M.~C.}\ \bibnamefont
  {Ba\~nuls}}, \bibinfo {author} {\bibfnamefont {D.}~\bibnamefont
  {P\'erez-Garc\'{\i}a}}, \bibinfo {author} {\bibfnamefont {M.~M.}\
  \bibnamefont {Wolf}}, \bibinfo {author} {\bibfnamefont {F.}~\bibnamefont
  {Verstraete}},\ and\ \bibinfo {author} {\bibfnamefont {J.~I.}\ \bibnamefont
  {Cirac}},\ }\bibfield  {title} {\bibinfo {title} {Sequentially generated
  states for the study of two-dimensional systems},\ }\href
  {https://doi.org/10.1103/PhysRevA.77.052306} {\bibfield  {journal} {\bibinfo
  {journal} {Phys. Rev. A}\ }\textbf {\bibinfo {volume} {77}},\ \bibinfo
  {pages} {052306} (\bibinfo {year} {2008})}\BibitemShut {NoStop}%
\bibitem [{\citenamefont {Satzinger}\ \emph {et~al.}(2021)\citenamefont
  {Satzinger}, \citenamefont {Liu}, \citenamefont {Smith}, \citenamefont
  {Knapp}, \citenamefont {Newman}, \citenamefont {Jones}, \citenamefont {Chen},
  \citenamefont {Quintana}, \citenamefont {Mi}, \citenamefont {Dunsworth},
  \citenamefont {Gidney}, \citenamefont {Aleiner}, \citenamefont {Arute},
  \citenamefont {Arya}, \citenamefont {Atalaya}, \citenamefont {Babbush},
  \citenamefont {Bardin}, \citenamefont {Barends}, \citenamefont {Basso},
  \citenamefont {Bengtsson}, \citenamefont {Bilmes}, \citenamefont {Broughton},
  \citenamefont {Buckley}, \citenamefont {Buell}, \citenamefont {Burkett},
  \citenamefont {Bushnell}, \citenamefont {Chiaro}, \citenamefont {Collins},
  \citenamefont {Courtney}, \citenamefont {Demura}, \citenamefont {Derk},
  \citenamefont {Eppens}, \citenamefont {Erickson}, \citenamefont {Faoro},
  \citenamefont {Farhi}, \citenamefont {Fowler}, \citenamefont {Foxen},
  \citenamefont {Giustina}, \citenamefont {Greene}, \citenamefont {Gross},
  \citenamefont {Harrigan}, \citenamefont {Harrington}, \citenamefont {Hilton},
  \citenamefont {Hong}, \citenamefont {Huang}, \citenamefont {Huggins},
  \citenamefont {Ioffe}, \citenamefont {Isakov}, \citenamefont {Jeffrey},
  \citenamefont {Jiang}, \citenamefont {Kafri}, \citenamefont {Kechedzhi},
  \citenamefont {Khattar}, \citenamefont {Kim}, \citenamefont {Klimov},
  \citenamefont {Korotkov}, \citenamefont {Kostritsa}, \citenamefont
  {Landhuis}, \citenamefont {Laptev}, \citenamefont {Locharla}, \citenamefont
  {Lucero}, \citenamefont {Martin}, \citenamefont {McClean}, \citenamefont
  {McEwen}, \citenamefont {Miao}, \citenamefont {Mohseni}, \citenamefont
  {Montazeri}, \citenamefont {Mruczkiewicz}, \citenamefont {Mutus},
  \citenamefont {Naaman}, \citenamefont {Neeley}, \citenamefont {Neill},
  \citenamefont {Niu}, \citenamefont {O’Brien}, \citenamefont {Opremcak},
  \citenamefont {Pató}, \citenamefont {Petukhov}, \citenamefont {Rubin},
  \citenamefont {Sank}, \citenamefont {Shvarts}, \citenamefont {Strain},
  \citenamefont {Szalay}, \citenamefont {Villalonga}, \citenamefont {White},
  \citenamefont {Yao}, \citenamefont {Yeh}, \citenamefont {Yoo}, \citenamefont
  {Zalcman}, \citenamefont {Neven}, \citenamefont {Boixo}, \citenamefont
  {Megrant}, \citenamefont {Chen}, \citenamefont {Kelly}, \citenamefont
  {Smelyanskiy}, \citenamefont {Kitaev}, \citenamefont {Knap}, \citenamefont
  {Pollmann},\ and\ \citenamefont {Roushan}}]{Roushan_2021}%
  \BibitemOpen
  \bibfield  {author} {\bibinfo {author} {\bibfnamefont {K.~J.}\ \bibnamefont
  {Satzinger}}, \bibinfo {author} {\bibfnamefont {Y.-J.}\ \bibnamefont {Liu}},
  \bibinfo {author} {\bibfnamefont {A.}~\bibnamefont {Smith}}, \bibinfo
  {author} {\bibfnamefont {C.}~\bibnamefont {Knapp}}, \bibinfo {author}
  {\bibfnamefont {M.}~\bibnamefont {Newman}}, \bibinfo {author} {\bibfnamefont
  {C.}~\bibnamefont {Jones}}, \bibinfo {author} {\bibfnamefont
  {Z.}~\bibnamefont {Chen}}, \bibinfo {author} {\bibfnamefont {C.}~\bibnamefont
  {Quintana}}, \bibinfo {author} {\bibfnamefont {X.}~\bibnamefont {Mi}},
  \bibinfo {author} {\bibfnamefont {A.}~\bibnamefont {Dunsworth}}, \bibinfo
  {author} {\bibfnamefont {C.}~\bibnamefont {Gidney}}, \bibinfo {author}
  {\bibfnamefont {I.}~\bibnamefont {Aleiner}}, \bibinfo {author} {\bibfnamefont
  {F.}~\bibnamefont {Arute}}, \bibinfo {author} {\bibfnamefont
  {K.}~\bibnamefont {Arya}}, \bibinfo {author} {\bibfnamefont {J.}~\bibnamefont
  {Atalaya}}, \bibinfo {author} {\bibfnamefont {R.}~\bibnamefont {Babbush}},
  \bibinfo {author} {\bibfnamefont {J.~C.}\ \bibnamefont {Bardin}}, \bibinfo
  {author} {\bibfnamefont {R.}~\bibnamefont {Barends}}, \bibinfo {author}
  {\bibfnamefont {J.}~\bibnamefont {Basso}}, \bibinfo {author} {\bibfnamefont
  {A.}~\bibnamefont {Bengtsson}}, \bibinfo {author} {\bibfnamefont
  {A.}~\bibnamefont {Bilmes}}, \bibinfo {author} {\bibfnamefont
  {M.}~\bibnamefont {Broughton}}, \bibinfo {author} {\bibfnamefont {B.~B.}\
  \bibnamefont {Buckley}}, \bibinfo {author} {\bibfnamefont {D.~A.}\
  \bibnamefont {Buell}}, \bibinfo {author} {\bibfnamefont {B.}~\bibnamefont
  {Burkett}}, \bibinfo {author} {\bibfnamefont {N.}~\bibnamefont {Bushnell}},
  \bibinfo {author} {\bibfnamefont {B.}~\bibnamefont {Chiaro}}, \bibinfo
  {author} {\bibfnamefont {R.}~\bibnamefont {Collins}}, \bibinfo {author}
  {\bibfnamefont {W.}~\bibnamefont {Courtney}}, \bibinfo {author}
  {\bibfnamefont {S.}~\bibnamefont {Demura}}, \bibinfo {author} {\bibfnamefont
  {A.~R.}\ \bibnamefont {Derk}}, \bibinfo {author} {\bibfnamefont
  {D.}~\bibnamefont {Eppens}}, \bibinfo {author} {\bibfnamefont
  {C.}~\bibnamefont {Erickson}}, \bibinfo {author} {\bibfnamefont
  {L.}~\bibnamefont {Faoro}}, \bibinfo {author} {\bibfnamefont
  {E.}~\bibnamefont {Farhi}}, \bibinfo {author} {\bibfnamefont {A.~G.}\
  \bibnamefont {Fowler}}, \bibinfo {author} {\bibfnamefont {B.}~\bibnamefont
  {Foxen}}, \bibinfo {author} {\bibfnamefont {M.}~\bibnamefont {Giustina}},
  \bibinfo {author} {\bibfnamefont {A.}~\bibnamefont {Greene}}, \bibinfo
  {author} {\bibfnamefont {J.~A.}\ \bibnamefont {Gross}}, \bibinfo {author}
  {\bibfnamefont {M.~P.}\ \bibnamefont {Harrigan}}, \bibinfo {author}
  {\bibfnamefont {S.~D.}\ \bibnamefont {Harrington}}, \bibinfo {author}
  {\bibfnamefont {J.}~\bibnamefont {Hilton}}, \bibinfo {author} {\bibfnamefont
  {S.}~\bibnamefont {Hong}}, \bibinfo {author} {\bibfnamefont {T.}~\bibnamefont
  {Huang}}, \bibinfo {author} {\bibfnamefont {W.~J.}\ \bibnamefont {Huggins}},
  \bibinfo {author} {\bibfnamefont {L.~B.}\ \bibnamefont {Ioffe}}, \bibinfo
  {author} {\bibfnamefont {S.~V.}\ \bibnamefont {Isakov}}, \bibinfo {author}
  {\bibfnamefont {E.}~\bibnamefont {Jeffrey}}, \bibinfo {author} {\bibfnamefont
  {Z.}~\bibnamefont {Jiang}}, \bibinfo {author} {\bibfnamefont
  {D.}~\bibnamefont {Kafri}}, \bibinfo {author} {\bibfnamefont
  {K.}~\bibnamefont {Kechedzhi}}, \bibinfo {author} {\bibfnamefont
  {T.}~\bibnamefont {Khattar}}, \bibinfo {author} {\bibfnamefont
  {S.}~\bibnamefont {Kim}}, \bibinfo {author} {\bibfnamefont {P.~V.}\
  \bibnamefont {Klimov}}, \bibinfo {author} {\bibfnamefont {A.~N.}\
  \bibnamefont {Korotkov}}, \bibinfo {author} {\bibfnamefont {F.}~\bibnamefont
  {Kostritsa}}, \bibinfo {author} {\bibfnamefont {D.}~\bibnamefont {Landhuis}},
  \bibinfo {author} {\bibfnamefont {P.}~\bibnamefont {Laptev}}, \bibinfo
  {author} {\bibfnamefont {A.}~\bibnamefont {Locharla}}, \bibinfo {author}
  {\bibfnamefont {E.}~\bibnamefont {Lucero}}, \bibinfo {author} {\bibfnamefont
  {O.}~\bibnamefont {Martin}}, \bibinfo {author} {\bibfnamefont {J.~R.}\
  \bibnamefont {McClean}}, \bibinfo {author} {\bibfnamefont {M.}~\bibnamefont
  {McEwen}}, \bibinfo {author} {\bibfnamefont {K.~C.}\ \bibnamefont {Miao}},
  \bibinfo {author} {\bibfnamefont {M.}~\bibnamefont {Mohseni}}, \bibinfo
  {author} {\bibfnamefont {S.}~\bibnamefont {Montazeri}}, \bibinfo {author}
  {\bibfnamefont {W.}~\bibnamefont {Mruczkiewicz}}, \bibinfo {author}
  {\bibfnamefont {J.}~\bibnamefont {Mutus}}, \bibinfo {author} {\bibfnamefont
  {O.}~\bibnamefont {Naaman}}, \bibinfo {author} {\bibfnamefont
  {M.}~\bibnamefont {Neeley}}, \bibinfo {author} {\bibfnamefont
  {C.}~\bibnamefont {Neill}}, \bibinfo {author} {\bibfnamefont {M.~Y.}\
  \bibnamefont {Niu}}, \bibinfo {author} {\bibfnamefont {T.~E.}\ \bibnamefont
  {O’Brien}}, \bibinfo {author} {\bibfnamefont {A.}~\bibnamefont {Opremcak}},
  \bibinfo {author} {\bibfnamefont {B.}~\bibnamefont {Pató}}, \bibinfo
  {author} {\bibfnamefont {A.}~\bibnamefont {Petukhov}}, \bibinfo {author}
  {\bibfnamefont {N.~C.}\ \bibnamefont {Rubin}}, \bibinfo {author}
  {\bibfnamefont {D.}~\bibnamefont {Sank}}, \bibinfo {author} {\bibfnamefont
  {V.}~\bibnamefont {Shvarts}}, \bibinfo {author} {\bibfnamefont
  {D.}~\bibnamefont {Strain}}, \bibinfo {author} {\bibfnamefont
  {M.}~\bibnamefont {Szalay}}, \bibinfo {author} {\bibfnamefont
  {B.}~\bibnamefont {Villalonga}}, \bibinfo {author} {\bibfnamefont {T.~C.}\
  \bibnamefont {White}}, \bibinfo {author} {\bibfnamefont {Z.}~\bibnamefont
  {Yao}}, \bibinfo {author} {\bibfnamefont {P.}~\bibnamefont {Yeh}}, \bibinfo
  {author} {\bibfnamefont {J.}~\bibnamefont {Yoo}}, \bibinfo {author}
  {\bibfnamefont {A.}~\bibnamefont {Zalcman}}, \bibinfo {author} {\bibfnamefont
  {H.}~\bibnamefont {Neven}}, \bibinfo {author} {\bibfnamefont
  {S.}~\bibnamefont {Boixo}}, \bibinfo {author} {\bibfnamefont
  {A.}~\bibnamefont {Megrant}}, \bibinfo {author} {\bibfnamefont
  {Y.}~\bibnamefont {Chen}}, \bibinfo {author} {\bibfnamefont {J.}~\bibnamefont
  {Kelly}}, \bibinfo {author} {\bibfnamefont {V.}~\bibnamefont {Smelyanskiy}},
  \bibinfo {author} {\bibfnamefont {A.}~\bibnamefont {Kitaev}}, \bibinfo
  {author} {\bibfnamefont {M.}~\bibnamefont {Knap}}, \bibinfo {author}
  {\bibfnamefont {F.}~\bibnamefont {Pollmann}},\ and\ \bibinfo {author}
  {\bibfnamefont {P.}~\bibnamefont {Roushan}},\ }\bibfield  {title} {\bibinfo
  {title} {Realizing topologically ordered states on a quantum processor},\
  }\href {https://doi.org/10.1126/science.abi8378} {\bibfield  {journal}
  {\bibinfo  {journal} {Science}\ }\textbf {\bibinfo {volume} {374}},\ \bibinfo
  {pages} {1237–1241} (\bibinfo {year} {2021})}\BibitemShut {NoStop}%
\bibitem [{\citenamefont {Liu}\ \emph {et~al.}(2022)\citenamefont {Liu},
  \citenamefont {Shtengel}, \citenamefont {Smith},\ and\ \citenamefont
  {Pollmann}}]{Liu_Yu-Jie2021_SQC_string_net}%
  \BibitemOpen
  \bibfield  {author} {\bibinfo {author} {\bibfnamefont {Y.-J.}\ \bibnamefont
  {Liu}}, \bibinfo {author} {\bibfnamefont {K.}~\bibnamefont {Shtengel}},
  \bibinfo {author} {\bibfnamefont {A.}~\bibnamefont {Smith}},\ and\ \bibinfo
  {author} {\bibfnamefont {F.}~\bibnamefont {Pollmann}},\ }\bibfield  {title}
  {\bibinfo {title} {Methods for simulating string-net states and anyons on a
  digital quantum computer},\ }\href
  {https://doi.org/10.1103/PRXQuantum.3.040315} {\bibfield  {journal} {\bibinfo
   {journal} {PRX Quantum}\ }\textbf {\bibinfo {volume} {3}},\ \bibinfo {pages}
  {040315} (\bibinfo {year} {2022})}\BibitemShut {NoStop}%
\bibitem [{\citenamefont {Wei}\ \emph {et~al.}(2022)\citenamefont {Wei},
  \citenamefont {Malz},\ and\ \citenamefont
  {Cirac}}]{Wei_Zhi-Yuan_2021_SQC_PEPS}%
  \BibitemOpen
  \bibfield  {author} {\bibinfo {author} {\bibfnamefont {Z.-Y.}\ \bibnamefont
  {Wei}}, \bibinfo {author} {\bibfnamefont {D.}~\bibnamefont {Malz}},\ and\
  \bibinfo {author} {\bibfnamefont {J.~I.}\ \bibnamefont {Cirac}},\ }\bibfield
  {title} {\bibinfo {title} {Sequential generation of projected entangled-pair
  states},\ }\href {https://doi.org/10.1103/PhysRevLett.128.010607} {\bibfield
  {journal} {\bibinfo  {journal} {Phys. Rev. Lett.}\ }\textbf {\bibinfo
  {volume} {128}},\ \bibinfo {pages} {010607} (\bibinfo {year}
  {2022})}\BibitemShut {NoStop}%
\bibitem [{\citenamefont {Lin}\ \emph {et~al.}(2021)\citenamefont {Lin},
  \citenamefont {Dilip}, \citenamefont {Green}, \citenamefont {Smith},\ and\
  \citenamefont {Pollmann}}]{SQC_time_evolution}%
  \BibitemOpen
  \bibfield  {author} {\bibinfo {author} {\bibfnamefont {S.-H.}\ \bibnamefont
  {Lin}}, \bibinfo {author} {\bibfnamefont {R.}~\bibnamefont {Dilip}}, \bibinfo
  {author} {\bibfnamefont {A.~G.}\ \bibnamefont {Green}}, \bibinfo {author}
  {\bibfnamefont {A.}~\bibnamefont {Smith}},\ and\ \bibinfo {author}
  {\bibfnamefont {F.}~\bibnamefont {Pollmann}},\ }\bibfield  {title} {\bibinfo
  {title} {Real- and imaginary-time evolution with compressed quantum
  circuits},\ }\href {https://doi.org/10.1103/PRXQuantum.2.010342} {\bibfield
  {journal} {\bibinfo  {journal} {PRX Quantum}\ }\textbf {\bibinfo {volume}
  {2}},\ \bibinfo {pages} {010342} (\bibinfo {year} {2021})}\BibitemShut
  {NoStop}%
\bibitem [{\citenamefont {Chen}\ \emph {et~al.}(2024)\citenamefont {Chen},
  \citenamefont {Dua}, \citenamefont {Hermele}, \citenamefont {Stephen},
  \citenamefont {Tantivasadakarn}, \citenamefont {Vanhove},\ and\ \citenamefont
  {Zhao}}]{chen2024sequential}%
  \BibitemOpen
  \bibfield  {author} {\bibinfo {author} {\bibfnamefont {X.}~\bibnamefont
  {Chen}}, \bibinfo {author} {\bibfnamefont {A.}~\bibnamefont {Dua}}, \bibinfo
  {author} {\bibfnamefont {M.}~\bibnamefont {Hermele}}, \bibinfo {author}
  {\bibfnamefont {D.~T.}\ \bibnamefont {Stephen}}, \bibinfo {author}
  {\bibfnamefont {N.}~\bibnamefont {Tantivasadakarn}}, \bibinfo {author}
  {\bibfnamefont {R.}~\bibnamefont {Vanhove}},\ and\ \bibinfo {author}
  {\bibfnamefont {J.-Y.}\ \bibnamefont {Zhao}},\ }\bibfield  {title} {\bibinfo
  {title} {Sequential quantum circuits as maps between gapped phases},\ }\href
  {https://doi.org/10.1103/PhysRevB.109.075116} {\bibfield  {journal} {\bibinfo
   {journal} {Phys. Rev. B}\ }\textbf {\bibinfo {volume} {109}},\ \bibinfo
  {pages} {075116} (\bibinfo {year} {2024})}\BibitemShut {NoStop}%
\bibitem [{\citenamefont {Vanhove}\ \emph {et~al.}(2025)\citenamefont
  {Vanhove}, \citenamefont {Ravindran}, \citenamefont {Stephen}, \citenamefont
  {Wen},\ and\ \citenamefont {Chen}}]{SQC_duality}%
  \BibitemOpen
  \bibfield  {author} {\bibinfo {author} {\bibfnamefont {R.}~\bibnamefont
  {Vanhove}}, \bibinfo {author} {\bibfnamefont {V.}~\bibnamefont {Ravindran}},
  \bibinfo {author} {\bibfnamefont {D.~T.}\ \bibnamefont {Stephen}}, \bibinfo
  {author} {\bibfnamefont {X.-G.}\ \bibnamefont {Wen}},\ and\ \bibinfo {author}
  {\bibfnamefont {X.}~\bibnamefont {Chen}},\ }\bibfield  {title} {\bibinfo
  {title} {Duality via sequential quantum circuit in the topological holography
  formalism},\ }\href {https://doi.org/10.1103/y9cd-dz5k} {\bibfield  {journal}
  {\bibinfo  {journal} {Phys. Rev. B}\ }\textbf {\bibinfo {volume} {112}},\
  \bibinfo {pages} {035173} (\bibinfo {year} {2025})}\BibitemShut {NoStop}%
\bibitem [{\citenamefont {{Pace}}\ \emph {et~al.}(2024)\citenamefont {{Pace}},
  \citenamefont {{Delfino}}, \citenamefont {{Lam}},\ and\ \citenamefont
  {{Aksoy}}}]{2024arXiv240612962P}%
  \BibitemOpen
  \bibfield  {author} {\bibinfo {author} {\bibfnamefont {S.~D.}\ \bibnamefont
  {{Pace}}}, \bibinfo {author} {\bibfnamefont {G.}~\bibnamefont {{Delfino}}},
  \bibinfo {author} {\bibfnamefont {H.~T.}\ \bibnamefont {{Lam}}},\ and\
  \bibinfo {author} {\bibfnamefont {{\"O}.~M.}\ \bibnamefont {{Aksoy}}},\
  }\bibfield  {title} {\bibinfo {title} {{Gauging modulated symmetries:
  Kramers-Wannier dualities and non-invertible reflections}},\ }\href
  {https://doi.org/10.48550/arXiv.2406.12962} {\bibfield  {journal} {\bibinfo
  {journal} {arXiv e-prints}\ ,\ \bibinfo {eid} {arXiv:2406.12962}} (\bibinfo
  {year} {2024})},\ \Eprint {https://arxiv.org/abs/2406.12962}
  {arXiv:2406.12962 [cond-mat.str-el]} \BibitemShut {NoStop}%
\bibitem [{\citenamefont {{Parayil Mana}}\ \emph {et~al.}(2024)\citenamefont
  {{Parayil Mana}}, \citenamefont {{Li}}, \citenamefont {{Sukeno}},\ and\
  \citenamefont {{Wei}}}]{SQC_non_invertible_symmetry_2}%
  \BibitemOpen
  \bibfield  {author} {\bibinfo {author} {\bibfnamefont {A.}~\bibnamefont
  {{Parayil Mana}}}, \bibinfo {author} {\bibfnamefont {Y.}~\bibnamefont
  {{Li}}}, \bibinfo {author} {\bibfnamefont {H.}~\bibnamefont {{Sukeno}}},\
  and\ \bibinfo {author} {\bibfnamefont {T.-C.}\ \bibnamefont {{Wei}}},\
  }\bibfield  {title} {\bibinfo {title} {{Kennedy-Tasaki transformation and
  noninvertible symmetry in lattice models beyond one dimension}},\ }\href
  {https://doi.org/10.1103/PhysRevB.109.245129} {\bibfield  {journal} {\bibinfo
   {journal} {\prb}\ }\textbf {\bibinfo {volume} {109}},\ \bibinfo {eid}
  {245129} (\bibinfo {year} {2024})},\ \Eprint
  {https://arxiv.org/abs/2402.09520} {arXiv:2402.09520 [cond-mat.str-el]}
  \BibitemShut {NoStop}%
\bibitem [{\citenamefont {{Tantivasadakarn}}\ and\ \citenamefont
  {{Chen}}(2024)}]{SQC_Cheshire_string}%
  \BibitemOpen
  \bibfield  {author} {\bibinfo {author} {\bibfnamefont {N.}~\bibnamefont
  {{Tantivasadakarn}}}\ and\ \bibinfo {author} {\bibfnamefont {X.}~\bibnamefont
  {{Chen}}},\ }\bibfield  {title} {\bibinfo {title} {{String operators for
  Cheshire strings in topological phases}},\ }\href
  {https://doi.org/10.1103/PhysRevB.109.165149} {\bibfield  {journal} {\bibinfo
   {journal} {\prb}\ }\textbf {\bibinfo {volume} {109}},\ \bibinfo {eid}
  {165149} (\bibinfo {year} {2024})},\ \Eprint
  {https://arxiv.org/abs/2307.03180} {arXiv:2307.03180 [cond-mat.str-el]}
  \BibitemShut {NoStop}%
\bibitem [{\citenamefont {{Lu}}\ \emph {et~al.}(2024)\citenamefont {{Lu}},
  \citenamefont {{Sun}},\ and\ \citenamefont {{You}}}]{SQC_triality}%
  \BibitemOpen
  \bibfield  {author} {\bibinfo {author} {\bibfnamefont {D.-C.}\ \bibnamefont
  {{Lu}}}, \bibinfo {author} {\bibfnamefont {Z.}~\bibnamefont {{Sun}}},\ and\
  \bibinfo {author} {\bibfnamefont {Y.-Z.}\ \bibnamefont {{You}}},\ }\bibfield
  {title} {\bibinfo {title} {{Realizing triality and p-ality by lattice twisted
  gauging in (1+1)d quantum spin systems}},\ }\href
  {https://doi.org/10.21468/SciPostPhys.17.5.136} {\bibfield  {journal}
  {\bibinfo  {journal} {SciPost Physics}\ }\textbf {\bibinfo {volume} {17}},\
  \bibinfo {eid} {136} (\bibinfo {year} {2024})},\ \Eprint
  {https://arxiv.org/abs/2405.14939} {arXiv:2405.14939 [cond-mat.str-el]}
  \BibitemShut {NoStop}%
\bibitem [{\citenamefont {{Lyons}}\ \emph {et~al.}(2024)\citenamefont
  {{Lyons}}, \citenamefont {{Bowen Lo}}, \citenamefont {{Tantivasadakarn}},
  \citenamefont {{Vishwanath}},\ and\ \citenamefont
  {{Verresen}}}]{SQC_ribbon_operator}%
  \BibitemOpen
  \bibfield  {author} {\bibinfo {author} {\bibfnamefont {A.}~\bibnamefont
  {{Lyons}}}, \bibinfo {author} {\bibfnamefont {C.~F.}\ \bibnamefont {{Bowen
  Lo}}}, \bibinfo {author} {\bibfnamefont {N.}~\bibnamefont
  {{Tantivasadakarn}}}, \bibinfo {author} {\bibfnamefont {A.}~\bibnamefont
  {{Vishwanath}}},\ and\ \bibinfo {author} {\bibfnamefont {R.}~\bibnamefont
  {{Verresen}}},\ }\bibfield  {title} {\bibinfo {title} {{Protocols for
  Creating Anyons and Defects via Gauging}},\ }\href
  {https://doi.org/10.48550/arXiv.2411.04181} {\bibfield  {journal} {\bibinfo
  {journal} {arXiv e-prints}\ ,\ \bibinfo {eid} {arXiv:2411.04181}} (\bibinfo
  {year} {2024})},\ \Eprint {https://arxiv.org/abs/2411.04181}
  {arXiv:2411.04181 [quant-ph]} \BibitemShut {NoStop}%
\bibitem [{\citenamefont {Zeng}\ \emph {et~al.}(2019)\citenamefont {Zeng},
  \citenamefont {Chen}, \citenamefont {Zhou}, \citenamefont {Wen} \emph
  {et~al.}}]{zeng2019quantum}%
  \BibitemOpen
  \bibfield  {author} {\bibinfo {author} {\bibfnamefont {B.}~\bibnamefont
  {Zeng}}, \bibinfo {author} {\bibfnamefont {X.}~\bibnamefont {Chen}}, \bibinfo
  {author} {\bibfnamefont {D.-L.}\ \bibnamefont {Zhou}}, \bibinfo {author}
  {\bibfnamefont {X.-G.}\ \bibnamefont {Wen}}, \emph {et~al.},\ }\href
  {https://doi.org/https://doi.org/10.1007/978-1-4939-9084-9} {\emph {\bibinfo
  {title} {Quantum information meets quantum matter}}}\ (\bibinfo  {publisher}
  {Springer},\ \bibinfo {year} {2019})\BibitemShut {NoStop}%
\bibitem [{\citenamefont {{Vidal}}(2007)}]{ERG_vidal}%
  \BibitemOpen
  \bibfield  {author} {\bibinfo {author} {\bibfnamefont {G.}~\bibnamefont
  {{Vidal}}},\ }\bibfield  {title} {\bibinfo {title} {{Entanglement
  Renormalization}},\ }\href {https://doi.org/10.1103/PhysRevLett.99.220405}
  {\bibfield  {journal} {\bibinfo  {journal} {\prl}\ }\textbf {\bibinfo
  {volume} {99}},\ \bibinfo {eid} {220405} (\bibinfo {year} {2007})},\ \Eprint
  {https://arxiv.org/abs/cond-mat/0512165} {arXiv:cond-mat/0512165
  [cond-mat.str-el]} \BibitemShut {NoStop}%
\bibitem [{\citenamefont {Shirley}\ \emph {et~al.}(2018)\citenamefont
  {Shirley}, \citenamefont {Slagle}, \citenamefont {Wang},\ and\ \citenamefont
  {Chen}}]{fracton16}%
  \BibitemOpen
  \bibfield  {author} {\bibinfo {author} {\bibfnamefont {W.}~\bibnamefont
  {Shirley}}, \bibinfo {author} {\bibfnamefont {K.}~\bibnamefont {Slagle}},
  \bibinfo {author} {\bibfnamefont {Z.}~\bibnamefont {Wang}},\ and\ \bibinfo
  {author} {\bibfnamefont {X.}~\bibnamefont {Chen}},\ }\bibfield  {title}
  {\bibinfo {title} {Fracton models on general three-dimensional manifolds},\
  }\href {https://doi.org/10.1103/PhysRevX.8.031051} {\bibfield  {journal}
  {\bibinfo  {journal} {Phys. Rev. X}\ }\textbf {\bibinfo {volume} {8}},\
  \bibinfo {pages} {031051} (\bibinfo {year} {2018})}\BibitemShut {NoStop}%
\bibitem [{\citenamefont {Li}\ and\ \citenamefont {Ye}(2023)}]{dddd_ERG}%
  \BibitemOpen
  \bibfield  {author} {\bibinfo {author} {\bibfnamefont {M.-Y.}\ \bibnamefont
  {Li}}\ and\ \bibinfo {author} {\bibfnamefont {P.}~\bibnamefont {Ye}},\
  }\bibfield  {title} {\bibinfo {title} {Hierarchy of entanglement
  renormalization and long-range entangled states},\ }\href
  {https://doi.org/10.1103/PhysRevB.107.115169} {\bibfield  {journal} {\bibinfo
   {journal} {Phys. Rev. B}\ }\textbf {\bibinfo {volume} {107}},\ \bibinfo
  {pages} {115169} (\bibinfo {year} {2023})}\BibitemShut {NoStop}%
\bibitem [{\citenamefont {{Kitaev}}(2003)}]{toric_code}%
  \BibitemOpen
  \bibfield  {author} {\bibinfo {author} {\bibfnamefont {A.~Y.}\ \bibnamefont
  {{Kitaev}}},\ }\bibfield  {title} {\bibinfo {title} {{Fault-tolerant quantum
  computation by anyons}},\ }\href
  {https://doi.org/10.1016/S0003-4916(02)00018-0} {\bibfield  {journal}
  {\bibinfo  {journal} {Annals of Physics}\ }\textbf {\bibinfo {volume}
  {303}},\ \bibinfo {pages} {2} (\bibinfo {year} {2003})},\ \Eprint
  {https://arxiv.org/abs/quant-ph/9707021} {arXiv:quant-ph/9707021 [quant-ph]}
  \BibitemShut {NoStop}%
\bibitem [{\citenamefont {{Ali}}\ \emph {et~al.}(2024)\citenamefont {{Ali}},
  \citenamefont {{Marques}}, \citenamefont {{Crawford}}, \citenamefont
  {{Majaniemi}}, \citenamefont {{Serra-Peralta}}, \citenamefont {{Byfield}},
  \citenamefont {{Varbanov}}, \citenamefont {{Terhal}}, \citenamefont
  {{DiCarlo}},\ and\ \citenamefont
  {{Campbell}}}]{toric_code_error_correction_experiment}%
  \BibitemOpen
  \bibfield  {author} {\bibinfo {author} {\bibfnamefont {H.}~\bibnamefont
  {{Ali}}}, \bibinfo {author} {\bibfnamefont {J.}~\bibnamefont {{Marques}}},
  \bibinfo {author} {\bibfnamefont {O.}~\bibnamefont {{Crawford}}}, \bibinfo
  {author} {\bibfnamefont {J.}~\bibnamefont {{Majaniemi}}}, \bibinfo {author}
  {\bibfnamefont {M.}~\bibnamefont {{Serra-Peralta}}}, \bibinfo {author}
  {\bibfnamefont {D.}~\bibnamefont {{Byfield}}}, \bibinfo {author}
  {\bibfnamefont {B.}~\bibnamefont {{Varbanov}}}, \bibinfo {author}
  {\bibfnamefont {B.~M.}\ \bibnamefont {{Terhal}}}, \bibinfo {author}
  {\bibfnamefont {L.}~\bibnamefont {{DiCarlo}}},\ and\ \bibinfo {author}
  {\bibfnamefont {E.~T.}\ \bibnamefont {{Campbell}}},\ }\bibfield  {title}
  {\bibinfo {title} {{Reducing the error rate of a superconducting logical
  qubit using analog readout information}},\ }\href
  {https://doi.org/10.1103/PhysRevApplied.22.044031} {\bibfield  {journal}
  {\bibinfo  {journal} {Physical Review Applied}\ }\textbf {\bibinfo {volume}
  {22}},\ \bibinfo {eid} {044031} (\bibinfo {year} {2024})},\ \Eprint
  {https://arxiv.org/abs/2403.00706} {arXiv:2403.00706 [quant-ph]} \BibitemShut
  {NoStop}%
\bibitem [{\citenamefont {{Cai}}\ \emph {et~al.}(2024)\citenamefont {{Cai}},
  \citenamefont {{Mu}}, \citenamefont {{Wang}}, \citenamefont {{Zhou}},
  \citenamefont {{Ma}}, \citenamefont {{Pan}}, \citenamefont {{Hua}},
  \citenamefont {{Liu}}, \citenamefont {{Xue}}, \citenamefont {{Yu}},
  \citenamefont {{Wang}}, \citenamefont {{Song}}, \citenamefont {{Zou}},\ and\
  \citenamefont {{Sun}}}]{logical_qubits_entanglement}%
  \BibitemOpen
  \bibfield  {author} {\bibinfo {author} {\bibfnamefont {W.}~\bibnamefont
  {{Cai}}}, \bibinfo {author} {\bibfnamefont {X.}~\bibnamefont {{Mu}}},
  \bibinfo {author} {\bibfnamefont {W.}~\bibnamefont {{Wang}}}, \bibinfo
  {author} {\bibfnamefont {J.}~\bibnamefont {{Zhou}}}, \bibinfo {author}
  {\bibfnamefont {Y.}~\bibnamefont {{Ma}}}, \bibinfo {author} {\bibfnamefont
  {X.}~\bibnamefont {{Pan}}}, \bibinfo {author} {\bibfnamefont
  {Z.}~\bibnamefont {{Hua}}}, \bibinfo {author} {\bibfnamefont
  {X.}~\bibnamefont {{Liu}}}, \bibinfo {author} {\bibfnamefont
  {G.}~\bibnamefont {{Xue}}}, \bibinfo {author} {\bibfnamefont
  {H.}~\bibnamefont {{Yu}}}, \bibinfo {author} {\bibfnamefont {H.}~\bibnamefont
  {{Wang}}}, \bibinfo {author} {\bibfnamefont {Y.}~\bibnamefont {{Song}}},
  \bibinfo {author} {\bibfnamefont {C.-L.}\ \bibnamefont {{Zou}}},\ and\
  \bibinfo {author} {\bibfnamefont {L.}~\bibnamefont {{Sun}}},\ }\bibfield
  {title} {\bibinfo {title} {{Protecting entanglement between logical qubits
  via quantum error correction}},\ }\href
  {https://doi.org/10.1038/s41567-024-02446-8} {\bibfield  {journal} {\bibinfo
  {journal} {Nature Physics}\ }\textbf {\bibinfo {volume} {20}},\ \bibinfo
  {pages} {1022} (\bibinfo {year} {2024})},\ \Eprint
  {https://arxiv.org/abs/2302.13027} {arXiv:2302.13027 [quant-ph]} \BibitemShut
  {NoStop}%
\bibitem [{\citenamefont {{Horsman}}\ \emph {et~al.}(2012)\citenamefont
  {{Horsman}}, \citenamefont {{Fowler}}, \citenamefont {{Devitt}},\ and\
  \citenamefont {{Van Meter}}}]{toric_code_lattice_surgery}%
  \BibitemOpen
  \bibfield  {author} {\bibinfo {author} {\bibfnamefont {C.}~\bibnamefont
  {{Horsman}}}, \bibinfo {author} {\bibfnamefont {A.~G.}\ \bibnamefont
  {{Fowler}}}, \bibinfo {author} {\bibfnamefont {S.}~\bibnamefont {{Devitt}}},\
  and\ \bibinfo {author} {\bibfnamefont {R.}~\bibnamefont {{Van Meter}}},\
  }\bibfield  {title} {\bibinfo {title} {{Surface code quantum computing by
  lattice surgery}},\ }\href {https://doi.org/10.1088/1367-2630/14/12/123011}
  {\bibfield  {journal} {\bibinfo  {journal} {New Journal of Physics}\ }\textbf
  {\bibinfo {volume} {14}},\ \bibinfo {eid} {123011} (\bibinfo {year}
  {2012})},\ \Eprint {https://arxiv.org/abs/1111.4022} {arXiv:1111.4022
  [quant-ph]} \BibitemShut {NoStop}%
\bibitem [{\citenamefont {{Fowler}}\ and\ \citenamefont
  {{Gidney}}(2018)}]{fowlow_2018_lattice_surgery}%
  \BibitemOpen
  \bibfield  {author} {\bibinfo {author} {\bibfnamefont {A.~G.}\ \bibnamefont
  {{Fowler}}}\ and\ \bibinfo {author} {\bibfnamefont {C.}~\bibnamefont
  {{Gidney}}},\ }\bibfield  {title} {\bibinfo {title} {{Low overhead quantum
  computation using lattice surgery}},\ }\href
  {https://doi.org/10.48550/arXiv.1808.06709} {\bibfield  {journal} {\bibinfo
  {journal} {arXiv e-prints}\ ,\ \bibinfo {eid} {arXiv:1808.06709}} (\bibinfo
  {year} {2018})},\ \Eprint {https://arxiv.org/abs/1808.06709}
  {arXiv:1808.06709 [quant-ph]} \BibitemShut {NoStop}%
\bibitem [{\citenamefont {{Litinski}}(2019)}]{game_lattice_surgery}%
  \BibitemOpen
  \bibfield  {author} {\bibinfo {author} {\bibfnamefont {D.}~\bibnamefont
  {{Litinski}}},\ }\bibfield  {title} {\bibinfo {title} {{A Game of Surface
  Codes: Large-Scale Quantum Computing with Lattice Surgery}},\ }\href
  {https://doi.org/10.22331/q-2019-03-05-128} {\bibfield  {journal} {\bibinfo
  {journal} {Quantum}\ }\textbf {\bibinfo {volume} {3}},\ \bibinfo {pages}
  {128} (\bibinfo {year} {2019})},\ \Eprint {https://arxiv.org/abs/1808.02892}
  {arXiv:1808.02892 [quant-ph]} \BibitemShut {NoStop}%
\bibitem [{\citenamefont {Li}\ and\ \citenamefont {Ye}(2020)}]{li2020fracton}%
  \BibitemOpen
  \bibfield  {author} {\bibinfo {author} {\bibfnamefont {M.-Y.}\ \bibnamefont
  {Li}}\ and\ \bibinfo {author} {\bibfnamefont {P.}~\bibnamefont {Ye}},\
  }\bibfield  {title} {\bibinfo {title} {Fracton physics of spatially extended
  excitations},\ }\href {https://doi.org/10.1103/PhysRevB.101.245134}
  {\bibfield  {journal} {\bibinfo  {journal} {Physical Review B}\ }\textbf
  {\bibinfo {volume} {101}},\ \bibinfo {pages} {245134} (\bibinfo {year}
  {2020})}\BibitemShut {NoStop}%
\bibitem [{\citenamefont {Li}\ and\ \citenamefont
  {Ye}(2021{\natexlab{a}})}]{li2021fracton}%
  \BibitemOpen
  \bibfield  {author} {\bibinfo {author} {\bibfnamefont {M.-Y.}\ \bibnamefont
  {Li}}\ and\ \bibinfo {author} {\bibfnamefont {P.}~\bibnamefont {Ye}},\
  }\bibfield  {title} {\bibinfo {title} {Fracton physics of spatially extended
  excitations. ii. polynomial ground state degeneracy of exactly solvable
  models},\ }\href {https://doi.org/10.1103/PhysRevB.104.235127} {\bibfield
  {journal} {\bibinfo  {journal} {Physical Review B}\ }\textbf {\bibinfo
  {volume} {104}},\ \bibinfo {pages} {235127} (\bibinfo {year}
  {2021}{\natexlab{a}})}\BibitemShut {NoStop}%
\bibitem [{\citenamefont {Chamon}(2005)}]{fractonorder1}%
  \BibitemOpen
  \bibfield  {author} {\bibinfo {author} {\bibfnamefont {C.}~\bibnamefont
  {Chamon}},\ }\bibfield  {title} {\bibinfo {title} {Quantum glassiness in
  strongly correlated clean systems: An example of topological
  overprotection},\ }\href {https://doi.org/10.1103/PhysRevLett.94.040402}
  {\bibfield  {journal} {\bibinfo  {journal} {Phys. Rev. Lett.}\ }\textbf
  {\bibinfo {volume} {94}},\ \bibinfo {pages} {040402} (\bibinfo {year}
  {2005})}\BibitemShut {NoStop}%
\bibitem [{\citenamefont {Vijay}\ \emph {et~al.}(2015)\citenamefont {Vijay},
  \citenamefont {Haah},\ and\ \citenamefont {Fu}}]{fractonorder2}%
  \BibitemOpen
  \bibfield  {author} {\bibinfo {author} {\bibfnamefont {S.}~\bibnamefont
  {Vijay}}, \bibinfo {author} {\bibfnamefont {J.}~\bibnamefont {Haah}},\ and\
  \bibinfo {author} {\bibfnamefont {L.}~\bibnamefont {Fu}},\ }\bibfield
  {title} {\bibinfo {title} {A new kind of topological quantum order: A
  dimensional hierarchy of quasiparticles built from stationary excitations},\
  }\href {https://doi.org/10.1103/PhysRevB.92.235136} {\bibfield  {journal}
  {\bibinfo  {journal} {Phys. Rev. B}\ }\textbf {\bibinfo {volume} {92}},\
  \bibinfo {pages} {235136} (\bibinfo {year} {2015})}\BibitemShut {NoStop}%
\bibitem [{\citenamefont {Vijay}\ \emph {et~al.}(2016)\citenamefont {Vijay},
  \citenamefont {Haah},\ and\ \citenamefont {Fu}}]{fractonorder3}%
  \BibitemOpen
  \bibfield  {author} {\bibinfo {author} {\bibfnamefont {S.}~\bibnamefont
  {Vijay}}, \bibinfo {author} {\bibfnamefont {J.}~\bibnamefont {Haah}},\ and\
  \bibinfo {author} {\bibfnamefont {L.}~\bibnamefont {Fu}},\ }\bibfield
  {title} {\bibinfo {title} {Fracton topological order, generalized lattice
  gauge theory, and duality},\ }\href
  {https://doi.org/10.1103/PhysRevB.94.235157} {\bibfield  {journal} {\bibinfo
  {journal} {Phys. Rev. B}\ }\textbf {\bibinfo {volume} {94}},\ \bibinfo
  {pages} {235157} (\bibinfo {year} {2016})}\BibitemShut {NoStop}%
\bibitem [{\citenamefont {Song}\ \emph {et~al.}(2022)\citenamefont {Song},
  \citenamefont {Sch\"onmeier-Kromer}, \citenamefont {Liu}, \citenamefont
  {Viyuela}, \citenamefont {Pollet},\ and\ \citenamefont
  {Martin-Delgado}}]{PhysRevLett.129.230502}%
  \BibitemOpen
  \bibfield  {author} {\bibinfo {author} {\bibfnamefont {H.}~\bibnamefont
  {Song}}, \bibinfo {author} {\bibfnamefont {J.}~\bibnamefont
  {Sch\"onmeier-Kromer}}, \bibinfo {author} {\bibfnamefont {K.}~\bibnamefont
  {Liu}}, \bibinfo {author} {\bibfnamefont {O.}~\bibnamefont {Viyuela}},
  \bibinfo {author} {\bibfnamefont {L.}~\bibnamefont {Pollet}},\ and\ \bibinfo
  {author} {\bibfnamefont {M.~A.}\ \bibnamefont {Martin-Delgado}},\ }\bibfield
  {title} {\bibinfo {title} {Optimal thresholds for fracton codes and random
  spin models with subsystem symmetry},\ }\href
  {https://doi.org/10.1103/PhysRevLett.129.230502} {\bibfield  {journal}
  {\bibinfo  {journal} {Phys. Rev. Lett.}\ }\textbf {\bibinfo {volume} {129}},\
  \bibinfo {pages} {230502} (\bibinfo {year} {2022})}\BibitemShut {NoStop}%
\bibitem [{\citenamefont {Ma}\ \emph {et~al.}(2017)\citenamefont {Ma},
  \citenamefont {Lake}, \citenamefont {Chen},\ and\ \citenamefont
  {Hermele}}]{fracton3}%
  \BibitemOpen
  \bibfield  {author} {\bibinfo {author} {\bibfnamefont {H.}~\bibnamefont
  {Ma}}, \bibinfo {author} {\bibfnamefont {E.}~\bibnamefont {Lake}}, \bibinfo
  {author} {\bibfnamefont {X.}~\bibnamefont {Chen}},\ and\ \bibinfo {author}
  {\bibfnamefont {M.}~\bibnamefont {Hermele}},\ }\bibfield  {title} {\bibinfo
  {title} {Fracton topological order via coupled layers},\ }\href
  {https://doi.org/10.1103/PhysRevB.95.245126} {\bibfield  {journal} {\bibinfo
  {journal} {Phys. Rev. B}\ }\textbf {\bibinfo {volume} {95}},\ \bibinfo
  {pages} {245126} (\bibinfo {year} {2017})}\BibitemShut {NoStop}%
\bibitem [{\citenamefont {Shirley}\ \emph
  {et~al.}(2019{\natexlab{a}})\citenamefont {Shirley}, \citenamefont {Slagle},\
  and\ \citenamefont {Chen}}]{fracton2}%
  \BibitemOpen
  \bibfield  {author} {\bibinfo {author} {\bibfnamefont {W.}~\bibnamefont
  {Shirley}}, \bibinfo {author} {\bibfnamefont {K.}~\bibnamefont {Slagle}},\
  and\ \bibinfo {author} {\bibfnamefont {X.}~\bibnamefont {Chen}},\ }\bibfield
  {title} {\bibinfo {title} {{Foliated fracton order from gauging subsystem
  symmetries}},\ }\href {https://doi.org/10.21468/SciPostPhys.6.4.041}
  {\bibfield  {journal} {\bibinfo  {journal} {SciPost Phys.}\ }\textbf
  {\bibinfo {volume} {6}},\ \bibinfo {pages} {041} (\bibinfo {year}
  {2019}{\natexlab{a}})}\BibitemShut {NoStop}%
\bibitem [{\citenamefont {Prem}\ \emph {et~al.}(2017)\citenamefont {Prem},
  \citenamefont {Haah},\ and\ \citenamefont {Nandkishore}}]{fracton1}%
  \BibitemOpen
  \bibfield  {author} {\bibinfo {author} {\bibfnamefont {A.}~\bibnamefont
  {Prem}}, \bibinfo {author} {\bibfnamefont {J.}~\bibnamefont {Haah}},\ and\
  \bibinfo {author} {\bibfnamefont {R.}~\bibnamefont {Nandkishore}},\
  }\bibfield  {title} {\bibinfo {title} {Glassy quantum dynamics in translation
  invariant fracton models},\ }\href
  {https://doi.org/10.1103/PhysRevB.95.155133} {\bibfield  {journal} {\bibinfo
  {journal} {Phys. Rev. B}\ }\textbf {\bibinfo {volume} {95}},\ \bibinfo
  {pages} {155133} (\bibinfo {year} {2017})}\BibitemShut {NoStop}%
\bibitem [{\citenamefont {Dua}\ \emph {et~al.}(2019)\citenamefont {Dua},
  \citenamefont {Kim}, \citenamefont {Cheng},\ and\ \citenamefont
  {Williamson}}]{fracton19}%
  \BibitemOpen
  \bibfield  {author} {\bibinfo {author} {\bibfnamefont {A.}~\bibnamefont
  {Dua}}, \bibinfo {author} {\bibfnamefont {I.~H.}\ \bibnamefont {Kim}},
  \bibinfo {author} {\bibfnamefont {M.}~\bibnamefont {Cheng}},\ and\ \bibinfo
  {author} {\bibfnamefont {D.~J.}\ \bibnamefont {Williamson}},\ }\bibfield
  {title} {\bibinfo {title} {Sorting topological stabilizer models in three
  dimensions},\ }\href {https://doi.org/10.1103/PhysRevB.100.155137} {\bibfield
   {journal} {\bibinfo  {journal} {Phys. Rev. B}\ }\textbf {\bibinfo {volume}
  {100}},\ \bibinfo {pages} {155137} (\bibinfo {year} {2019})}\BibitemShut
  {NoStop}%
\bibitem [{\citenamefont {Nandkishore}\ and\ \citenamefont
  {Hermele}(2019)}]{fracton}%
  \BibitemOpen
  \bibfield  {author} {\bibinfo {author} {\bibfnamefont {R.~M.}\ \bibnamefont
  {Nandkishore}}\ and\ \bibinfo {author} {\bibfnamefont {M.}~\bibnamefont
  {Hermele}},\ }\bibfield  {title} {\bibinfo {title} {Fractons},\ }\href
  {https://doi.org/https://doi.org/10.1146/annurev-conmatphys-031218-013604}
  {\bibfield  {journal} {\bibinfo  {journal} {Annual Review of Condensed Matter
  Physics}\ }\textbf {\bibinfo {volume} {10}},\ \bibinfo {pages} {295}
  (\bibinfo {year} {2019})}\BibitemShut {NoStop}%
\bibitem [{\citenamefont {Bulmash}\ and\ \citenamefont
  {Barkeshli}(2019)}]{fracton4}%
  \BibitemOpen
  \bibfield  {author} {\bibinfo {author} {\bibfnamefont {D.}~\bibnamefont
  {Bulmash}}\ and\ \bibinfo {author} {\bibfnamefont {M.}~\bibnamefont
  {Barkeshli}},\ }\bibfield  {title} {\bibinfo {title} {Gauging fractons:
  Immobile non-abelian quasiparticles, fractals, and position-dependent
  degeneracies},\ }\href {https://doi.org/10.1103/PhysRevB.100.155146}
  {\bibfield  {journal} {\bibinfo  {journal} {Phys. Rev. B}\ }\textbf {\bibinfo
  {volume} {100}},\ \bibinfo {pages} {155146} (\bibinfo {year}
  {2019})}\BibitemShut {NoStop}%
\bibitem [{\citenamefont {Prem}\ \emph {et~al.}(2019)\citenamefont {Prem},
  \citenamefont {Huang}, \citenamefont {Song},\ and\ \citenamefont
  {Hermele}}]{fracton17}%
  \BibitemOpen
  \bibfield  {author} {\bibinfo {author} {\bibfnamefont {A.}~\bibnamefont
  {Prem}}, \bibinfo {author} {\bibfnamefont {S.-J.}\ \bibnamefont {Huang}},
  \bibinfo {author} {\bibfnamefont {H.}~\bibnamefont {Song}},\ and\ \bibinfo
  {author} {\bibfnamefont {M.}~\bibnamefont {Hermele}},\ }\bibfield  {title}
  {\bibinfo {title} {Cage-net fracton models},\ }\href
  {https://doi.org/10.1103/PhysRevX.9.021010} {\bibfield  {journal} {\bibinfo
  {journal} {Phys. Rev. X}\ }\textbf {\bibinfo {volume} {9}},\ \bibinfo {pages}
  {021010} (\bibinfo {year} {2019})}\BibitemShut {NoStop}%
\bibitem [{\citenamefont {Slagle}(2021)}]{fracton27}%
  \BibitemOpen
  \bibfield  {author} {\bibinfo {author} {\bibfnamefont {K.}~\bibnamefont
  {Slagle}},\ }\bibfield  {title} {\bibinfo {title} {Foliated quantum field
  theory of fracton order},\ }\href
  {https://doi.org/10.1103/PhysRevLett.126.101603} {\bibfield  {journal}
  {\bibinfo  {journal} {Phys. Rev. Lett.}\ }\textbf {\bibinfo {volume} {126}},\
  \bibinfo {pages} {101603} (\bibinfo {year} {2021})}\BibitemShut {NoStop}%
\bibitem [{\citenamefont {Zhou}\ \emph
  {et~al.}(2022{\natexlab{a}})\citenamefont {Zhou}, \citenamefont {Li},
  \citenamefont {Yan}, \citenamefont {Ye},\ and\ \citenamefont
  {Meng}}]{fracton53}%
  \BibitemOpen
  \bibfield  {author} {\bibinfo {author} {\bibfnamefont {C.}~\bibnamefont
  {Zhou}}, \bibinfo {author} {\bibfnamefont {M.-Y.}\ \bibnamefont {Li}},
  \bibinfo {author} {\bibfnamefont {Z.}~\bibnamefont {Yan}}, \bibinfo {author}
  {\bibfnamefont {P.}~\bibnamefont {Ye}},\ and\ \bibinfo {author}
  {\bibfnamefont {Z.~Y.}\ \bibnamefont {Meng}},\ }\bibfield  {title} {\bibinfo
  {title} {Evolution of dynamical signature in the x-cube fracton topological
  order},\ }\href {https://doi.org/10.1103/PhysRevResearch.4.033111} {\bibfield
   {journal} {\bibinfo  {journal} {Phys. Rev. Res.}\ }\textbf {\bibinfo
  {volume} {4}},\ \bibinfo {pages} {033111} (\bibinfo {year}
  {2022}{\natexlab{a}})}\BibitemShut {NoStop}%
\bibitem [{\citenamefont {Zhu}\ \emph {et~al.}(2023)\citenamefont {Zhu},
  \citenamefont {Chen}, \citenamefont {Ye},\ and\ \citenamefont
  {Trebst}}]{fracton54}%
  \BibitemOpen
  \bibfield  {author} {\bibinfo {author} {\bibfnamefont {G.-Y.}\ \bibnamefont
  {Zhu}}, \bibinfo {author} {\bibfnamefont {J.-Y.}\ \bibnamefont {Chen}},
  \bibinfo {author} {\bibfnamefont {P.}~\bibnamefont {Ye}},\ and\ \bibinfo
  {author} {\bibfnamefont {S.}~\bibnamefont {Trebst}},\ }\bibfield  {title}
  {\bibinfo {title} {Topological fracton quantum phase transitions by tuning
  exact tensor network states},\ }\href
  {https://doi.org/10.1103/PhysRevLett.130.216704} {\bibfield  {journal}
  {\bibinfo  {journal} {Phys. Rev. Lett.}\ }\textbf {\bibinfo {volume} {130}},\
  \bibinfo {pages} {216704} (\bibinfo {year} {2023})}\BibitemShut {NoStop}%
\bibitem [{\citenamefont {Canossa}\ \emph {et~al.}(2024)\citenamefont
  {Canossa}, \citenamefont {Pollet}, \citenamefont {Martin-Delgado},
  \citenamefont {Song},\ and\ \citenamefont {Liu}}]{fracton15}%
  \BibitemOpen
  \bibfield  {author} {\bibinfo {author} {\bibfnamefont {G.}~\bibnamefont
  {Canossa}}, \bibinfo {author} {\bibfnamefont {L.}~\bibnamefont {Pollet}},
  \bibinfo {author} {\bibfnamefont {M.~A.}\ \bibnamefont {Martin-Delgado}},
  \bibinfo {author} {\bibfnamefont {H.}~\bibnamefont {Song}},\ and\ \bibinfo
  {author} {\bibfnamefont {K.}~\bibnamefont {Liu}},\ }\bibfield  {title}
  {\bibinfo {title} {Exotic symmetry breaking properties of self-dual fracton
  spin models},\ }\href {https://doi.org/10.1103/PhysRevResearch.6.013304}
  {\bibfield  {journal} {\bibinfo  {journal} {Phys. Rev. Res.}\ }\textbf
  {\bibinfo {volume} {6}},\ \bibinfo {pages} {013304} (\bibinfo {year}
  {2024})}\BibitemShut {NoStop}%
\bibitem [{\citenamefont {Li}\ \emph {et~al.}(2024)\citenamefont {Li},
  \citenamefont {Zhou},\ and\ \citenamefont {Ye}}]{PhysRevB.110.205108}%
  \BibitemOpen
  \bibfield  {author} {\bibinfo {author} {\bibfnamefont {B.-X.}\ \bibnamefont
  {Li}}, \bibinfo {author} {\bibfnamefont {Y.}~\bibnamefont {Zhou}},\ and\
  \bibinfo {author} {\bibfnamefont {P.}~\bibnamefont {Ye}},\ }\bibfield
  {title} {\bibinfo {title} {Three-dimensional fracton topological orders with
  boundary toeplitz braiding},\ }\href
  {https://doi.org/10.1103/PhysRevB.110.205108} {\bibfield  {journal} {\bibinfo
   {journal} {Phys. Rev. B}\ }\textbf {\bibinfo {volume} {110}},\ \bibinfo
  {pages} {205108} (\bibinfo {year} {2024})}\BibitemShut {NoStop}%
\bibitem [{\citenamefont {Hansson}\ \emph {et~al.}(2004)\citenamefont
  {Hansson}, \citenamefont {Oganesyan},\ and\ \citenamefont
  {Sondhi}}]{hansson_superconductors_2004}%
  \BibitemOpen
  \bibfield  {author} {\bibinfo {author} {\bibfnamefont {T.}~\bibnamefont
  {Hansson}}, \bibinfo {author} {\bibfnamefont {V.}~\bibnamefont {Oganesyan}},\
  and\ \bibinfo {author} {\bibfnamefont {S.}~\bibnamefont {Sondhi}},\
  }\bibfield  {title} {\bibinfo {title} {Superconductors are topologically
  ordered},\ }\href {https://doi.org/10.1016/j.aop.2004.05.006} {\bibfield
  {journal} {\bibinfo  {journal} {Annals of Physics}\ }\textbf {\bibinfo
  {volume} {313}},\ \bibinfo {pages} {497} (\bibinfo {year}
  {2004})}\BibitemShut {NoStop}%
\bibitem [{\citenamefont {Preskill}\ and\ \citenamefont
  {Krauss}(1990)}]{PRESKILL199050}%
  \BibitemOpen
  \bibfield  {author} {\bibinfo {author} {\bibfnamefont {J.}~\bibnamefont
  {Preskill}}\ and\ \bibinfo {author} {\bibfnamefont {L.~M.}\ \bibnamefont
  {Krauss}},\ }\bibfield  {title} {\bibinfo {title} {Local discrete symmetry
  and quantum-mechanical hair},\ }\href
  {https://doi.org/https://doi.org/10.1016/0550-3213(90)90262-C} {\bibfield
  {journal} {\bibinfo  {journal} {Nuclear Physics B}\ }\textbf {\bibinfo
  {volume} {341}},\ \bibinfo {pages} {50 } (\bibinfo {year}
  {1990})}\BibitemShut {NoStop}%
\bibitem [{\citenamefont {Alford}\ and\ \citenamefont
  {Wilczek}(1989)}]{PhysRevLett.62.1071}%
  \BibitemOpen
  \bibfield  {author} {\bibinfo {author} {\bibfnamefont {M.~G.}\ \bibnamefont
  {Alford}}\ and\ \bibinfo {author} {\bibfnamefont {F.}~\bibnamefont
  {Wilczek}},\ }\bibfield  {title} {\bibinfo {title} {Aharonov-bohm interaction
  of cosmic strings with matter},\ }\href
  {https://doi.org/10.1103/PhysRevLett.62.1071} {\bibfield  {journal} {\bibinfo
   {journal} {Phys. Rev. Lett.}\ }\textbf {\bibinfo {volume} {62}},\ \bibinfo
  {pages} {1071} (\bibinfo {year} {1989})}\BibitemShut {NoStop}%
\bibitem [{\citenamefont {Krauss}\ and\ \citenamefont
  {Wilczek}(1989)}]{PhysRevLett.62.1221}%
  \BibitemOpen
  \bibfield  {author} {\bibinfo {author} {\bibfnamefont {L.~M.}\ \bibnamefont
  {Krauss}}\ and\ \bibinfo {author} {\bibfnamefont {F.}~\bibnamefont
  {Wilczek}},\ }\bibfield  {title} {\bibinfo {title} {Discrete gauge symmetry
  in continuum theories},\ }\href {https://doi.org/10.1103/PhysRevLett.62.1221}
  {\bibfield  {journal} {\bibinfo  {journal} {Phys. Rev. Lett.}\ }\textbf
  {\bibinfo {volume} {62}},\ \bibinfo {pages} {1221} (\bibinfo {year}
  {1989})}\BibitemShut {NoStop}%
\bibitem [{\citenamefont {Alford}\ \emph {et~al.}(1992)\citenamefont {Alford},
  \citenamefont {Lee}, \citenamefont {March-Russell},\ and\ \citenamefont
  {Preskill}}]{ALFORD1992251}%
  \BibitemOpen
  \bibfield  {author} {\bibinfo {author} {\bibfnamefont {M.~G.}\ \bibnamefont
  {Alford}}, \bibinfo {author} {\bibfnamefont {K.-M.}\ \bibnamefont {Lee}},
  \bibinfo {author} {\bibfnamefont {J.}~\bibnamefont {March-Russell}},\ and\
  \bibinfo {author} {\bibfnamefont {J.}~\bibnamefont {Preskill}},\ }\bibfield
  {title} {\bibinfo {title} {Quantum field theory of non-abelian strings and
  vortices},\ }\href
  {https://doi.org/https://doi.org/10.1016/0550-3213(92)90468-Q} {\bibfield
  {journal} {\bibinfo  {journal} {Nuclear Physics B}\ }\textbf {\bibinfo
  {volume} {384}},\ \bibinfo {pages} {251 } (\bibinfo {year}
  {1992})}\BibitemShut {NoStop}%
\bibitem [{\citenamefont {Wang}\ and\ \citenamefont
  {Levin}(2014)}]{wang_levin1}%
  \BibitemOpen
  \bibfield  {author} {\bibinfo {author} {\bibfnamefont {C.}~\bibnamefont
  {Wang}}\ and\ \bibinfo {author} {\bibfnamefont {M.}~\bibnamefont {Levin}},\
  }\bibfield  {title} {\bibinfo {title} {Braiding statistics of loop
  excitations in three dimensions},\ }\href
  {https://doi.org/10.1103/PhysRevLett.113.080403} {\bibfield  {journal}
  {\bibinfo  {journal} {Phys. Rev. Lett.}\ }\textbf {\bibinfo {volume} {113}},\
  \bibinfo {pages} {080403} (\bibinfo {year} {2014})}\BibitemShut {NoStop}%
\bibitem [{\citenamefont {Wang}\ \emph {et~al.}(2015)\citenamefont {Wang},
  \citenamefont {Gu},\ and\ \citenamefont {Wen}}]{PhysRevLett.114.031601}%
  \BibitemOpen
  \bibfield  {author} {\bibinfo {author} {\bibfnamefont {J.~C.}\ \bibnamefont
  {Wang}}, \bibinfo {author} {\bibfnamefont {Z.-C.}\ \bibnamefont {Gu}},\ and\
  \bibinfo {author} {\bibfnamefont {X.-G.}\ \bibnamefont {Wen}},\ }\bibfield
  {title} {\bibinfo {title} {Field-theory representation of gauge-gravity
  symmetry-protected topological invariants, group cohomology, and beyond},\
  }\href {https://doi.org/10.1103/PhysRevLett.114.031601} {\bibfield  {journal}
  {\bibinfo  {journal} {Phys. Rev. Lett.}\ }\textbf {\bibinfo {volume} {114}},\
  \bibinfo {pages} {031601} (\bibinfo {year} {2015})}\BibitemShut {NoStop}%
\bibitem [{\citenamefont {Putrov}\ \emph {et~al.}(2017)\citenamefont {Putrov},
  \citenamefont {Wang},\ and\ \citenamefont {Yau}}]{2016arXiv161209298P}%
  \BibitemOpen
  \bibfield  {author} {\bibinfo {author} {\bibfnamefont {P.}~\bibnamefont
  {Putrov}}, \bibinfo {author} {\bibfnamefont {J.}~\bibnamefont {Wang}},\ and\
  \bibinfo {author} {\bibfnamefont {S.-T.}\ \bibnamefont {Yau}},\ }\bibfield
  {title} {\bibinfo {title} {Braiding statistics and link invariants of
  bosonic/fermionic topological quantum matter in 2+1 and 3+1 dimensions},\
  }\href {https://doi.org/https://doi.org/10.1016/j.aop.2017.06.019} {\bibfield
   {journal} {\bibinfo  {journal} {Annals of Physics}\ }\textbf {\bibinfo
  {volume} {384}},\ \bibinfo {pages} {254 } (\bibinfo {year}
  {2017})}\BibitemShut {NoStop}%
\bibitem [{\citenamefont {Wang}\ and\ \citenamefont {Wen}(2015)}]{string4}%
  \BibitemOpen
  \bibfield  {author} {\bibinfo {author} {\bibfnamefont {J.~C.}\ \bibnamefont
  {Wang}}\ and\ \bibinfo {author} {\bibfnamefont {X.-G.}\ \bibnamefont {Wen}},\
  }\bibfield  {title} {\bibinfo {title} {Non-abelian string and particle
  braiding in topological order: Modular $\mathrm{SL}(3,\mathbb{Z})$
  representation and $(3+1)$-dimensional twisted gauge theory},\ }\href
  {https://doi.org/10.1103/PhysRevB.91.035134} {\bibfield  {journal} {\bibinfo
  {journal} {Phys. Rev. B}\ }\textbf {\bibinfo {volume} {91}},\ \bibinfo
  {pages} {035134} (\bibinfo {year} {2015})}\BibitemShut {NoStop}%
\bibitem [{\citenamefont {Jian}\ and\ \citenamefont {Qi}(2014)}]{jian_qi_14}%
  \BibitemOpen
  \bibfield  {author} {\bibinfo {author} {\bibfnamefont {C.-M.}\ \bibnamefont
  {Jian}}\ and\ \bibinfo {author} {\bibfnamefont {X.-L.}\ \bibnamefont {Qi}},\
  }\bibfield  {title} {\bibinfo {title} {Layer construction of 3d topological
  states and string braiding statistics},\ }\href
  {https://doi.org/10.1103/PhysRevX.4.041043} {\bibfield  {journal} {\bibinfo
  {journal} {Phys. Rev. X}\ }\textbf {\bibinfo {volume} {4}},\ \bibinfo {pages}
  {041043} (\bibinfo {year} {2014})}\BibitemShut {NoStop}%
\bibitem [{\citenamefont {Jiang}\ \emph {et~al.}(2014)\citenamefont {Jiang},
  \citenamefont {Mesaros},\ and\ \citenamefont {Ran}}]{string5}%
  \BibitemOpen
  \bibfield  {author} {\bibinfo {author} {\bibfnamefont {S.}~\bibnamefont
  {Jiang}}, \bibinfo {author} {\bibfnamefont {A.}~\bibnamefont {Mesaros}},\
  and\ \bibinfo {author} {\bibfnamefont {Y.}~\bibnamefont {Ran}},\ }\bibfield
  {title} {\bibinfo {title} {Generalized modular transformations in
  $(3+1)\mathrm{D}$ topologically ordered phases and triple linking invariant
  of loop braiding},\ }\href {https://doi.org/10.1103/PhysRevX.4.031048}
  {\bibfield  {journal} {\bibinfo  {journal} {Phys. Rev. X}\ }\textbf {\bibinfo
  {volume} {4}},\ \bibinfo {pages} {031048} (\bibinfo {year}
  {2014})}\BibitemShut {NoStop}%
\bibitem [{\citenamefont {Wang}\ \emph {et~al.}(2016)\citenamefont {Wang},
  \citenamefont {Lin},\ and\ \citenamefont {Levin}}]{PhysRevX.6.021015}%
  \BibitemOpen
  \bibfield  {author} {\bibinfo {author} {\bibfnamefont {C.}~\bibnamefont
  {Wang}}, \bibinfo {author} {\bibfnamefont {C.-H.}\ \bibnamefont {Lin}},\ and\
  \bibinfo {author} {\bibfnamefont {M.}~\bibnamefont {Levin}},\ }\bibfield
  {title} {\bibinfo {title} {Bulk-boundary correspondence for three-dimensional
  symmetry-protected topological phases},\ }\href
  {https://doi.org/10.1103/PhysRevX.6.021015} {\bibfield  {journal} {\bibinfo
  {journal} {Phys. Rev. X}\ }\textbf {\bibinfo {volume} {6}},\ \bibinfo {pages}
  {021015} (\bibinfo {year} {2016})}\BibitemShut {NoStop}%
\bibitem [{\citenamefont {Tiwari}\ \emph {et~al.}(2017)\citenamefont {Tiwari},
  \citenamefont {Chen},\ and\ \citenamefont {Ryu}}]{Tiwari:2016aa}%
  \BibitemOpen
  \bibfield  {author} {\bibinfo {author} {\bibfnamefont {A.}~\bibnamefont
  {Tiwari}}, \bibinfo {author} {\bibfnamefont {X.}~\bibnamefont {Chen}},\ and\
  \bibinfo {author} {\bibfnamefont {S.}~\bibnamefont {Ryu}},\ }\bibfield
  {title} {\bibinfo {title} {Wilson operator algebras and ground states of
  coupled $\mathit{BF}$ theories},\ }\href
  {https://doi.org/10.1103/PhysRevB.95.245124} {\bibfield  {journal} {\bibinfo
  {journal} {Phys. Rev. B}\ }\textbf {\bibinfo {volume} {95}},\ \bibinfo
  {pages} {245124} (\bibinfo {year} {2017})}\BibitemShut {NoStop}%
\bibitem [{\citenamefont {{Kapustin}}\ and\ \citenamefont
  {{Thorngren}}(2014)}]{corbodism3}%
  \BibitemOpen
  \bibfield  {author} {\bibinfo {author} {\bibfnamefont {A.}~\bibnamefont
  {{Kapustin}}}\ and\ \bibinfo {author} {\bibfnamefont {R.}~\bibnamefont
  {{Thorngren}}},\ }\bibfield  {title} {\bibinfo {title} {{Anomalies of
  discrete symmetries in various dimensions and group cohomology}},\
  }\href@noop {} {\bibfield  {journal} {\bibinfo  {journal} {ArXiv e-prints}\ }
  (\bibinfo {year} {2014})},\ \Eprint {https://arxiv.org/abs/1404.3230}
  {arXiv:1404.3230 [hep-th]} \BibitemShut {NoStop}%
\bibitem [{\citenamefont {Wan}\ \emph {et~al.}(2015)\citenamefont {Wan},
  \citenamefont {Wang},\ and\ \citenamefont {He}}]{string6}%
  \BibitemOpen
  \bibfield  {author} {\bibinfo {author} {\bibfnamefont {Y.}~\bibnamefont
  {Wan}}, \bibinfo {author} {\bibfnamefont {J.~C.}\ \bibnamefont {Wang}},\ and\
  \bibinfo {author} {\bibfnamefont {H.}~\bibnamefont {He}},\ }\bibfield
  {title} {\bibinfo {title} {Twisted gauge theory model of topological phases
  in three dimensions},\ }\href {https://doi.org/10.1103/PhysRevB.92.045101}
  {\bibfield  {journal} {\bibinfo  {journal} {Phys. Rev. B}\ }\textbf {\bibinfo
  {volume} {92}},\ \bibinfo {pages} {045101} (\bibinfo {year}
  {2015})}\BibitemShut {NoStop}%
\bibitem [{\citenamefont {Chen}\ \emph {et~al.}(2016)\citenamefont {Chen},
  \citenamefont {Tiwari},\ and\ \citenamefont {Ryu}}]{3loop_ryu}%
  \BibitemOpen
  \bibfield  {author} {\bibinfo {author} {\bibfnamefont {X.}~\bibnamefont
  {Chen}}, \bibinfo {author} {\bibfnamefont {A.}~\bibnamefont {Tiwari}},\ and\
  \bibinfo {author} {\bibfnamefont {S.}~\bibnamefont {Ryu}},\ }\bibfield
  {title} {\bibinfo {title} {Bulk-boundary correspondence in (3+1)-dimensional
  topological phases},\ }\href {https://doi.org/10.1103/PhysRevB.94.045113}
  {\bibfield  {journal} {\bibinfo  {journal} {Phys. Rev. B}\ }\textbf {\bibinfo
  {volume} {94}},\ \bibinfo {pages} {045113} (\bibinfo {year}
  {2016})}\BibitemShut {NoStop}%
\bibitem [{\citenamefont {Chan}\ \emph {et~al.}(2018)\citenamefont {Chan},
  \citenamefont {Ye},\ and\ \citenamefont {Ryu}}]{PhysRevLett.121.061601}%
  \BibitemOpen
  \bibfield  {author} {\bibinfo {author} {\bibfnamefont {A.~P.~O.}\
  \bibnamefont {Chan}}, \bibinfo {author} {\bibfnamefont {P.}~\bibnamefont
  {Ye}},\ and\ \bibinfo {author} {\bibfnamefont {S.}~\bibnamefont {Ryu}},\
  }\bibfield  {title} {\bibinfo {title} {Braiding with borromean rings in
  ($3+1$)-dimensional spacetime},\ }\href
  {https://doi.org/10.1103/PhysRevLett.121.061601} {\bibfield  {journal}
  {\bibinfo  {journal} {Phys. Rev. Lett.}\ }\textbf {\bibinfo {volume} {121}},\
  \bibinfo {pages} {061601} (\bibinfo {year} {2018})}\BibitemShut {NoStop}%
\bibitem [{\citenamefont {Kapustin}\ and\ \citenamefont
  {Seiberg}(2014)}]{Kapustin:2014gua}%
  \BibitemOpen
  \bibfield  {author} {\bibinfo {author} {\bibfnamefont {A.}~\bibnamefont
  {Kapustin}}\ and\ \bibinfo {author} {\bibfnamefont {N.}~\bibnamefont
  {Seiberg}},\ }\bibfield  {title} {\bibinfo {title} {{Coupling a QFT to a TQFT
  and Duality}},\ }\href {https://doi.org/10.1007/JHEP04(2014)001} {\bibfield
  {journal} {\bibinfo  {journal} {JHEP}\ }\textbf {\bibinfo {volume} {04}},\
  \bibinfo {pages} {001}},\ \Eprint {https://arxiv.org/abs/1401.0740}
  {arXiv:1401.0740 [hep-th]} \BibitemShut {NoStop}%
\bibitem [{\citenamefont {Ye}\ and\ \citenamefont {Gu}(2015)}]{bti2}%
  \BibitemOpen
  \bibfield  {author} {\bibinfo {author} {\bibfnamefont {P.}~\bibnamefont
  {Ye}}\ and\ \bibinfo {author} {\bibfnamefont {Z.-C.}\ \bibnamefont {Gu}},\
  }\bibfield  {title} {\bibinfo {title} {Vortex-line condensation in three
  dimensions: A physical mechanism for bosonic topological insulators},\ }\href
  {https://doi.org/10.1103/PhysRevX.5.021029} {\bibfield  {journal} {\bibinfo
  {journal} {Phys. Rev. X}\ }\textbf {\bibinfo {volume} {5}},\ \bibinfo {pages}
  {021029} (\bibinfo {year} {2015})}\BibitemShut {NoStop}%
\bibitem [{\citenamefont {Wang}\ \emph {et~al.}(2019)\citenamefont {Wang},
  \citenamefont {Cheng}, \citenamefont {Wang},\ and\ \citenamefont
  {Gu}}]{PhysRevB.99.235137}%
  \BibitemOpen
  \bibfield  {author} {\bibinfo {author} {\bibfnamefont {Q.-R.}\ \bibnamefont
  {Wang}}, \bibinfo {author} {\bibfnamefont {M.}~\bibnamefont {Cheng}},
  \bibinfo {author} {\bibfnamefont {C.}~\bibnamefont {Wang}},\ and\ \bibinfo
  {author} {\bibfnamefont {Z.-C.}\ \bibnamefont {Gu}},\ }\bibfield  {title}
  {\bibinfo {title} {Topological quantum field theory for abelian topological
  phases and loop braiding statistics in $(3+1)$-dimensions},\ }\href
  {https://doi.org/10.1103/PhysRevB.99.235137} {\bibfield  {journal} {\bibinfo
  {journal} {Phys. Rev. B}\ }\textbf {\bibinfo {volume} {99}},\ \bibinfo
  {pages} {235137} (\bibinfo {year} {2019})}\BibitemShut {NoStop}%
\bibitem [{\citenamefont {Chen}\ and\ \citenamefont {Hsin}(2023)}]{cyaloop1}%
  \BibitemOpen
  \bibfield  {author} {\bibinfo {author} {\bibfnamefont {Y.-A.}\ \bibnamefont
  {Chen}}\ and\ \bibinfo {author} {\bibfnamefont {P.-S.}\ \bibnamefont
  {Hsin}},\ }\bibfield  {title} {\bibinfo {title} {{Exactly solvable lattice
  Hamiltonians and gravitational anomalies}},\ }\href
  {https://doi.org/10.21468/SciPostPhys.14.5.089} {\bibfield  {journal}
  {\bibinfo  {journal} {SciPost Phys.}\ }\textbf {\bibinfo {volume} {14}},\
  \bibinfo {pages} {089} (\bibinfo {year} {2023})}\BibitemShut {NoStop}%
\bibitem [{\citenamefont {Barkeshli}\ \emph {et~al.}(2023)\citenamefont
  {Barkeshli}, \citenamefont {Chen}, \citenamefont {Huang}, \citenamefont
  {Kobayashi}, \citenamefont {Tantivasadakarn},\ and\ \citenamefont
  {Zhu}}]{cyaloop2}%
  \BibitemOpen
  \bibfield  {author} {\bibinfo {author} {\bibfnamefont {M.}~\bibnamefont
  {Barkeshli}}, \bibinfo {author} {\bibfnamefont {Y.-A.}\ \bibnamefont {Chen}},
  \bibinfo {author} {\bibfnamefont {S.-J.}\ \bibnamefont {Huang}}, \bibinfo
  {author} {\bibfnamefont {R.}~\bibnamefont {Kobayashi}}, \bibinfo {author}
  {\bibfnamefont {N.}~\bibnamefont {Tantivasadakarn}},\ and\ \bibinfo {author}
  {\bibfnamefont {G.}~\bibnamefont {Zhu}},\ }\bibfield  {title} {\bibinfo
  {title} {{Codimension-2 defects and higher symmetries in (3+1)D topological
  phases}},\ }\href {https://doi.org/10.21468/SciPostPhys.14.4.065} {\bibfield
  {journal} {\bibinfo  {journal} {SciPost Phys.}\ }\textbf {\bibinfo {volume}
  {14}},\ \bibinfo {pages} {065} (\bibinfo {year} {2023})}\BibitemShut
  {NoStop}%
\bibitem [{\citenamefont {Barkeshli}\ \emph {et~al.}(2024)\citenamefont
  {Barkeshli}, \citenamefont {Chen}, \citenamefont {Hsin},\ and\ \citenamefont
  {Kobayashi}}]{cyaloop3}%
  \BibitemOpen
  \bibfield  {author} {\bibinfo {author} {\bibfnamefont {M.}~\bibnamefont
  {Barkeshli}}, \bibinfo {author} {\bibfnamefont {Y.-A.}\ \bibnamefont {Chen}},
  \bibinfo {author} {\bibfnamefont {P.-S.}\ \bibnamefont {Hsin}},\ and\
  \bibinfo {author} {\bibfnamefont {R.}~\bibnamefont {Kobayashi}},\ }\bibfield
  {title} {\bibinfo {title} {{Higher-group symmetry in finite gauge theory and
  stabilizer codes}},\ }\href {https://doi.org/10.21468/SciPostPhys.16.4.089}
  {\bibfield  {journal} {\bibinfo  {journal} {SciPost Phys.}\ }\textbf
  {\bibinfo {volume} {16}},\ \bibinfo {pages} {089} (\bibinfo {year}
  {2024})}\BibitemShut {NoStop}%
\bibitem [{\citenamefont {{Kobayashi}}\ \emph {et~al.}(2024)\citenamefont
  {{Kobayashi}}, \citenamefont {{Li}}, \citenamefont {{Xue}}, \citenamefont
  {{Hsin}},\ and\ \citenamefont {{Chen}}}]{2024arXiv241201886K}%
  \BibitemOpen
  \bibfield  {author} {\bibinfo {author} {\bibfnamefont {R.}~\bibnamefont
  {{Kobayashi}}}, \bibinfo {author} {\bibfnamefont {Y.}~\bibnamefont {{Li}}},
  \bibinfo {author} {\bibfnamefont {H.}~\bibnamefont {{Xue}}}, \bibinfo
  {author} {\bibfnamefont {P.-S.}\ \bibnamefont {{Hsin}}},\ and\ \bibinfo
  {author} {\bibfnamefont {Y.-A.}\ \bibnamefont {{Chen}}},\ }\bibfield  {title}
  {\bibinfo {title} {{Universal microscopic descriptions for statistics of
  particles and extended excitations}},\ }\href
  {https://doi.org/10.48550/arXiv.2412.01886} {\bibfield  {journal} {\bibinfo
  {journal} {arXiv e-prints}\ ,\ \bibinfo {eid} {arXiv:2412.01886}} (\bibinfo
  {year} {2024})},\ \Eprint {https://arxiv.org/abs/2412.01886}
  {arXiv:2412.01886 [quant-ph]} \BibitemShut {NoStop}%
\bibitem [{\citenamefont {Zhang}\ and\ \citenamefont
  {Ye}(2021)}]{zhang_compatible_2021}%
  \BibitemOpen
  \bibfield  {author} {\bibinfo {author} {\bibfnamefont {Z.-F.}\ \bibnamefont
  {Zhang}}\ and\ \bibinfo {author} {\bibfnamefont {P.}~\bibnamefont {Ye}},\
  }\bibfield  {title} {\bibinfo {title} {Compatible braidings with {Hopf}
  links, multi-loop, and {Borromean} rings in (3+1)-dimensional spacetime},\
  }\href {https://doi.org/10.1103/PhysRevResearch.3.023132} {\bibfield
  {journal} {\bibinfo  {journal} {Phys. Rev. Research}\ }\textbf {\bibinfo
  {volume} {3}},\ \bibinfo {pages} {023132} (\bibinfo {year}
  {2021})}\BibitemShut {NoStop}%
\bibitem [{\citenamefont {Zhang}\ and\ \citenamefont
  {Ye}(2022)}]{zhang_topological_2022}%
  \BibitemOpen
  \bibfield  {author} {\bibinfo {author} {\bibfnamefont {Z.-F.}\ \bibnamefont
  {Zhang}}\ and\ \bibinfo {author} {\bibfnamefont {P.}~\bibnamefont {Ye}},\
  }\bibfield  {title} {\bibinfo {title} {Topological orders, braiding
  statistics, and mixture of two types of twisted {BF} theories in five
  dimensions},\ }\href {https://doi.org/10.1007/JHEP04(2022)138} {\bibfield
  {journal} {\bibinfo  {journal} {J. High Energ. Phys.}\ }\textbf {\bibinfo
  {volume} {2022}}\bibinfo  {number} { (4)},\ \bibinfo {pages}
  {138}}\BibitemShut {NoStop}%
\bibitem [{\citenamefont {Zhang}\ \emph
  {et~al.}(2023{\natexlab{a}})\citenamefont {Zhang}, \citenamefont {Wang},\
  and\ \citenamefont {Ye}}]{Zhang2023fusion}%
  \BibitemOpen
\bibfield  {number} {  }\bibfield  {author} {\bibinfo {author} {\bibfnamefont
  {Z.-F.}\ \bibnamefont {Zhang}}, \bibinfo {author} {\bibfnamefont {Q.-R.}\
  \bibnamefont {Wang}},\ and\ \bibinfo {author} {\bibfnamefont
  {P.}~\bibnamefont {Ye}},\ }\bibfield  {title} {\bibinfo {title} {Non-abelian
  fusion, shrinking, and quantum dimensions of abelian gauge fluxes},\ }\href
  {https://doi.org/10.1103/PhysRevB.107.165117} {\bibfield  {journal} {\bibinfo
   {journal} {Phys. Rev. B}\ }\textbf {\bibinfo {volume} {107}},\ \bibinfo
  {pages} {165117} (\bibinfo {year} {2023}{\natexlab{a}})}\BibitemShut
  {NoStop}%
\bibitem [{\citenamefont {Zhang}\ \emph
  {et~al.}(2023{\natexlab{b}})\citenamefont {Zhang}, \citenamefont {Wang},\
  and\ \citenamefont {Ye}}]{Zhang2023Continuum}%
  \BibitemOpen
  \bibfield  {author} {\bibinfo {author} {\bibfnamefont {Z.-F.}\ \bibnamefont
  {Zhang}}, \bibinfo {author} {\bibfnamefont {Q.-R.}\ \bibnamefont {Wang}},\
  and\ \bibinfo {author} {\bibfnamefont {P.}~\bibnamefont {Ye}},\ }\bibfield
  {title} {\bibinfo {title} {Continuum field theory of three-dimensional
  topological orders with emergent fermions and braiding statistics},\ }\href
  {https://doi.org/10.1103/PhysRevResearch.5.043111} {\bibfield  {journal}
  {\bibinfo  {journal} {Phys. Rev. Res.}\ }\textbf {\bibinfo {volume} {5}},\
  \bibinfo {pages} {043111} (\bibinfo {year} {2023}{\natexlab{b}})}\BibitemShut
  {NoStop}%
\bibitem [{\citenamefont {{Huang}}\ \emph {et~al.}(2023)\citenamefont
  {{Huang}}, \citenamefont {{Zhang}},\ and\ \citenamefont {{Ye}}}]{Huang2023}%
  \BibitemOpen
  \bibfield  {author} {\bibinfo {author} {\bibfnamefont {Y.}~\bibnamefont
  {{Huang}}}, \bibinfo {author} {\bibfnamefont {Z.-F.}\ \bibnamefont
  {{Zhang}}},\ and\ \bibinfo {author} {\bibfnamefont {P.}~\bibnamefont
  {{Ye}}},\ }\bibfield  {title} {\bibinfo {title} {{Fusion rules and shrinking
  rules of topological orders in five dimensions}},\ }\href
  {https://doi.org/10.1007/JHEP11(2023)210} {\bibfield  {journal} {\bibinfo
  {journal} {Journal of High Energy Physics}\ }\textbf {\bibinfo {volume}
  {2023}},\ \bibinfo {eid} {210} (\bibinfo {year} {2023})},\ \Eprint
  {https://arxiv.org/abs/2306.14611} {arXiv:2306.14611 [hep-th]} \BibitemShut
  {NoStop}%
\bibitem [{\citenamefont {{Huang}}\ \emph {et~al.}(2025)\citenamefont
  {{Huang}}, \citenamefont {{Zhang}},\ and\ \citenamefont
  {{Ye}}}]{huang2025diagrammatics}%
  \BibitemOpen
  \bibfield  {author} {\bibinfo {author} {\bibfnamefont {Y.}~\bibnamefont
  {{Huang}}}, \bibinfo {author} {\bibfnamefont {Z.-F.}\ \bibnamefont
  {{Zhang}}},\ and\ \bibinfo {author} {\bibfnamefont {P.}~\bibnamefont
  {{Ye}}},\ }\bibfield  {title} {\bibinfo {title} {{Diagrammatics, pentagon
  equations, and hexagon equations of topological orders with loop- and
  membrane-like excitations}},\ }\href
  {https://doi.org/10.1007/JHEP06(2025)238} {\bibfield  {journal} {\bibinfo
  {journal} {Journal of High Energy Physics}\ }\textbf {\bibinfo {volume}
  {2025}},\ \bibinfo {eid} {238} (\bibinfo {year} {2025})},\ \Eprint
  {https://arxiv.org/abs/2405.19077} {arXiv:2405.19077 [hep-th]} \BibitemShut
  {NoStop}%
\bibitem [{\citenamefont {Lapa}\ \emph {et~al.}(2017)\citenamefont {Lapa},
  \citenamefont {Jian}, \citenamefont {Ye},\ and\ \citenamefont
  {Hughes}}]{lapa17}%
  \BibitemOpen
  \bibfield  {author} {\bibinfo {author} {\bibfnamefont {M.~F.}\ \bibnamefont
  {Lapa}}, \bibinfo {author} {\bibfnamefont {C.-M.}\ \bibnamefont {Jian}},
  \bibinfo {author} {\bibfnamefont {P.}~\bibnamefont {Ye}},\ and\ \bibinfo
  {author} {\bibfnamefont {T.~L.}\ \bibnamefont {Hughes}},\ }\bibfield  {title}
  {\bibinfo {title} {Topological electromagnetic responses of bosonic quantum
  hall, topological insulator, and chiral semimetal phases in all dimensions},\
  }\href {https://doi.org/10.1103/PhysRevB.95.035149} {\bibfield  {journal}
  {\bibinfo  {journal} {Phys. Rev. B}\ }\textbf {\bibinfo {volume} {95}},\
  \bibinfo {pages} {035149} (\bibinfo {year} {2017})}\BibitemShut {NoStop}%
\bibitem [{\citenamefont {Ye}\ and\ \citenamefont {Wen}(2014)}]{YW13a}%
  \BibitemOpen
  \bibfield  {author} {\bibinfo {author} {\bibfnamefont {P.}~\bibnamefont
  {Ye}}\ and\ \bibinfo {author} {\bibfnamefont {X.-G.}\ \bibnamefont {Wen}},\
  }\bibfield  {title} {\bibinfo {title} {Constructing symmetric topological
  phases of bosons in three dimensions via fermionic projective construction
  and dyon condensation},\ }\href {https://doi.org/10.1103/PhysRevB.89.045127}
  {\bibfield  {journal} {\bibinfo  {journal} {Phys. Rev. B}\ }\textbf {\bibinfo
  {volume} {89}},\ \bibinfo {pages} {045127} (\bibinfo {year}
  {2014})}\BibitemShut {NoStop}%
\bibitem [{\citenamefont {Ye}\ \emph {et~al.}(2016)\citenamefont {Ye},
  \citenamefont {Hughes}, \citenamefont {Maciejko},\ and\ \citenamefont
  {Fradkin}}]{ye16a}%
  \BibitemOpen
  \bibfield  {author} {\bibinfo {author} {\bibfnamefont {P.}~\bibnamefont
  {Ye}}, \bibinfo {author} {\bibfnamefont {T.~L.}\ \bibnamefont {Hughes}},
  \bibinfo {author} {\bibfnamefont {J.}~\bibnamefont {Maciejko}},\ and\
  \bibinfo {author} {\bibfnamefont {E.}~\bibnamefont {Fradkin}},\ }\bibfield
  {title} {\bibinfo {title} {Composite particle theory of three-dimensional
  gapped fermionic phases: Fractional topological insulators and charge-loop
  excitation symmetry},\ }\href {https://doi.org/10.1103/PhysRevB.94.115104}
  {\bibfield  {journal} {\bibinfo  {journal} {Phys. Rev. B}\ }\textbf {\bibinfo
  {volume} {94}},\ \bibinfo {pages} {115104} (\bibinfo {year}
  {2016})}\BibitemShut {NoStop}%
\bibitem [{\citenamefont {Ye}\ \emph {et~al.}(2017)\citenamefont {Ye},
  \citenamefont {Cheng},\ and\ \citenamefont {Fradkin}}]{Ye:2017aa}%
  \BibitemOpen
  \bibfield  {author} {\bibinfo {author} {\bibfnamefont {P.}~\bibnamefont
  {Ye}}, \bibinfo {author} {\bibfnamefont {M.}~\bibnamefont {Cheng}},\ and\
  \bibinfo {author} {\bibfnamefont {E.}~\bibnamefont {Fradkin}},\ }\bibfield
  {title} {\bibinfo {title} {Fractional $s$-duality, classification of
  fractional topological insulators, and surface topological order},\ }\href
  {https://doi.org/10.1103/PhysRevB.96.085125} {\bibfield  {journal} {\bibinfo
  {journal} {Phys. Rev. B}\ }\textbf {\bibinfo {volume} {96}},\ \bibinfo
  {pages} {085125} (\bibinfo {year} {2017})}\BibitemShut {NoStop}%
\bibitem [{\citenamefont {Ye}\ and\ \citenamefont {Wang}(2013)}]{bti6}%
  \BibitemOpen
  \bibfield  {author} {\bibinfo {author} {\bibfnamefont {P.}~\bibnamefont
  {Ye}}\ and\ \bibinfo {author} {\bibfnamefont {J.}~\bibnamefont {Wang}},\
  }\bibfield  {title} {\bibinfo {title} {Symmetry-protected topological phases
  with charge and spin symmetries: Response theory and dynamical gauge theory
  in two and three dimensions},\ }\href
  {https://doi.org/10.1103/PhysRevB.88.235109} {\bibfield  {journal} {\bibinfo
  {journal} {Phys. Rev. B}\ }\textbf {\bibinfo {volume} {88}},\ \bibinfo
  {pages} {235109} (\bibinfo {year} {2013})}\BibitemShut {NoStop}%
\bibitem [{\citenamefont {Witten}(2016)}]{RevModPhys.88.035001}%
  \BibitemOpen
  \bibfield  {author} {\bibinfo {author} {\bibfnamefont {E.}~\bibnamefont
  {Witten}},\ }\bibfield  {title} {\bibinfo {title} {Fermion path integrals and
  topological phases},\ }\href {https://doi.org/10.1103/RevModPhys.88.035001}
  {\bibfield  {journal} {\bibinfo  {journal} {Rev. Mod. Phys.}\ }\textbf
  {\bibinfo {volume} {88}},\ \bibinfo {pages} {035001} (\bibinfo {year}
  {2016})}\BibitemShut {NoStop}%
\bibitem [{\citenamefont {Han}\ \emph {et~al.}(2019)\citenamefont {Han},
  \citenamefont {Wang},\ and\ \citenamefont {Ye}}]{PhysRevB.99.205120}%
  \BibitemOpen
  \bibfield  {author} {\bibinfo {author} {\bibfnamefont {B.}~\bibnamefont
  {Han}}, \bibinfo {author} {\bibfnamefont {H.}~\bibnamefont {Wang}},\ and\
  \bibinfo {author} {\bibfnamefont {P.}~\bibnamefont {Ye}},\ }\bibfield
  {title} {\bibinfo {title} {Generalized wen-zee terms},\ }\href
  {https://doi.org/10.1103/PhysRevB.99.205120} {\bibfield  {journal} {\bibinfo
  {journal} {Phys. Rev. B}\ }\textbf {\bibinfo {volume} {99}},\ \bibinfo
  {pages} {205120} (\bibinfo {year} {2019})}\BibitemShut {NoStop}%
\bibitem [{\citenamefont {Ning}\ \emph {et~al.}(2022)\citenamefont {Ning},
  \citenamefont {Liu},\ and\ \citenamefont {Ye}}]{Ning2018prb}%
  \BibitemOpen
  \bibfield  {author} {\bibinfo {author} {\bibfnamefont {S.-Q.}\ \bibnamefont
  {Ning}}, \bibinfo {author} {\bibfnamefont {Z.-X.}\ \bibnamefont {Liu}},\ and\
  \bibinfo {author} {\bibfnamefont {P.}~\bibnamefont {Ye}},\ }\bibfield
  {title} {\bibinfo {title} {Fractionalizing global symmetry on looplike
  topological excitations},\ }\href
  {https://doi.org/10.1103/PhysRevB.105.205137} {\bibfield  {journal} {\bibinfo
   {journal} {Phys. Rev. B}\ }\textbf {\bibinfo {volume} {105}},\ \bibinfo
  {pages} {205137} (\bibinfo {year} {2022})}\BibitemShut {NoStop}%
\bibitem [{\citenamefont {Ning}\ \emph {et~al.}(2016)\citenamefont {Ning},
  \citenamefont {Liu},\ and\ \citenamefont {Ye}}]{ye16_set}%
  \BibitemOpen
  \bibfield  {author} {\bibinfo {author} {\bibfnamefont {S.-Q.}\ \bibnamefont
  {Ning}}, \bibinfo {author} {\bibfnamefont {Z.-X.}\ \bibnamefont {Liu}},\ and\
  \bibinfo {author} {\bibfnamefont {P.}~\bibnamefont {Ye}},\ }\bibfield
  {title} {\bibinfo {title} {Symmetry enrichment in three-dimensional
  topological phases},\ }\href {https://doi.org/10.1103/PhysRevB.94.245120}
  {\bibfield  {journal} {\bibinfo  {journal} {Phys. Rev. B}\ }\textbf {\bibinfo
  {volume} {94}},\ \bibinfo {pages} {245120} (\bibinfo {year}
  {2016})}\BibitemShut {NoStop}%
\bibitem [{\citenamefont {Ye}(2018)}]{2016arXiv161008645Y}%
  \BibitemOpen
  \bibfield  {author} {\bibinfo {author} {\bibfnamefont {P.}~\bibnamefont
  {Ye}},\ }\bibfield  {title} {\bibinfo {title} {Three-dimensional anomalous
  twisted gauge theories with global symmetry: Implications for quantum spin
  liquids},\ }\href {https://doi.org/10.1103/PhysRevB.97.125127} {\bibfield
  {journal} {\bibinfo  {journal} {Phys. Rev. B}\ }\textbf {\bibinfo {volume}
  {97}},\ \bibinfo {pages} {125127} (\bibinfo {year} {2018})}\BibitemShut
  {NoStop}%
\bibitem [{\citenamefont {Pretko}(2017{\natexlab{a}})}]{fracton58}%
  \BibitemOpen
  \bibfield  {author} {\bibinfo {author} {\bibfnamefont {M.}~\bibnamefont
  {Pretko}},\ }\bibfield  {title} {\bibinfo {title} {Generalized
  electromagnetism of subdimensional particles: A spin liquid story},\ }\href
  {https://doi.org/10.1103/PhysRevB.96.035119} {\bibfield  {journal} {\bibinfo
  {journal} {Phys. Rev. B}\ }\textbf {\bibinfo {volume} {96}},\ \bibinfo
  {pages} {035119} (\bibinfo {year} {2017}{\natexlab{a}})}\BibitemShut
  {NoStop}%
\bibitem [{\citenamefont {Pretko}(2017{\natexlab{b}})}]{fracton59}%
  \BibitemOpen
  \bibfield  {author} {\bibinfo {author} {\bibfnamefont {M.}~\bibnamefont
  {Pretko}},\ }\bibfield  {title} {\bibinfo {title} {Subdimensional particle
  structure of higher rank $u(1)$ spin liquids},\ }\href
  {https://doi.org/10.1103/PhysRevB.95.115139} {\bibfield  {journal} {\bibinfo
  {journal} {Phys. Rev. B}\ }\textbf {\bibinfo {volume} {95}},\ \bibinfo
  {pages} {115139} (\bibinfo {year} {2017}{\natexlab{b}})}\BibitemShut
  {NoStop}%
\bibitem [{\citenamefont {Pretko}(2018)}]{Fractongauge}%
  \BibitemOpen
  \bibfield  {author} {\bibinfo {author} {\bibfnamefont {M.}~\bibnamefont
  {Pretko}},\ }\bibfield  {title} {\bibinfo {title} {The fracton gauge
  principle},\ }\href {https://doi.org/10.1103/PhysRevB.98.115134} {\bibfield
  {journal} {\bibinfo  {journal} {Phys. Rev. B}\ }\textbf {\bibinfo {volume}
  {98}},\ \bibinfo {pages} {115134} (\bibinfo {year} {2018})}\BibitemShut
  {NoStop}%
\bibitem [{\citenamefont {Pai}\ and\ \citenamefont {Pretko}(2019)}]{fracton55}%
  \BibitemOpen
  \bibfield  {author} {\bibinfo {author} {\bibfnamefont {S.}~\bibnamefont
  {Pai}}\ and\ \bibinfo {author} {\bibfnamefont {M.}~\bibnamefont {Pretko}},\
  }\bibfield  {title} {\bibinfo {title} {Dynamical scar states in driven
  fracton systems},\ }\href {https://doi.org/10.1103/PhysRevLett.123.136401}
  {\bibfield  {journal} {\bibinfo  {journal} {Phys. Rev. Lett.}\ }\textbf
  {\bibinfo {volume} {123}},\ \bibinfo {pages} {136401} (\bibinfo {year}
  {2019})}\BibitemShut {NoStop}%
\bibitem [{\citenamefont {Ma}\ and\ \citenamefont {Pretko}(2018)}]{fracton56}%
  \BibitemOpen
  \bibfield  {author} {\bibinfo {author} {\bibfnamefont {H.}~\bibnamefont
  {Ma}}\ and\ \bibinfo {author} {\bibfnamefont {M.}~\bibnamefont {Pretko}},\
  }\bibfield  {title} {\bibinfo {title} {Higher-rank deconfined quantum
  criticality at the lifshitz transition and the exciton bose condensate},\
  }\href {https://doi.org/10.1103/PhysRevB.98.125105} {\bibfield  {journal}
  {\bibinfo  {journal} {Phys. Rev. B}\ }\textbf {\bibinfo {volume} {98}},\
  \bibinfo {pages} {125105} (\bibinfo {year} {2018})}\BibitemShut {NoStop}%
\bibitem [{\citenamefont {Pai}\ and\ \citenamefont {Pretko}(2018)}]{fracton57}%
  \BibitemOpen
  \bibfield  {author} {\bibinfo {author} {\bibfnamefont {S.}~\bibnamefont
  {Pai}}\ and\ \bibinfo {author} {\bibfnamefont {M.}~\bibnamefont {Pretko}},\
  }\bibfield  {title} {\bibinfo {title} {Fractonic line excitations: An inroad
  from three-dimensional elasticity theory},\ }\href
  {https://doi.org/10.1103/PhysRevB.97.235102} {\bibfield  {journal} {\bibinfo
  {journal} {Phys. Rev. B}\ }\textbf {\bibinfo {volume} {97}},\ \bibinfo
  {pages} {235102} (\bibinfo {year} {2018})}\BibitemShut {NoStop}%
\bibitem [{\citenamefont {Pai}\ \emph {et~al.}(2019)\citenamefont {Pai},
  \citenamefont {Pretko},\ and\ \citenamefont {Nandkishore}}]{fracton60}%
  \BibitemOpen
  \bibfield  {author} {\bibinfo {author} {\bibfnamefont {S.}~\bibnamefont
  {Pai}}, \bibinfo {author} {\bibfnamefont {M.}~\bibnamefont {Pretko}},\ and\
  \bibinfo {author} {\bibfnamefont {R.~M.}\ \bibnamefont {Nandkishore}},\
  }\bibfield  {title} {\bibinfo {title} {Localization in fractonic random
  circuits},\ }\href {https://doi.org/10.1103/PhysRevX.9.021003} {\bibfield
  {journal} {\bibinfo  {journal} {Phys. Rev. X}\ }\textbf {\bibinfo {volume}
  {9}},\ \bibinfo {pages} {021003} (\bibinfo {year} {2019})}\BibitemShut
  {NoStop}%
\bibitem [{\citenamefont {Gromov}(2019)}]{fracton20}%
  \BibitemOpen
  \bibfield  {author} {\bibinfo {author} {\bibfnamefont {A.}~\bibnamefont
  {Gromov}},\ }\bibfield  {title} {\bibinfo {title} {Towards classification of
  fracton phases: The multipole algebra},\ }\href
  {https://doi.org/10.1103/PhysRevX.9.031035} {\bibfield  {journal} {\bibinfo
  {journal} {Phys. Rev. X}\ }\textbf {\bibinfo {volume} {9}},\ \bibinfo {pages}
  {031035} (\bibinfo {year} {2019})}\BibitemShut {NoStop}%
\bibitem [{\citenamefont {Wang}\ and\ \citenamefont {Yau}(2020)}]{fracton26}%
  \BibitemOpen
  \bibfield  {author} {\bibinfo {author} {\bibfnamefont {J.}~\bibnamefont
  {Wang}}\ and\ \bibinfo {author} {\bibfnamefont {S.-T.}\ \bibnamefont {Yau}},\
  }\bibfield  {title} {\bibinfo {title} {Non-abelian gauged fracton matter
  field theory: Sigma models, superfluids, and vortices},\ }\href
  {https://doi.org/10.1103/PhysRevResearch.2.043219} {\bibfield  {journal}
  {\bibinfo  {journal} {Phys. Rev. Res.}\ }\textbf {\bibinfo {volume} {2}},\
  \bibinfo {pages} {043219} (\bibinfo {year} {2020})}\BibitemShut {NoStop}%
\bibitem [{\citenamefont {Gromov}\ \emph {et~al.}(2020)\citenamefont {Gromov},
  \citenamefont {Lucas},\ and\ \citenamefont {Nandkishore}}]{fracton5}%
  \BibitemOpen
  \bibfield  {author} {\bibinfo {author} {\bibfnamefont {A.}~\bibnamefont
  {Gromov}}, \bibinfo {author} {\bibfnamefont {A.}~\bibnamefont {Lucas}},\ and\
  \bibinfo {author} {\bibfnamefont {R.~M.}\ \bibnamefont {Nandkishore}},\
  }\bibfield  {title} {\bibinfo {title} {Fracton hydrodynamics},\ }\href
  {https://doi.org/10.1103/PhysRevResearch.2.033124} {\bibfield  {journal}
  {\bibinfo  {journal} {Phys. Rev. Res.}\ }\textbf {\bibinfo {volume} {2}},\
  \bibinfo {pages} {033124} (\bibinfo {year} {2020})}\BibitemShut {NoStop}%
\bibitem [{\citenamefont {Feldmeier}\ \emph {et~al.}(2020)\citenamefont
  {Feldmeier}, \citenamefont {Sala}, \citenamefont {De~Tomasi}, \citenamefont
  {Pollmann},\ and\ \citenamefont {Knap}}]{diffusionofhigher-moment}%
  \BibitemOpen
  \bibfield  {author} {\bibinfo {author} {\bibfnamefont {J.}~\bibnamefont
  {Feldmeier}}, \bibinfo {author} {\bibfnamefont {P.}~\bibnamefont {Sala}},
  \bibinfo {author} {\bibfnamefont {G.}~\bibnamefont {De~Tomasi}}, \bibinfo
  {author} {\bibfnamefont {F.}~\bibnamefont {Pollmann}},\ and\ \bibinfo
  {author} {\bibfnamefont {M.}~\bibnamefont {Knap}},\ }\bibfield  {title}
  {\bibinfo {title} {Anomalous diffusion in dipole- and
  higher-moment-conserving systems},\ }\href
  {https://doi.org/10.1103/PhysRevLett.125.245303} {\bibfield  {journal}
  {\bibinfo  {journal} {Phys. Rev. Lett.}\ }\textbf {\bibinfo {volume} {125}},\
  \bibinfo {pages} {245303} (\bibinfo {year} {2020})}\BibitemShut {NoStop}%
\bibitem [{\citenamefont {Yuan}\ \emph {et~al.}(2020)\citenamefont {Yuan},
  \citenamefont {Chen},\ and\ \citenamefont {Ye}}]{Fractonicsuperfluids1}%
  \BibitemOpen
  \bibfield  {author} {\bibinfo {author} {\bibfnamefont {J.-K.}\ \bibnamefont
  {Yuan}}, \bibinfo {author} {\bibfnamefont {S.~A.}\ \bibnamefont {Chen}},\
  and\ \bibinfo {author} {\bibfnamefont {P.}~\bibnamefont {Ye}},\ }\bibfield
  {title} {\bibinfo {title} {Fractonic superfluids},\ }\href
  {https://doi.org/10.1103/PhysRevResearch.2.023267} {\bibfield  {journal}
  {\bibinfo  {journal} {Phys. Rev. Res.}\ }\textbf {\bibinfo {volume} {2}},\
  \bibinfo {pages} {023267} (\bibinfo {year} {2020})}\BibitemShut {NoStop}%
\bibitem [{\citenamefont {Chen}\ \emph {et~al.}(2021)\citenamefont {Chen},
  \citenamefont {Yuan},\ and\ \citenamefont {Ye}}]{Fractonicsuperfluids2}%
  \BibitemOpen
  \bibfield  {author} {\bibinfo {author} {\bibfnamefont {S.~A.}\ \bibnamefont
  {Chen}}, \bibinfo {author} {\bibfnamefont {J.-K.}\ \bibnamefont {Yuan}},\
  and\ \bibinfo {author} {\bibfnamefont {P.}~\bibnamefont {Ye}},\ }\bibfield
  {title} {\bibinfo {title} {Fractonic superfluids. ii. condensing
  subdimensional particles},\ }\href
  {https://doi.org/10.1103/PhysRevResearch.3.013226} {\bibfield  {journal}
  {\bibinfo  {journal} {Phys. Rev. Res.}\ }\textbf {\bibinfo {volume} {3}},\
  \bibinfo {pages} {013226} (\bibinfo {year} {2021})}\BibitemShut {NoStop}%
\bibitem [{\citenamefont {Wang}\ \emph {et~al.}(2025)\citenamefont {Wang},
  \citenamefont {Chen},\ and\ \citenamefont {Ye}}]{wanghanxie}%
  \BibitemOpen
  \bibfield  {author} {\bibinfo {author} {\bibfnamefont {H.-X.}\ \bibnamefont
  {Wang}}, \bibinfo {author} {\bibfnamefont {S.~A.}\ \bibnamefont {Chen}},\
  and\ \bibinfo {author} {\bibfnamefont {P.}~\bibnamefont {Ye}},\ }\bibfield
  {title} {\bibinfo {title} {Fractonic superfluids. iii. hybridizing higher
  moments},\ }\href {https://doi.org/10.1103/jvs9-6qkn} {\bibfield  {journal}
  {\bibinfo  {journal} {Phys. Rev. Res.}\ }\textbf {\bibinfo {volume} {7}},\
  \bibinfo {pages} {033118} (\bibinfo {year} {2025})}\BibitemShut {NoStop}%
\bibitem [{\citenamefont {Yuan}\ \emph {et~al.}(2022)\citenamefont {Yuan},
  \citenamefont {Chen},\ and\ \citenamefont {Ye}}]{NSofFractonicsuperfluids}%
  \BibitemOpen
  \bibfield  {author} {\bibinfo {author} {\bibfnamefont {J.-K.}\ \bibnamefont
  {Yuan}}, \bibinfo {author} {\bibfnamefont {S.~A.}\ \bibnamefont {Chen}},\
  and\ \bibinfo {author} {\bibfnamefont {P.}~\bibnamefont {Ye}},\ }\bibfield
  {title} {\bibinfo {title} {Quantum hydrodynamics of fractonic superfluids
  with lineon condensate: From navier--stokes-like equations to landau-like
  criterion},\ }\href {https://doi.org/10.1088/0256-307X/39/5/057101}
  {\bibfield  {journal} {\bibinfo  {journal} {Chin. Phys. Lett.}\ }\textbf
  {\bibinfo {volume} {39}},\ \bibinfo {pages} {057101} (\bibinfo {year}
  {2022})}\BibitemShut {NoStop}%
\bibitem [{\citenamefont {Li}\ and\ \citenamefont
  {Ye}(2021{\natexlab{b}})}]{fracton23}%
  \BibitemOpen
  \bibfield  {author} {\bibinfo {author} {\bibfnamefont {H.}~\bibnamefont
  {Li}}\ and\ \bibinfo {author} {\bibfnamefont {P.}~\bibnamefont {Ye}},\
  }\bibfield  {title} {\bibinfo {title} {Renormalization group analysis on
  emergence of higher rank symmetry and higher moment conservation},\ }\href
  {https://doi.org/10.1103/PhysRevResearch.3.043176} {\bibfield  {journal}
  {\bibinfo  {journal} {Phys. Rev. Res.}\ }\textbf {\bibinfo {volume} {3}},\
  \bibinfo {pages} {043176} (\bibinfo {year} {2021}{\natexlab{b}})}\BibitemShut
  {NoStop}%
\bibitem [{\citenamefont {Yuan}\ \emph {et~al.}(2023)\citenamefont {Yuan},
  \citenamefont {Chen},\ and\ \citenamefont {Ye}}]{Fractonicsuperfluidsdefect}%
  \BibitemOpen
  \bibfield  {author} {\bibinfo {author} {\bibfnamefont {J.-K.}\ \bibnamefont
  {Yuan}}, \bibinfo {author} {\bibfnamefont {S.~A.}\ \bibnamefont {Chen}},\
  and\ \bibinfo {author} {\bibfnamefont {P.}~\bibnamefont {Ye}},\ }\bibfield
  {title} {\bibinfo {title} {Hierarchical proliferation of higher-rank symmetry
  defects in fractonic superfluids},\ }\href
  {https://doi.org/10.1103/PhysRevB.107.205134} {\bibfield  {journal} {\bibinfo
   {journal} {Phys. Rev. B}\ }\textbf {\bibinfo {volume} {107}},\ \bibinfo
  {pages} {205134} (\bibinfo {year} {2023})}\BibitemShut {NoStop}%
\bibitem [{\citenamefont {Chen}\ and\ \citenamefont
  {Ye}(2023)}]{reviewofFractonicsuperfluids}%
  \BibitemOpen
  \bibfield  {author} {\bibinfo {author} {\bibfnamefont {S.~A.}\ \bibnamefont
  {Chen}}\ and\ \bibinfo {author} {\bibfnamefont {P.}~\bibnamefont {Ye}},\
  }\href {https://arxiv.org/abs/2305.00941} {\bibinfo {title} {Many-body
  physics of spontaneously broken higher-rank symmetry: from fractonic
  superfluids to dipolar hubbard model}} (\bibinfo {year} {2023}),\ \Eprint
  {https://arxiv.org/abs/2305.00941} {arXiv:2305.00941 [cond-mat.str-el]}
  \BibitemShut {NoStop}%
\bibitem [{\citenamefont {Doshi}\ and\ \citenamefont
  {Gromov}(2021)}]{fracton28}%
  \BibitemOpen
  \bibfield  {author} {\bibinfo {author} {\bibfnamefont {D.}~\bibnamefont
  {Doshi}}\ and\ \bibinfo {author} {\bibfnamefont {A.}~\bibnamefont {Gromov}},\
  }\bibfield  {title} {\bibinfo {title} {Vortices as fractons},\ }\href
  {https://doi.org/10.1038/s42005-021-00540-4} {\bibfield  {journal} {\bibinfo
  {journal} {Communications Physics}\ }\textbf {\bibinfo {volume} {4}},\
  \bibinfo {pages} {44} (\bibinfo {year} {2021})}\BibitemShut {NoStop}%
\bibitem [{\citenamefont {Grosvenor}\ \emph {et~al.}(2021)\citenamefont
  {Grosvenor}, \citenamefont {Hoyos}, \citenamefont {Pe\~na Benitez},\ and\
  \citenamefont {Sur\'owka}}]{fracton7}%
  \BibitemOpen
  \bibfield  {author} {\bibinfo {author} {\bibfnamefont {K.~T.}\ \bibnamefont
  {Grosvenor}}, \bibinfo {author} {\bibfnamefont {C.}~\bibnamefont {Hoyos}},
  \bibinfo {author} {\bibfnamefont {F.}~\bibnamefont {Pe\~na Benitez}},\ and\
  \bibinfo {author} {\bibfnamefont {P.}~\bibnamefont {Sur\'owka}},\ }\bibfield
  {title} {\bibinfo {title} {Hydrodynamics of ideal fracton fluids},\ }\href
  {https://doi.org/10.1103/PhysRevResearch.3.043186} {\bibfield  {journal}
  {\bibinfo  {journal} {Phys. Rev. Res.}\ }\textbf {\bibinfo {volume} {3}},\
  \bibinfo {pages} {043186} (\bibinfo {year} {2021})}\BibitemShut {NoStop}%
\bibitem [{\citenamefont {Wang}\ \emph {et~al.}(2021)\citenamefont {Wang},
  \citenamefont {Xu},\ and\ \citenamefont {Yau}}]{fracton21}%
  \BibitemOpen
  \bibfield  {author} {\bibinfo {author} {\bibfnamefont {J.}~\bibnamefont
  {Wang}}, \bibinfo {author} {\bibfnamefont {K.}~\bibnamefont {Xu}},\ and\
  \bibinfo {author} {\bibfnamefont {S.-T.}\ \bibnamefont {Yau}},\ }\bibfield
  {title} {\bibinfo {title} {Higher-rank tensor non-abelian field theory:
  Higher-moment or subdimensional polynomial global symmetry, algebraic
  variety, noether's theorem, and gauging},\ }\href
  {https://doi.org/10.1103/PhysRevResearch.3.013185} {\bibfield  {journal}
  {\bibinfo  {journal} {Phys. Rev. Res.}\ }\textbf {\bibinfo {volume} {3}},\
  \bibinfo {pages} {013185} (\bibinfo {year} {2021})}\BibitemShut {NoStop}%
\bibitem [{\citenamefont {Argurio}\ \emph {et~al.}(2021)\citenamefont
  {Argurio}, \citenamefont {Hoyos}, \citenamefont {Musso},\ and\ \citenamefont
  {Naegels}}]{fracton14}%
  \BibitemOpen
  \bibfield  {author} {\bibinfo {author} {\bibfnamefont {R.}~\bibnamefont
  {Argurio}}, \bibinfo {author} {\bibfnamefont {C.}~\bibnamefont {Hoyos}},
  \bibinfo {author} {\bibfnamefont {D.}~\bibnamefont {Musso}},\ and\ \bibinfo
  {author} {\bibfnamefont {D.}~\bibnamefont {Naegels}},\ }\bibfield  {title}
  {\bibinfo {title} {Fractons in effective field theories for spontaneously
  broken translations},\ }\href {https://doi.org/10.1103/PhysRevD.104.105001}
  {\bibfield  {journal} {\bibinfo  {journal} {Phys. Rev. D}\ }\textbf {\bibinfo
  {volume} {104}},\ \bibinfo {pages} {105001} (\bibinfo {year}
  {2021})}\BibitemShut {NoStop}%
\bibitem [{\citenamefont {Iaconis}\ \emph {et~al.}(2021)\citenamefont
  {Iaconis}, \citenamefont {Lucas},\ and\ \citenamefont
  {Nandkishore}}]{fracton22}%
  \BibitemOpen
  \bibfield  {author} {\bibinfo {author} {\bibfnamefont {J.}~\bibnamefont
  {Iaconis}}, \bibinfo {author} {\bibfnamefont {A.}~\bibnamefont {Lucas}},\
  and\ \bibinfo {author} {\bibfnamefont {R.}~\bibnamefont {Nandkishore}},\
  }\bibfield  {title} {\bibinfo {title} {Multipole conservation laws and
  subdiffusion in any dimension},\ }\href
  {https://doi.org/10.1103/PhysRevE.103.022142} {\bibfield  {journal} {\bibinfo
   {journal} {Phys. Rev. E}\ }\textbf {\bibinfo {volume} {103}},\ \bibinfo
  {pages} {022142} (\bibinfo {year} {2021})}\BibitemShut {NoStop}%
\bibitem [{\citenamefont {Stahl}\ \emph {et~al.}(2022)\citenamefont {Stahl},
  \citenamefont {Lake},\ and\ \citenamefont {Nandkishore}}]{fracton30}%
  \BibitemOpen
  \bibfield  {author} {\bibinfo {author} {\bibfnamefont {C.}~\bibnamefont
  {Stahl}}, \bibinfo {author} {\bibfnamefont {E.}~\bibnamefont {Lake}},\ and\
  \bibinfo {author} {\bibfnamefont {R.}~\bibnamefont {Nandkishore}},\
  }\bibfield  {title} {\bibinfo {title} {Spontaneous breaking of multipole
  symmetries},\ }\href {https://doi.org/10.1103/PhysRevB.105.155107} {\bibfield
   {journal} {\bibinfo  {journal} {Phys. Rev. B}\ }\textbf {\bibinfo {volume}
  {105}},\ \bibinfo {pages} {155107} (\bibinfo {year} {2022})}\BibitemShut
  {NoStop}%
\bibitem [{\citenamefont {Radzihovsky}(2022)}]{Lifshitzduality}%
  \BibitemOpen
  \bibfield  {author} {\bibinfo {author} {\bibfnamefont {L.}~\bibnamefont
  {Radzihovsky}},\ }\bibfield  {title} {\bibinfo {title} {Lifshitz gauge
  duality},\ }\href {https://doi.org/10.1103/PhysRevB.106.224510} {\bibfield
  {journal} {\bibinfo  {journal} {Phys. Rev. B}\ }\textbf {\bibinfo {volume}
  {106}},\ \bibinfo {pages} {224510} (\bibinfo {year} {2022})}\BibitemShut
  {NoStop}%
\bibitem [{\citenamefont {Osborne}\ and\ \citenamefont
  {Lucas}(2022)}]{fracton11}%
  \BibitemOpen
  \bibfield  {author} {\bibinfo {author} {\bibfnamefont {A.}~\bibnamefont
  {Osborne}}\ and\ \bibinfo {author} {\bibfnamefont {A.}~\bibnamefont
  {Lucas}},\ }\bibfield  {title} {\bibinfo {title} {Infinite families of
  fracton fluids with momentum conservation},\ }\href
  {https://doi.org/10.1103/PhysRevB.105.024311} {\bibfield  {journal} {\bibinfo
   {journal} {Phys. Rev. B}\ }\textbf {\bibinfo {volume} {105}},\ \bibinfo
  {pages} {024311} (\bibinfo {year} {2022})}\BibitemShut {NoStop}%
\bibitem [{\citenamefont {Kapustin}\ and\ \citenamefont
  {Spodyneiko}(2022)}]{HMWT}%
  \BibitemOpen
  \bibfield  {author} {\bibinfo {author} {\bibfnamefont {A.}~\bibnamefont
  {Kapustin}}\ and\ \bibinfo {author} {\bibfnamefont {L.}~\bibnamefont
  {Spodyneiko}},\ }\bibfield  {title} {\bibinfo {title}
  {Hohenberg-mermin-wagner-type theorems and dipole symmetry},\ }\href
  {https://doi.org/10.1103/PhysRevB.106.245125} {\bibfield  {journal} {\bibinfo
   {journal} {Phys. Rev. B}\ }\textbf {\bibinfo {volume} {106}},\ \bibinfo
  {pages} {245125} (\bibinfo {year} {2022})}\BibitemShut {NoStop}%
\bibitem [{\citenamefont {Lake}\ \emph {et~al.}(2022)\citenamefont {Lake},
  \citenamefont {Hermele},\ and\ \citenamefont {Senthil}}]{DBHM}%
  \BibitemOpen
  \bibfield  {author} {\bibinfo {author} {\bibfnamefont {E.}~\bibnamefont
  {Lake}}, \bibinfo {author} {\bibfnamefont {M.}~\bibnamefont {Hermele}},\ and\
  \bibinfo {author} {\bibfnamefont {T.}~\bibnamefont {Senthil}},\ }\bibfield
  {title} {\bibinfo {title} {Dipolar bose-hubbard model},\ }\href
  {https://doi.org/10.1103/PhysRevB.106.064511} {\bibfield  {journal} {\bibinfo
   {journal} {Phys. Rev. B}\ }\textbf {\bibinfo {volume} {106}},\ \bibinfo
  {pages} {064511} (\bibinfo {year} {2022})}\BibitemShut {NoStop}%
\bibitem [{\citenamefont {Lake}\ \emph {et~al.}(2023)\citenamefont {Lake},
  \citenamefont {Lee}, \citenamefont {Han},\ and\ \citenamefont
  {Senthil}}]{DBHM2}%
  \BibitemOpen
  \bibfield  {author} {\bibinfo {author} {\bibfnamefont {E.}~\bibnamefont
  {Lake}}, \bibinfo {author} {\bibfnamefont {H.-Y.}\ \bibnamefont {Lee}},
  \bibinfo {author} {\bibfnamefont {J.~H.}\ \bibnamefont {Han}},\ and\ \bibinfo
  {author} {\bibfnamefont {T.}~\bibnamefont {Senthil}},\ }\bibfield  {title}
  {\bibinfo {title} {Dipole condensates in tilted bose-hubbard chains},\ }\href
  {https://doi.org/10.1103/PhysRevB.107.195132} {\bibfield  {journal} {\bibinfo
   {journal} {Phys. Rev. B}\ }\textbf {\bibinfo {volume} {107}},\ \bibinfo
  {pages} {195132} (\bibinfo {year} {2023})}\BibitemShut {NoStop}%
\bibitem [{\citenamefont {Giergiel}\ \emph {et~al.}(2022)\citenamefont
  {Giergiel}, \citenamefont {Lier}, \citenamefont {Sur\'owka},\ and\
  \citenamefont {Kosior}}]{fracton31}%
  \BibitemOpen
  \bibfield  {author} {\bibinfo {author} {\bibfnamefont {K.}~\bibnamefont
  {Giergiel}}, \bibinfo {author} {\bibfnamefont {R.}~\bibnamefont {Lier}},
  \bibinfo {author} {\bibfnamefont {P.}~\bibnamefont {Sur\'owka}},\ and\
  \bibinfo {author} {\bibfnamefont {A.}~\bibnamefont {Kosior}},\ }\bibfield
  {title} {\bibinfo {title} {Bose-hubbard realization of fracton defects},\
  }\href {https://doi.org/10.1103/PhysRevResearch.4.023151} {\bibfield
  {journal} {\bibinfo  {journal} {Phys. Rev. Res.}\ }\textbf {\bibinfo {volume}
  {4}},\ \bibinfo {pages} {023151} (\bibinfo {year} {2022})}\BibitemShut
  {NoStop}%
\bibitem [{\citenamefont {Zechmann}\ \emph {et~al.}(2023)\citenamefont
  {Zechmann}, \citenamefont {Altman}, \citenamefont {Knap},\ and\ \citenamefont
  {Feldmeier}}]{fracton35}%
  \BibitemOpen
  \bibfield  {author} {\bibinfo {author} {\bibfnamefont {P.}~\bibnamefont
  {Zechmann}}, \bibinfo {author} {\bibfnamefont {E.}~\bibnamefont {Altman}},
  \bibinfo {author} {\bibfnamefont {M.}~\bibnamefont {Knap}},\ and\ \bibinfo
  {author} {\bibfnamefont {J.}~\bibnamefont {Feldmeier}},\ }\bibfield  {title}
  {\bibinfo {title} {Fractonic luttinger liquids and supersolids in a
  constrained bose-hubbard model},\ }\href
  {https://doi.org/10.1103/PhysRevB.107.195131} {\bibfield  {journal} {\bibinfo
   {journal} {Phys. Rev. B}\ }\textbf {\bibinfo {volume} {107}},\ \bibinfo
  {pages} {195131} (\bibinfo {year} {2023})}\BibitemShut {NoStop}%
\bibitem [{\citenamefont {Boesl}\ \emph {et~al.}(2024)\citenamefont {Boesl},
  \citenamefont {Zechmann}, \citenamefont {Feldmeier},\ and\ \citenamefont
  {Knap}}]{fracton44}%
  \BibitemOpen
  \bibfield  {author} {\bibinfo {author} {\bibfnamefont {J.}~\bibnamefont
  {Boesl}}, \bibinfo {author} {\bibfnamefont {P.}~\bibnamefont {Zechmann}},
  \bibinfo {author} {\bibfnamefont {J.}~\bibnamefont {Feldmeier}},\ and\
  \bibinfo {author} {\bibfnamefont {M.}~\bibnamefont {Knap}},\ }\bibfield
  {title} {\bibinfo {title} {Deconfinement dynamics of fractons in tilted
  bose-hubbard chains},\ }\href
  {https://doi.org/10.1103/PhysRevLett.132.143401} {\bibfield  {journal}
  {\bibinfo  {journal} {Phys. Rev. Lett.}\ }\textbf {\bibinfo {volume} {132}},\
  \bibinfo {pages} {143401} (\bibinfo {year} {2024})}\BibitemShut {NoStop}%
\bibitem [{\citenamefont {Xu}\ \emph {et~al.}(2024)\citenamefont {Xu},
  \citenamefont {Lv},\ and\ \citenamefont {Zhou}}]{fracton45}%
  \BibitemOpen
  \bibfield  {author} {\bibinfo {author} {\bibfnamefont {W.}~\bibnamefont
  {Xu}}, \bibinfo {author} {\bibfnamefont {C.}~\bibnamefont {Lv}},\ and\
  \bibinfo {author} {\bibfnamefont {Q.}~\bibnamefont {Zhou}},\ }\bibfield
  {title} {\bibinfo {title} {Multipolar condensates and multipolar {Josephson}
  effects},\ }\href {https://doi.org/10.1038/s41467-024-48907-9} {\bibfield
  {journal} {\bibinfo  {journal} {Nature Communications}\ }\textbf {\bibinfo
  {volume} {15}},\ \bibinfo {pages} {4786} (\bibinfo {year}
  {2024})}\BibitemShut {NoStop}%
\bibitem [{\citenamefont {Gorantla}\ \emph {et~al.}(2022)\citenamefont
  {Gorantla}, \citenamefont {Lam}, \citenamefont {Seiberg},\ and\ \citenamefont
  {Shao}}]{fracton6}%
  \BibitemOpen
  \bibfield  {author} {\bibinfo {author} {\bibfnamefont {P.}~\bibnamefont
  {Gorantla}}, \bibinfo {author} {\bibfnamefont {H.~T.}\ \bibnamefont {Lam}},
  \bibinfo {author} {\bibfnamefont {N.}~\bibnamefont {Seiberg}},\ and\ \bibinfo
  {author} {\bibfnamefont {S.-H.}\ \bibnamefont {Shao}},\ }\bibfield  {title}
  {\bibinfo {title} {Global dipole symmetry, compact lifshitz theory, tensor
  gauge theory, and fractons},\ }\href
  {https://doi.org/10.1103/PhysRevB.106.045112} {\bibfield  {journal} {\bibinfo
   {journal} {Phys. Rev. B}\ }\textbf {\bibinfo {volume} {106}},\ \bibinfo
  {pages} {045112} (\bibinfo {year} {2022})}\BibitemShut {NoStop}%
\bibitem [{\citenamefont {Grosvenor}\ \emph {et~al.}(2022)\citenamefont
  {Grosvenor}, \citenamefont {Hoyos}, \citenamefont {Peña-Benítez},\ and\
  \citenamefont {Surówka}}]{fracton29}%
  \BibitemOpen
  \bibfield  {author} {\bibinfo {author} {\bibfnamefont {K.~T.}\ \bibnamefont
  {Grosvenor}}, \bibinfo {author} {\bibfnamefont {C.}~\bibnamefont {Hoyos}},
  \bibinfo {author} {\bibfnamefont {F.}~\bibnamefont {Peña-Benítez}},\ and\
  \bibinfo {author} {\bibfnamefont {P.}~\bibnamefont {Surówka}},\ }\bibfield
  {title} {\bibinfo {title} {Space-dependent symmetries and fractons},\
  }\bibfield  {journal} {\bibinfo  {journal} {Frontiers in Physics}\ }\textbf
  {\bibinfo {volume} {9}},\ \href {https://doi.org/10.3389/fphy.2021.792621}
  {10.3389/fphy.2021.792621} (\bibinfo {year} {2022})\BibitemShut {NoStop}%
\bibitem [{\citenamefont {Lake}\ and\ \citenamefont
  {Senthil}(2023)}]{fracton24}%
  \BibitemOpen
  \bibfield  {author} {\bibinfo {author} {\bibfnamefont {E.}~\bibnamefont
  {Lake}}\ and\ \bibinfo {author} {\bibfnamefont {T.}~\bibnamefont {Senthil}},\
  }\bibfield  {title} {\bibinfo {title} {Non-fermi liquids from kinetic
  constraints in tilted optical lattices},\ }\href
  {https://doi.org/10.1103/PhysRevLett.131.043403} {\bibfield  {journal}
  {\bibinfo  {journal} {Phys. Rev. Lett.}\ }\textbf {\bibinfo {volume} {131}},\
  \bibinfo {pages} {043403} (\bibinfo {year} {2023})}\BibitemShut {NoStop}%
\bibitem [{\citenamefont {Molina-Vilaplana}(2023)}]{fracton34}%
  \BibitemOpen
  \bibfield  {author} {\bibinfo {author} {\bibfnamefont {J.}~\bibnamefont
  {Molina-Vilaplana}},\ }\bibfield  {title} {\bibinfo {title} {A
  post-{Gaussian} approach to dipole symmetries and interacting fractons},\
  }\href {https://doi.org/10.1007/JHEP08(2023)065} {\bibfield  {journal}
  {\bibinfo  {journal} {Journal of High Energy Physics}\ }\textbf {\bibinfo
  {volume} {2023}},\ \bibinfo {pages} {65} (\bibinfo {year}
  {2023})}\BibitemShut {NoStop}%
\bibitem [{\citenamefont {Gorantla}\ \emph {et~al.}(2023)\citenamefont
  {Gorantla}, \citenamefont {Lam}, \citenamefont {Seiberg},\ and\ \citenamefont
  {Shao}}]{fracton36}%
  \BibitemOpen
  \bibfield  {author} {\bibinfo {author} {\bibfnamefont {P.}~\bibnamefont
  {Gorantla}}, \bibinfo {author} {\bibfnamefont {H.~T.}\ \bibnamefont {Lam}},
  \bibinfo {author} {\bibfnamefont {N.}~\bibnamefont {Seiberg}},\ and\ \bibinfo
  {author} {\bibfnamefont {S.-H.}\ \bibnamefont {Shao}},\ }\bibfield  {title}
  {\bibinfo {title} {(2+1)-dimensional compact lifshitz theory, tensor gauge
  theory, and fractons},\ }\href {https://doi.org/10.1103/PhysRevB.108.075106}
  {\bibfield  {journal} {\bibinfo  {journal} {Phys. Rev. B}\ }\textbf {\bibinfo
  {volume} {108}},\ \bibinfo {pages} {075106} (\bibinfo {year}
  {2023})}\BibitemShut {NoStop}%
\bibitem [{\citenamefont {Anakru}\ and\ \citenamefont {Bi}(2023)}]{fracton37}%
  \BibitemOpen
  \bibfield  {author} {\bibinfo {author} {\bibfnamefont {A.}~\bibnamefont
  {Anakru}}\ and\ \bibinfo {author} {\bibfnamefont {Z.}~\bibnamefont {Bi}},\
  }\bibfield  {title} {\bibinfo {title} {Non-fermi liquids from dipolar
  symmetry breaking},\ }\href {https://doi.org/10.1103/PhysRevB.108.165112}
  {\bibfield  {journal} {\bibinfo  {journal} {Phys. Rev. B}\ }\textbf {\bibinfo
  {volume} {108}},\ \bibinfo {pages} {165112} (\bibinfo {year}
  {2023})}\BibitemShut {NoStop}%
\bibitem [{\citenamefont {Glorioso}\ \emph {et~al.}(2023)\citenamefont
  {Glorioso}, \citenamefont {Huang}, \citenamefont {Guo}, \citenamefont
  {Rodriguez-Nieva},\ and\ \citenamefont {Lucas}}]{fracton33}%
  \BibitemOpen
  \bibfield  {author} {\bibinfo {author} {\bibfnamefont {P.}~\bibnamefont
  {Glorioso}}, \bibinfo {author} {\bibfnamefont {X.}~\bibnamefont {Huang}},
  \bibinfo {author} {\bibfnamefont {J.}~\bibnamefont {Guo}}, \bibinfo {author}
  {\bibfnamefont {J.~F.}\ \bibnamefont {Rodriguez-Nieva}},\ and\ \bibinfo
  {author} {\bibfnamefont {A.}~\bibnamefont {Lucas}},\ }\bibfield  {title}
  {\bibinfo {title} {Goldstone bosons and fluctuating hydrodynamics with dipole
  and momentum conservation},\ }\href {https://doi.org/10.1007/JHEP05(2023)022}
  {\bibfield  {journal} {\bibinfo  {journal} {Journal of High Energy Physics}\
  }\textbf {\bibinfo {volume} {2023}},\ \bibinfo {pages} {22} (\bibinfo {year}
  {2023})}\BibitemShut {NoStop}%
\bibitem [{\citenamefont {Morningstar}\ \emph {et~al.}(2023)\citenamefont
  {Morningstar}, \citenamefont {O'Dea},\ and\ \citenamefont
  {Richter}}]{fracton38}%
  \BibitemOpen
  \bibfield  {author} {\bibinfo {author} {\bibfnamefont {A.}~\bibnamefont
  {Morningstar}}, \bibinfo {author} {\bibfnamefont {N.}~\bibnamefont {O'Dea}},\
  and\ \bibinfo {author} {\bibfnamefont {J.}~\bibnamefont {Richter}},\
  }\bibfield  {title} {\bibinfo {title} {Hydrodynamics in long-range
  interacting systems with center-of-mass conservation},\ }\href
  {https://doi.org/10.1103/PhysRevB.108.L020304} {\bibfield  {journal}
  {\bibinfo  {journal} {Phys. Rev. B}\ }\textbf {\bibinfo {volume} {108}},\
  \bibinfo {pages} {L020304} (\bibinfo {year} {2023})}\BibitemShut {NoStop}%
\bibitem [{\citenamefont {Huang}(2023)}]{fracton39}%
  \BibitemOpen
  \bibfield  {author} {\bibinfo {author} {\bibfnamefont {X.}~\bibnamefont
  {Huang}},\ }\bibfield  {title} {\bibinfo {title} {{A Chern-Simons theory for
  dipole symmetry}},\ }\href {https://doi.org/10.21468/SciPostPhys.15.4.153}
  {\bibfield  {journal} {\bibinfo  {journal} {SciPost Phys.}\ }\textbf
  {\bibinfo {volume} {15}},\ \bibinfo {pages} {153} (\bibinfo {year}
  {2023})}\BibitemShut {NoStop}%
\bibitem [{\citenamefont {Afxonidis}\ \emph {et~al.}(2023)\citenamefont
  {Afxonidis}, \citenamefont {Caddeo}, \citenamefont {Hoyos},\ and\
  \citenamefont {Musso}}]{fracton40}%
  \BibitemOpen
  \bibfield  {author} {\bibinfo {author} {\bibfnamefont {E.}~\bibnamefont
  {Afxonidis}}, \bibinfo {author} {\bibfnamefont {A.}~\bibnamefont {Caddeo}},
  \bibinfo {author} {\bibfnamefont {C.}~\bibnamefont {Hoyos}},\ and\ \bibinfo
  {author} {\bibfnamefont {D.}~\bibnamefont {Musso}},\ }\bibfield  {title}
  {\bibinfo {title} {{Dipole symmetry breaking and fractonic Nambu-Goldstone
  mode}},\ }\href {https://doi.org/10.21468/SciPostPhysCore.6.4.082} {\bibfield
   {journal} {\bibinfo  {journal} {SciPost Phys. Core}\ }\textbf {\bibinfo
  {volume} {6}},\ \bibinfo {pages} {082} (\bibinfo {year} {2023})}\BibitemShut
  {NoStop}%
\bibitem [{\citenamefont {Stahl}\ \emph {et~al.}(2023)\citenamefont {Stahl},
  \citenamefont {Qi}, \citenamefont {Glorioso}, \citenamefont {Lucas},\ and\
  \citenamefont {Nandkishore}}]{fracton9}%
  \BibitemOpen
  \bibfield  {author} {\bibinfo {author} {\bibfnamefont {C.}~\bibnamefont
  {Stahl}}, \bibinfo {author} {\bibfnamefont {M.}~\bibnamefont {Qi}}, \bibinfo
  {author} {\bibfnamefont {P.}~\bibnamefont {Glorioso}}, \bibinfo {author}
  {\bibfnamefont {A.}~\bibnamefont {Lucas}},\ and\ \bibinfo {author}
  {\bibfnamefont {R.}~\bibnamefont {Nandkishore}},\ }\bibfield  {title}
  {\bibinfo {title} {Fracton superfluid hydrodynamics},\ }\href
  {https://doi.org/10.1103/PhysRevB.108.144509} {\bibfield  {journal} {\bibinfo
   {journal} {Phys. Rev. B}\ }\textbf {\bibinfo {volume} {108}},\ \bibinfo
  {pages} {144509} (\bibinfo {year} {2023})}\BibitemShut {NoStop}%
\bibitem [{\citenamefont {Lam}(2024)}]{fracton41}%
  \BibitemOpen
  \bibfield  {author} {\bibinfo {author} {\bibfnamefont {H.~T.}\ \bibnamefont
  {Lam}},\ }\bibfield  {title} {\bibinfo {title} {Classification of dipolar
  symmetry-protected topological phases: Matrix product states, stabilizer
  hamiltonians, and finite tensor gauge theories},\ }\href
  {https://doi.org/10.1103/PhysRevB.109.115142} {\bibfield  {journal} {\bibinfo
   {journal} {Phys. Rev. B}\ }\textbf {\bibinfo {volume} {109}},\ \bibinfo
  {pages} {115142} (\bibinfo {year} {2024})}\BibitemShut {NoStop}%
\bibitem [{\citenamefont {Han}\ \emph {et~al.}(2024)\citenamefont {Han},
  \citenamefont {Lake},\ and\ \citenamefont {Ro}}]{fracton32}%
  \BibitemOpen
  \bibfield  {author} {\bibinfo {author} {\bibfnamefont {J.~H.}\ \bibnamefont
  {Han}}, \bibinfo {author} {\bibfnamefont {E.}~\bibnamefont {Lake}},\ and\
  \bibinfo {author} {\bibfnamefont {S.}~\bibnamefont {Ro}},\ }\bibfield
  {title} {\bibinfo {title} {Scaling and localization in multipole-conserving
  diffusion},\ }\href {https://doi.org/10.1103/PhysRevLett.132.137102}
  {\bibfield  {journal} {\bibinfo  {journal} {Phys. Rev. Lett.}\ }\textbf
  {\bibinfo {volume} {132}},\ \bibinfo {pages} {137102} (\bibinfo {year}
  {2024})}\BibitemShut {NoStop}%
\bibitem [{\citenamefont {Burnell}\ \emph {et~al.}(2024)\citenamefont
  {Burnell}, \citenamefont {Moudgalya},\ and\ \citenamefont
  {Prem}}]{fracton43}%
  \BibitemOpen
  \bibfield  {author} {\bibinfo {author} {\bibfnamefont {F.~J.}\ \bibnamefont
  {Burnell}}, \bibinfo {author} {\bibfnamefont {S.}~\bibnamefont {Moudgalya}},\
  and\ \bibinfo {author} {\bibfnamefont {A.}~\bibnamefont {Prem}},\ }\bibfield
  {title} {\bibinfo {title} {Filling constraints on translation invariant
  dipole conserving systems},\ }\href
  {https://doi.org/10.1103/PhysRevB.110.L121113} {\bibfield  {journal}
  {\bibinfo  {journal} {Phys. Rev. B}\ }\textbf {\bibinfo {volume} {110}},\
  \bibinfo {pages} {L121113} (\bibinfo {year} {2024})}\BibitemShut {NoStop}%
\bibitem [{\citenamefont {Chavda}\ \emph {et~al.}(2025)\citenamefont {Chavda},
  \citenamefont {Naegels},\ and\ \citenamefont {Staunton}}]{fracton48}%
  \BibitemOpen
  \bibfield  {author} {\bibinfo {author} {\bibfnamefont {A.}~\bibnamefont
  {Chavda}}, \bibinfo {author} {\bibfnamefont {D.}~\bibnamefont {Naegels}},\
  and\ \bibinfo {author} {\bibfnamefont {J.}~\bibnamefont {Staunton}},\
  }\bibfield  {title} {\bibinfo {title} {Fractonic coset construction for
  spontaneously broken translations},\ }\href
  {https://doi.org/10.1103/PhysRevD.111.085018} {\bibfield  {journal} {\bibinfo
   {journal} {Phys. Rev. D}\ }\textbf {\bibinfo {volume} {111}},\ \bibinfo
  {pages} {085018} (\bibinfo {year} {2025})}\BibitemShut {NoStop}%
\bibitem [{\citenamefont {Armas}\ and\ \citenamefont {Have}(2024)}]{fracton8}%
  \BibitemOpen
  \bibfield  {author} {\bibinfo {author} {\bibfnamefont {J.}~\bibnamefont
  {Armas}}\ and\ \bibinfo {author} {\bibfnamefont {E.}~\bibnamefont {Have}},\
  }\bibfield  {title} {\bibinfo {title} {{Ideal fracton superfluids}},\ }\href
  {https://doi.org/10.21468/SciPostPhys.16.1.039} {\bibfield  {journal}
  {\bibinfo  {journal} {SciPost Phys.}\ }\textbf {\bibinfo {volume} {16}},\
  \bibinfo {pages} {039} (\bibinfo {year} {2024})}\BibitemShut {NoStop}%
\bibitem [{\citenamefont {Zechmann}\ \emph {et~al.}(2024)\citenamefont
  {Zechmann}, \citenamefont {Boesl}, \citenamefont {Feldmeier},\ and\
  \citenamefont {Knap}}]{fracton42}%
  \BibitemOpen
  \bibfield  {author} {\bibinfo {author} {\bibfnamefont {P.}~\bibnamefont
  {Zechmann}}, \bibinfo {author} {\bibfnamefont {J.}~\bibnamefont {Boesl}},
  \bibinfo {author} {\bibfnamefont {J.}~\bibnamefont {Feldmeier}},\ and\
  \bibinfo {author} {\bibfnamefont {M.}~\bibnamefont {Knap}},\ }\bibfield
  {title} {\bibinfo {title} {Dynamical spectral response of fractonic quantum
  matter},\ }\href {https://doi.org/10.1103/PhysRevB.109.125137} {\bibfield
  {journal} {\bibinfo  {journal} {Phys. Rev. B}\ }\textbf {\bibinfo {volume}
  {109}},\ \bibinfo {pages} {125137} (\bibinfo {year} {2024})}\BibitemShut
  {NoStop}%
\bibitem [{\citenamefont {Głódkowski}\ \emph {et~al.}(2024)\citenamefont
  {Głódkowski}, \citenamefont {Peña-Benítez},\ and\ \citenamefont
  {Surówka}}]{fracton46}%
  \BibitemOpen
  \bibfield  {author} {\bibinfo {author} {\bibfnamefont {A.}~\bibnamefont
  {Głódkowski}}, \bibinfo {author} {\bibfnamefont {F.}~\bibnamefont
  {Peña-Benítez}},\ and\ \bibinfo {author} {\bibfnamefont {P.}~\bibnamefont
  {Surówka}},\ }\bibfield  {title} {\bibinfo {title} {Dissipative fracton
  superfluids},\ }\href {https://doi.org/10.1007/JHEP07(2024)285} {\bibfield
  {journal} {\bibinfo  {journal} {Journal of High Energy Physics}\ }\textbf
  {\bibinfo {volume} {2024}},\ \bibinfo {pages} {285} (\bibinfo {year}
  {2024})}\BibitemShut {NoStop}%
\bibitem [{\citenamefont {Angus}\ \emph {et~al.}(2022)\citenamefont {Angus},
  \citenamefont {Kim},\ and\ \citenamefont {Park}}]{fracton25}%
  \BibitemOpen
  \bibfield  {author} {\bibinfo {author} {\bibfnamefont {S.}~\bibnamefont
  {Angus}}, \bibinfo {author} {\bibfnamefont {M.}~\bibnamefont {Kim}},\ and\
  \bibinfo {author} {\bibfnamefont {J.-H.}\ \bibnamefont {Park}},\ }\bibfield
  {title} {\bibinfo {title} {Fractons, non-riemannian geometry, and double
  field theory},\ }\href {https://doi.org/10.1103/PhysRevResearch.4.033186}
  {\bibfield  {journal} {\bibinfo  {journal} {Phys. Rev. Res.}\ }\textbf
  {\bibinfo {volume} {4}},\ \bibinfo {pages} {033186} (\bibinfo {year}
  {2022})}\BibitemShut {NoStop}%
\bibitem [{\citenamefont {Pe\~na Ben\'{\i}tez}(2023)}]{fracton10}%
  \BibitemOpen
  \bibfield  {author} {\bibinfo {author} {\bibfnamefont {F.}~\bibnamefont
  {Pe\~na Ben\'{\i}tez}},\ }\bibfield  {title} {\bibinfo {title} {Fractons,
  symmetric gauge fields and geometry},\ }\href
  {https://doi.org/10.1103/PhysRevResearch.5.013101} {\bibfield  {journal}
  {\bibinfo  {journal} {Phys. Rev. Res.}\ }\textbf {\bibinfo {volume} {5}},\
  \bibinfo {pages} {013101} (\bibinfo {year} {2023})}\BibitemShut {NoStop}%
\bibitem [{\citenamefont {Bertolini}\ \emph {et~al.}(2023)\citenamefont
  {Bertolini}, \citenamefont {Maggiore},\ and\ \citenamefont
  {Palumbo}}]{fracton12}%
  \BibitemOpen
  \bibfield  {author} {\bibinfo {author} {\bibfnamefont {E.}~\bibnamefont
  {Bertolini}}, \bibinfo {author} {\bibfnamefont {N.}~\bibnamefont
  {Maggiore}},\ and\ \bibinfo {author} {\bibfnamefont {G.}~\bibnamefont
  {Palumbo}},\ }\bibfield  {title} {\bibinfo {title} {Covariant fracton gauge
  theory with boundary},\ }\href {https://doi.org/10.1103/PhysRevD.108.025009}
  {\bibfield  {journal} {\bibinfo  {journal} {Phys. Rev. D}\ }\textbf {\bibinfo
  {volume} {108}},\ \bibinfo {pages} {025009} (\bibinfo {year}
  {2023})}\BibitemShut {NoStop}%
\bibitem [{\citenamefont {Bertolini}\ and\ \citenamefont
  {Kim}(2025)}]{fracton49}%
  \BibitemOpen
  \bibfield  {author} {\bibinfo {author} {\bibfnamefont {E.}~\bibnamefont
  {Bertolini}}\ and\ \bibinfo {author} {\bibfnamefont {H.}~\bibnamefont
  {Kim}},\ }\bibfield  {title} {\bibinfo {title} {Covariant interacting
  fractons},\ }\href {https://doi.org/10.1103/PhysRevD.111.025006} {\bibfield
  {journal} {\bibinfo  {journal} {Phys. Rev. D}\ }\textbf {\bibinfo {volume}
  {111}},\ \bibinfo {pages} {025006} (\bibinfo {year} {2025})}\BibitemShut
  {NoStop}%
\bibitem [{\citenamefont {{Bertolini}}\ and\ \citenamefont
  {{Kim}}(2024)}]{fracton50}%
  \BibitemOpen
  \bibfield  {author} {\bibinfo {author} {\bibfnamefont {E.}~\bibnamefont
  {{Bertolini}}}\ and\ \bibinfo {author} {\bibfnamefont {H.}~\bibnamefont
  {{Kim}}},\ }\bibfield  {title} {\bibinfo {title} {{Strings as
  Hyper-Fractons}},\ }\href {https://doi.org/10.48550/arXiv.2410.11678}
  {\bibfield  {journal} {\bibinfo  {journal} {arXiv e-prints}\ ,\ \bibinfo
  {eid} {arXiv:2410.11678}} (\bibinfo {year} {2024})},\ \Eprint
  {https://arxiv.org/abs/2410.11678} {arXiv:2410.11678 [hep-th]} \BibitemShut
  {NoStop}%
\bibitem [{\citenamefont {Hu}\ and\ \citenamefont {Lian}(2025)}]{fracton47}%
  \BibitemOpen
  \bibfield  {author} {\bibinfo {author} {\bibfnamefont {Y.-M.}\ \bibnamefont
  {Hu}}\ and\ \bibinfo {author} {\bibfnamefont {B.}~\bibnamefont {Lian}},\
  }\bibfield  {title} {\bibinfo {title} {Bosonic quantum breakdown hubbard
  model},\ }\href {https://doi.org/10.1103/1r4m-7psy} {\bibfield  {journal}
  {\bibinfo  {journal} {Phys. Rev. B}\ }\textbf {\bibinfo {volume} {112}},\
  \bibinfo {pages} {L100504} (\bibinfo {year} {2025})}\BibitemShut {NoStop}%
\bibitem [{\citenamefont {Oh}\ \emph {et~al.}(2022)\citenamefont {Oh},
  \citenamefont {Kim}, \citenamefont {Moon},\ and\ \citenamefont
  {Han}}]{rank_two_TC}%
  \BibitemOpen
  \bibfield  {author} {\bibinfo {author} {\bibfnamefont {Y.-T.}\ \bibnamefont
  {Oh}}, \bibinfo {author} {\bibfnamefont {J.}~\bibnamefont {Kim}}, \bibinfo
  {author} {\bibfnamefont {E.-G.}\ \bibnamefont {Moon}},\ and\ \bibinfo
  {author} {\bibfnamefont {J.~H.}\ \bibnamefont {Han}},\ }\bibfield  {title}
  {\bibinfo {title} {Rank-2 toric code in two dimensions},\ }\href
  {https://doi.org/10.1103/PhysRevB.105.045128} {\bibfield  {journal} {\bibinfo
   {journal} {Phys. Rev. B}\ }\textbf {\bibinfo {volume} {105}},\ \bibinfo
  {pages} {045128} (\bibinfo {year} {2022})}\BibitemShut {NoStop}%
\bibitem [{\citenamefont {Pace}\ and\ \citenamefont
  {Wen}(2022)}]{rank_two_TC_Z_N}%
  \BibitemOpen
  \bibfield  {author} {\bibinfo {author} {\bibfnamefont {S.~D.}\ \bibnamefont
  {Pace}}\ and\ \bibinfo {author} {\bibfnamefont {X.-G.}\ \bibnamefont {Wen}},\
  }\bibfield  {title} {\bibinfo {title} {Position-dependent excitations and
  uv/ir mixing in the ${\mathbb{z}}_{N}$ rank-2 toric code and its low-energy
  effective field theory},\ }\href
  {https://doi.org/10.1103/PhysRevB.106.045145} {\bibfield  {journal} {\bibinfo
   {journal} {Phys. Rev. B}\ }\textbf {\bibinfo {volume} {106}},\ \bibinfo
  {pages} {045145} (\bibinfo {year} {2022})}\BibitemShut {NoStop}%
\bibitem [{\citenamefont {{Delfino}}\ \emph {et~al.}(2023)\citenamefont
  {{Delfino}}, \citenamefont {{Fontana}}, \citenamefont {{Gomes}},\ and\
  \citenamefont {{Chamon}}}]{effective_2d_fracton}%
  \BibitemOpen
  \bibfield  {author} {\bibinfo {author} {\bibfnamefont {G.}~\bibnamefont
  {{Delfino}}}, \bibinfo {author} {\bibfnamefont {W.~B.}\ \bibnamefont
  {{Fontana}}}, \bibinfo {author} {\bibfnamefont {P.~R.~S.}\ \bibnamefont
  {{Gomes}}},\ and\ \bibinfo {author} {\bibfnamefont {C.}~\bibnamefont
  {{Chamon}}},\ }\bibfield  {title} {\bibinfo {title} {{Effective fractonic
  behavior in a two-dimensional exactly solvable spin liquid}},\ }\href
  {https://doi.org/10.21468/SciPostPhys.14.1.002} {\bibfield  {journal}
  {\bibinfo  {journal} {SciPost Physics}\ }\textbf {\bibinfo {volume} {14}},\
  \bibinfo {eid} {002} (\bibinfo {year} {2023})},\ \Eprint
  {https://arxiv.org/abs/2207.00409} {arXiv:2207.00409 [cond-mat.str-el]}
  \BibitemShut {NoStop}%
\bibitem [{\citenamefont {{Delfino}}\ and\ \citenamefont
  {{You}}(2024)}]{2024PhRvB.109t5146D}%
  \BibitemOpen
  \bibfield  {author} {\bibinfo {author} {\bibfnamefont {G.}~\bibnamefont
  {{Delfino}}}\ and\ \bibinfo {author} {\bibfnamefont {Y.}~\bibnamefont
  {{You}}},\ }\bibfield  {title} {\bibinfo {title} {{Anyon condensation web and
  multipartite entanglement in two-dimensional modulated gauge theories}},\
  }\href {https://doi.org/10.1103/PhysRevB.109.205146} {\bibfield  {journal}
  {\bibinfo  {journal} {\prb}\ }\textbf {\bibinfo {volume} {109}},\ \bibinfo
  {eid} {205146} (\bibinfo {year} {2024})},\ \Eprint
  {https://arxiv.org/abs/2310.09490} {arXiv:2310.09490 [cond-mat.str-el]}
  \BibitemShut {NoStop}%
\bibitem [{\citenamefont {Shen}\ \emph {et~al.}(2022)\citenamefont {Shen},
  \citenamefont {Wu}, \citenamefont {Li}, \citenamefont {Qin},\ and\
  \citenamefont {Yao}}]{Shen_2022}%
  \BibitemOpen
  \bibfield  {author} {\bibinfo {author} {\bibfnamefont {X.}~\bibnamefont
  {Shen}}, \bibinfo {author} {\bibfnamefont {Z.}~\bibnamefont {Wu}}, \bibinfo
  {author} {\bibfnamefont {L.}~\bibnamefont {Li}}, \bibinfo {author}
  {\bibfnamefont {Z.}~\bibnamefont {Qin}},\ and\ \bibinfo {author}
  {\bibfnamefont {H.}~\bibnamefont {Yao}},\ }\bibfield  {title} {\bibinfo
  {title} {Fracton topological order at finite temperature},\ }\bibfield
  {journal} {\bibinfo  {journal} {Physical Review Research}\ }\textbf {\bibinfo
  {volume} {4}},\ \href {https://doi.org/10.1103/physrevresearch.4.l032008}
  {10.1103/physrevresearch.4.l032008} (\bibinfo {year} {2022})\BibitemShut
  {NoStop}%
\bibitem [{\citenamefont {Boada}\ \emph {et~al.}(2012)\citenamefont {Boada},
  \citenamefont {Celi}, \citenamefont {Latorre},\ and\ \citenamefont
  {Lewenstein}}]{synthetic_dimension_1}%
  \BibitemOpen
  \bibfield  {author} {\bibinfo {author} {\bibfnamefont {O.}~\bibnamefont
  {Boada}}, \bibinfo {author} {\bibfnamefont {A.}~\bibnamefont {Celi}},
  \bibinfo {author} {\bibfnamefont {J.~I.}\ \bibnamefont {Latorre}},\ and\
  \bibinfo {author} {\bibfnamefont {M.}~\bibnamefont {Lewenstein}},\ }\bibfield
   {title} {\bibinfo {title} {Quantum simulation of an extra dimension},\
  }\href {https://doi.org/10.1103/PhysRevLett.108.133001} {\bibfield  {journal}
  {\bibinfo  {journal} {Phys. Rev. Lett.}\ }\textbf {\bibinfo {volume} {108}},\
  \bibinfo {pages} {133001} (\bibinfo {year} {2012})}\BibitemShut {NoStop}%
\bibitem [{\citenamefont {{Hazzard}}\ and\ \citenamefont
  {{Gadway}}(2023)}]{synthetic_dimension_2}%
  \BibitemOpen
  \bibfield  {author} {\bibinfo {author} {\bibfnamefont {K.~R.~A.}\
  \bibnamefont {{Hazzard}}}\ and\ \bibinfo {author} {\bibfnamefont
  {B.}~\bibnamefont {{Gadway}}},\ }\bibfield  {title} {\bibinfo {title}
  {{Synthetic dimensions}},\ }\href {https://doi.org/10.1063/PT.3.5225}
  {\bibfield  {journal} {\bibinfo  {journal} {Physics Today}\ }\textbf
  {\bibinfo {volume} {76}},\ \bibinfo {pages} {62} (\bibinfo {year} {2023})},\
  \Eprint {https://arxiv.org/abs/2306.13658} {arXiv:2306.13658
  [physics.pop-ph]} \BibitemShut {NoStop}%
\bibitem [{\citenamefont {{Arg{\"u}ello-Luengo}}\ \emph
  {et~al.}(2024)\citenamefont {{Arg{\"u}ello-Luengo}}, \citenamefont
  {{Bhattacharya}}, \citenamefont {{Celi}}, \citenamefont {{Chhajlany}},
  \citenamefont {{Grass}}, \citenamefont {{P{\l}odzie{\'n}}}, \citenamefont
  {{Rakshit}}, \citenamefont {{Salamon}}, \citenamefont {{Stornati}},
  \citenamefont {{Tarruell}},\ and\ \citenamefont
  {{Lewenstein}}}]{synthetic_dimension_3}%
  \BibitemOpen
  \bibfield  {author} {\bibinfo {author} {\bibfnamefont {J.}~\bibnamefont
  {{Arg{\"u}ello-Luengo}}}, \bibinfo {author} {\bibfnamefont {U.}~\bibnamefont
  {{Bhattacharya}}}, \bibinfo {author} {\bibfnamefont {A.}~\bibnamefont
  {{Celi}}}, \bibinfo {author} {\bibfnamefont {R.~W.}\ \bibnamefont
  {{Chhajlany}}}, \bibinfo {author} {\bibfnamefont {T.}~\bibnamefont
  {{Grass}}}, \bibinfo {author} {\bibfnamefont {M.}~\bibnamefont
  {{P{\l}odzie{\'n}}}}, \bibinfo {author} {\bibfnamefont {D.}~\bibnamefont
  {{Rakshit}}}, \bibinfo {author} {\bibfnamefont {T.}~\bibnamefont
  {{Salamon}}}, \bibinfo {author} {\bibfnamefont {P.}~\bibnamefont
  {{Stornati}}}, \bibinfo {author} {\bibfnamefont {L.}~\bibnamefont
  {{Tarruell}}},\ and\ \bibinfo {author} {\bibfnamefont {M.}~\bibnamefont
  {{Lewenstein}}},\ }\bibfield  {title} {\bibinfo {title} {{Synthetic
  dimensions for topological and quantum phases}},\ }\href
  {https://doi.org/10.1038/s42005-024-01636-3} {\bibfield  {journal} {\bibinfo
  {journal} {Communications Physics}\ }\textbf {\bibinfo {volume} {7}},\
  \bibinfo {eid} {143} (\bibinfo {year} {2024})},\ \Eprint
  {https://arxiv.org/abs/2310.19549} {arXiv:2310.19549 [quant-ph]} \BibitemShut
  {NoStop}%
\bibitem [{\citenamefont {{Joyce}}(2009)}]{manifolds_with_corners}%
  \BibitemOpen
  \bibfield  {author} {\bibinfo {author} {\bibfnamefont {D.}~\bibnamefont
  {{Joyce}}},\ }\bibfield  {title} {\bibinfo {title} {{On manifolds with
  corners}},\ }\href {https://doi.org/10.48550/arXiv.0910.3518} {\bibfield
  {journal} {\bibinfo  {journal} {arXiv e-prints}\ ,\ \bibinfo {eid}
  {arXiv:0910.3518}} (\bibinfo {year} {2009})},\ \Eprint
  {https://arxiv.org/abs/0910.3518} {arXiv:0910.3518 [math.DG]} \BibitemShut
  {NoStop}%
\bibitem [{\citenamefont {Candel}\ and\ \citenamefont
  {Conlon}(2000)}]{foliation_1}%
  \BibitemOpen
  \bibfield  {author} {\bibinfo {author} {\bibfnamefont {A.}~\bibnamefont
  {Candel}}\ and\ \bibinfo {author} {\bibfnamefont {L.}~\bibnamefont
  {Conlon}},\ }\href {https://books.google.com/books?id=7x8SCgAAQBAJ} {\emph
  {\bibinfo {title} {Foliations I}}},\ Foliations\ (\bibinfo  {publisher}
  {American Mathematical Society},\ \bibinfo {year} {2000})\BibitemShut
  {NoStop}%
\bibitem [{\citenamefont {{Vijay}}\ \emph {et~al.}(2016)\citenamefont
  {{Vijay}}, \citenamefont {{Haah}},\ and\ \citenamefont
  {{Fu}}}]{fuliang_xcube}%
  \BibitemOpen
  \bibfield  {author} {\bibinfo {author} {\bibfnamefont {S.}~\bibnamefont
  {{Vijay}}}, \bibinfo {author} {\bibfnamefont {J.}~\bibnamefont {{Haah}}},\
  and\ \bibinfo {author} {\bibfnamefont {L.}~\bibnamefont {{Fu}}},\ }\bibfield
  {title} {\bibinfo {title} {{Fracton topological order, generalized lattice
  gauge theory, and duality}},\ }\href
  {https://doi.org/10.1103/PhysRevB.94.235157} {\bibfield  {journal} {\bibinfo
  {journal} {\prb}\ }\textbf {\bibinfo {volume} {94}},\ \bibinfo {eid} {235157}
  (\bibinfo {year} {2016})},\ \Eprint {https://arxiv.org/abs/1603.04442}
  {arXiv:1603.04442 [cond-mat.str-el]} \BibitemShut {NoStop}%
\bibitem [{\citenamefont {{Pai}}\ and\ \citenamefont
  {{Hermele}}(2019)}]{fracton_fusion}%
  \BibitemOpen
  \bibfield  {author} {\bibinfo {author} {\bibfnamefont {S.}~\bibnamefont
  {{Pai}}}\ and\ \bibinfo {author} {\bibfnamefont {M.}~\bibnamefont
  {{Hermele}}},\ }\bibfield  {title} {\bibinfo {title} {{Fracton fusion and
  statistics}},\ }\href {https://doi.org/10.1103/PhysRevB.100.195136}
  {\bibfield  {journal} {\bibinfo  {journal} {\prb}\ }\textbf {\bibinfo
  {volume} {100}},\ \bibinfo {eid} {195136} (\bibinfo {year} {2019})},\ \Eprint
  {https://arxiv.org/abs/1903.11625} {arXiv:1903.11625 [cond-mat.str-el]}
  \BibitemShut {NoStop}%
\bibitem [{\citenamefont {{Dennis}}\ \emph {et~al.}(2002)\citenamefont
  {{Dennis}}, \citenamefont {{Kitaev}}, \citenamefont {{Landahl}},\ and\
  \citenamefont {{Preskill}}}]{Kitaev_topo_quantum_memory}%
  \BibitemOpen
  \bibfield  {author} {\bibinfo {author} {\bibfnamefont {E.}~\bibnamefont
  {{Dennis}}}, \bibinfo {author} {\bibfnamefont {A.}~\bibnamefont {{Kitaev}}},
  \bibinfo {author} {\bibfnamefont {A.}~\bibnamefont {{Landahl}}},\ and\
  \bibinfo {author} {\bibfnamefont {J.}~\bibnamefont {{Preskill}}},\ }\bibfield
   {title} {\bibinfo {title} {{Topological quantum memory}},\ }\href
  {https://doi.org/10.1063/1.1499754} {\bibfield  {journal} {\bibinfo
  {journal} {Journal of Mathematical Physics}\ }\textbf {\bibinfo {volume}
  {43}},\ \bibinfo {pages} {4452} (\bibinfo {year} {2002})},\ \Eprint
  {https://arxiv.org/abs/quant-ph/0110143} {arXiv:quant-ph/0110143 [quant-ph]}
  \BibitemShut {NoStop}%
\bibitem [{\citenamefont {{Jochym-O'Connor}}\ and\ \citenamefont
  {{Yoder}}(2021)}]{octaplex_4d_toric_code}%
  \BibitemOpen
  \bibfield  {author} {\bibinfo {author} {\bibfnamefont {T.}~\bibnamefont
  {{Jochym-O'Connor}}}\ and\ \bibinfo {author} {\bibfnamefont {T.~J.}\
  \bibnamefont {{Yoder}}},\ }\bibfield  {title} {\bibinfo {title}
  {{Four-dimensional toric code with non-Clifford transversal gates}},\ }\href
  {https://doi.org/10.1103/PhysRevResearch.3.013118} {\bibfield  {journal}
  {\bibinfo  {journal} {Physical Review Research}\ }\textbf {\bibinfo {volume}
  {3}},\ \bibinfo {eid} {013118} (\bibinfo {year} {2021})},\ \Eprint
  {https://arxiv.org/abs/2010.02238} {arXiv:2010.02238 [quant-ph]} \BibitemShut
  {NoStop}%
\bibitem [{\citenamefont {Shirley}\ \emph
  {et~al.}(2019{\natexlab{b}})\citenamefont {Shirley}, \citenamefont {Slagle},\
  and\ \citenamefont {Chen}}]{chen_2018_foliated_fracton_phase_classification}%
  \BibitemOpen
  \bibfield  {author} {\bibinfo {author} {\bibfnamefont {W.}~\bibnamefont
  {Shirley}}, \bibinfo {author} {\bibfnamefont {K.}~\bibnamefont {Slagle}},\
  and\ \bibinfo {author} {\bibfnamefont {X.}~\bibnamefont {Chen}},\ }\bibfield
  {title} {\bibinfo {title} {{Universal entanglement signatures of foliated
  fracton phases}},\ }\href {https://doi.org/10.21468/SciPostPhys.6.1.015}
  {\bibfield  {journal} {\bibinfo  {journal} {SciPost Phys.}\ }\textbf
  {\bibinfo {volume} {6}},\ \bibinfo {pages} {015} (\bibinfo {year}
  {2019}{\natexlab{b}})}\BibitemShut {NoStop}%
\bibitem [{\citenamefont {Cardy}(1984)}]{cardy1984conformal}%
  \BibitemOpen
  \bibfield  {author} {\bibinfo {author} {\bibfnamefont {J.~L.}\ \bibnamefont
  {Cardy}},\ }\bibfield  {title} {\bibinfo {title} {Conformal invariance and
  universality in finite-size scaling},\ }\href
  {https://doi.org/10.1088/0305-4470/17/7/003} {\bibfield  {journal} {\bibinfo
  {journal} {Journal of Physics A: Mathematical and General}\ }\textbf
  {\bibinfo {volume} {17}},\ \bibinfo {pages} {L385} (\bibinfo {year}
  {1984})}\BibitemShut {NoStop}%
\bibitem [{\citenamefont {Oshikawa}(2019)}]{oshikawa2019universal}%
  \BibitemOpen
  \bibfield  {author} {\bibinfo {author} {\bibfnamefont {M.}~\bibnamefont
  {Oshikawa}},\ }\bibfield  {title} {\bibinfo {title} {Universal finite-size
  gap scaling of the quantum ising chain},\ }\href
  {https://inspirehep.net/literature/1759287} {\bibfield  {journal} {\bibinfo
  {journal} {arXiv preprint arXiv:1910.06353}\ } (\bibinfo {year}
  {2019})}\BibitemShut {NoStop}%
\bibitem [{\citenamefont {{{\L}odyga}}\ \emph {et~al.}(2015)\citenamefont
  {{{\L}odyga}}, \citenamefont {{Mazurek}}, \citenamefont {{Grudka}},\ and\
  \citenamefont {{Horodecki}}}]{any_single_qubit_logical_state}%
  \BibitemOpen
  \bibfield  {author} {\bibinfo {author} {\bibfnamefont {J.}~\bibnamefont
  {{{\L}odyga}}}, \bibinfo {author} {\bibfnamefont {P.}~\bibnamefont
  {{Mazurek}}}, \bibinfo {author} {\bibfnamefont {A.}~\bibnamefont
  {{Grudka}}},\ and\ \bibinfo {author} {\bibfnamefont {M.}~\bibnamefont
  {{Horodecki}}},\ }\bibfield  {title} {\bibinfo {title} {{Simple scheme for
  encoding and decoding a qubit in unknown state for various topological
  codes}},\ }\href {https://doi.org/10.1038/srep08975} {\bibfield  {journal}
  {\bibinfo  {journal} {Scientific Reports}\ }\textbf {\bibinfo {volume} {5}},\
  \bibinfo {eid} {8975} (\bibinfo {year} {2015})},\ \Eprint
  {https://arxiv.org/abs/1404.2495} {arXiv:1404.2495 [quant-ph]} \BibitemShut
  {NoStop}%
\bibitem [{\citenamefont {Tantivasadakarn}\ \emph {et~al.}(2024)\citenamefont
  {Tantivasadakarn}, \citenamefont {Thorngren}, \citenamefont {Vishwanath},\
  and\ \citenamefont {Verresen}}]{long_range_entanglement_from_measuring_SPT}%
  \BibitemOpen
  \bibfield  {author} {\bibinfo {author} {\bibfnamefont {N.}~\bibnamefont
  {Tantivasadakarn}}, \bibinfo {author} {\bibfnamefont {R.}~\bibnamefont
  {Thorngren}}, \bibinfo {author} {\bibfnamefont {A.}~\bibnamefont
  {Vishwanath}},\ and\ \bibinfo {author} {\bibfnamefont {R.}~\bibnamefont
  {Verresen}},\ }\bibfield  {title} {\bibinfo {title} {Long-range entanglement
  from measuring symmetry-protected topological phases},\ }\href
  {https://doi.org/10.1103/PhysRevX.14.021040} {\bibfield  {journal} {\bibinfo
  {journal} {Phys. Rev. X}\ }\textbf {\bibinfo {volume} {14}},\ \bibinfo
  {pages} {021040} (\bibinfo {year} {2024})}\BibitemShut {NoStop}%
\bibitem [{\citenamefont {Zhang}\ \emph {et~al.}(2024)\citenamefont {Zhang},
  \citenamefont {Li},\ and\ \citenamefont {Ye}}]{PRXquantum.5.030342}%
  \BibitemOpen
  \bibfield  {author} {\bibinfo {author} {\bibfnamefont {J.-Y.}\ \bibnamefont
  {Zhang}}, \bibinfo {author} {\bibfnamefont {M.-Y.}\ \bibnamefont {Li}},\ and\
  \bibinfo {author} {\bibfnamefont {P.}~\bibnamefont {Ye}},\ }\bibfield
  {title} {\bibinfo {title} {Higher-order cellular automata generated
  symmetry-protected topological phases and detection through multi point
  strange correlators},\ }\href {https://doi.org/10.1103/PRXQuantum.5.030342}
  {\bibfield  {journal} {\bibinfo  {journal} {PRX Quantum}\ }\textbf {\bibinfo
  {volume} {5}},\ \bibinfo {pages} {030342} (\bibinfo {year}
  {2024})}\BibitemShut {NoStop}%
\bibitem [{\citenamefont {Zhou}\ \emph
  {et~al.}(2022{\natexlab{b}})\citenamefont {Zhou}, \citenamefont {Li},
  \citenamefont {Yan}, \citenamefont {Ye},\ and\ \citenamefont {Meng}}]{SSPT1}%
  \BibitemOpen
  \bibfield  {author} {\bibinfo {author} {\bibfnamefont {C.}~\bibnamefont
  {Zhou}}, \bibinfo {author} {\bibfnamefont {M.-Y.}\ \bibnamefont {Li}},
  \bibinfo {author} {\bibfnamefont {Z.}~\bibnamefont {Yan}}, \bibinfo {author}
  {\bibfnamefont {P.}~\bibnamefont {Ye}},\ and\ \bibinfo {author}
  {\bibfnamefont {Z.~Y.}\ \bibnamefont {Meng}},\ }\bibfield  {title} {\bibinfo
  {title} {Detecting subsystem symmetry protected topological order through
  strange correlators},\ }\href {https://doi.org/10.1103/PhysRevB.106.214428}
  {\bibfield  {journal} {\bibinfo  {journal} {Phys. Rev. B}\ }\textbf {\bibinfo
  {volume} {106}},\ \bibinfo {pages} {214428} (\bibinfo {year}
  {2022}{\natexlab{b}})}\BibitemShut {NoStop}%
\bibitem [{\citenamefont {Zhang}\ and\ \citenamefont
  {Ye}(2025)}]{zhang2025programmable}%
  \BibitemOpen
  \bibfield  {author} {\bibinfo {author} {\bibfnamefont {J.-Y.}\ \bibnamefont
  {Zhang}}\ and\ \bibinfo {author} {\bibfnamefont {P.}~\bibnamefont {Ye}},\
  }\bibfield  {title} {\bibinfo {title} {Programmable anyon mobility through
  higher order cellular automata},\ }\href@noop {} {\bibfield  {journal}
  {\bibinfo  {journal} {arXiv preprint arXiv:2508.13961}\ } (\bibinfo {year}
  {2025})}\BibitemShut {NoStop}%
\bibitem [{\citenamefont {Liang}\ \emph {et~al.}(2024)\citenamefont {Liang},
  \citenamefont {Xu}, \citenamefont {Iosue},\ and\ \citenamefont
  {Chen}}]{PRXquantum.5.030328}%
  \BibitemOpen
  \bibfield  {author} {\bibinfo {author} {\bibfnamefont {Z.}~\bibnamefont
  {Liang}}, \bibinfo {author} {\bibfnamefont {Y.}~\bibnamefont {Xu}}, \bibinfo
  {author} {\bibfnamefont {J.~T.}\ \bibnamefont {Iosue}},\ and\ \bibinfo
  {author} {\bibfnamefont {Y.-A.}\ \bibnamefont {Chen}},\ }\bibfield  {title}
  {\bibinfo {title} {Extracting topological orders of generalized pauli
  stabilizer codes in two dimensions},\ }\href
  {https://doi.org/10.1103/PRXQuantum.5.030328} {\bibfield  {journal} {\bibinfo
   {journal} {PRX Quantum}\ }\textbf {\bibinfo {volume} {5}},\ \bibinfo {pages}
  {030328} (\bibinfo {year} {2024})}\BibitemShut {NoStop}%
\bibitem [{\citenamefont {{Arute}}\ \emph {et~al.}(2019)\citenamefont
  {{Arute}}, \citenamefont {{Arya}}, \citenamefont {{Babbush}}, \citenamefont
  {{Bacon}}, \citenamefont {{Bardin}}, \citenamefont {{Barends}}, \citenamefont
  {{Biswas}}, \citenamefont {{Boixo}}, \citenamefont {{Brandao}}, \citenamefont
  {{Buell}}, \citenamefont {{Burkett}}, \citenamefont {{Chen}}, \citenamefont
  {{Chen}}, \citenamefont {{Chiaro}}, \citenamefont {{Collins}}, \citenamefont
  {{Courtney}}, \citenamefont {{Dunsworth}}, \citenamefont {{Farhi}},
  \citenamefont {{Foxen}}, \citenamefont {{Fowler}}, \citenamefont {{Gidney}},
  \citenamefont {{Giustina}}, \citenamefont {{Graff}}, \citenamefont
  {{Guerin}}, \citenamefont {{Habegger}}, \citenamefont {{Harrigan}},
  \citenamefont {{Hartmann}}, \citenamefont {{Ho}}, \citenamefont {{Hoffmann}},
  \citenamefont {{Huang}}, \citenamefont {{Humble}}, \citenamefont {{Isakov}},
  \citenamefont {{Jeffrey}}, \citenamefont {{Jiang}}, \citenamefont {{Kafri}},
  \citenamefont {{Kechedzhi}}, \citenamefont {{Kelly}}, \citenamefont
  {{Klimov}}, \citenamefont {{Knysh}}, \citenamefont {{Korotkov}},
  \citenamefont {{Kostritsa}}, \citenamefont {{Landhuis}}, \citenamefont
  {{Lindmark}}, \citenamefont {{Lucero}}, \citenamefont {{Lyakh}},
  \citenamefont {{Mandr{\`a}}}, \citenamefont {{McClean}}, \citenamefont
  {{McEwen}}, \citenamefont {{Megrant}}, \citenamefont {{Mi}}, \citenamefont
  {{Michielsen}}, \citenamefont {{Mohseni}}, \citenamefont {{Mutus}},
  \citenamefont {{Naaman}}, \citenamefont {{Neeley}}, \citenamefont {{Neill}},
  \citenamefont {{Niu}}, \citenamefont {{Ostby}}, \citenamefont {{Petukhov}},
  \citenamefont {{Platt}}, \citenamefont {{Quintana}}, \citenamefont
  {{Rieffel}}, \citenamefont {{Roushan}}, \citenamefont {{Rubin}},
  \citenamefont {{Sank}}, \citenamefont {{Satzinger}}, \citenamefont
  {{Smelyanskiy}}, \citenamefont {{Sung}}, \citenamefont {{Trevithick}},
  \citenamefont {{Vainsencher}}, \citenamefont {{Villalonga}}, \citenamefont
  {{White}}, \citenamefont {{Yao}}, \citenamefont {{Yeh}}, \citenamefont
  {{Zalcman}}, \citenamefont {{Neven}},\ and\ \citenamefont
  {{Martinis}}}]{2019Nature_programmable_superconducting_qubits}%
  \BibitemOpen
  \bibfield  {author} {\bibinfo {author} {\bibfnamefont {F.}~\bibnamefont
  {{Arute}}}, \bibinfo {author} {\bibfnamefont {K.}~\bibnamefont {{Arya}}},
  \bibinfo {author} {\bibfnamefont {R.}~\bibnamefont {{Babbush}}}, \bibinfo
  {author} {\bibfnamefont {D.}~\bibnamefont {{Bacon}}}, \bibinfo {author}
  {\bibfnamefont {J.~C.}\ \bibnamefont {{Bardin}}}, \bibinfo {author}
  {\bibfnamefont {R.}~\bibnamefont {{Barends}}}, \bibinfo {author}
  {\bibfnamefont {R.}~\bibnamefont {{Biswas}}}, \bibinfo {author}
  {\bibfnamefont {S.}~\bibnamefont {{Boixo}}}, \bibinfo {author} {\bibfnamefont
  {F.~G.~S.~L.}\ \bibnamefont {{Brandao}}}, \bibinfo {author} {\bibfnamefont
  {D.~A.}\ \bibnamefont {{Buell}}}, \bibinfo {author} {\bibfnamefont
  {B.}~\bibnamefont {{Burkett}}}, \bibinfo {author} {\bibfnamefont
  {Y.}~\bibnamefont {{Chen}}}, \bibinfo {author} {\bibfnamefont
  {Z.}~\bibnamefont {{Chen}}}, \bibinfo {author} {\bibfnamefont
  {B.}~\bibnamefont {{Chiaro}}}, \bibinfo {author} {\bibfnamefont
  {R.}~\bibnamefont {{Collins}}}, \bibinfo {author} {\bibfnamefont
  {W.}~\bibnamefont {{Courtney}}}, \bibinfo {author} {\bibfnamefont
  {A.}~\bibnamefont {{Dunsworth}}}, \bibinfo {author} {\bibfnamefont
  {E.}~\bibnamefont {{Farhi}}}, \bibinfo {author} {\bibfnamefont
  {B.}~\bibnamefont {{Foxen}}}, \bibinfo {author} {\bibfnamefont
  {A.}~\bibnamefont {{Fowler}}}, \bibinfo {author} {\bibfnamefont
  {C.}~\bibnamefont {{Gidney}}}, \bibinfo {author} {\bibfnamefont
  {M.}~\bibnamefont {{Giustina}}}, \bibinfo {author} {\bibfnamefont
  {R.}~\bibnamefont {{Graff}}}, \bibinfo {author} {\bibfnamefont
  {K.}~\bibnamefont {{Guerin}}}, \bibinfo {author} {\bibfnamefont
  {S.}~\bibnamefont {{Habegger}}}, \bibinfo {author} {\bibfnamefont {M.~P.}\
  \bibnamefont {{Harrigan}}}, \bibinfo {author} {\bibfnamefont {M.~J.}\
  \bibnamefont {{Hartmann}}}, \bibinfo {author} {\bibfnamefont
  {A.}~\bibnamefont {{Ho}}}, \bibinfo {author} {\bibfnamefont {M.}~\bibnamefont
  {{Hoffmann}}}, \bibinfo {author} {\bibfnamefont {T.}~\bibnamefont {{Huang}}},
  \bibinfo {author} {\bibfnamefont {T.~S.}\ \bibnamefont {{Humble}}}, \bibinfo
  {author} {\bibfnamefont {S.~V.}\ \bibnamefont {{Isakov}}}, \bibinfo {author}
  {\bibfnamefont {E.}~\bibnamefont {{Jeffrey}}}, \bibinfo {author}
  {\bibfnamefont {Z.}~\bibnamefont {{Jiang}}}, \bibinfo {author} {\bibfnamefont
  {D.}~\bibnamefont {{Kafri}}}, \bibinfo {author} {\bibfnamefont
  {K.}~\bibnamefont {{Kechedzhi}}}, \bibinfo {author} {\bibfnamefont
  {J.}~\bibnamefont {{Kelly}}}, \bibinfo {author} {\bibfnamefont {P.~V.}\
  \bibnamefont {{Klimov}}}, \bibinfo {author} {\bibfnamefont {S.}~\bibnamefont
  {{Knysh}}}, \bibinfo {author} {\bibfnamefont {A.}~\bibnamefont {{Korotkov}}},
  \bibinfo {author} {\bibfnamefont {F.}~\bibnamefont {{Kostritsa}}}, \bibinfo
  {author} {\bibfnamefont {D.}~\bibnamefont {{Landhuis}}}, \bibinfo {author}
  {\bibfnamefont {M.}~\bibnamefont {{Lindmark}}}, \bibinfo {author}
  {\bibfnamefont {E.}~\bibnamefont {{Lucero}}}, \bibinfo {author}
  {\bibfnamefont {D.}~\bibnamefont {{Lyakh}}}, \bibinfo {author} {\bibfnamefont
  {S.}~\bibnamefont {{Mandr{\`a}}}}, \bibinfo {author} {\bibfnamefont {J.~R.}\
  \bibnamefont {{McClean}}}, \bibinfo {author} {\bibfnamefont {M.}~\bibnamefont
  {{McEwen}}}, \bibinfo {author} {\bibfnamefont {A.}~\bibnamefont {{Megrant}}},
  \bibinfo {author} {\bibfnamefont {X.}~\bibnamefont {{Mi}}}, \bibinfo {author}
  {\bibfnamefont {K.}~\bibnamefont {{Michielsen}}}, \bibinfo {author}
  {\bibfnamefont {M.}~\bibnamefont {{Mohseni}}}, \bibinfo {author}
  {\bibfnamefont {J.}~\bibnamefont {{Mutus}}}, \bibinfo {author} {\bibfnamefont
  {O.}~\bibnamefont {{Naaman}}}, \bibinfo {author} {\bibfnamefont
  {M.}~\bibnamefont {{Neeley}}}, \bibinfo {author} {\bibfnamefont
  {C.}~\bibnamefont {{Neill}}}, \bibinfo {author} {\bibfnamefont {M.~Y.}\
  \bibnamefont {{Niu}}}, \bibinfo {author} {\bibfnamefont {E.}~\bibnamefont
  {{Ostby}}}, \bibinfo {author} {\bibfnamefont {A.}~\bibnamefont {{Petukhov}}},
  \bibinfo {author} {\bibfnamefont {J.~C.}\ \bibnamefont {{Platt}}}, \bibinfo
  {author} {\bibfnamefont {C.}~\bibnamefont {{Quintana}}}, \bibinfo {author}
  {\bibfnamefont {E.~G.}\ \bibnamefont {{Rieffel}}}, \bibinfo {author}
  {\bibfnamefont {P.}~\bibnamefont {{Roushan}}}, \bibinfo {author}
  {\bibfnamefont {N.~C.}\ \bibnamefont {{Rubin}}}, \bibinfo {author}
  {\bibfnamefont {D.}~\bibnamefont {{Sank}}}, \bibinfo {author} {\bibfnamefont
  {K.~J.}\ \bibnamefont {{Satzinger}}}, \bibinfo {author} {\bibfnamefont
  {V.}~\bibnamefont {{Smelyanskiy}}}, \bibinfo {author} {\bibfnamefont {K.~J.}\
  \bibnamefont {{Sung}}}, \bibinfo {author} {\bibfnamefont {M.~D.}\
  \bibnamefont {{Trevithick}}}, \bibinfo {author} {\bibfnamefont
  {A.}~\bibnamefont {{Vainsencher}}}, \bibinfo {author} {\bibfnamefont
  {B.}~\bibnamefont {{Villalonga}}}, \bibinfo {author} {\bibfnamefont
  {T.}~\bibnamefont {{White}}}, \bibinfo {author} {\bibfnamefont {Z.~J.}\
  \bibnamefont {{Yao}}}, \bibinfo {author} {\bibfnamefont {P.}~\bibnamefont
  {{Yeh}}}, \bibinfo {author} {\bibfnamefont {A.}~\bibnamefont {{Zalcman}}},
  \bibinfo {author} {\bibfnamefont {H.}~\bibnamefont {{Neven}}},\ and\ \bibinfo
  {author} {\bibfnamefont {J.~M.}\ \bibnamefont {{Martinis}}},\ }\bibfield
  {title} {\bibinfo {title} {{Quantum supremacy using a programmable
  superconducting processor}},\ }\href
  {https://doi.org/10.1038/s41586-019-1666-5} {\bibfield  {journal} {\bibinfo
  {journal} {\nat}\ }\textbf {\bibinfo {volume} {574}},\ \bibinfo {pages} {505}
  (\bibinfo {year} {2019})},\ \Eprint {https://arxiv.org/abs/1910.11333}
  {arXiv:1910.11333 [quant-ph]} \BibitemShut {NoStop}%
\bibitem [{\citenamefont {Semeghini}\ \emph {et~al.}(2021)\citenamefont
  {Semeghini}, \citenamefont {Levine}, \citenamefont {Keesling}, \citenamefont
  {Ebadi}, \citenamefont {Wang}, \citenamefont {Bluvstein}, \citenamefont
  {Verresen}, \citenamefont {Pichler}, \citenamefont {Kalinowski},
  \citenamefont {Samajdar}, \citenamefont {Omran}, \citenamefont {Sachdev},
  \citenamefont {Vishwanath}, \citenamefont {Greiner}, \citenamefont
  {Vuletić},\ and\ \citenamefont {Lukin}}]{Rydberg_atom_toric_code_1}%
  \BibitemOpen
  \bibfield  {author} {\bibinfo {author} {\bibfnamefont {G.}~\bibnamefont
  {Semeghini}}, \bibinfo {author} {\bibfnamefont {H.}~\bibnamefont {Levine}},
  \bibinfo {author} {\bibfnamefont {A.}~\bibnamefont {Keesling}}, \bibinfo
  {author} {\bibfnamefont {S.}~\bibnamefont {Ebadi}}, \bibinfo {author}
  {\bibfnamefont {T.~T.}\ \bibnamefont {Wang}}, \bibinfo {author}
  {\bibfnamefont {D.}~\bibnamefont {Bluvstein}}, \bibinfo {author}
  {\bibfnamefont {R.}~\bibnamefont {Verresen}}, \bibinfo {author}
  {\bibfnamefont {H.}~\bibnamefont {Pichler}}, \bibinfo {author} {\bibfnamefont
  {M.}~\bibnamefont {Kalinowski}}, \bibinfo {author} {\bibfnamefont
  {R.}~\bibnamefont {Samajdar}}, \bibinfo {author} {\bibfnamefont
  {A.}~\bibnamefont {Omran}}, \bibinfo {author} {\bibfnamefont
  {S.}~\bibnamefont {Sachdev}}, \bibinfo {author} {\bibfnamefont
  {A.}~\bibnamefont {Vishwanath}}, \bibinfo {author} {\bibfnamefont
  {M.}~\bibnamefont {Greiner}}, \bibinfo {author} {\bibfnamefont
  {V.}~\bibnamefont {Vuletić}},\ and\ \bibinfo {author} {\bibfnamefont
  {M.~D.}\ \bibnamefont {Lukin}},\ }\bibfield  {title} {\bibinfo {title}
  {Probing topological spin liquids on a programmable quantum simulator},\
  }\href {https://doi.org/10.1126/science.abi8794} {\bibfield  {journal}
  {\bibinfo  {journal} {Science}\ }\textbf {\bibinfo {volume} {374}},\ \bibinfo
  {pages} {1242} (\bibinfo {year} {2021})},\ \Eprint
  {https://arxiv.org/abs/https://www.science.org/doi/pdf/10.1126/science.abi8794}
  {https://www.science.org/doi/pdf/10.1126/science.abi8794} \BibitemShut
  {NoStop}%
\bibitem [{\citenamefont {Verresen}\ \emph
  {et~al.}(2021{\natexlab{a}})\citenamefont {Verresen}, \citenamefont {Lukin},\
  and\ \citenamefont {Vishwanath}}]{Rydberg_atom_toric_code_2}%
  \BibitemOpen
  \bibfield  {author} {\bibinfo {author} {\bibfnamefont {R.}~\bibnamefont
  {Verresen}}, \bibinfo {author} {\bibfnamefont {M.~D.}\ \bibnamefont
  {Lukin}},\ and\ \bibinfo {author} {\bibfnamefont {A.}~\bibnamefont
  {Vishwanath}},\ }\bibfield  {title} {\bibinfo {title} {Prediction of toric
  code topological order from rydberg blockade},\ }\href
  {https://doi.org/10.1103/PhysRevX.11.031005} {\bibfield  {journal} {\bibinfo
  {journal} {Phys. Rev. X}\ }\textbf {\bibinfo {volume} {11}},\ \bibinfo
  {pages} {031005} (\bibinfo {year} {2021}{\natexlab{a}})}\BibitemShut
  {NoStop}%
\bibitem [{\citenamefont {Verresen}\ \emph
  {et~al.}(2021{\natexlab{b}})\citenamefont {Verresen}, \citenamefont {Lukin},\
  and\ \citenamefont {Vishwanath}}]{Rydberg_atom_toric_code_3}%
  \BibitemOpen
  \bibfield  {author} {\bibinfo {author} {\bibfnamefont {R.}~\bibnamefont
  {Verresen}}, \bibinfo {author} {\bibfnamefont {M.~D.}\ \bibnamefont
  {Lukin}},\ and\ \bibinfo {author} {\bibfnamefont {A.}~\bibnamefont
  {Vishwanath}},\ }\bibfield  {title} {\bibinfo {title} {Prediction of toric
  code topological order from rydberg blockade},\ }\href
  {https://doi.org/10.1103/PhysRevX.11.031005} {\bibfield  {journal} {\bibinfo
  {journal} {Phys. Rev. X}\ }\textbf {\bibinfo {volume} {11}},\ \bibinfo
  {pages} {031005} (\bibinfo {year} {2021}{\natexlab{b}})}\BibitemShut
  {NoStop}%
\bibitem [{\citenamefont {{Bluvstein}}\ \emph {et~al.}(2022)\citenamefont
  {{Bluvstein}}, \citenamefont {{Levine}}, \citenamefont {{Semeghini}},
  \citenamefont {{Wang}}, \citenamefont {{Ebadi}}, \citenamefont
  {{Kalinowski}}, \citenamefont {{Keesling}}, \citenamefont {{Maskara}},
  \citenamefont {{Pichler}}, \citenamefont {{Greiner}}, \citenamefont
  {{Vuleti{\'c}}},\ and\ \citenamefont
  {{Lukin}}}]{quantum_simulation_with_movable_qubits_2}%
  \BibitemOpen
  \bibfield  {author} {\bibinfo {author} {\bibfnamefont {D.}~\bibnamefont
  {{Bluvstein}}}, \bibinfo {author} {\bibfnamefont {H.}~\bibnamefont
  {{Levine}}}, \bibinfo {author} {\bibfnamefont {G.}~\bibnamefont
  {{Semeghini}}}, \bibinfo {author} {\bibfnamefont {T.~T.}\ \bibnamefont
  {{Wang}}}, \bibinfo {author} {\bibfnamefont {S.}~\bibnamefont {{Ebadi}}},
  \bibinfo {author} {\bibfnamefont {M.}~\bibnamefont {{Kalinowski}}}, \bibinfo
  {author} {\bibfnamefont {A.}~\bibnamefont {{Keesling}}}, \bibinfo {author}
  {\bibfnamefont {N.}~\bibnamefont {{Maskara}}}, \bibinfo {author}
  {\bibfnamefont {H.}~\bibnamefont {{Pichler}}}, \bibinfo {author}
  {\bibfnamefont {M.}~\bibnamefont {{Greiner}}}, \bibinfo {author}
  {\bibfnamefont {V.}~\bibnamefont {{Vuleti{\'c}}}},\ and\ \bibinfo {author}
  {\bibfnamefont {M.~D.}\ \bibnamefont {{Lukin}}},\ }\bibfield  {title}
  {\bibinfo {title} {{A quantum processor based on coherent transport of
  entangled atom arrays}},\ }\href {https://doi.org/10.1038/s41586-022-04592-6}
  {\bibfield  {journal} {\bibinfo  {journal} {\nat}\ }\textbf {\bibinfo
  {volume} {604}},\ \bibinfo {pages} {451} (\bibinfo {year} {2022})},\ \Eprint
  {https://arxiv.org/abs/2112.03923} {arXiv:2112.03923 [quant-ph]} \BibitemShut
  {NoStop}%
\bibitem [{\citenamefont {{Maskara}}\ \emph {et~al.}(2023)\citenamefont
  {{Maskara}}, \citenamefont {{Ostermann}}, \citenamefont {{Shee}},
  \citenamefont {{Kalinowski}}, \citenamefont {{McClain Gomez}}, \citenamefont
  {{Araiza Bravo}}, \citenamefont {{Wang}}, \citenamefont {{Krylov}},
  \citenamefont {{Yao}}, \citenamefont {{Head-Gordon}}, \citenamefont
  {{Lukin}},\ and\ \citenamefont
  {{Yelin}}}]{quantum_simulation_with_movable_qubits_3}%
  \BibitemOpen
  \bibfield  {author} {\bibinfo {author} {\bibfnamefont {N.}~\bibnamefont
  {{Maskara}}}, \bibinfo {author} {\bibfnamefont {S.}~\bibnamefont
  {{Ostermann}}}, \bibinfo {author} {\bibfnamefont {J.}~\bibnamefont {{Shee}}},
  \bibinfo {author} {\bibfnamefont {M.}~\bibnamefont {{Kalinowski}}}, \bibinfo
  {author} {\bibfnamefont {A.}~\bibnamefont {{McClain Gomez}}}, \bibinfo
  {author} {\bibfnamefont {R.}~\bibnamefont {{Araiza Bravo}}}, \bibinfo
  {author} {\bibfnamefont {D.~S.}\ \bibnamefont {{Wang}}}, \bibinfo {author}
  {\bibfnamefont {A.~I.}\ \bibnamefont {{Krylov}}}, \bibinfo {author}
  {\bibfnamefont {N.~Y.}\ \bibnamefont {{Yao}}}, \bibinfo {author}
  {\bibfnamefont {M.}~\bibnamefont {{Head-Gordon}}}, \bibinfo {author}
  {\bibfnamefont {M.~D.}\ \bibnamefont {{Lukin}}},\ and\ \bibinfo {author}
  {\bibfnamefont {S.~F.}\ \bibnamefont {{Yelin}}},\ }\bibfield  {title}
  {\bibinfo {title} {{Programmable Simulations of Molecules and Materials with
  Reconfigurable Quantum Processors}},\ }\href
  {https://doi.org/10.48550/arXiv.2312.02265} {\bibfield  {journal} {\bibinfo
  {journal} {arXiv e-prints}\ ,\ \bibinfo {eid} {arXiv:2312.02265}} (\bibinfo
  {year} {2023})},\ \Eprint {https://arxiv.org/abs/2312.02265}
  {arXiv:2312.02265 [quant-ph]} \BibitemShut {NoStop}%
\bibitem [{\citenamefont
  {Schabbauer}(2023)}]{quantum_simulation_with_movable_qubits_4}%
  \BibitemOpen
  \bibfield  {author} {\bibinfo {author} {\bibfnamefont {J.}~\bibnamefont
  {Schabbauer}},\ }\emph {\bibinfo {title} {Experiment control for quantum
  simulation with non-local interactions}},\ \href
  {https://doi.org/https://doi.org/10.34726/hss.2023.104001} {Ph.D. thesis},\
  \bibinfo  {school} {Wien} (\bibinfo {year} {2023})\BibitemShut {NoStop}%
\bibitem [{\citenamefont {Moses}\ \emph {et~al.}(2023)\citenamefont {Moses},
  \citenamefont {Baldwin}, \citenamefont {Allman}, \citenamefont {Ancona},
  \citenamefont {Ascarrunz}, \citenamefont {Barnes}, \citenamefont
  {Bartolotta}, \citenamefont {Bjork}, \citenamefont {Blanchard}, \citenamefont
  {Bohn}, \citenamefont {Bohnet}, \citenamefont {Brown}, \citenamefont
  {Burdick}, \citenamefont {Burton}, \citenamefont {Campbell}, \citenamefont
  {Campora}, \citenamefont {Carron}, \citenamefont {Chambers}, \citenamefont
  {Chan}, \citenamefont {Chen}, \citenamefont {Chernoguzov}, \citenamefont
  {Chertkov}, \citenamefont {Colina}, \citenamefont {Curtis}, \citenamefont
  {Daniel}, \citenamefont {DeCross}, \citenamefont {Deen}, \citenamefont
  {Delaney}, \citenamefont {Dreiling}, \citenamefont {Ertsgaard}, \citenamefont
  {Esposito}, \citenamefont {Estey}, \citenamefont {Fabrikant}, \citenamefont
  {Figgatt}, \citenamefont {Foltz}, \citenamefont {Foss-Feig}, \citenamefont
  {Francois}, \citenamefont {Gaebler}, \citenamefont {Gatterman}, \citenamefont
  {Gilbreth}, \citenamefont {Giles}, \citenamefont {Glynn}, \citenamefont
  {Hall}, \citenamefont {Hankin}, \citenamefont {Hansen}, \citenamefont
  {Hayes}, \citenamefont {Higashi}, \citenamefont {Hoffman}, \citenamefont
  {Horning}, \citenamefont {Hout}, \citenamefont {Jacobs}, \citenamefont
  {Johansen}, \citenamefont {Jones}, \citenamefont {Karcz}, \citenamefont
  {Klein}, \citenamefont {Lauria}, \citenamefont {Lee}, \citenamefont {Liefer},
  \citenamefont {Lu}, \citenamefont {Lucchetti}, \citenamefont {Lytle},
  \citenamefont {Malm}, \citenamefont {Matheny}, \citenamefont {Mathewson},
  \citenamefont {Mayer}, \citenamefont {Miller}, \citenamefont {Mills},
  \citenamefont {Neyenhuis}, \citenamefont {Nugent}, \citenamefont {Olson},
  \citenamefont {Parks}, \citenamefont {Price}, \citenamefont {Price},
  \citenamefont {Pugh}, \citenamefont {Ransford}, \citenamefont {Reed},
  \citenamefont {Roman}, \citenamefont {Rowe}, \citenamefont {Ryan-Anderson},
  \citenamefont {Sanders}, \citenamefont {Sedlacek}, \citenamefont {Shevchuk},
  \citenamefont {Siegfried}, \citenamefont {Skripka}, \citenamefont {Spaun},
  \citenamefont {Sprenkle}, \citenamefont {Stutz}, \citenamefont {Swallows},
  \citenamefont {Tobey}, \citenamefont {Tran}, \citenamefont {Tran},
  \citenamefont {Vogt}, \citenamefont {Volin}, \citenamefont {Walker},
  \citenamefont {Zolot},\ and\ \citenamefont
  {Pino}}]{quantum_simulation_with_movable_qubits_1}%
  \BibitemOpen
  \bibfield  {author} {\bibinfo {author} {\bibfnamefont {S.~A.}\ \bibnamefont
  {Moses}}, \bibinfo {author} {\bibfnamefont {C.~H.}\ \bibnamefont {Baldwin}},
  \bibinfo {author} {\bibfnamefont {M.~S.}\ \bibnamefont {Allman}}, \bibinfo
  {author} {\bibfnamefont {R.}~\bibnamefont {Ancona}}, \bibinfo {author}
  {\bibfnamefont {L.}~\bibnamefont {Ascarrunz}}, \bibinfo {author}
  {\bibfnamefont {C.}~\bibnamefont {Barnes}}, \bibinfo {author} {\bibfnamefont
  {J.}~\bibnamefont {Bartolotta}}, \bibinfo {author} {\bibfnamefont
  {B.}~\bibnamefont {Bjork}}, \bibinfo {author} {\bibfnamefont
  {P.}~\bibnamefont {Blanchard}}, \bibinfo {author} {\bibfnamefont
  {M.}~\bibnamefont {Bohn}}, \bibinfo {author} {\bibfnamefont {J.~G.}\
  \bibnamefont {Bohnet}}, \bibinfo {author} {\bibfnamefont {N.~C.}\
  \bibnamefont {Brown}}, \bibinfo {author} {\bibfnamefont {N.~Q.}\ \bibnamefont
  {Burdick}}, \bibinfo {author} {\bibfnamefont {W.~C.}\ \bibnamefont {Burton}},
  \bibinfo {author} {\bibfnamefont {S.~L.}\ \bibnamefont {Campbell}}, \bibinfo
  {author} {\bibfnamefont {J.~P.}\ \bibnamefont {Campora}}, \bibinfo {author}
  {\bibfnamefont {C.}~\bibnamefont {Carron}}, \bibinfo {author} {\bibfnamefont
  {J.}~\bibnamefont {Chambers}}, \bibinfo {author} {\bibfnamefont {J.~W.}\
  \bibnamefont {Chan}}, \bibinfo {author} {\bibfnamefont {Y.~H.}\ \bibnamefont
  {Chen}}, \bibinfo {author} {\bibfnamefont {A.}~\bibnamefont {Chernoguzov}},
  \bibinfo {author} {\bibfnamefont {E.}~\bibnamefont {Chertkov}}, \bibinfo
  {author} {\bibfnamefont {J.}~\bibnamefont {Colina}}, \bibinfo {author}
  {\bibfnamefont {J.~P.}\ \bibnamefont {Curtis}}, \bibinfo {author}
  {\bibfnamefont {R.}~\bibnamefont {Daniel}}, \bibinfo {author} {\bibfnamefont
  {M.}~\bibnamefont {DeCross}}, \bibinfo {author} {\bibfnamefont
  {D.}~\bibnamefont {Deen}}, \bibinfo {author} {\bibfnamefont {C.}~\bibnamefont
  {Delaney}}, \bibinfo {author} {\bibfnamefont {J.~M.}\ \bibnamefont
  {Dreiling}}, \bibinfo {author} {\bibfnamefont {C.~T.}\ \bibnamefont
  {Ertsgaard}}, \bibinfo {author} {\bibfnamefont {J.}~\bibnamefont {Esposito}},
  \bibinfo {author} {\bibfnamefont {B.}~\bibnamefont {Estey}}, \bibinfo
  {author} {\bibfnamefont {M.}~\bibnamefont {Fabrikant}}, \bibinfo {author}
  {\bibfnamefont {C.}~\bibnamefont {Figgatt}}, \bibinfo {author} {\bibfnamefont
  {C.}~\bibnamefont {Foltz}}, \bibinfo {author} {\bibfnamefont
  {M.}~\bibnamefont {Foss-Feig}}, \bibinfo {author} {\bibfnamefont
  {D.}~\bibnamefont {Francois}}, \bibinfo {author} {\bibfnamefont {J.~P.}\
  \bibnamefont {Gaebler}}, \bibinfo {author} {\bibfnamefont {T.~M.}\
  \bibnamefont {Gatterman}}, \bibinfo {author} {\bibfnamefont {C.~N.}\
  \bibnamefont {Gilbreth}}, \bibinfo {author} {\bibfnamefont {J.}~\bibnamefont
  {Giles}}, \bibinfo {author} {\bibfnamefont {E.}~\bibnamefont {Glynn}},
  \bibinfo {author} {\bibfnamefont {A.}~\bibnamefont {Hall}}, \bibinfo {author}
  {\bibfnamefont {A.~M.}\ \bibnamefont {Hankin}}, \bibinfo {author}
  {\bibfnamefont {A.}~\bibnamefont {Hansen}}, \bibinfo {author} {\bibfnamefont
  {D.}~\bibnamefont {Hayes}}, \bibinfo {author} {\bibfnamefont
  {B.}~\bibnamefont {Higashi}}, \bibinfo {author} {\bibfnamefont {I.~M.}\
  \bibnamefont {Hoffman}}, \bibinfo {author} {\bibfnamefont {B.}~\bibnamefont
  {Horning}}, \bibinfo {author} {\bibfnamefont {J.~J.}\ \bibnamefont {Hout}},
  \bibinfo {author} {\bibfnamefont {R.}~\bibnamefont {Jacobs}}, \bibinfo
  {author} {\bibfnamefont {J.}~\bibnamefont {Johansen}}, \bibinfo {author}
  {\bibfnamefont {L.}~\bibnamefont {Jones}}, \bibinfo {author} {\bibfnamefont
  {J.}~\bibnamefont {Karcz}}, \bibinfo {author} {\bibfnamefont
  {T.}~\bibnamefont {Klein}}, \bibinfo {author} {\bibfnamefont
  {P.}~\bibnamefont {Lauria}}, \bibinfo {author} {\bibfnamefont
  {P.}~\bibnamefont {Lee}}, \bibinfo {author} {\bibfnamefont {D.}~\bibnamefont
  {Liefer}}, \bibinfo {author} {\bibfnamefont {S.~T.}\ \bibnamefont {Lu}},
  \bibinfo {author} {\bibfnamefont {D.}~\bibnamefont {Lucchetti}}, \bibinfo
  {author} {\bibfnamefont {C.}~\bibnamefont {Lytle}}, \bibinfo {author}
  {\bibfnamefont {A.}~\bibnamefont {Malm}}, \bibinfo {author} {\bibfnamefont
  {M.}~\bibnamefont {Matheny}}, \bibinfo {author} {\bibfnamefont
  {B.}~\bibnamefont {Mathewson}}, \bibinfo {author} {\bibfnamefont
  {K.}~\bibnamefont {Mayer}}, \bibinfo {author} {\bibfnamefont {D.~B.}\
  \bibnamefont {Miller}}, \bibinfo {author} {\bibfnamefont {M.}~\bibnamefont
  {Mills}}, \bibinfo {author} {\bibfnamefont {B.}~\bibnamefont {Neyenhuis}},
  \bibinfo {author} {\bibfnamefont {L.}~\bibnamefont {Nugent}}, \bibinfo
  {author} {\bibfnamefont {S.}~\bibnamefont {Olson}}, \bibinfo {author}
  {\bibfnamefont {J.}~\bibnamefont {Parks}}, \bibinfo {author} {\bibfnamefont
  {G.~N.}\ \bibnamefont {Price}}, \bibinfo {author} {\bibfnamefont
  {Z.}~\bibnamefont {Price}}, \bibinfo {author} {\bibfnamefont
  {M.}~\bibnamefont {Pugh}}, \bibinfo {author} {\bibfnamefont {A.}~\bibnamefont
  {Ransford}}, \bibinfo {author} {\bibfnamefont {A.~P.}\ \bibnamefont {Reed}},
  \bibinfo {author} {\bibfnamefont {C.}~\bibnamefont {Roman}}, \bibinfo
  {author} {\bibfnamefont {M.}~\bibnamefont {Rowe}}, \bibinfo {author}
  {\bibfnamefont {C.}~\bibnamefont {Ryan-Anderson}}, \bibinfo {author}
  {\bibfnamefont {S.}~\bibnamefont {Sanders}}, \bibinfo {author} {\bibfnamefont
  {J.}~\bibnamefont {Sedlacek}}, \bibinfo {author} {\bibfnamefont
  {P.}~\bibnamefont {Shevchuk}}, \bibinfo {author} {\bibfnamefont
  {P.}~\bibnamefont {Siegfried}}, \bibinfo {author} {\bibfnamefont
  {T.}~\bibnamefont {Skripka}}, \bibinfo {author} {\bibfnamefont
  {B.}~\bibnamefont {Spaun}}, \bibinfo {author} {\bibfnamefont {R.~T.}\
  \bibnamefont {Sprenkle}}, \bibinfo {author} {\bibfnamefont {R.~P.}\
  \bibnamefont {Stutz}}, \bibinfo {author} {\bibfnamefont {M.}~\bibnamefont
  {Swallows}}, \bibinfo {author} {\bibfnamefont {R.~I.}\ \bibnamefont {Tobey}},
  \bibinfo {author} {\bibfnamefont {A.}~\bibnamefont {Tran}}, \bibinfo {author}
  {\bibfnamefont {T.}~\bibnamefont {Tran}}, \bibinfo {author} {\bibfnamefont
  {E.}~\bibnamefont {Vogt}}, \bibinfo {author} {\bibfnamefont {C.}~\bibnamefont
  {Volin}}, \bibinfo {author} {\bibfnamefont {J.}~\bibnamefont {Walker}},
  \bibinfo {author} {\bibfnamefont {A.~M.}\ \bibnamefont {Zolot}},\ and\
  \bibinfo {author} {\bibfnamefont {J.~M.}\ \bibnamefont {Pino}},\ }\bibfield
  {title} {\bibinfo {title} {A race-track trapped-ion quantum processor},\
  }\href {https://doi.org/10.1103/PhysRevX.13.041052} {\bibfield  {journal}
  {\bibinfo  {journal} {Phys. Rev. X}\ }\textbf {\bibinfo {volume} {13}},\
  \bibinfo {pages} {041052} (\bibinfo {year} {2023})}\BibitemShut {NoStop}%
\bibitem [{\citenamefont {{Yuan}}\ \emph {et~al.}(2018)\citenamefont {{Yuan}},
  \citenamefont {{Lin}}, \citenamefont {{Xiao}},\ and\ \citenamefont
  {{Fan}}}]{photon_synthetic_dimension_1}%
  \BibitemOpen
  \bibfield  {author} {\bibinfo {author} {\bibfnamefont {L.}~\bibnamefont
  {{Yuan}}}, \bibinfo {author} {\bibfnamefont {Q.}~\bibnamefont {{Lin}}},
  \bibinfo {author} {\bibfnamefont {M.}~\bibnamefont {{Xiao}}},\ and\ \bibinfo
  {author} {\bibfnamefont {S.}~\bibnamefont {{Fan}}},\ }\bibfield  {title}
  {\bibinfo {title} {{Synthetic dimension in photonics}},\ }\href
  {https://doi.org/10.1364/OPTICA.5.001396} {\bibfield  {journal} {\bibinfo
  {journal} {Optica}\ }\textbf {\bibinfo {volume} {5}},\ \bibinfo {pages}
  {1396} (\bibinfo {year} {2018})},\ \Eprint {https://arxiv.org/abs/1807.11468}
  {arXiv:1807.11468 [physics.optics]} \BibitemShut {NoStop}%
\bibitem [{\citenamefont {{Lustig}}\ \emph {et~al.}(2019)\citenamefont
  {{Lustig}}, \citenamefont {{Weimann}}, \citenamefont {{Plotnik}},
  \citenamefont {{Lumer}}, \citenamefont {{Bandres}}, \citenamefont
  {{Szameit}},\ and\ \citenamefont {{Segev}}}]{photon_synthetic_dimension_2}%
  \BibitemOpen
  \bibfield  {author} {\bibinfo {author} {\bibfnamefont {E.}~\bibnamefont
  {{Lustig}}}, \bibinfo {author} {\bibfnamefont {S.}~\bibnamefont {{Weimann}}},
  \bibinfo {author} {\bibfnamefont {Y.}~\bibnamefont {{Plotnik}}}, \bibinfo
  {author} {\bibfnamefont {Y.}~\bibnamefont {{Lumer}}}, \bibinfo {author}
  {\bibfnamefont {M.~A.}\ \bibnamefont {{Bandres}}}, \bibinfo {author}
  {\bibfnamefont {A.}~\bibnamefont {{Szameit}}},\ and\ \bibinfo {author}
  {\bibfnamefont {M.}~\bibnamefont {{Segev}}},\ }\bibfield  {title} {\bibinfo
  {title} {{Photonic topological insulator in synthetic dimensions}},\ }\href
  {https://doi.org/10.1038/s41586-019-0943-7} {\bibfield  {journal} {\bibinfo
  {journal} {\nat}\ }\textbf {\bibinfo {volume} {567}},\ \bibinfo {pages} {356}
  (\bibinfo {year} {2019})},\ \Eprint {https://arxiv.org/abs/1807.01983}
  {arXiv:1807.01983 [physics.optics]} \BibitemShut {NoStop}%
\bibitem [{\citenamefont {Ozawa}\ \emph {et~al.}(2016)\citenamefont {Ozawa},
  \citenamefont {Price}, \citenamefont {Goldman}, \citenamefont {Zilberberg},\
  and\ \citenamefont {Carusotto}}]{photon_synthetic_dimension_3}%
  \BibitemOpen
  \bibfield  {author} {\bibinfo {author} {\bibfnamefont {T.}~\bibnamefont
  {Ozawa}}, \bibinfo {author} {\bibfnamefont {H.~M.}\ \bibnamefont {Price}},
  \bibinfo {author} {\bibfnamefont {N.}~\bibnamefont {Goldman}}, \bibinfo
  {author} {\bibfnamefont {O.}~\bibnamefont {Zilberberg}},\ and\ \bibinfo
  {author} {\bibfnamefont {I.}~\bibnamefont {Carusotto}},\ }\bibfield  {title}
  {\bibinfo {title} {Synthetic dimensions in integrated photonics: From optical
  isolation to four-dimensional quantum hall physics},\ }\href
  {https://doi.org/10.1103/PhysRevA.93.043827} {\bibfield  {journal} {\bibinfo
  {journal} {Phys. Rev. A}\ }\textbf {\bibinfo {volume} {93}},\ \bibinfo
  {pages} {043827} (\bibinfo {year} {2016})}\BibitemShut {NoStop}%
\bibitem [{\citenamefont {{Dutt}}\ \emph {et~al.}(2020)\citenamefont {{Dutt}},
  \citenamefont {{Lin}}, \citenamefont {{Yuan}}, \citenamefont {{Minkov}},
  \citenamefont {{Xiao}},\ and\ \citenamefont
  {{Fan}}}]{photon_synthetic_dimension_4}%
  \BibitemOpen
  \bibfield  {author} {\bibinfo {author} {\bibfnamefont {A.}~\bibnamefont
  {{Dutt}}}, \bibinfo {author} {\bibfnamefont {Q.}~\bibnamefont {{Lin}}},
  \bibinfo {author} {\bibfnamefont {L.}~\bibnamefont {{Yuan}}}, \bibinfo
  {author} {\bibfnamefont {M.}~\bibnamefont {{Minkov}}}, \bibinfo {author}
  {\bibfnamefont {M.}~\bibnamefont {{Xiao}}},\ and\ \bibinfo {author}
  {\bibfnamefont {S.}~\bibnamefont {{Fan}}},\ }\bibfield  {title} {\bibinfo
  {title} {{A single photonic cavity with two independent physical synthetic
  dimensions}},\ }\href {https://doi.org/10.1126/science.aaz3071} {\bibfield
  {journal} {\bibinfo  {journal} {Science}\ }\textbf {\bibinfo {volume}
  {367}},\ \bibinfo {pages} {59} (\bibinfo {year} {2020})},\ \Eprint
  {https://arxiv.org/abs/1909.04828} {arXiv:1909.04828 [physics.optics]}
  \BibitemShut {NoStop}%
\end{thebibliography}
\end{document}